\documentclass[12pt,a4paper,twoside]{book}
\usepackage [T1] {fontenc}
\usepackage [utf8] {inputenc}
\usepackage[english]{babel}
\usepackage[dvipsnames]{xcolor}
\usepackage{amsmath}
\usepackage{amsfonts}
\usepackage{amssymb}
\usepackage{makeidx}
\usepackage{graphicx}
\usepackage[left=3.5cm,right=2.5cm,top=2.5cm,bottom=2.5cm, includefoot, headheight=25.6pt]{geometry}
\usepackage[explicit]{titlesec}
\usepackage{titletoc}
\usepackage{epigraph}
\usepackage{xpatch}
\usepackage{lmodern}
\usepackage{fancyhdr}
\usepackage{tocbibind}
\usepackage[urlcolor=green]{hyperref}
\usepackage{physics}
\usepackage{setspace}
\usepackage[colorinlistoftodos]{todonotes}
\usepackage{listings}
\usepackage{subfiles}
\usepackage{accents}
\usepackage{url}
\usepackage{subcaption}
\usepackage{cleveref}
\crefname{figure}{Fig.}{Fig.}
\Crefname{figure}{Fig.}{Fig.}
\crefname{table}{Table}{Table}
\Crefname{table}{Table}{Table}
\crefname{equation}{Eq.}{Eq.}
\Crefname{equation}{Eq.}{Eq.}
\crefname{section}{Sec.}{Sec.}
\crefname{section}{Sec.}{Sec.}
\crefname{subsection}{Sec.}{Sec.}
\crefname{chapter}{Chapter}{Chapter}
\crefname{appendix}{Appendix}{Appendix}

\usepackage{makecell} 
\usepackage{arydshln} 
\usepackage{multirow} 
\usepackage{booktabs}
\usepackage{cite}
\usepackage{bm}
\usepackage{resizegather} 

\newcommand{\Kompost}{\textsc{K{\o}MP{\o}ST}}

\usepackage[font=small, labelfont=bf, margin=15pt, tableposition=top]{caption}
\usepackage{tikz}
\usepackage{pgfplots}
\pgfplotsset{compat=newest}
\pgfplotsset{compat=1.10}
\usepgfplotslibrary{fillbetween} 
\usetikzlibrary{decorations.text,patterns,calc,arrows,positioning,snakes,decorations.pathmorphing,decorations.markings,decorations.pathreplacing,shapes.misc}

\let\oldFootnote\footnote
\newcommand\nextToken\relax

\renewcommand\footnote[1]{%
    \oldFootnote{#1}\futurelet\nextToken\isFootnote}

\newcommand\isFootnote{%
    \ifx\footnote\nextToken\textsuperscript{,}\fi}

\fancypagestyle{plain}{%
	\fancyhf{}
	\fancyfoot[LE,RO]{\thepage}
		
	\renewcommand{\headrulewidth}{0pt}	
	}
	
\renewcommand{\headrulewidth}{1.5pt}

\renewcommand{\headrule}{\hbox to\headwidth{\color{UniRed}\leaders\hrule height \headrulewidth\hfill}}
\renewcommand{\footrule}{\hbox to\headwidth{\color{UniRed}\leaders\hrule height \headrulewidth\hfill}}

\newlength\ChapWd
\definecolor{UniRed}{RGB}{128, 0, 32}

\titleformat{\chapter}[display]
  {\normalfont\filcenter\sffamily}
  {\tikz[remember picture,overlay]
    {
    \node[fill=UniRed,font=\fontsize{60}{72}\selectfont\color{white},anchor=north east,minimum size=\ChapWd] 
      at ([xshift=-40pt,yshift=-40pt]current page.north east) 
      (numb) {\thechapter};
    \node[rotate=90,anchor=south,inner sep=0pt,font=\huge] at (numb.west) {\chaptertitlename};
    }
  }{0pt}{\fontsize{33}{40}\selectfont\color{UniRed}#1}[\vskip10pt]
\titlespacing*{\chapter}
  {0pt}{50pt}{10pt}

\makeatletter
\xpatchcmd{\ttl@printlist}{\endgroup}{{\noindent\color{UniRed}\rule{\textwidth}{1.5pt}}\vskip30pt\endgroup}{}{}
\makeatother

\newcommand\DoPToC{%
\startcontents[chapters]
\printcontents[chapters]{}{1}{\noindent{\color{UniRed}\rule{\textwidth}{1.5pt}}\par\medskip}%
}

\newcommand\placedate[2]{
\noindent\begin{minipage}{5cm}
	\noindent\vspace{3cm}\par
    \noindent\text{#1}\par
    \noindent\rule{5cm}{1pt}\par
    \noindent\textbf{#2}
\end{minipage}}

\newcommand\signature[1]{
\noindent\begin{minipage}{5cm}
    \noindent\vspace{3cm}\par
    \noindent\rule{5cm}{1pt}\par
    \noindent\textbf{#1}
\end{minipage}}

\newcounter{Rcomment}

\newcounter{Rinline}

\begin{document}
\pagenumbering{gobble}
\begin{titlepage}
\begin{center}
\vspace*{.06\textheight}
\textsc{\Large Dissertation}\\[0.5cm] 

\noindent {\color{UniRed} \rule{\linewidth}{1.5pt} } \\[0.4cm] 
{\huge \bfseries Fluctuation modes in the initial state of heavy-ion collisions: from 2D to 3D hybrid simulations \par}\vspace{0.4cm} 
\noindent {\color{UniRed} \rule{\linewidth}{1.5pt} } \\[1.5cm] 
\end{center}

\begin{center} \large
{\huge\textbf{Renata Krupczak}} \\[3cm] 

\emph{Supervisor and 1$^{st}$ Reviewer:} \\
{Prof. Dr. N. Borghini} \\[0.5cm] 

\emph{2$^{\,nd}$ Reviewer:} \\
{Prof. Dr. S. Schlichting} \\[1.5cm] 
\end{center}

\begin{center}
\includegraphics[width=0.3\linewidth]{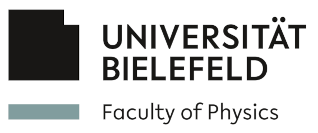}\\[0.1cm]
\textsc{Universität Bielefeld}\\[0.1cm]
\textsc{Fakultät für Physik}\\[0.5cm]
\vfill

{\large October 2, 2026}\\[3cm] 
\end{center}

\end{titlepage}
\cleardoublepage
\newpage
~
\vspace*{\fill}
\vspace{0.8\textheight}
\begin{flushright}
    \emph{In memory of my father.}
\end{flushright}

\chapter*{Abstract}

This dissertation investigates the role of the initial stages of heavy-ion collisions and their impact on final observables. The hybrid description of heavy-ion collisions relies on several effective models, and the largest theoretical uncertainties arise from the initial state and pre-equilibrium evolution, where first-principles descriptions remain limited. The objective of this work is to contribute to a more systematic understanding of these early stages and to clarify how they influence the extraction of the physical properties of the quark–gluon plasma (QGP).

The first part of the dissertation establishes the theoretical framework by reviewing the properties of the QGP and its dynamical evolution within state-of-the-art hybrid simulations. The theoretical foundations of each stage of the hybrid description are discussed, together with the numerical tools used in the phenomenological analyses presented throughout this dissertation.

The second part focuses on the matching between the pre-equilibrium stage and viscous hydrodynamics. While the necessity of pre-equilibrium evolution is well established, current models typically rely on conformal kinetic-theory descriptions, which are inconsistent with the non-conformal equation of state employed in hydrodynamics. This conceptual tension is analyzed, and a new entropy-density-based matching prescription is proposed. The main outcome of this approach is a reduction in the artificial bulk pressure generated at the switching time, thereby improving the consistency of the hybrid framework.

The third part introduces a mode-by-mode decomposition of the initial energy-density profiles to analyze the role of initial fluctuations. This framework enables the construction of an uncorrelated basis of fluctuation modes and the systematic study of their evolution and contributions to final observables. The method is applied to Pb-Pb collisions in different centrality classes and extended to a fully three-dimensional scenario for oxygen-oxygen collisions. The results demonstrate that initial fluctuations survive the hydrodynamic evolution and leave measurable imprints on final observables. In intermediate-sized systems such as O-O collisions, fluctuations dominate over purely geometric effects, indicating that these systems are not simply scaled-down versions of large heavy-ion collisions.

Overall, this dissertation provides new tools and insights for constraining the initial stages of heavy-ion collisions and for understanding how fluctuations propagate throughout the hybrid evolution.

\newpage
\thispagestyle{empty}
\setcounter{secnumdepth}{3}
\setcounter{tocdepth}{3}
\tableofcontents
\cleardoublepage

\pagenumbering{arabic}

\chapter{Introduction to the physics of the quark-gluon plasma}
\DoPToC
\noindent
\label{chap:introd}

With the goal of facilitating the understanding of the analysis presented in this work, it is necessary to introduce the research field and its current status. The state of matter explored in this dissertation is known as the quark–gluon plasma (QGP), and this chapter provides the basic background required to understand the context and the main properties of QGP physics.

This chapter is divided into four sections to build this introductory framework. \Cref{sec:QCD} presents the main properties of quantum chromodynamics (QCD), the theory responsible for describing hadronic matter. \Cref{sec:phase_diagram} explores the properties of hadronic matter under different conditions of temperature and baryonic potential. Particular attention is given to extreme temperatures and vanishing baryonic potential, where the QGP can be produced through heavy-ion collisions, which are discussed in \cref{sec:ultrarelativistic_HIC}. Finally, \cref{sec:importance} discusses the relevance of this research field and how this work contributes to advancing the current frontier of knowledge in the area.

It is important to emphasize that this dissertation uses natural units. Therefore, the fundamental physical constants, such as the reduced Planck constant $\hbar$, the speed of light $c$, and the Boltzmann constant $k_B$, are set equal to 1. Consequently, for convenience, physical quantities are expressed in units of energy and length.

\section{Quantum chromodynamics (QCD)}
\label{sec:QCD}

Throughout the history of humanity, there has been a persistent effort to better understand nature and its fundamental laws. In this pursuit, the identification of the most elementary constituents of matter became a central objective. As a result, the Standard Model of particle physics was established, providing a successful theoretical framework that describes matter as being composed of 12 fermions and 5 bosons responsible for the fundamental interactions \cite{griffiths2020introduction}.

There are four fundamental interactions in nature: the gravitational force, originating from the dynamics of massive bodies; the electromagnetic force, generated by electric charges; the weak nuclear force, responsible for atomic decays; and the strong nuclear force, generated by the color charge of quarks and gluons \cite{griffiths2020introduction}. Among these, the Standard Model is still not able to associate a boson that explains gravitation within a quantum framework.

In this dissertation, the focus is on hadronic matter, which is composed of quarks and is subject to the strong interaction. Although quarks can also interact through the other forces, the relevant energy scale for studying internal nuclear structure is sufficiently high that these contributions become negligible compared to the strong force, as shown in \cref{tab:forces}.

\begin{table}[ht]
    \centering
    \begin{tabular}{c|c|c|c}
    \hline
       \textbf{Force} & \textbf{Relative intensity} & \textbf{Boson mediator} & \textbf{Force range} \\
    \hline
       Strong nucelar & $1$ & Gluon & $10^{-15}$ m \\
       Eletromagnetic & $10^{-3}$ & Photon & $\infty$ \\
       Weak nuclear & $10^{-8}$ & W$^{\pm}$ and Z & $10^{-13}$ m \\
       Gravitational & $10^{-37}$ & ? & $\infty$\\
    \hline
    \end{tabular}
    \caption{Information about the fundamental interactions. The relative intensity of each force is approximately given for the interaction between two fundamental particles separated by a distance on the order of the atomic radius $1$ fm $=  10^{-15}$ m. These values are estimations and should not be interpreted as exact. The table is adapted from Ref. \cite{thomson2013modern}.}
\label{tab:forces}
\end{table}

The strong nuclear interaction was proposed to explain atomic stability. Since atoms contain positively charged protons concentrated within a very small nucleus, one would expect a strong electrostatic repulsion among them. The existence of nuclei therefore indicates the presence of an interaction stronger than the electromagnetic repulsion and limited to nuclear dimensions, which explains the origin of its name.

The first attempt to describe this interaction was the Yukawa model, which proposed that it is mediated by mesons, namely pions. Because these mediators are massive, the interaction is naturally limited to a short range compatible with nuclear scales. This description relies on isospin symmetry, in which protons and neutrons are understood as two different states of the same particle. However, this symmetry is not exact, since protons and neutrons present a small mass difference.

In 1964, the quark model was independently proposed by Gell-Mann and Zweig, explaining that protons and neutrons are composed of more fundamental particles named quarks. A few years later, experimental evidence for quarks was obtained through deep inelastic scattering experiments \cite{halzen2008quark}.

The modern description of the strong interaction is given by QCD \cite{greiner2007quantum, griffiths2020introduction}, with quarks interacting via gluons. The allowed bound states include three quarks or three antiquarks, forming baryons such as protons and neutrons, and quark–antiquark pairs, forming mesons such as pions and kaons. In all cases, the resulting particles are neutral with respect to color charge due to confinement. This property also allows exotic states, such as tetraquarks and pentaquarks, which combine four and five quarks/antiquarks, respectively, while still preserving overall color neutrality. Consequently, color charge is not observed at macroscopic scales.

Confinement is not directly derived from the QCD Lagrangian density; instead, it is incorporated phenomenologically to explain why isolated color charges are not observed. For this reason, its formal mathematical description is still not fully understood. Current phenomenological models associate confinement with a linear potential \cite{thomson2013modern}, analogous to the restoring force of an elastic spring.

\begin{figure}[ht]
    \centering
    \includegraphics[width=0.78\linewidth]{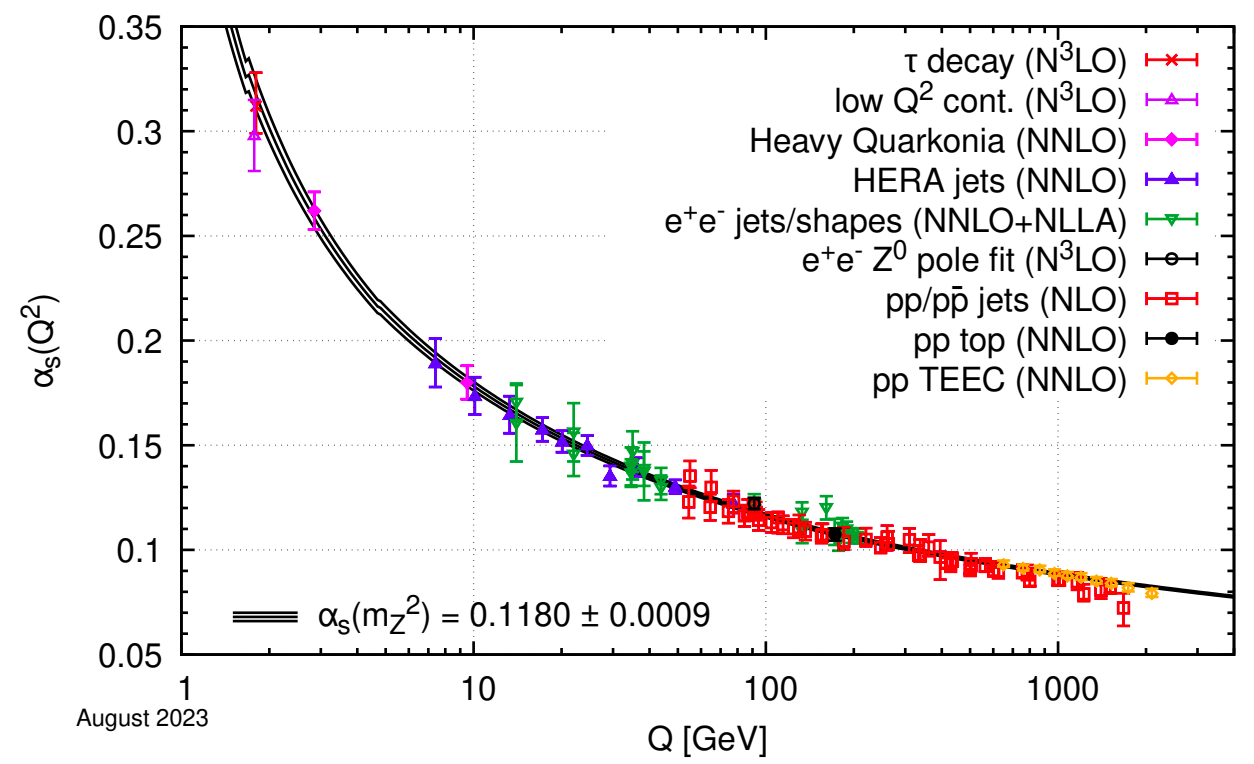}
    \caption{QCD coupling constant $\alpha_s$ as a function of the energy scale $Q$. This behavior reflects a fundamental property of QCD: a strong coupling at low energies and a weaker coupling at high energies. Figure from Ref. \cite{ParticleDataGroup:2024cfk}.}
\label{fig:alpha_s}
\end{figure}

Another fundamental property of QCD is asymptotic freedom. This feature refers to the energy dependence of the strong coupling constant, as shown in \cref{fig:alpha_s}. In QCD, the gauge bosons that mediate the interaction, namely gluons, also carry color charge and can therefore interact among themselves. This self-interaction distinguishes QCD from quantum electrodynamics (QED). As a consequence, the coupling constant decreases at high energies, allowing QCD to be treated perturbatively in this regime. At low energies, however, the coupling becomes large and perturbative methods are no longer applicable. In this case, a first-principles numerical approach known as lattice QCD is employed, in which space-time is discretized in order to evaluate the QCD path integrals. Nevertheless, this method is limited to vanishing chemical potential due to the QCD sign problem \cite{Giacosa:2024scx}, although extensions to small chemical potentials are possible as an approximation.

The properties of confinement and asymptotic freedom are central to QCD. At sufficiently high energies, the coupling constant becomes small enough to significantly weaken the interaction between quarks and gluons. Under such conditions, it is expected that a new state of matter can be formed, in which quarks and gluons are no longer confined inside hadrons but become deconfined in the QGP, which will be discussed in the next section.

\section{The QCD phase diagram}
\label{sec:phase_diagram}

A phase diagram is a representation of the relevant thermodynamic variables that, when varied, modify the microscopic state of matter and may lead to phase transitions. According to statistical mechanics \cite{nishimori2010elements}, a phase transition can be first order, characterized by a discontinuity in the first derivative of the Helmholtz free energy; second order, when the discontinuity appears in the second or higher-order derivative; or a crossover, when the microscopic structure of matter changes smoothly without any discontinuity. A simple everyday example is water at atmospheric pressure: when heated from the solid phase, its molecular structure gradually reorganizes until it reaches $0\,^\circ$C, where the transition to the liquid phase occurs. This first-order transformation is easily identified because it is accompanied by a clear macroscopic change in the system.

\begin{figure}[ht]
    \centering
    \includegraphics[width=0.65\linewidth]{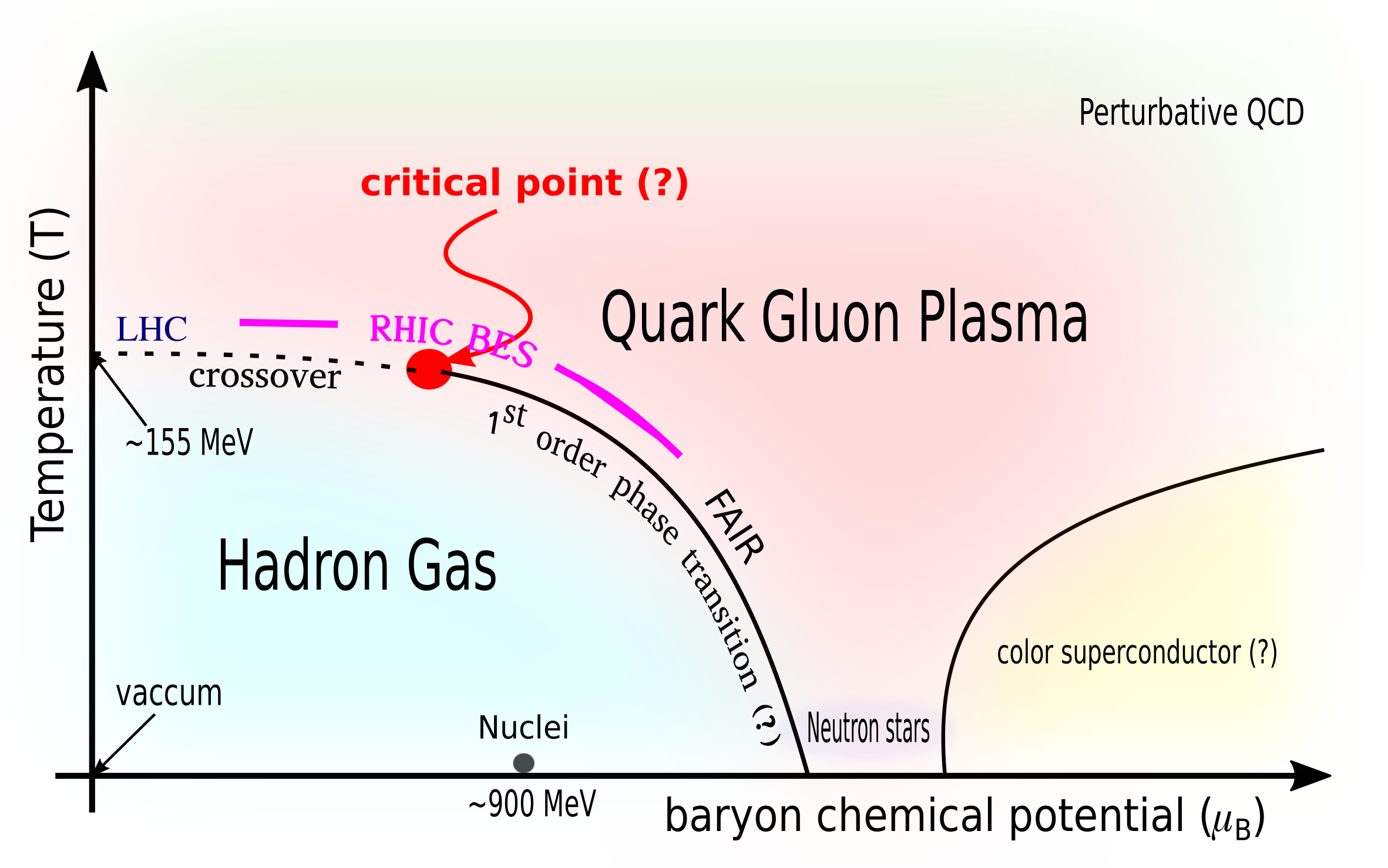}
    \caption{Schematic phase diagram of QCD matter in terms of the relevant thermodynamic variables, temperature $T$ and baryonic chemical potential $\mu_B$. Figure from Ref.~\cite{parida2025phenomenology}.}
\label{fig:phase_diagram}
\end{figure}

The phase diagram of hadronic matter is shown in \cref{fig:phase_diagram}. It should be understood as a schematic representation, since several regions of the diagram have not yet been experimentally accessed. Nuclear matter under ordinary conditions is represented by the black circle, located at finite baryonic chemical potential and low temperature. In this region of low to moderate temperature and baryonic chemical potential, quarks are confined within bound states, as described in \cref{sec:QCD}. As suggested by the behavior of the QCD coupling constant shown in \cref{fig:alpha_s}, the transition from nuclear matter to other phases is expected to occur only under conditions of extremely high energy.

One way to increase the energy of the system is by increasing the baryonic chemical potential. In this direction of the diagram, a first-order phase transition is expected, which may lead to a color superconducting phase. In this phase, quarks form correlated states that allow color charge to be transported analogously to electric charge in an electrical superconductor. This region of the phase diagram is believed to be realized in the cores of neutron stars, where the baryonic density is sufficiently high \cite{Fraga:2015xha}.

Alternatively, the energy of nuclear matter can be increased by raising the temperature while keeping the baryonic chemical potential close to zero. In this regime, matter undergoes a crossover transition to a new phase known as the QGP, in which quarks and gluons are deconfined within a strongly interacting medium. The temperatures reached in this transition are of the same order as those of the early universe, which is why the QGP is often described as the matter that filled the primordial universe shortly after the Big Bang. The possibility of such a phase was first discussed theoretically in 1975 by Collins and Perry \cite{Collins:1974ky}, who proposed the existence of superdense matter, and independently by Cabibbo and Parisi \cite{Cabibbo:1975ig}. Experimental evidence for the QGP, however, was only reported decades later, when CERN announced strong indications for the formation of a new state of matter in heavy-ion collisions \cite{Heinz:2000bk}.

Because of the extreme conditions involved, neither high baryonic density nor extremely high temperature is easily accessible. Therefore, it is necessary to recreate such conditions in the laboratory in order to study these phase transitions. One way to achieve this is through heavy-ion collisions, in which highly energetic nuclei are accelerated and collide, producing a small and short-lived system that can reach the required energy densities, as discussed in the next section.

In this work, we focus exclusively on the region of vanishing baryonic chemical potential and high temperature. Nevertheless, other regions of the QCD phase diagram remain active areas of research, including studies of the internal structure of neutron stars \cite{Baym:2017whm,Fraga:2015xha} and the search for a possible critical point \cite{Pandav:2022xxx,Borsanyi:2025dyp}.

\section{Ultrarelativistic heavy ion collisions}
\label{sec:ultrarelativistic_HIC}

As briefly introduced in the previous section, the QGP cannot be accessed under ordinary natural conditions. Therefore, it is necessary to consider how this state of matter can be produced in the laboratory. In this context, ultrarelativistic heavy ion collisions are used to create the QGP, since they provide a dense hadronic environment at sufficiently high energy to undergo the crossover phase transition of QCD matter, as discussed previously. Currently, two main colliders collect data from such collisions: the Large Hadron Collider (LHC) at CERN, located on the border between France and Switzerland, where the principal heavy ion results are obtained by the ALICE detector; and the Relativistic Heavy Ion Collider (RHIC), located at Brookhaven National Laboratory in New York.

\begin{figure}[ht]
    \centering
    \includegraphics[width=0.98\linewidth]{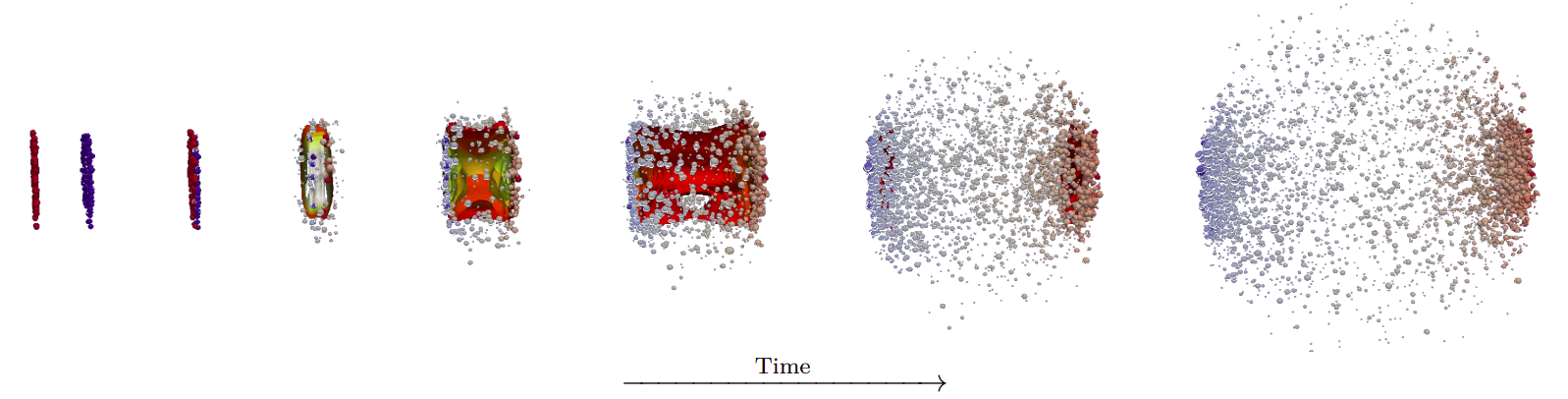}
    \caption{Illustrative schema of the time evolution during a heavy ion collision. Initially, the nuclei are Lorentz contracted into a pancake shape. After the collision, the QGP is produced as a continuous medium. As the system expands, the temperature decreases and hadronization occurs. Figure from Ref. \cite{Bernhard:2018hnz}.}
\label{fig:HIC}
\end{figure}

The basic idea is to accelerate heavy nuclei, such as Pb at the LHC and Au at RHIC, to velocities close to the speed of light. Along the direction of motion, Lorentz contraction occurs, and the collision can be described as a collision between two pancakes,\footnote{The Lorentz factor is approximately 100 at RHIC and about 2500 at the LHC \cite{Busza:2018rrf}. This implies that, if a lead nucleus is approximated as a sphere, its diameter is reduced from $14$ fm to $0.14$ fm at RHIC and to $0.0056$ fm at the LHC.} as shown in \cref{fig:HIC}. The plasma created in such collisions \cite{Busza:2018rrf} is an extremely small system, with a characteristic size on the order of $10^{-14}$ m, and highly short-lived, surviving for only about $10^{-23}$ s. The small size results from the challenge of concentrating hadronic matter at such high energies, since the ion beams must be precisely aligned. The short lifetime of the QGP is a consequence of the rapid expansion that follows the collision, which lowers the temperature and drives the system back to a confined state when $T<155$ MeV \cite{Busza:2018rrf}. Consequently, only the final confined particles are detected experimentally, as illustrated in the time evolution shown in \cref{fig:HIC}.

Since the QGP produced in heavy ion collisions cannot be observed directly, its study relies on reconstructing the evolution of the system from the final state particles. Collision simulations provide a valuable tool for this purpose, allowing theoretical models to be tested by verifying whether they reproduce the experimentally measured observables under comparable initial conditions \cite{Heinz:2024jwu}. The theoretical description of the full-time evolution shown in \cref{fig:HIC} is complex and involves multiple physical concepts; a more detailed discussion will be presented in the next chapter.

\begin{figure}[ht]
    \centering
    \includegraphics[width=0.7\linewidth]{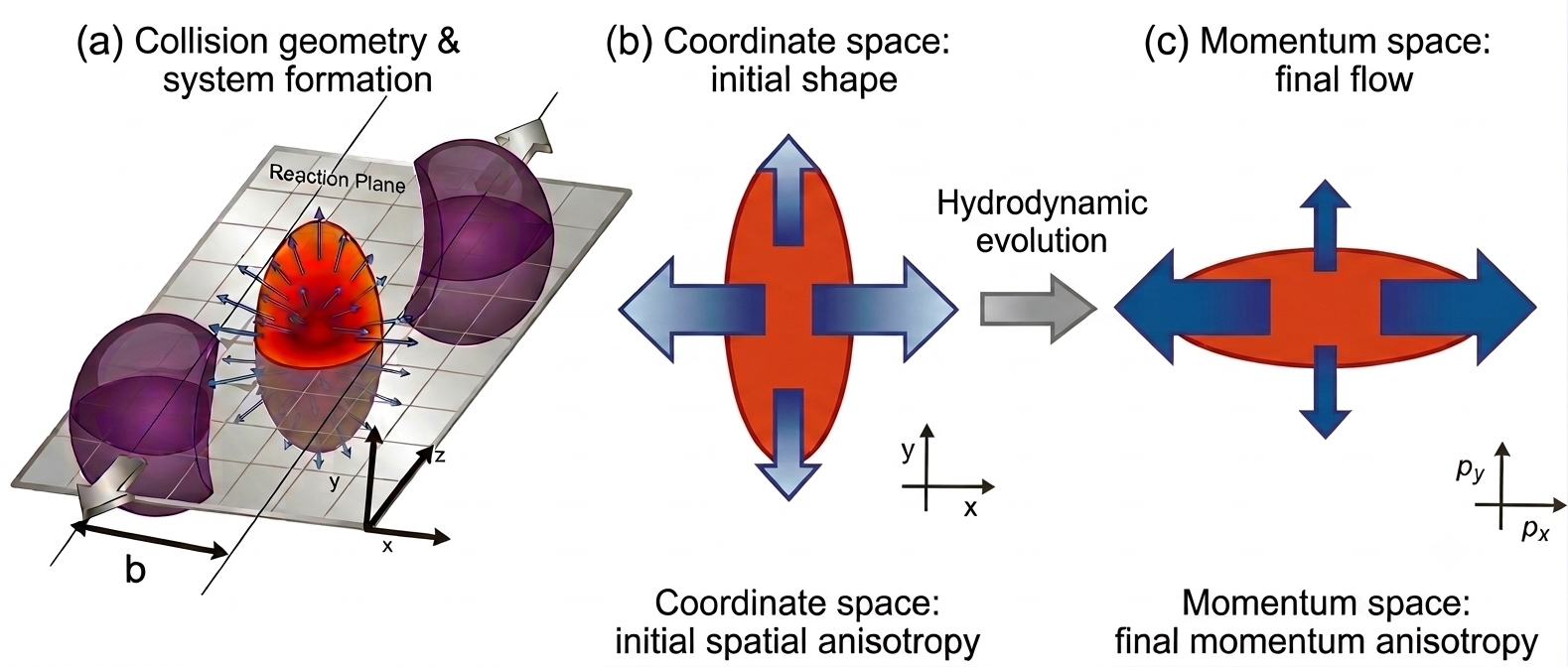}
    \caption{a) Illustration of the initial elliptical shape produced by the superposition of two nuclei that are not aligned. b) Emphasis on the initial spatial anisotropy, which generates different pressure gradients. c) During the time evolution, this spatial anisotropy is transformed into momentum anisotropy. This results in an enhanced production of final-state particles in a specific direction, which can serve as indirect evidence of QGP formation. Figure adapted from Ref. \cite{Nouicer:2015jrf}.}
\label{fig:v2_schema}
\end{figure}

However, since the QGP cannot be measured directly, how can we determine whether it is indeed produced? When a collision occurs and the two nuclei are not perfectly aligned, they typically overlap in an elliptical region, as illustrated in \cref{fig:v2_schema}a. In \cref{fig:v2_schema}b, the resulting spatial anisotropy from this elliptical geometry is shown. If the produced matter exhibits collective behavior, its time evolution converts this spatial anisotropy into momentum anisotropy, as depicted in \cref{fig:v2_schema}c. This conversion arises from the anisotropic pressure gradients generated by the initial elliptical shape of the plasma, represented by the blue arrows in \cref{fig:v2_schema}b.

Experimentally, the number of hadrons produced in these collisions is measured as a function of the azimuthal angle. A larger particle yield observed along a preferred direction provides evidence that collective behavior is present. If no QGP were formed, one would expect, statistically, an isotropic particle distribution in azimuthal angle, independent of the initial geometric configuration. In that case, the initial elliptical overlap would not influence the directions of the emitted particles.

Several final state observables are used to quantify this collective behavior. One of the most important is the set of flow coefficients \cite{ALICE:2018rtz, Borghini:2001vi}. These coefficients arise from expressing the azimuthal distribution of final state particles as a Fourier series \cite{Voloshin:1994mz}: 
\begin{equation} 
    \frac{\mathrm{d}N_\mathrm{ch}}{\mathrm{d}\phi} \propto 1 + 2 \sum_{n=1}^\infty v_n \cos{n(\phi - \Phi_n)}, 
\label{eq:flow_coefficients} 
\end{equation} 
where $\phi$ is the emission angle, $\Phi_n$ is the event plane angle, and 
$v_n$ are the flow coefficients, which constitute key final observables in heavy ion collisions. In particular, the experimental observation of a nonzero $v_2$, known as elliptic flow, provides strong evidence of collectivity in these systems \cite{Heinz:2001xi}.

For small collision systems \cite{Grosse-Oetringhaus:2024bwr, Krupczak:2026byr}, the situation is somewhat different. In these cases, the initial geometry is closer to spherical, and anisotropies arise predominantly from event-by-event fluctuations rather than from a well-defined elliptical overlap region, as discussed later in this dissertation.

\section{Why is this work important?}
\label{sec:importance}

After this brief introduction to the field, the reader should now have a general understanding of the context of heavy-ion physics and how heavy-ion collisions can be used to investigate QCD matter under extreme conditions.

The primary objective of this dissertation is to study the early stages of heavy-ion collisions, a stage of the system evolution where most of the model uncertainties are concentrated. To achieve this, a phenomenological approach is employed to investigate fluctuations in the initial energy density profiles and their evolution toward the final observables. Through this analysis, it is possible to gain a deeper understanding of QCD properties in extreme regimes.

This dissertation begins in \cref{chap:HIC} with a detailed description of the standard hybrid framework used in heavy-ion collision simulations, including the numerical codes employed throughout this work. The following chapters then present the original analyses developed in this dissertation. \Cref{chap:entropy_matching} proposes a new matching procedure between the initial state and the pre-equilibrium phase based on entropy density. \Cref{chap:modesPb} introduces the mode formalism to investigate initial fluctuations as a function of centrality. \Cref{chap:3d} extends the framework to three-dimensional analyses in smaller collision systems. Finally, \cref{chap:Conclusion} interconnects the main ideas presented in this dissertation and provides an outlook for future research.

In this context, this dissertation focuses on the dynamics of QCD matter, particularly in out-of-equilibrium regimes and at high energies, contributing to the advancement of knowledge at the frontier of nuclear physics.

\chapter{Hybrid model for heavy-ion collisions}
\label{chap:HIC}
\DoPToC
\noindent

After the brief introduction to the properties of QCD and the discussion of QGP production, it becomes necessary to better understand ultrarelativistic heavy-ion collisions, since they provide a valuable tool for studying this state of matter. As discussed in \cref{sec:ultrarelativistic_HIC}, due to the indirect experimental measurements, it is essential to develop theoretical frameworks and test them through numerical simulations of such collisions. In this context, this chapter reviews the standard hybrid approach used to describe these collisions numerically. Since different physical regimes are reached throughout the system evolution, distinct theoretical models must be employed. As a consequence, different numerical codes need to be consistently connected.

\begin{figure}
    \centering
    \includegraphics[width=0.8\linewidth]{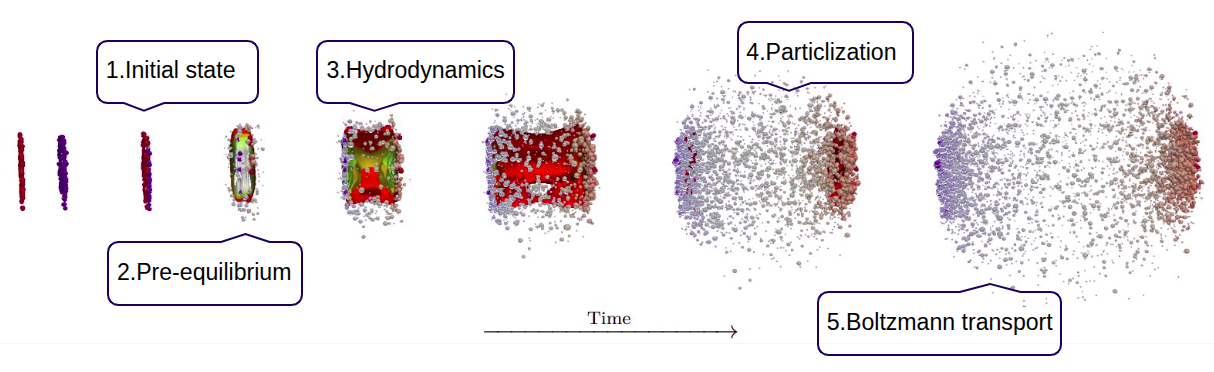}
    \caption{Illustrative schema of the standard hybrid evolution for heavy ion collisions, including the initial state, pre-equilibrium phase, hydrodynamic evolution, particlization, and Boltzmann transport. Figure adapted from Ref. \cite{Bernhard:2018hnz}.}
\label{fig:HIC_numerical}
\end{figure}

We reproduce \cref{fig:HIC} from \cref{sec:ultrarelativistic_HIC}, now emphasizing the numerical steps represented by the bubbles in \cref{fig:HIC_numerical}. From a physical perspective, this process is described as a collision between two Lorentz-contracted nuclei, followed by QGP production and collective evolution. The subsequent rapid expansion leads to a decrease in temperature until hadronization takes place, followed by final-state particle interactions that continue until detection in the experimental apparatus.

This physical evolution is now decomposed into the numerical steps shown in \cref{fig:HIC_numerical}, each corresponding to a different underlying physical theory. The collision between the two nuclei that leads to QGP formation is referred to as the initial state or initial condition. Because the produced matter is far from equilibrium, a pre-equilibrium model is required to drive the system to near-equilibrium, where a hydrodynamic description becomes valid. The hydrodynamic evolution then describes the collective behavior of the matter, covering the QGP phase. However, at a certain stage, determined by the equation of state and freeze-out conditions, the continuous fluid description must be replaced by a microscopic approach. The particlization procedure is the computational step that converts continuous fluid degrees of freedom into discrete particles. Finally, particle interactions continue through a Boltzmann transport approach until the system reaches the detectors.

This chapter is organized such that each section describes in detail the physical theory underlying each numerical step shown in \cref{fig:HIC_numerical}. Within each section, subsections present the specific codes used in the analyses throughout this dissertation, thereby providing both the necessary background for their understanding and a concise reference for the model parameters discussed in the following chapters.

\section{Initial Condition}
\label{sec:IC}

The initial condition model aims to take specific nuclear configurations and collision energy as input and provide, as output, an energy or entropy density profile. This profile can be determined in either two or three dimensions. By convention, the $z$-axis is defined as the collision axis. In the two-dimensional case, the energy or entropy density deposition is studied in the transverse plane $\mathbf{x_T} = (x, y)$, and in the three-dimensional case, the deposition happens on $(x, y, \eta_s)$, where $\eta_s$ is the longitudinal variable explained later in this section.

\begin{figure}[h]
    \centering
    \includegraphics[width=0.6\linewidth]{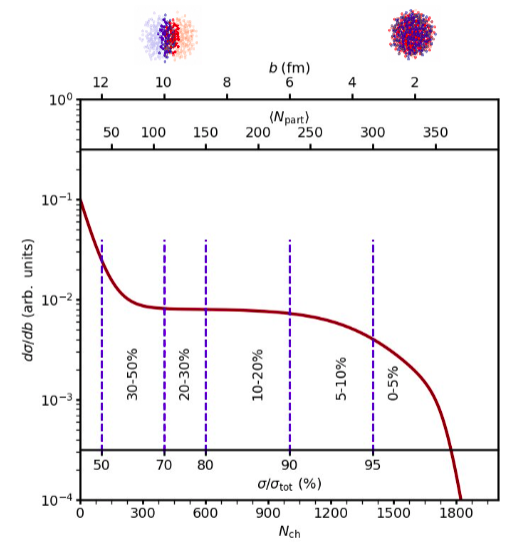}
    \caption{Visual connection between impact parameter $b$, number of participants $N_{\mathrm{part}}$, centrality bins, the fraction of the total cross section $\sigma/\sigma_{\mathrm{tot}}$, and the final charged particle multiplicity $N_{\mathrm{ch}}$. More peripheral events correspond to larger centrality values and larger $b$, whereas central events are associated with smaller centrality bins and smaller values of $b$. The numerical values shown on the scales are illustrative, as they depend on the collision system and beam energy. Figure adapted from Ref. \cite{Miller:2007ri}.}
\label{fig:impact_b}
\end{figure}

The first step is to arrange the $A$ nucleons inside each nucleus, since the collision dynamics depend on the positions of these sources. The nucleon positions can be described using a parametric nucleon density distribution \cite{Miller:2007ri}, or obtained from \emph{ab initio} nuclear structure calculations \cite{Zhang:2024vkh}. Then, the geometry of the collision must be determined. This geometry is characterized by the impact parameter b, defined as the transverse distance between the centers of the two colliding nuclei. A schematic representation is shown in \cref{fig:impact_b}, where the darker overlapping region corresponds to the interaction zone.

An event is referred to as central when $b$ is close to zero, corresponding to an almost complete overlap of the nuclei. Conversely, an event is called peripheral when $b$ is larger and the nuclei overlap only partially. In principle, the impact parameter can vary from zero up to approximately twice the nuclear radius.\footnote{The impact parameter can be larger than $2R$, corresponding to ultra-peripheral collisions \cite{Bertulani:2024hmb}. These are the focus of many research groups studying electromagnetic interactions; however, this topic is not explored in this work.} However, this quantity cannot be measured directly in experiments. Instead, experimentalists measure the final number of produced particles and their angular distribution. From these measurements, a histogram can be constructed and divided into centrality classes. Consequently, collisions are categorized according to centrality percentiles, ranging from $0\%$ for the most central events to $100\%$ for the most peripheral events. The relation between the impact parameter and the centrality classes is illustrated in \cref{fig:impact_b}. This procedure further highlights the importance of performing minimum bias simulations to capture all geometries for comparison with experimental data \cite{Qiu:2011iv}.

The inelastic nucleon–nucleon cross section is used to identify the nucleons that interact during the collision and subsequently contribute to the initial energy density deposition. These are called participants, $N_\mathrm{part}$, while those that pass through without interacting are referred to as spectators. From $N_\mathrm{part}$, the thickness function of each nucleus is calculated, followed by the reduced thickness function, which is proportional to the energy deposition in the collision.

Every initial-state model contains a normalization constant that controls the overall amount of energy deposited after the collision and must be calibrated to reproduce the observed charged-particle multiplicity. In this dissertation, this parameter is referred to as the 
$K$-factor. Its physical interpretation depends on the model under consideration: in more physically motivated approaches, it accounts for small corrections arising from approximations in the theoretical calculations, whereas in parametric models it acts simply as an overall normalization parameter. A more detailed discussion of its calibration and the procedure used to match the final charged multiplicity will be presented in \cref{chap:entropy_matching}.

When comparing theory, simulations, and experimental data, it is essential to adopt a consistent coordinate system. In high-energy collisions, Milne coordinates are commonly used because they naturally incorporate longitudinal expansion and simplify the description under the assumption of boost invariance. The Milne coordinates $(\tau, x, y, \eta_s)$ are related to Minkowski coordinates $(t, x, y, z)$ through 
\begin{equation} 
    \tau = \sqrt{t^2 - z^2}, \hspace{1.0cm} x = x, \hspace{1.0cm} y = y, \hspace{1.0cm} \eta_s = \frac{1}{2}\ln{\frac{t+z}{t-z}},
\label{eq:Milne_coord} 
\end{equation} 
where $\tau$ is the proper time and $\eta_s$ is the space-time rapidity. The metric in these coordinates is given by $g_{\mu\nu} = \mathrm{diag}(1, -1, -1, -\tau^2)$.

Two additional important kinematic variables are the rapidity $\mathrm{y}$ and the pseudorapidity $\eta$: 
\begin{equation} 
    \mathrm{y} = \frac{1}{2}\ln{\frac{E+p_z}{E-p_z}}, \hspace{1.0cm}    \eta = - \ln{\left(\tan{\frac{\theta}{2}}\right)} = \frac{1}{2}\ln{\frac{|\vec{p}|+p_z}{|\vec{p}|-p_z}}, 
\end{equation} 
where $E$ is the energy, $p$ is the momentum of the particle, and $p_z$ is the $z$ component of the momentum. The pseudorapidity is defined in terms of the polar angle $\theta$ with respect to the beam axis. This variable is particularly useful in experiments, since $\theta$ corresponds directly to the angle measured in detectors. For this reason, final observables are commonly expressed as functions of $\eta$, for example, the charged particle multiplicity $\mathrm{d}N_\mathrm{ch}/\mathrm{d}\eta$. It is important to emphasize that quantities in the initial state are functions of the space-time rapidity $\eta_s$, while final-state observables depend on the pseudorapidity $\eta$, which encodes information about both the particle momentum and the measurable emission angle. It is also useful to present the transformation \cite{wong1994introduction} between pseudorapidity $\eta$ and rapidity $\mathrm{y}$
\begin{equation} 
    \eta = \frac{1}{2}\ln{\left(\frac{\sqrt{m_\mathrm{T}^2 \cosh^2\mathrm{y} - m^2} + m_\mathrm{T} \sinh \mathrm{y}}{\sqrt{m_\mathrm{T}^2 \cosh^2\mathrm{y} - m^2} - m_\mathrm{T} \sinh \mathrm{y}}\right)}, 
\label{eq:jacobian_yeta} 
\end{equation} 
where $m$ is the rest mass and $m_\mathrm{T} = \sqrt{m^2+p_T^2}$ is the transverse mass of the particle.

When two-dimensional profiles are considered, the boost-invariance approximation is typically employed. In this case, the energy density is assumed to be independent of space-time rapidity, such that $e(\tau, x, y, \eta_s) = e(\tau, x, y)$. Boost invariance implies that the total transverse energy per unit space-time rapidity is flat. Experimentally, this approximate boost invariance is reflected in the charged-particle multiplicity distribution, which exhibits an almost plateau around midrapidity $\eta \sim 0$. This behavior characterizes the Bjorken flow \cite{Bjorken:1982qr}.

In this dissertation, the characterization of initial quantities from the initial energy density is referred to as ``initial observables''. In addition to the total energy, an important set of quantities is given by the spatial eccentricities: 
\begin{equation} 
\begin{aligned} 
    \varepsilon_1 e^{i\Psi_1} &\equiv - \frac{\int r^3 e^{i\theta} f(r, \theta) \, r \, \mathrm{d}r \, \mathrm{d}\theta}{\int r^3 f(r, \theta) \, r \, \mathrm{d}r \, \mathrm{d}\theta}, \\ 
    \varepsilon_n e^{in\Psi_n} &\equiv - \frac{\int r^n e^{in\theta} f(r, \theta) \, r \, \mathrm{d}r \, \mathrm{d}\theta}{\int r^n f(r, \theta) \, r \, \mathrm{d}r \, \mathrm{d}\theta}, \hspace{0.5cm} \mathrm{for}\ n > 1, 
\end{aligned}
\end{equation} 
where $\varepsilon_n$ is the magnitude of the spatial anisotropy (eccentricity), $\Psi_n$ is the participant-plane angle with respect to the reaction plane, and $f(r, \theta)$ represents a weight function, typically the energy or entropy density profile. These quantities characterize the initial geometry and fluctuations of the system and are related to the final flow coefficients \cite{Gardim:2014tya, Ollitrault:1992bk}, due to the mechanism discussed previously in \cref{fig:v2_schema}.

Although we refer to the eccentricities as initial observables in this dissertation, it is important to emphasize that they are not directly measured in detectors. Instead, they are initial-state quantities estimated from the energy or entropy density profile, which influence measurable final-state observables such as the flow coefficients.

Finally, there are two main classes of codes used to model initial conditions in heavy ion collisions: parametric and non-parametric models. The latter are based on physical descriptions inspired by the color glass condensate (CGC) framework, which is a QCD-based effective theory applicable in the high-energy regime where gluon saturation effects become important \cite{Frankfurt:2005mc}. Examples of such models include \textsc{IP-Glasma} \cite{Schenke:2014jza,Schenke:2016ksl} and \textsc{MC-Dipper} \cite{Garcia-Montero:2023gex}, the latter of which will be discussed in \cref{subsec:mcdipper}. In contrast, parametric models employ a set of parameters to describe the initial energy deposition and are often combined with Bayesian analyses to constrain these parameters efficiently via experimental data. Examples include $\mathrm{T}_{\mathrm{R}}\mathrm{ENTo}$ \cite{Moreland:2014oya} and \textsc{MC-Glauber} \cite{Miller:2007ri}, the latter of which will be presented in \cref{subsec:glauber}.

\subsection{\textsc{Monte Carlo Glauber}}
\label{subsec:glauber}

As a parametric model, the Monte Carlo Glauber (\textsc{MC-Glauber}) approach \cite{Miller:2007ri,dEnterria:2020dwq} uses phenomenological input to model initial energy deposition in a collision based on the eikonal approximation, which assumes straight-line propagation of the colliding nuclei. The model begins by sampling the positions of nucleons inside each nucleus, using a nucleon density described by the Woods–Saxon distribution,
\begin{equation}
    \rho(\vec{r}) = \frac{\rho_0}{1+\exp{\frac{r-R}{a}}},
\label{eq:woods-saxon}
\end{equation}
where $\rho_0$ is the nucleon density in the center of the nucleus, $R$ is the nuclear radius, and $a$ is the skin depth, two parameters specific to each nucleus.

In the \textsc{MC-Glauber} approach, a nucleon position is randomly sampled within $0 \le r \le 2R$, and the Woods–Saxon distribution is used to determine whether that position is accepted. A minimum inter-nucleon distance of $0.4$ fm is enforced. After sampling the nucleon positions, an impact parameter is also sampled as a random number within $0 \le b \le 2.5R$. In this dissertation, the impact parameter is fixed along the $x$-direction.\footnote{This choice is adopted for the analysis in this work for reasons discussed later. In general, $b$ can have any orientation.} The two nuclei are then shifted such that their centers are located at $\pm b/2$.

The \textsc{MC-Glauber} model uses the geometric criterion $d \leq \sqrt{\sigma^\mathrm{NN}_\mathrm{inel}/\pi}$ to determine whether a collision occurs between two nucleons. This criterion treats each nucleon–nucleon interaction independently and relies solely on the experimental value of the inelastic nucleon–nucleon cross section $\sigma^\mathrm{NN}_\mathrm{inel}$. Based on this condition, the number of participants $N_\mathrm{part}(\mathbf{x_T})$ and the number of binary collisions $N_\mathrm{coll}(\mathbf{\bar{x}_T})$ are calculated, where $\mathbf{\bar{x}_T}$ denotes the average position of the two nucleons involved in each collision. Since these quantities are initially defined as discrete sources at specific spatial points, each source is smeared using a Gaussian profile with width $\sigma = 0.4$ fm in order to obtain continuous density distributions. The deposited energy density is then taken to be proportional to a linear combination of the participant and collision densities \cite{Strickland:2011aa}
\begin{equation}
    e(\mathbf{x_T}) = \frac{1}{2}(1-\alpha)N_\mathrm{part}(\mathbf{x_T}) + \alpha N_\mathrm{coll}(\mathbf{x_T}),
\label{eq:MCGlauber_energy}
\end{equation}
with $\alpha$ which depends on the energy collision, in this dissertation $\alpha=0.2$. 

This model does not include any internal substructure of the nucleons. After this procedure, the energy density is multiplied by the normalization $K$-factor, which, as discussed previously, is a global multiplicative constant used to reproduce the observed final multiplicity. Finally, the energy density profile is recentered on the computational grid by shifting the center of mass to the center of the grid.

One of the focuses of this dissertation is the study of event-by-event fluctuations in the initial energy density defined above. These fluctuations arise naturally from the random sampling of nucleon positions \cite{Luzum:2013yya}. In this context, it is also important to introduce the optical limit of the Glauber model. In this approach, individual nucleon positions are not sampled, and therefore, the type of initial-state fluctuations studied in this work is absent. For this reason, the optical model provides a useful baseline for comparison, as will be discussed in \cref{chap:modesPb}.

The optical Glauber model \cite{Miller:2007ri,Strickland:2011aa} is also based on the eikonal approximation. In this framework, the projectile sees the target as a smooth continuous density, and the discrete nucleon structure is replaced by continuous nuclear densities. The model employs the nuclear thickness function for each nucleus,
\begin{equation}
    \hat{T}_{A, B}(x, y) = \int \rho_{A, B}(x, y, z) \, \mathrm{d}z,
\end{equation}
where the thickness function represents the probability of finding a nucleon in the target flux tube of interaction. It is obtained by integrating the nuclear density $\rho_{A, B}$, typically given by the Woods–Saxon distribution, along the beam axis $z$. The indices $A$ and $B$ denote the projectile and the target nuclei. The reduced thickness \cite{Moreland:2014oya}, which characterizes the overlap region between nuclei $A$ and $B$ at a given impact parameter $b$, is defined as
\begin{equation}
    \hat{T}_{AB}(b) = \int \hat{T}_A(x+b/2, y) \, \hat{T}_B(x-b/2, y) \, \mathrm{d}^2 r.
\end{equation}
From the reduced thickness, it is possible to determine the number of participants, the number of collisions, and the corresponding energy deposition, all based on the interaction probability $\hat{T}_{AB}(b)\sigma^\mathrm{NN}_\mathrm{inel}$.

\subsection{\textsc{McDipper}}
\label{subsec:mcdipper}

\textsc{McDipper} is a three-dimensional model \cite{Garcia-Montero:2023gex,Garcia-Montero:2025bpn} for initial states in heavy ion collisions based on the $k_\perp$-factorization framework of the CGC.\footnote{The code is available at \url{https://github.com/Non-Equilibrium-QCD/McDipper}.} It can generate initial profiles of energy, electric charge, or baryon densities; however, in this dissertation, we focus exclusively on energy deposition, which is the quantity used throughout the analysis.

Extending a two-dimensional model to three-dimensional is not straightforward \cite{Schenke:2016ksl}, since the inclusion of longitudinal dynamics leads to a larger gluon production cross section compared to that of valence quarks. For this reason, \textsc{McDipper} separates the energy deposition into gluon and quark contributions. The gluon component arises from radiative processes between two color glass condensates. In contrast, the quark contribution originates from the stopping of a quark or antiquark from each nucleus. The longitudinal dependence \cite{Garcia-Montero:2023gex} enters through the unintegrated gluon distribution or the adjoint dipole, both of which depend on the momentum fraction $x = \frac{|\mathbf{p}|}{\sqrt{s_\mathrm{NN}}} e^{\pm \mathrm{y}}$,  where $\mathrm{y}$ is the momentum rapidity.

\textsc{McDipper} provides two different modes for modeling the fundamental dipole gluon distribution. The first, known as the GBW (Golec–Biernat–Wüsthoff) model, assumes a Gaussian form with phenomenological input for the saturation scale. The second, referred to as the IP-Sat mode, incorporates a more realistic distribution that includes a tail at large transverse momentum. In this dissertation, the analysis is restricted to the GBW mode, since it is computationally faster. This is particularly important because the present analysis requires to produce a large number of event, as will become clear in the following chapters.

The code begins by sampling nucleon positions according to the Woods–Saxon distribution given in \cref{eq:woods-saxon}. Alternatively, it can read nuclear configurations from input files generated by nuclear structure calculations \cite{Zhang:2024vkh}. Subsequently, the internal structure of the nucleon is introduced in terms of subnucleonic degrees of freedom, referred to as hotspots \cite{Garcia-Montero:2025bpn}. These hotspots are described by a two-dimensional Gaussian profile 
\begin{equation}
    T_q(\mathbf{r}_\perp) = \frac{1}{2\pi B_q} \exp{-\frac{\mathbf{r}_\perp^2}{2B_q}},
\label{eq:hotspots_fluc}
\end{equation}
where the quark width is $B_q = 0.04$ fm$^2$, and $\mathbf{r}_\perp$ denotes the hotspot position in the transverse plane. The hotspot positions are sampled from a Gaussian distribution with width $B_{qc} = \frac{B_G - B_q}{1 - 1/N_q}$, typically, $N_q = 3$ is used to represent the three valence quarks and $B_G = 0.156$ fm$^2$. As in the \textsc{MC-Glauber} approach, the impact parameter is sampled along the fixed $x$-direction, and participants and spectators are determined using the geometric criterion $d \le \sqrt{\sigma^\mathrm{NN}_\mathrm{inel}/\pi}$.

In addition to the inclusion of hotspots, \textsc{McDipper} accounts for fluctuations \cite{Garcia-Montero:2025bpn} in the participant thickness function 
\begin{equation}
    T_{\mathrm{p/n}} = \frac{1}{N_q} \sum_{i=1}^{N_q} p_i T_q, 
\label{eq:thickness_fluc}
\end{equation}
where the weights $p_i$ are sampled from a log-normal distribution \cite{Mantysaari:2022ffw}. Using the proton and neutron thickness functions as input, the gluon and quark spectra are computed. The total deposited energy corresponds to the sum of these two contributions \cite{Garcia-Montero:2023gex}: 
\begin{equation}
    (e\tau)_0 = \int \mathrm{d}^2 \mathbf{p} \, |\mathbf{p}| \left[ K_g \frac{\mathrm{d}N_g}{\mathrm{d}^2\mathbf{x}\, \mathrm{d}^2\mathbf{p}\, \mathrm{d}\mathrm{y}} + \sum_{f, \bar{f}} \frac{\mathrm{d}N_{q_f}}{\mathrm{d}^2\mathbf{x}\, \mathrm{d}^2\mathbf{p}\, \mathrm{d}\mathrm{y}} \right].
\label{eq:energy_mcdipper}
\end{equation}

The difference between the GBW and IP-Sat modes lies primarily in how the gluon and quark spectra are computed from the thickness functions. The only parameter requiring calibration is the $K_g$-factor, which accounts for higher-order corrections in the gluon sector. Notably, this $K_g$-factor multiplies only the gluon contribution and does not act as a global normalization for the total energy deposition. Since it compensates for higher-order effects in the gluon production relative to the quark contribution, it is expected to be of order unity. Its precise value is determined by analyzing the amount of initial energy required to reproduce the experimentally measured final multiplicity, which will be discussed in more detail in the next chapter.

\section{Pre-equilibrium}
\label{sec:pre_eq}

After the collision, the QGP is formed and, as described in \cref{chap:introd} (see \cref{fig:v2_schema}), it exhibits collective behavior that can be described within a hydrodynamic framework. However, hydrodynamics is applicable only when the system is in or close to thermodynamic equilibrium. Immediately after such high-energy collisions, the system is far from this regime. Therefore, a pre-equilibrium or pre-hydrodynamic phase is introduced to evolve the system toward conditions under which a hydrodynamic description becomes valid.

\begin{figure}[h]
    \centering
    \includegraphics[width=0.6\linewidth]{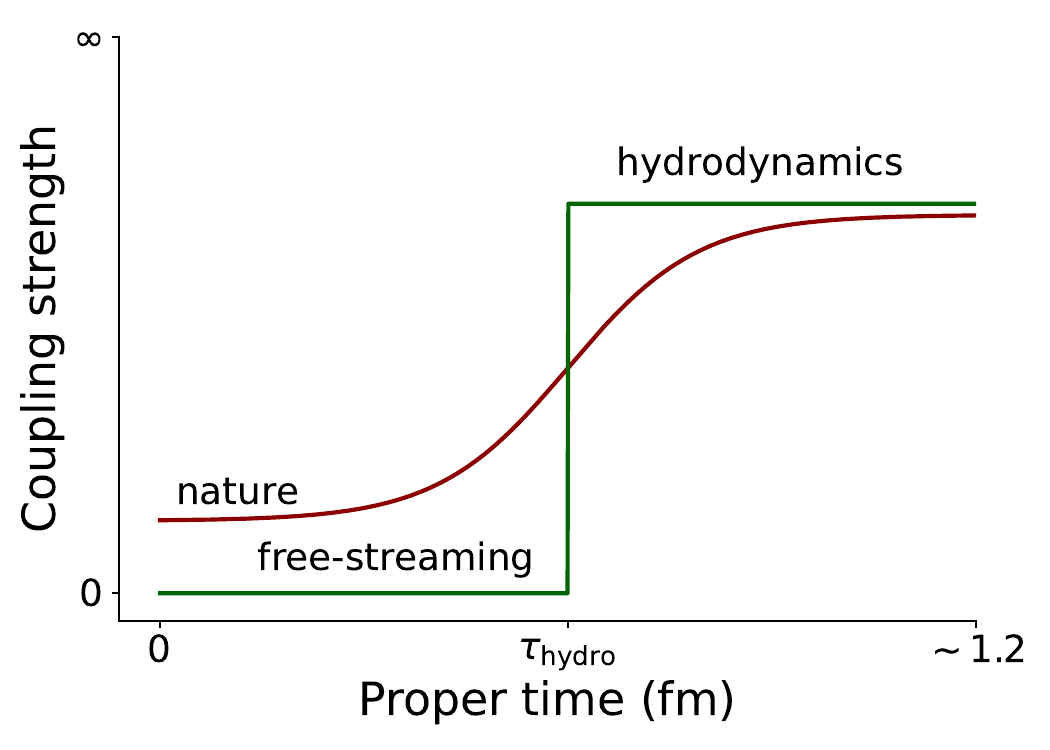}
    \caption{Schematic relation between the coupling strength and the proper time of the QGP. Naturally, the system evolves smoothly from a weakly coupled regime to a strongly coupled one. In the free-streaming approximation, this early stage is described as a zero-coupling evolution up to a switching time $\tau_\mathrm{hydro}$, after which a hydrodynamic description becomes applicable. Image adapted from Ref. \cite{Moreland:2018gsh}.}
\label{fig:free-streaming}
\end{figure}

The purpose of this phase is to use kinetic theory to evolve the QGP for a short time, leading to a smoother and more uniform profile. The simplest implementation of this evolution is free streaming, which consists of solving the Boltzmann equation without including a collision term \cite{Liu:2015nwa,Broniowski:2008qk}. Free streaming is a useful approximation because, shortly after the collision, the system is more likely to undergo rapid expansion than to interact strongly with neighboring cells. As a result, the matter behaves as a weakly coupled system at early times and gradually evolves toward a regime of stronger effective coupling, where hydrodynamics becomes applicable. \Cref{fig:free-streaming} illustrates how the coupling strength varies with proper time. It is important to note that free streaming effectively corresponds to an abrupt transition from zero to finite coupling. Nevertheless, this approximation can be practical when a smooth description of the transition is not available.

The advantage of the free-streaming approximation lies in its simplicity. It provides a straightforward evolution in which the system expands without modifying the particle momenta. However, the model strongly depends on the choice of the switching time $\tau_\mathrm{fs}$, since the approximation becomes unreliable for times larger than approximately $1$ fm \cite{Liu:2015nwa,Moreland:2018gsh}, as illustrated in \cref{fig:free-streaming}. In practice, this parameter $\tau_\mathrm{fs}$ is often constrained through Bayesian analyses, and it may be treated either as independent \cite{Bernhard:2018hnz} or dependent \cite{JETSCAPE:2020mzn} of the initial energy density.

An alternative approach to describing this out-of-equilibrium stage is to employ an effective kinetic theory \cite{Kurkela:2018vqr}, as implemented in the code \Kompost, which is the framework adopted in this dissertation. The central idea is to decompose the initial energy–momentum tensor as 
\begin{equation}
    T^{\mu\nu}(\tau_0, \mathbf{x}') = \bar{T}^{\mu\nu}_{\mathbf{x}}(\tau_0) + \delta T^{\mu\nu}_{\mathbf{x}}(\tau_0, \mathbf{x}'),
\end{equation}
where the homogeneous background energy–momentum tensor $\bar{T}^{\mu\nu}_{\mathbf{x}}(\tau_0)$ contains the dominant contribution to the energy density and pressure. Its evolution is governed by a nonlinear equilibration process characterized by attractor behavior \cite{Kurkela:2018vqr}. The second term, $\delta T^{\mu\nu}_{\mathbf{x}}(\tau_0, \mathbf{x}')$, represents perturbations around the background and is treated using linear response theory. With this decomposition, the system can be evolved smoothly toward the hydrodynamic regime, avoiding the strong sensitivity to a specific switching time that appears in the free-streaming approximation. Further details of this model are presented in the next subsection.

The pre-equilibrium phase, together with the initial conditions, constitutes one of the most uncertain stages of heavy ion collisions, primarily due to the absence of a first-principles theory applicable in this regime. These stages, therefore, play a central role in the analyses presented in this dissertation. In particular, the transition from the pre-equilibrium stage to hydrodynamics remains an open problem, with several conceptual and practical challenges still under investigation by the community. A dedicated analysis of this transition is provided in \cref{chap:entropy_matching}.

\subsection{\Kompost}
\label{subsec:kompost}

\begin{figure}[h]
    \centering
    \includegraphics[width=0.4\linewidth]{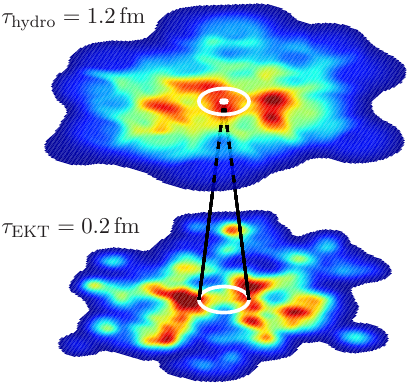}
    \caption{Schematic representation of the evolution of the transverse energy profile. In the figure $\tau_{\mathrm{EKT}}$ corresponds to what we denote as $\tau_0$, the initial time for \Kompost, while $\tau_{\mathrm{hydro}}$ represents the final evolution time in the code. Notice that the points contributing to the background are always located within the past light cone (white circle). Figure from Ref. \cite{Kurkela:2018vqr}.}
\label{fig:kompost_evolution}
\end{figure}

\Kompost~is the code that implements effective kinetic theory (EKT) to evolve the system from an out-of-equilibrium state toward a near-equilibrium regime \cite{Kurkela:2018vqr}.\footnote{The code is available at \url{https://github.com/KMPST/KoMPoST}.} Taking the energy–momentum tensor produced in the collision as input, the code performs a decomposition into background and perturbation components according to
\begin{equation}
    T^{\mu\nu}(\tau_0, \mathbf{x}') = \underbrace{\bar{T}^{\mu\nu}_{\mathbf{x}}(\tau_0)}_{\mathrm{background}} + \underbrace{T^{\mu\nu}(\tau_0, \mathbf{x}') - \bar{T}^{\mu\nu}_{\mathbf{x}}(\tau_0)}_{\equiv \delta T^{\mu\nu}_{\mathbf{x}}(\tau_0, \mathbf{x}')}.
\end{equation}
Here, the background is defined as an average over neighboring points $\mathbf{x}'$ within a causal region of the future $\mathbf{x}$ point defined as $|\mathbf{x} - \mathbf{x}'| < c (\tau_\mathrm{hydro} - \tau_0)$, as illustrated in the white circle of \cref{fig:kompost_evolution}.

After this decomposition, the evolved energy–momentum tensor at the hydrodynamization time can be written as
\begin{equation}
    T^{\mu\nu}(\tau_\mathrm{hydro}, \mathbf{x}) = \bar{T}^{\mu\nu}_{\mathbf{x}} (\tau_\mathrm{hydro}) + \frac{\bar{T}^{\tau\tau}_{\mathbf{x}}(\tau_\mathrm{hydro})}{\bar{T}^{\tau\tau}_{\mathbf{x}}(\tau_0)} \int \mathrm{d}^2 {\mathbf{x}'}\, G^{\mu\nu}_{\alpha\beta}(\mathbf{x}, \mathbf{x}', \tau_\mathrm{hydro}, \tau_0) \delta T^{\alpha\beta}_{\mathbf{x}}(\tau_0, \mathbf{x}'),
\label{eq:Tmunu_tauhydro}
\end{equation}
where $\tau_0$ and $\mathbf{x}'$ denote the initial coordinates, while $\tau_\mathrm{hydro}$ and $\mathbf{x}$ represent the final ones. In the present implementation, the code assumes boost invariance, so the vectors $\mathbf{x}$ lie entirely in the transverse plane. Although the framework was recently extended to include rapidity dependence in Ref. \cite{Du:2025hyk}.

A central challenge \cite{Kurkela:2018vqr} is determining the evolution of the background term $\bar{T}^{\mu\nu}_{\mathbf{x}}(\tau_\mathrm{hydro})$. This evolution is obtained through a universal scaling function derived from attractor solutions, which relates the energy density to an idealized parameterized ``temperature'' according to $T \propto e^{1/4}$.\footnote{Strictly speaking, temperature is a thermodynamic quantity defined only in equilibrium. For this reason, we use quotation marks.}

The second term in the evolution equation involves the determination of the Green’s function $G^{\mu\nu}_{\alpha\beta}(\mathbf{x}, \mathbf{x}', \tau_\mathrm{hydro}, \tau_0)$ in \cref{eq:Tmunu_tauhydro}. In Fourier space, this requires solving the Boltzmann equation for a linear perturbation of the distribution function 
\begin{equation}
    f_{\mathbf{x}, \mathbf{p}} = \bar{f}_{\mathbf{p}} + \int \frac{\mathrm{d}^2 \mathbf{k}}{(2\pi^2)} \delta f_{\mathbf{k}, \mathbf{p}} e^{i\mathbf{k} \cdot \mathbf{x}},
\end{equation}
which leads to coupled equations for the background distribution $\bar{f}_{\mathbf{p}}$ and the perturbation $\delta f_{\mathbf{k}, \mathbf{p}}$. The collision integral is treated within the relaxation time approximation (RTA) \cite{Heller:2016rtz,denicol2022microscopic}, where the relaxation time is defined in terms of the specific shear viscosity and the temperature 
\begin{equation}
    \tau_R \equiv 5\frac{\eta/s}{T}.
\end{equation}

After solving these equations, the evolved energy–momentum tensor is reconstructed according to \cref{eq:Tmunu_tauhydro}. Finally, in order to interface with the hydrodynamics code, a standard hydrodynamic decomposition is applied to extract the relevant tensor components, as will be described in the next section.

It has been shown \cite{Kurkela:2018vqr} that the effective kinetic evolution is less sensitive to the switching time $\tau_\mathrm{hydro}$ than the free-streaming approximation, since it provides a more physical evolution to the system. In this sense, although $\tau_\mathrm{hydro}$ is an input parameter, variations around values of order $1\,\mathrm{fm}$ do not significantly modify the evolution. In addition, the specific shear viscosity $\eta/s$ is also an input parameter. In this dissertation, a constant value of $\eta/s = 0.16$ is adopted throughout the analysis.\footnote{Note the notation used throughout this dissertation: space-time rapidity $\eta_s$, pseudorapidity $\eta$, and specific shear viscosity $\eta/s$.}

\Kompost~can be executed either in free-streaming mode or using EKT with energy and/or momentum perturbations, which determine the components of $G^{\mu\nu}_{\alpha\beta}$ included in the calculation. In this work, momentum perturbations are not relevant for the energy-related observables under investigation and are therefore neglected.

\section{Hydrodynamics}
\label{sec:hydro}

As mentioned previously, after the collision and the subsequent deconfinement of quarks and gluons into a plasma, the system exhibits collective behavior \cite{Nouicer:2015jrf}. Therefore, it is necessary to develop a hydrodynamic framework to describe this ultrarelativistic system, which is driven close to local equilibrium by the pre-equilibrium phase. This section introduces the ideal and viscous formulations of relativistic fluid dynamics. The next subsection \ref{subsec:music} provides details on the numerical implementation of this theory in the code \textsc{MUSIC}.

The ideal hydrodynamic theory \cite{landau1987fluid,denicol2022microscopic,Heinz:2024jwu} is based on the assumption of local thermodynamic equilibrium, in which all fluid properties, such as temperature $T(\vec{x})$, pressure $P(\vec{x})$, and fluid velocity $\vec{u}(\vec{x})$, are well defined at each space-time point. In this limit, dissipative effects are absent. In relativistic formulations, it is convenient to describe the fluid in its local rest frame, meaning that the reference frame moves together with the fluid element.

The central quantity in relativistic hydrodynamics is the energy–momentum tensor:
\begin{equation}
    T_{\mu\nu} =
    \begin{pmatrix}
    \colorbox{yellow!50}{$T_{00}$} & \colorbox{blue!40}{$T_{01}$} & \colorbox{blue!40}{$T_{02}$} & \colorbox{blue!40}{$T_{03}$} \\
    \colorbox{cyan!30}{$T_{10}$} & \colorbox{purple!50}{$T_{11}$} & \colorbox{green!50}{$T_{12}$} & \colorbox{green!50}{$T_{13}$} \\
    \colorbox{cyan!30}{$T_{20}$} & \colorbox{green!50}{$T_{21}$} & \colorbox{purple!50}{$T_{22}$} & \colorbox{green!50}{$T_{23}$} \\
    \colorbox{cyan!30}{$T_{30}$} & \colorbox{green!50}{$T_{31}$} & \colorbox{green!50}{$T_{32}$} & \colorbox{purple!50}{$T_{33}$}
    \end{pmatrix},
\label{eq:Tmunu_color}
\end{equation}
where the yellow component corresponds to the energy density, the purple components represent the energy flux, the blue components represent the momentum density, the red diagonal spatial components are related to the pressure, and the green off-diagonal spatial components are associated with shear stresses. For an ideal fluid in the local rest frame, only the diagonal components are nonzero: 
\begin{equation}
    T^{\mu\nu}_0 = \begin{pmatrix}
        \epsilon & 0 & 0 & 0 \\
        0 & P & 0 & 0 \\
        0 & 0 & P & 0 \\
        0 & 0 & 0 & P \\
    \end{pmatrix},
\label{eq:Tmunu_ideal}
\end{equation}
where $\epsilon$ is the energy density and $P$ is the pressure. The subscript $0$ in the energy-momentum tensor denotes the ideal contribution. This form assumes spatial isotropy, meaning that the pressure is identical in all spatial directions. More compactly, the ideal energy–momentum tensor and particle four-current can be written as \cite{denicol2022microscopic} 
\begin{equation}
\begin{aligned}
    T^{\mu\nu}_0 &= \epsilon u^\mu u^\nu - \Delta^{\mu\nu} P, \\
    N^\mu_0 &= n u^\mu,
\end{aligned}
\end{equation}
where $u^\mu = (\gamma, \vec{u})^T$ is the four-velocity, $\Delta^{\mu\nu} = g^{\mu\nu} - u^\mu u^\nu$ is the projector onto the three-space orthogonal to $u^\mu$ (such that $\Delta^{\mu\nu} u_\mu = 0$) and $g^{\mu\nu}$ is the space-time metric. The quantity $n$ denotes the particle density.

The equations of motion follow from conservation of particle number, energy, and momentum \cite{Heinz:2024jwu}:\footnote{It is also possible to include additional conserved charge currents, $\partial_\mu J_i^\mu = 0$, where the index $i$ refers to conserved quantum numbers such as net baryon number, strangeness, or electric charge. These are not considered in this dissertation.}
\begin{equation}
\begin{aligned}
    \partial_\mu N^\mu_0 &= 0,  \\
    \partial_\mu T^{\mu\nu}_0 &= 0.
\end{aligned}
\label{eq:conservation_idealTN}
\end{equation}
A common way to express the equations of motion is to contract the conservation laws with the four-velocity and the projector \cite{denicol2022microscopic}:
\begin{equation}
\begin{aligned}
    \partial_\mu N^\mu_0 &= \dot{n} + n \theta = 0, \\
    u_\nu \partial_\mu T^{\mu\nu}_0 &= \dot{\epsilon} + \epsilon\theta + P\theta = 0, \\
    \Delta^\alpha_\nu \partial_\mu T^{\mu\nu}_0 &= (\epsilon + P) \dot{u}^\alpha - \nabla^\alpha P =0,
\end{aligned}
\end{equation}
where the shorthand notation $\dot{A} = u^\alpha \partial_\alpha A$ and $\theta \equiv \partial_\alpha u^\alpha$ has been introduced. These five equations determine six variables ($n$, $\epsilon$, $P$, and the three independent components of $u^\mu$, subject to the normalization condition $u^\mu u_\mu = 1$). Therefore, to close the system, it is necessary to specify an equation of state (EOS), $P = P(\epsilon, n)$, which encodes the thermodynamic properties of the fluid under consideration.

In ideal hydrodynamics, only the conservation of particle number, energy, and momentum is imposed. However, it can be shown that the entropy four-current can be derived from the first law of Thermodynamics and written as
\begin{equation}
    \partial_\mu S^\mu_0 = \beta_\nu \partial_\mu T_0^{\mu\nu} - \alpha \partial_\mu N_0^\mu = 0~,
\label{eq:entropy_conservation}
\end{equation}
where $\beta^\nu = u^\nu/T$ and $\alpha = \mu/T$. Therefore, entropy is naturally conserved once the conservation equations \ref{eq:conservation_idealTN} are imposed.

\begin{figure}[h]
    \centering
    \includegraphics[width=0.63\linewidth]{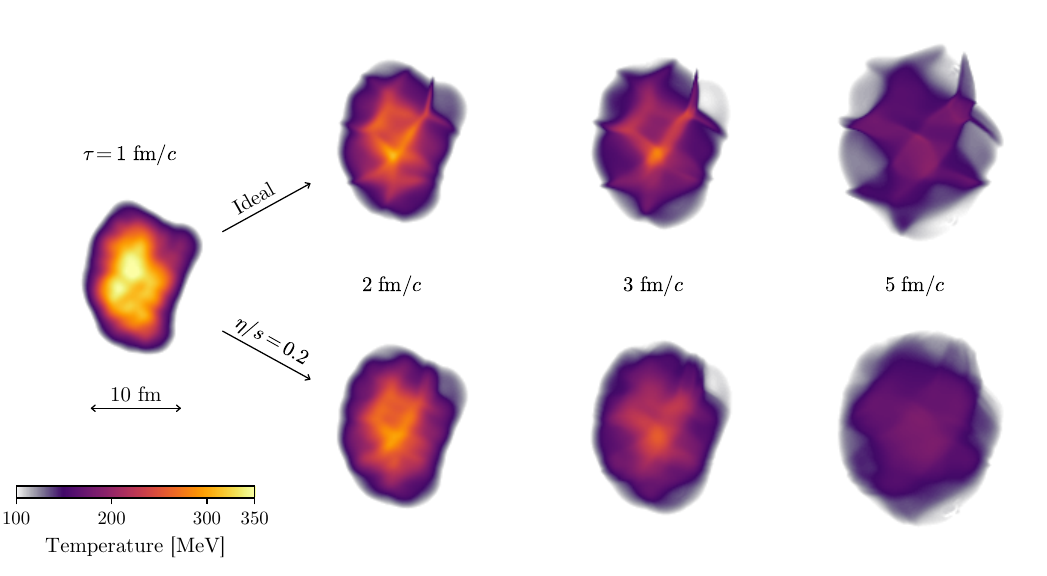}
    \caption{Evolution of a temperature profile in ideal hydrodynamics (top) and viscous hydrodynamics with $\eta/s = 0.2$ (bottom). Due to viscous effects, the temperature distribution after $4\,\mathrm{fm}$ differs significantly, even though the initial profile is the same in both cases. Figure from Ref. \cite{Bernhard:2018hnz}.}
\label{fig:hydro_evolve}
\end{figure}

Nevertheless, although the QGP is characterized by a very small shear viscosity \cite{Bernhard:2018hnz,Nouicer:2015jrf,Heinz:2024jwu}, it remains a viscous system and is not in exact local thermodynamic equilibrium. \Cref{fig:hydro_evolve} illustrates the difference between the temperature evolution in ideal and viscous hydrodynamics. Even for a relatively small value of the specific shear viscosity $\eta/s = 0.2$, which is significantly lower than that of liquid helium $\eta/s \sim 1$ or water $\eta/s \sim 2$, viscous effects still play a crucial role in accurately describing the dynamics and the final observables in heavy-ion collisions~\cite{Ryu:2015vwa,Baier:2006um}.

To develop a viscous relativistic hydrodynamic theory \cite{landau1987fluid,denicol2022microscopic}, the procedure follows the same general structure as in the ideal case, but additional dissipative currents are introduced. In particular, the particle four-current and the energy–momentum tensor are extended as
\begin{equation}
\begin{aligned}
    N^\mu &= N^\mu_0 + n^\mu, \\
    T^{\mu\nu} &= T_0^{\mu\nu} + W^{\mu\nu}.
\end{aligned}
\end{equation}
In dissipative hydrodynamics, the energy–momentum tensor in \cref{eq:Tmunu_color} now contains not only diagonal components but also off-diagonal terms, which account for fluxes and dissipative effects. A convenient decomposition of $W^{\mu\nu}$ introduces a scalar $\Pi$, known as the bulk pressure; a vector $h^\mu$, referred to as the heat flow;\footnote{Parentheses denote symmetrization over Lorentz indices: $a^{(\mu\nu)} = (a^{\mu\nu} + a^{\nu\mu})/2$.} and a traceless symmetric tensor $\pi^{\mu\nu}$, known as the shear stress tensor
\begin{equation}
    W^{\mu\nu} \equiv -\Pi \Delta^{\mu\nu} + 2 u^{(\mu}h^{\nu)} + \pi^{\mu\nu}~.
\end{equation}
With the inclusion of these additional dissipative quantities, the number of dynamical variables increases. Consequently, extra equations of motion are required to determine the evolution of the bulk pressure, heat flow, and shear stress tensor.

For ideal fluids, the local rest frame is defined as the frame in which there is no flow of particles or energy. In viscous fluids, however, energy and particle diffusion are generally present, and therefore, the fluid velocity can be defined in different ways. In this work, we define the velocity through the energy flow, which corresponds to the Landau frame \cite{landau1987fluid,denicol2022microscopic}:
\begin{equation}
    u_\mu T^\mu_\nu \equiv \epsilon u_\nu~.
\label{eq:landau_matching}
\end{equation}
This is the most commonly used frame in heavy-ion collisions. For the Landau frame be valid the energy diffusion current $h^\mu$ is set to zero, which simplify the equations. Another possible choice is the Eckart frame, in which the velocity is defined through the particle flow $N^\mu \equiv n u^\mu$. The choice of frame affects the form of the dissipative currents, but give the same evolution.

Once the frame is specified, the conserved currents take the form 
\begin{equation}
\begin{aligned}
    N^\mu &= n u^\mu + n^\nu~, \\
    T^{\mu\nu} &= \epsilon u^\mu u^\nu - (P + \Pi) \Delta^{\mu\nu}+ \pi^{\mu\nu}.
\end{aligned}
\label{eq:tmunu_hydro}
\end{equation}
The projection of the conservation equations follows the same procedure as in the ideal case \cite{denicol2022microscopic} 
\begin{equation}
\begin{aligned}
    \partial_\mu N^\mu &= \dot{n} + n\theta + \partial_\mu n^\mu = 0, \\
    u_\nu \partial_\mu T^{\mu\nu} &= \dot{\epsilon} + (\epsilon + P + \Pi) \theta - \pi^{\mu\nu} \sigma_{\mu\nu} = 0, \\
    \Delta^\alpha_\nu \partial_\mu T^{\mu\nu} &= (\epsilon + P + \Pi) \dot{u}^\alpha - \nabla^\alpha (P + \Pi) + \Delta^\alpha_\nu \partial_\mu \pi^{\mu\nu} = 0,
\end{aligned}
\label{eq:generic_dissipative_hydro}
\end{equation}
where $\sigma^{\mu\nu} \equiv \frac{1}{2}(\nabla^\mu u^\nu + \nabla^\nu u^\mu) - \frac{1}{3}\Delta^{\mu\nu}\theta$. Including the equation of state, there are six equations; however, in addition to the usual variables $n$, $u^\mu$, $\epsilon$, and $P$, there are now the dissipative quantities $n^\mu$, $\Pi$, and $\pi^{\mu\nu}$, which introduce additional independent degrees of freedom. These quantities satisfy the constraints 
\begin{equation}
\begin{aligned}
    u_\mu n^\mu & = 0, \\
    u_\mu \pi^{\mu\nu} & = 0, \\
    \pi^\mu_\mu & = 0, \\
    \pi^{\mu\nu} & = \pi^{\nu\mu}.
\end{aligned}
\end{equation}

To close the system, evolution equations for the dissipative terms are required. These can be derived from the entropy current. In the ideal case, entropy is naturally conserved, as shown in \cref{eq:entropy_conservation}. In a dissipative system, however, entropy is no longer conserved; according to the second law of thermodynamics, dissipation must generate entropy:
\begin{equation}
    \partial_\mu S^\mu = -n^\mu \nabla_\mu \alpha_0 - \beta_0 \Pi \theta + \beta_0 \pi^{\mu\nu} \sigma_{\mu\nu} \equiv Q,
\end{equation}
where the non-dissipative contributions are contained in $S^\mu_0$, and the entropy four-current is defined as $S^\mu = S^\mu_0 - \alpha_0 n^\mu$. The term $Q$ represents the entropy production due to dissipative effects.

Imposing the condition $Q \ge 0$, the linear relations that satisfy this requirement are \cite{denicol2022microscopic} 
\begin{equation}
\begin{aligned}
    \Pi &= - \zeta\theta, \\
    n^\mu &= \chi \nabla^\mu \alpha_0, \\
    \pi^{\mu\nu} &= 2\eta\sigma^{\mu\nu}.
\end{aligned}
\label{eq:linear_NS}
\end{equation}
The proportionality coefficients $\zeta$, $\chi$, and $\eta$ correspond to the bulk viscosity, particle diffusion coefficient, and shear viscosity, respectively. Substituting these relations into \cref{eq:generic_dissipative_hydro} yields the relativistic Navier–Stokes equations 
\begin{eqnarray}
    \dot{n} &=& - n\theta - \partial_\mu (\chi \nabla^\mu \alpha_0), \\
    \dot{\epsilon} &=& -(\epsilon + P - \zeta\theta) \theta + 2\eta\sigma_{\mu\nu}\sigma^{\mu\nu}, \\
    (\epsilon + P - \zeta \theta) \dot{u}^\alpha &=&  \nabla^\alpha (P - \zeta\theta) - 2\Delta^\alpha_\nu \partial_\mu (\eta\sigma^{\mu\nu}). 
\end{eqnarray}

An alternative derivation of the Navier–Stokes equations can be obtained through a gradient expansion in terms of the Knudsen number $\mathrm{Kn} \equiv \lambda/L$, which relates a typical microscopic scale $\lambda$ to a macroscopic scale $L$. When $\mathrm{Kn} \ll 1$, the system exhibits collective behavior, and the expansion in powers of $\mathrm{Kn}$ is expected to converge.

The main issue with the relativistic Navier–Stokes equations is that they are parabolic, which leads to acausal behavior. In particular, one can compute the group velocity of a perturbation in the fluid as $v_g \equiv \left|\frac{d\omega}{dk}\right| \propto k$ where the wave number $k$ is, in principle, unbounded. As a consequence, arbitrarily large propagation velocities are allowed, violating causality. In nonrelativistic systems, this problem is not apparent because the relevant wave numbers are small, and the Navier–Stokes theory successfully describes everyday fluids. However, in relativistic systems such as the QGP, this acausal behavior becomes physically unacceptable.

Similar issues with parabolic equations have been studied in other areas of physics, for example, in the diffusion equation
\begin{eqnarray}
    \partial_t A = D \nabla^2 A.
\end{eqnarray}
A common remedy is to introduce a second-order time derivative, thereby modifying the equation to \cite{denicol2022microscopic,karim2020effect} 
\begin{equation}
    \tau_R \partial_t^2 A + \partial_t A = D \nabla^2 A,
\end{equation}
where $\tau_R$ is a relaxation time of the variable $A$. This additional term limits the maximal signal velocity to $v_{\mathrm{max}} = \sqrt{D/\tau_R} \le 1$, thus restoring causality. This modification is known as the Maxwell–Cattaneo equation \cite{Romatschke:2017ejr}.

In causal relativistic hydrodynamics, the same idea is applied. The constitutive relations in \cref{eq:linear_NS} are replaced by relaxation-type equations, known as the Israel–Stewart equations \cite{Israel:1976tn,Israel:1979wp}:
\begin{equation}
\begin{aligned}
    \tau_\Pi \dot{\Pi} + \Pi &= - \zeta\theta + ...~,  \\
    \tau_n \Delta^\mu_\nu n^\nu + n^\mu &= \chi \nabla^\mu \alpha_0 + ...~,  \\
    \tau_\pi \Delta^{\mu\nu}_{\alpha\beta}\dot{\pi}^{\alpha\beta} +\pi^{\mu\nu} &= 2\eta\sigma^{\mu\nu} + ...~.
\end{aligned}
\label{eq:IS}
\end{equation}
The additional terms on the right side are second-order contributions, which can be derived using different theoretical approaches. The next section presents the specific formulation adopted in this dissertation.

The main codes that implement relativistic hydrodynamic theory in the context of heavy-ion collisions: \textsc{MUSIC} \cite{Schenke:2010nt}, \textsc{iEBE-VISHNU} \cite{Shen:2014vra} and \textsc{CLVisc} \cite{Pang:2018zzo}. The next section provides a more detailed discussion of Eq.\ref{eq:IS} within the DNMR framework \cite{denicol2022microscopic,Denicol:2012yr}, as well as its numerical implementation in \textsc{MUSIC}.

\subsection{\textsc{MUSIC}}
\label{subsec:music}

This subsection presents the numerical implementation of the hydrodynamic theory introduced in the previous section, as realized in the code \textsc{MUSIC} \cite{Schenke:2010nt,Schenke:2010rr,Paquet:2015lta}, which is used throughout this dissertation.\footnote{The code is available at \url{https://github.com/MUSIC-fluid/MUSIC}. A useful reference for understanding the code is the online manual: \url{https://webhome.phy.duke.edu/jp401/music_manual/index.html}.}

Here, the focus is restricted to the evolution of the energy–momentum tensor. Conserved charges, vorticity, and other related effects are not included, since they are beyond the scope of the present analysis.

The equations of motion in \ref{eq:IS} are implemented in \textsc{MUSIC} according to the DNMR (Denicol–Niemi–Molnár–Rischke) framework \cite{Denicol:2012yr, Denicol:2012cn}. Within the 14-moment approximation \cite{Denicol:2014vaa}, the dissipative equations can be written as follows:
\begin{equation}
\begin{aligned}
    \tau_\pi \dot{\pi}^{\langle \mu\nu \rangle} + \pi^{\mu\nu} &= 2\eta \sigma^{\mu\nu} + 2\tau_\pi \pi_\alpha^{\langle \mu}\omega^{\nu\rangle\alpha} - \delta_{\pi\pi} \pi^{\mu\nu} \theta + \phi_7\pi^{\langle \mu}_\alpha \pi^{\nu \rangle \alpha} \\
    &- \tau_{\pi\pi} \pi_\alpha^{\langle \mu} \sigma^{\nu\rangle\alpha} + \lambda_{\pi\Pi} \Pi \sigma^{\mu\nu} + \phi_6 \Pi \pi^{\mu\nu},
\end{aligned}
\end{equation}
for the shear stress tensor and 
\begin{equation}
    \tau_\Pi \dot{\Pi} + \Pi = -\zeta \theta - \delta_{\Pi\Pi}\Pi\theta + \phi_1 \Pi^2 + \lambda_{\Pi\pi} \pi^{\mu\nu}\sigma_{\mu\nu} + \phi_3 \pi^{\mu\nu} \pi_{\mu\nu},
\end{equation}
for the bulk pressure. The DNMR framework introduces 11 second-order transport coefficients ($\tau_\pi$, $\delta_{\pi\pi}$, $\phi_7$, $\tau_{\pi\pi}$, $\lambda_{\pi\Pi}$, $\phi_6$, $\tau_\Pi$, $\delta_{\Pi\Pi}$, $\phi_1$, $\lambda_{\Pi\pi}$, and $\phi_3$). These coefficients are given by
\begin{equation}
\begin{aligned}
    \tau_\pi &= \frac{5\eta}{(\epsilon + P)}, & \qquad
    \delta_{\pi\pi} &= \frac{4}{3}\tau_\pi, \\
    \phi_7 &= \frac{9}{70 P}, & \qquad
    \tau_{\pi\pi} &= \frac{10}{7}\tau_\pi, \\
    \lambda_{\pi\Pi} &= \frac{6}{5} \tau_\pi, & \qquad
    \tau_\Pi &= \frac{\zeta}{15 (1/3 - c_s^2)^2 (\epsilon + P)}, \\
    \delta_{\Pi\Pi} &= \frac{2}{3}\tau_\Pi, & \qquad
    \lambda_{\Pi\pi} &= \frac{8}{5}\left(\frac{1}{3} - c_s^2\right) \tau_\Pi,
\end{aligned}
\end{equation}
with $\phi_1 = \phi_3 = \phi_6 = 0$. 

In summary, the DNMR model introduces second-order transport coefficients that are derived within the 14-moment approximation \cite{Denicol:2014vaa}. These coefficients represent properties of the QGP and they depend exclusively on the first-order transport coefficients, which must be determined either from first-principles calculations or from effective models of QCD. 

To close the hydrodynamic system, an EOS must be specified. In this work, we employ the EOS from the HotQCD collaboration \cite{HotQCD:2014kol}, matched to the \textsc{SMASH} hadron resonance gas (HRG) at low temperatures. The matching procedure ensures consistency in the hadronic degrees of freedom by including the same particle species in the HRG. In the vicinity of the critical temperature, there exists an overlap region where the HotQCD and HRG descriptions are in a good agreement.

The numerical evolution in \textsc{MUSIC} is performed using a Runge–Kutta time-integration scheme combined with operator splitting. The evolution is typically separated into a conservation step and a relaxation step. In the first step, the conservation equations for the total-momentum tensor are solved, which can be written as $\partial_a T_0^{a\mu} = - \partial_a W^{a\mu}$, where $T^{\mu\nu}_0$ denotes the ideal contribution and $W^{\mu\nu}$ the dissipative part. In the second step, the dissipative quantities are evolved according to their relaxation-type equations. Since the evolution is time-dependent, each step requires information from the previous time slice, which makes the initialization of hydrodynamic simulations nontrivial and remains an active area of research \cite{Gavassino:2025mcq}. The code enforces the Landau frame condition as given in \cref{eq:landau_matching}.

\textsc{MUSIC} provides several operational modes. The standard configuration is the evolution mode, in which the hydrodynamic fields obtained from the decomposition are used as input, and the system is evolved until a specified switching temperature is reached. Additional modes allow the calculation of observables directly from the freeze-out hypersurface, including full particle spectra and resonance decays, without incorporating subsequent hadronic transport evolution.

\section{Particlization}
\label{sec:particlization}

Particlization is only a numerical step in heavy-ion collision simulations. Physically, the system is initially described as a fluid which, upon reaching the critical temperature, undergoes confinement of quarks and gluons into hadrons, a process known as hadronization. After this phase transition, the resulting hot gas of hadrons can still exhibit collective behavior and may continue to be described within a hydrodynamic framework using a hadron resonance gas equation of state.

As the system expands and cools further, it eventually reaches the switching temperature, defined as the temperature at which the hydrodynamic description is terminated and a microscopic hadronic approach is initiated. At this stage, particlization is performed. This procedure converts the continuous fluid fields into discrete particle degrees of freedom. It is important to emphasize that particlization itself does not describe further dynamical evolution; rather, it provides the initial conditions for the subsequent hadronic transport stage. In this dissertation, the switching temperature is set to $155\,\mathrm{MeV}$.

It is important to note that some hydrodynamic codes refer to the switching temperature as the freeze-out temperature. This terminology originates from earlier approaches, developed before hybrid models became widely used. In modern hybrid descriptions, however, the actual freeze-out occurs within the hadronic transport stage. There are two types of freeze-out processes: the first is chemical freeze-out, which takes place when inelastic interactions cease, thereby fixing the particle yields, and the second is kinetic freeze-out, which occurs at a later stage, when elastic collisions also stop as the system becomes sufficiently dilute. In general, when the term freeze-out is used without specification, it typically refers to chemical freeze-out.

The theory used to translate the fluid into discrete particles is based on the Cooper–Frye formula \cite{Cooper:1974mv} 
\begin{equation}
    E_p \frac{\mathrm{d}N_n}{\mathrm{d}^3p} = \frac{1}{(2\pi\hbar)^3} \int_\Sigma f_n(x, p) \, p^\mu \, \mathrm{d}^3 \sigma_\mu,
\label{eq:cooper-frye}
\end{equation}
which gives the probability that a particle with a given momentum is emitted from a fluid cell. The integral is performed over the so-called particlization hypersurface,\footnote{The particlization hypersurface is often referred to in the literature as the freeze-out hypersurface, although there is no actual freeze-out at this stage, as discussed previously.} which is an isothermal surface in space-time containing the fluid properties at the instant each fluid cell reaches the switching temperature.

In this expression, $f_n(x, p)$ is the distribution function for a particle species $n$. In local equilibrium, this distribution corresponds to the Maxwell–Jüttner form \cite{McNelis:2019auj} 
\begin{equation}
    f_{\text{eq},n}(x, p) = g_n \left[\exp{\frac{p \cdot u(x)}{T(x)} - b_n \alpha_B(x)} + \Theta_n \right]^{-1},
\end{equation}
where $g_n$ is the spin degeneracy, $u(x)$ is the fluid four-velocity, $T(x)$ is the temperature, $b_n$ is the baryon number of species $n$, and $\alpha_B(x) = \mu_B(x)/T(x)$ is the ratio between the baryon chemical potential and the temperature. The parameter $\Theta_n = -1$ for the quantum statistics of bosons and $\Theta_n = 1$ for fermions.

However, since the viscous fluid is not in exact local equilibrium, a small correction must be included
\begin{equation}
    f_n(x, p) = f_{\text{eq}, n}(x, p) + \delta f_n(x, p).    
\end{equation}
To determine the correction $\delta f_n(x, p)$, one must match the hydrodynamic definition of the energy–momentum tensor given in \cref{eq:tmunu_hydro} with its kinetic theory expression, 
\begin{equation}
    T^{\mu\nu} = \sum_n \int \frac{\mathrm{d}^3 p}{(2\pi\hbar)^3 E_p} \, p^\mu \, p^\nu \, f_n(x, p).
\end{equation}
The most common choices for $\delta f_n$ are obtained by solving the Boltzmann equation via RTA, which is the one used in this dissertation. Alternatives include solving Boltzmann within the 14-moment approximation or through the first-order Chapman–Enskog expansion, both of which retain only linear contributions from the viscous corrections. Or one may employ modified equilibrium distributions. Two commonly used approaches are the Pratt–Torrieri–Bernhard (PTB) distribution and a related variant known as the PTM distribution \cite{McNelis:2019auj}.

Physically, the Cooper–Frye formula \cref{eq:cooper-frye} provides the average particle spectrum obtained by integrating the distribution function over the entire particlization hypersurface. Converting this continuous hypersurface into discrete particles, with well-defined space-time coordinates $x$ and momenta $p$, is referred to as sampling. Since the sampling procedure introduces statistical fluctuations, multiple oversampled events are typically generated in order to reduce statistical noise and obtain a more accurate estimate of the average spectrum.

\subsection{\textsc{iSS}}
\label{subsec:iss}

\textsc{iSS}, short for iSpectraSampler, is a Monte Carlo sampler for particle distributions based on the Cooper–Frye procedure \cite{Shen:2014vra}.\footnote{The code used is available at \url{https://github.com/chunshen1987/iSS}.} The code takes as input the hydrodynamic variables in each fluid cell, including the viscous contributions, and then samples the particle species and their corresponding momenta according to the Cooper–Frye formula \cite{Shen:2014vra,McNelis:2019auj}.

It is important to ensure that the list of hadronic species used in \textsc{iSS} is consistent with the one implemented in the hadronic transport code \textsc{SMASH}, which will be described in the following section. The sampling is performed over the entire particlization hypersurface and includes multiple oversamplings. The code provides two main options: either fixing the total number of produced particles or specifying the number of oversamplings. The choice depends on the statistical precision required for the analysis.

Once the sampling procedure is completed, the output consists of a list of particles with their kinematic information, which serves as the input for the subsequent transport evolution. Alternatively, iSS can be executed in a mode where particle decays are performed immediately after sampling. Similar to the corresponding option in \textsc{MUSIC}, this mode is useful for calculating final observables that include resonance decays but do not account for the full hadronic transport evolution.

\section{Boltzmann transport}
\label{sec:transport}

The hadronic stage constitutes the final phase of the hybrid description of relativistic heavy-ion collisions, as shown in \cref{fig:HIC_numerical}. After particlization, the system is no longer well described by relativistic hydrodynamics, since the decreasing particle density leads to a growing mean free path, and eventually microscopic interactions become too infrequent to maintain local thermal equilibrium. In this dilute regime, a microscopic description based on kinetic theory becomes necessary.

The space-time evolution of the hadronic system is therefore modeled using a relativistic Boltzmann equation \cite{pottier2009nonequilibrium} for each hadronic species $i$,
\begin{equation}
    p^\mu \partial_\mu f_i(x,p) + m_i F^\alpha \partial^p_\alpha f_i(x,p) = C^i_{\mathrm{coll}},
\end{equation}
where $f_i(x,p)$ denotes the phase-space distribution function of particles of species $i$ with mass $m_i$. In the context of ultra-relativistic heavy-ion collisions, long-range mean-field forces are neglected, and thus the force term $F^\alpha$ vanishes. The collision term $C^i_{\mathrm{coll}}$ encodes all microscopic interactions among hadrons and represents the most nontrivial \cite{pottier2009nonequilibrium} component of the transport equation.

The collision kernel includes elastic and inelastic two-body scatterings, as well as resonance formation and decay processes governed by the strong interaction. Weak decays are typically neglected because their lifetimes are much longer than the duration of the hadronic phase, while electromagnetic processes are either ignored or treated perturbatively. Under these assumptions, the Boltzmann equation describes the non-equilibrium evolution of a dilute hadron gas, based on binary collisions and the assumption of molecular chaos.

Two widely used transport codes for this approach are \textsc{UrQMD} \cite{Bleicher:1999xi} and \textsc{SMASH} \cite{SMASH:2016zqf}, the latter of which is described in the next subsection. Both codes implement similar underlying physics. The choice of \textsc{SMASH} in this work is motivated by its more recent implementation in C++ and its focus on binary collisions and resonance dynamics, rather than on hadronic mean-field potentials.

\subsection{\textsc{SMASH}}
\label{subsec:smash}

The Simulating Many Accelerated Strongly-interacting Hadrons (\textsc{SMASH}) code \cite{SMASH:2016zqf} is a relativistic hadronic transport model designed to solve the Boltzmann equation through a microscopic cascade approach. In \textsc{SMASH},\footnote{The code used is available at \url{https://github.com/smash-transport/smash.git}.} hadrons propagate along classical trajectories between interactions, and collisions are implemented using experimentally constrained cross sections.

Binary interactions between particles are determined using a geometric collision criterion. Two particles are assumed to interact if their minimum transverse distance $d$ in the center-of-mass frame satisfies
\begin{equation}
    d < \sqrt{\frac{\sigma_{\mathrm{tot}}}{\pi}},
\end{equation}
where $\sigma_{\mathrm{tot}}$ is the total cross section for the considered reaction channel. The squared distance of closest approach is computed as 
\begin{equation}
    d^2 = (\vec{r}_1 - \vec{r}_2)^2 - \frac{\left[(\vec{r}_1 - \vec{r}_2) \cdot (\vec{p}_1 - \vec{p}_2)\right]^2} {(\vec{p}_1 - \vec{p}_2)^2}.
\end{equation}

Once a collision occurs, \textsc{SMASH} stochastically selects the reaction channel according to the relevant partial cross sections. Inelastic interactions are predominantly mediated by resonance excitation, followed by resonance decays. This procedure ensures detailed balance and exact conservation of energy, momentum, and quantum numbers on an event-by-event basis. Through repeated interactions, the system naturally evolves from chemical freeze-out, when inelastic collisions cease, to kinetic freeze-out, when elastic scatterings also become negligible and the free-streaming regime is reached as the hadron gas becomes dilute and particles stream toward the detectors.

\textsc{SMASH} includes a comprehensive hadronic particle list and interaction database based on the Review of Particle Physics \cite{ParticleDataGroup:2024cfk}, covering most well-established mesons and baryons. The particle content must be consistent with that used at the particlization stage, which in this dissertation is performed with \textsc{iSS}. 

Although \textsc{SMASH} is also capable of generating initial conditions and simulating the full space-time evolution of heavy-ion collisions at low energies (up to the GeV scale), in this work it is used exclusively as a hadronic afterburner for high-energy collisions at LHC energies (TeV scale).

Due to the oversampling procedure applied during particlization, each hydrodynamic event produces multiple hadronic realizations, all of which must be evolved independently through the transport stage. This significantly increases the computational cost of the hadronic evolution. The final output of \textsc{SMASH} consists of a list of stable hadrons, primarily protons, pions, and kaons, which serve as input for subsequent analysis. This list contains information about the particle species as well as their kinematic properties, such as energy, momentum, and emission angle. These quantities can then be used to compute final observables, including the charged particle multiplicity per unit pseudorapidity $\mathrm{d}N_\mathrm{ch}/\mathrm{d}\eta$, the flow coefficients $v_n$, and the average transverse momentum $\langle p_T \rangle$.

In this work, the final determination of observables from \textsc{SMASH} are performed using \textsc{SPARKX} \cite{Sass:2025opk}, a Python-based framework that calculates observables using different methods.\footnote{The code was developed during this PhD and is available at \url{https://github.com/smash-transport/sparkx}.} This tool facilitates the selection of particle species and kinematic ranges in order to perform direct comparisons with experimental data.

After completing this chapter, we now have a solid foundation for understanding a hybrid simulation of heavy-ion collisions, as summarized in \cref{fig:HIC_numerical}, as well as the main codes employed throughout the analysis. This chapter will serve as a reference for the subsequent discussions presented in the following chapters.\footnote{The correct version of each code used in this dissertation, including the hybrid evolution implementation, is available at \url{https://github.com/Hendrik1704/EBE-PREEQ-HYDRO-TRANSPORT} and \url{https://github.com/Hendrik1704/CRONOS}, depending on the chapter under consideration. These branches were developed during this PhD in collaboration with co-authors.}

\chapter{Matching prescriptions between pre-equilibrium and hydrodynamic stages}
\label{chap:entropy_matching}
\DoPToC
\noindent

This chapter is primarily based on the published article \cite{Borghini:2024kll}, ``Comparing matching prescriptions between pre-equilibrium and hydrodynamic models in high-energy nuclear collisions'' European Physical Journal C (2024), and includes additional adaptations and extensions.

As discussed in the previous chapter, the simulation of heavy-ion collisions relies on a hybrid framework composed of different physical descriptions. Although each stage of the hybrid framework is based on a well-established theory, it is crucial to understand its regime of validity. Hydrodynamics successfully describes the QGP when the system is close to local equilibrium, but it should not be applied immediately after the collision, nor once the system becomes too dilute. Determining when to switch between descriptions is therefore essential.

The question of how to consistently connect different stages of the evolution remains an active topic of research. In this context, the present chapter aims to clarify the matching procedure between the pre-equilibrium and hydrodynamic phases. A new method, referred to as entropy matching, is proposed and compared with the conventional energy matching procedure.

The chapter is organized as follows. A brief introduction to the topic is provided in \cref{sec:introduction_s}. The conventional energy matching procedure and the proposed entropy matching method are discussed in \cref{sec:matching}. The numerical model description is presented in \cref{sec:Model_s}, followed by the procedure for calibrating the $K$-factor in \cref{subsec:calibration_kfactor}, which is also relevant for the next chapters of this dissertation. The comparison between the two approaches is presented in \cref{sec:results_s}, and a concise conclusion is given in \cref{sec:summary_matching}.

\section{From pre-equilibrium to hydrodynamics}
\label{sec:introduction_s}

As discussed in the previous chapter, initial-state models typically rely on the saturation regime of QCD to describe the production of the initial energy density profile after the collision. In other approaches, parametric models are employed to achieve the same objective. Since the system is far from equilibrium immediately after the collision, a pre-equilibrium phase is introduced. This stage evolves the system using kinetic theory in order to reduce large gradients and drive the matter closer to local equilibrium, after which a hydrodynamic description becomes applicable. Beyond facilitating the applicability of hydrodynamics, the pre-equilibrium phase also plays an important role in improving the description of final observables \cite{NunesdaSilva:2020bfs,Liyanage:2022nua}.

The models commonly used to describe the pre-equilibrium phase, such as free-streaming and \Kompost~(see \cref{sec:pre_eq} for details), assume a conformal EOS $P = \epsilon/3$. This assumption reflects the treatment of the system as composed of effectively massless particles, implying the absence of an intrinsic scale. In contrast, the hydrodynamic equation of state employed in the subsequent evolution is based on lattice QCD calculations and explicitly incorporates the breaking of conformal symmetry. As shown in \cref{fig:alpha_s}, the QCD coupling constant runs with the energy scale, introducing a characteristic scale $\Lambda_\mathrm{QCD}$. Consequently, the lattice EOS is intrinsically nonconformal \cite{HotQCD:2014kol} and the trace of the energy–momentum tensor is no longer zero, $T^\mu_{\ \mu} \neq 0$, and therefore $P - 3\epsilon \neq 0$ \cite{daSilva:2022xwu}.

\begin{figure}[h]
    \centering
    \includegraphics[width=0.65\linewidth]{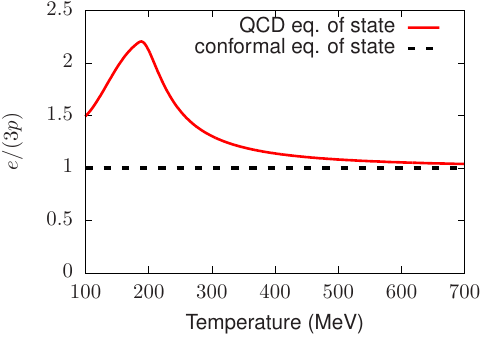}
    \caption{Comparison between the conformal EOS used in pre-equilibrium models and the lattice QCD EOS employed in hydrodynamics simulations. The largest discrepancy in the pressure occurs in the temperature range $\sim 100-300$ MeV, which is precisely the region most relevant for heavy-ion collisions. Figure from Ref. \cite{Kurkela:2018vqr}.}
\label{fig:EOS_differences}
\end{figure}

This mismatch in conformal symmetry leads to a nontrivial issue when matching the pre-equilibrium stage to hydrodynamics, since the pressures in the two descriptions generally differ $P_\mathrm{conformal} \neq P_\mathrm{LQCD}$. To address this, it is customary to impose continuity of the total pressure, defined as the sum of the thermodynamic pressure $P$ and the bulk pressure $\Pi$. Because the pre-equilibrium stage is conformal, bulk effects are absent in that phase. Therefore, the matching condition between pre-equilibrium and hydrodynamics is written as 
\begin{equation}
    P_\mathrm{conformal} \equiv P_\mathrm{QCD} + \Pi.
\label{eq:total_pressure}
\end{equation}
In this prescription, the bulk pressure is initialized as the difference between the conformal and lattice QCD EOS. However, this procedure can lead to sizable bulk contributions, particularly in the temperature range relevant for heavy-ion collisions, where the difference between the two equations of state is significant, as illustrated in \cref{fig:EOS_differences}.

This sizable difference introduces ``artificial'' contributions to the bulk pressure \cite{daSilva:2022xwu}. According to hydrodynamic theory, viscous corrections are expected to remain small \cite{denicol2022microscopic}, since many theoretical expansions rely on the assumption $\Pi / P \ll 1$. However, this condition is not necessarily satisfied when the matching prescription in \cref{eq:total_pressure} is applied.

A possible way to address this issue would be to formulate a pre-equilibrium framework based on a nonconformal equation of state that can be consistently matched to the lattice QCD EOS. For example, introducing massive particles in the pre-equilibrium stage would explicitly break conformal symmetry. Such a model, however, is not yet available and remains an open topic of research.

In the absence of a definitive solution, one may instead attempt to reduce the impact of the artificially large bulk pressure. A commonly discussed approach is to implement free-streaming with a velocity smaller than the speed of light. The optimal value of $v_\mathrm{fs}$ can be determined through Bayesian analysis, as performed for Pb–Pb and p–Pb collisions in Ref. \cite{Nijs:2020roc} by the Trajectum collaboration. Alternatively, one may tune $v_\mathrm{fs}$ to minimize the deviation from the lattice equation of state, as done by the ExTrEMe collaboration in Ref. \cite{daSilva:2022xwu}.

It is important to emphasize that modifying $v_\mathrm{fs}$ also requires a recalibration of the initial energy normalization, referred to in this dissertation as the $K$-factor, in order to reproduce the correct final observables. The approaches in Ref. \cite{Nijs:2020roc} and Ref. \cite{daSilva:2022xwu} indicate that a reduced free-streaming velocity, $v_\mathrm{fs} \simeq 0.85c$, improves the agreement within a conformal framework by decreasing the pressure mismatch and consequently reducing the artificial bulk contribution. In all these studies, $v_\mathrm{fs}$ is treated as independent of the local energy density.

However, this procedure is applicable primarily to the free-streaming scenario, which neglects interactions. In contrast, the \Kompost~model is based on effective kinetic theory and incorporates shear effects during the pre-equilibrium evolution. In this framework, there is limited flexibility in the choice of parameters. Therefore, instead of modifying model parameters, the strategy consists of changing the matching prescription between the pre-equilibrium and hydrodynamic stages, without altering the numerical implementation of the codes.

The main objective of this alternative matching procedure, described in the following sections, is to reduce the artificial bulk pressure generated at the transition between the two stages.

\section{Matching prescriptions}
\label{sec:matching}

In the present section, we present the usual energy matching in \cref{subsec:energy_matching} between pre-equilibrium and hydrodynamic. In \cref{subsec:entropy_matching}, we develop the generic approach for entropy matching.

\subsection{Energy matching}
\label{subsec:energy_matching}

The conventional procedure for matching the pre-equilibrium stage to the hydrodynamic phase is referred to here as energy matching \cite{Borghini:2024kll}. After running a pre-equilibrium evolution model, the output consists of the local energy density $\epsilon_\mathrm{pre}(\vec{x})$, fluid velocity $u^\mu(\vec{x})$, and shear stress tensor $\pi^{\mu\nu}_\mathrm{pre}(\vec{x})$. In energy matching, the energy density is passed directly to the hydrodynamic code, enforcing
\begin{equation}
    \epsilon_\mathrm{pre}(\vec{x}) = \epsilon_\mathrm{hydro}(\vec{x}),
\end{equation}
where the subscript ``$\mathrm{pre}$'' makes reference to a conformal pre-equilibrium model and ``$\mathrm{hydro}$'' makes reference to a hydrodynamic model.

Since the pre-equilibrium stage is conformal, the bulk pressure vanishes $\Pi_\mathrm{pre} \equiv 0$. Therefore, the bulk pressure in hydrodynamics must be initialized explicitly. To preserve the continuity of the total pressure $P + \Pi$, one imposes 
\begin{equation}
    \Pi_\mathrm{hydro}(\vec{x}) = P_\mathrm{pre}(\vec{x}) - P_\mathrm{LQCD}(\vec{x}) = \frac{\epsilon_\mathrm{pre}(\vec{x})}{3} - P_\mathrm{LQCD}(\epsilon_\mathrm{pre}(\vec{x})),
\label{eq:bulkM}
\end{equation}
which corresponds to the difference between the conformal and lattice QCD pressures shown in \cref{fig:EOS_differences}. If the pre-equilibrium model includes shear effects (as \Kompost~described in the previous chapter), the shear stress tensor is transferred to hydrodynamic model without modification $\pi^{\mu\nu}_\mathrm{pre}(\vec{x}) = \pi^{\mu\nu}_\mathrm{hydro}(\vec{x})$.

\subsection{Entropy matching}
\label{subsec:entropy_matching}

The alternative prescription proposed in this work is entropy matching \cite{Borghini:2024kll}. In this case, the matching condition is imposed on the equilibrium entropy density 
\begin{equation}
    s_\mathrm{pre}^\mathrm{eq}(\vec{x}) = s_\mathrm{hydro}^\mathrm{eq}(\vec{x}).
\end{equation}

Since both pre-equilibrium and hydrodynamical codes evolve the energy density, the output of the pre-equilibrium at the switching time is still the energy density $\epsilon_\mathrm{pre}(\vec{x})$, that is translated in the effective equilibrium entropy density via \cite{Giacalone:2019ldn}
\begin{equation}
    s_\mathrm{pre}^\mathrm{eq}(\vec{x}) \equiv \frac{4}{3} \left(\frac{\pi^2}{30} \nu_\mathrm{eff}\right)^{1/4} \epsilon_\mathrm{pre}(\vec{x})^{3/4}.
\label{eq:entropy_kompost}
\end{equation}
This is referred to as an effective equilibrium entropy because it does not include viscous corrections present in the full entropy current. The matched entropy is then converted into the corresponding energy density using the lattice QCD EOS
\begin{equation}
    \epsilon_\mathrm{hydro}(\vec{x}) = \epsilon_\mathrm{LQCD}(s_\mathrm{hydro}^\mathrm{eq}(\vec{x})).
\label{eq:energy_music}
\end{equation}
In this prescription, the energies before and after matching are not equal $\epsilon_\mathrm{pre}(\vec{x}) \neq \epsilon_\mathrm{hydro}(\vec{x})$.

The bulk pressure is initialized in the same manner as in energy matching, but using the appropriate variables
\begin{equation}
    \Pi_\mathrm{hydro}(\vec{x}) = P_\mathrm{pre}(\vec{x}) - P_\mathrm{LQCD}(\vec{x}) = \frac{\epsilon_\mathrm{pre}(\vec{x})}{3} - P_\mathrm{LQCD}(\epsilon_\mathrm{hydro}(\vec{x})).
\label{eq:bulk_s_matching}
\end{equation}
In this way, the total pressure remains continuous across the transition. Note that \cref{eq:bulkM} gives a different bulk pressure than \cref{eq:bulk_s_matching}. As in the energy matching case, the shear stress tensor is passed unchanged between the codes.

In our numerical implementation, the matching prescription can be chosen. For energy matching, the hydrodynamic stage receives the energy density directly from the pre-equilibrium code. For entropy matching, the energy density is first converted into entropy via \cref{eq:entropy_kompost}, and then mapped to the corresponding lattice-QCD energy density via \cref{eq:energy_music}. The bulk pressure must be computed externally, since the hydrodynamic code does not have access to the original energy $\epsilon_\mathrm{pre}(\vec{x})$ before the entropy matching procedure.

The motivation for entropy matching is that final particle multiplicities are more directly related to entropy production than to energy density. Therefore, entropy may be the more relevant quantity for describing observables sensitive to particle yields. Nevertheless, this prescription represents only an alternative possibility. Within a conformal pre-equilibrium framework, it is impossible to match both energy and entropy simultaneously; a discontinuity in thermodynamic variables is therefore unavoidable.

In energy matching, the local energy density is conserved across the transition. In entropy matching, the equilibrium entropy density is conserved.\footnote{A full entropy current $S^\mu$ would include viscous corrections and cannot be consistently matched due to the different equations of state.} In both prescriptions, momentum conservation is preserved, since only the scalar component $T^{00}$ of the energy–momentum tensor is modified, while the fluid velocity $u^\mu$ remains unchanged. The quantitative differences between these two approaches are analyzed in \cref{sec:results_s} for the specific numerical chain described in the next section.

\section{Numerical model}
\label{sec:Model_s}

Based on the framework introduced in \cref{chap:HIC}, we now describe the specific setup used for the present analysis. The goal is to simulate Pb–Pb collisions at $\sqrt{s_\mathrm{NN}} = 5.02$ TeV in two centrality classes, $0-10\%$ and $40-50\%$, generating $5000$ events per centrality.

For the initial condition, a Monte Carlo Glauber model is employed with a Gaussian smearing radius of $0.4$ fm, corresponding to a typical hot-spot size (see \cref{subsec:glauber}). The geometric collision criterion uses an inelastic nucleon–nucleon cross section of $\sigma_\mathrm{inel}^\mathrm{NN} = 67.6$ mb. The initial energy density profile is constructed according to \cref{eq:MCGlauber_energy} and subsequently normalized by a $K$-factor. The determination of this normalization is discussed in more detail in the next subsection, where the centrality definition is also clarified.

After the energy density is produced, normalized, and recentered on a transverse grid of total size $30$ fm with a spatial resolution of $0.1$ fm, the local energy–momentum tensor is initialized as \cite{Kurkela:2018vqr} 
\begin{equation}
    T^{\mu\nu}(\vec{x}) = \operatorname{diag}\left(\epsilon(\vec{x}), P_T(\vec{x}), P_T(\vec{x}), \frac{P_L(\vec{x})}{\tau^2}\right) = \operatorname{diag}\left(\epsilon(\vec{x}), \frac{\epsilon(\vec{x})}{2}, \frac{\epsilon(\vec{x})}{2}, 0\right),
\label{eq:Tmunu_input_kompost}
\end{equation}
This choice corresponds to a boost-invariant energy–momentum tensor with vanishing longitudinal pressure $P_L = 0$ and transverse pressure $P_T = \epsilon/2$, ensuring that the tensor remains traceless. It is important to note that this configuration is anisotropic and therefore differs from the ideal isotropic tensor defined in \cref{eq:Tmunu_ideal}. The deviation from isotropy is encoded in the shear stress tensor $\pi^{\mu\nu}$.

This energy–momentum tensor is provided as input to \Kompost~(see \cref{subsec:kompost}), which is run with a constant shear viscosity $\eta/s = 0.16$ from $\tau_0 = 0.2$ fm to $\tau_\mathrm{hydro} = 1.0$ fm, where the transition to \textsc{MUSIC} occurs. At the end of the \Kompost~evolution, the energy–momentum tensor is decomposed into hydrodynamic variables: the energy density $\epsilon$, the four-velocity $u^\mu$, and the ten independent components of the shear stress tensor $\pi^{\mu\nu}$. At this stage, the Landau matching condition $T^\mu_\nu u_\mu = \epsilon u_\nu$ must be satisfied, providing the necessary input for the hydrodynamic evolution in \textsc{MUSIC} (see \cref{subsec:music}).

The hydrodynamic evolution is performed under the assumption of boost invariance and uses the lattice QCD EOS. The shear viscosity remains constant at $\eta/s = 0.16$, consistent with the pre-equilibrium stage. The bulk viscosity is implemented using the parametrization from \cite{Denicol:2009am}: 
\begin{equation}
    \frac{\zeta}{s}= \begin{cases}
    - 13.77 x^2 +27.55 x - 13.45 & \left(0.995 T_c \geqslant T \geqslant 1.05 T_c\right)\\ 
    0.9 \exp \left[-(x-1)/0.025\right] + 0.25 \exp\left[-(x-1)/0.13\right]+0.001 & \left(T>1.05 T_c\right)\\ 
    0.9 \exp\left[(x-1)/0.0025\right] + 0.22 \exp\left[(x-1)/0.022\right]+0.03 & \left(T<0.995 T_c\right)
\end{cases},
\label{eq:bulk_viscosity_parametrization}
\end{equation}
where $x = T/T_c$ and $T_c = 200$ MeV. Hydrodynamics proceeds until the switching temperature $T_\mathrm{switch} = 155$ MeV is reached. At this point, \textsc{iSS} is used to perform particlization (see \cref{subsec:iss}), including shear and bulk corrections in $\delta f$, with $20$ oversamplings per event. The subsequent hadronic evolution is carried out with \textsc{SMASH} (see \cref{subsec:smash}).

\subsection{Calibration of the $K$-factor based on the estimator equation}
\label{subsec:calibration_kfactor}

Before explaining how the $K$-factor of an initial model is calibrated, it is necessary to introduce the theoretical relation that encodes this calibration, referred to in this dissertation as the estimator equation.

Experimental data are reported in terms of centrality, a quantity that for large systems is related to the collision geometry, as illustrated in \cref{fig:impact_b}. Since experiments cannot directly measure the impact parameter, they instead measure the number of charged particles produced in each event. From millions of recorded events, a multiplicity distribution is constructed and ordered, allowing the definition of centrality classes according to percentile ranges \cite{Miller:2007ri,ALICE:2015juo}.

In simulations, minimum-bias events are generated, where the impact parameter varies from $0$ to approximately $2.5R$. After producing and evolving these events, one can classify them according to their multiplicity using the experimental centrality bins. This approach is useful when the goal is to compute observables across all centrality classes. Nevertheless, if the objective is to study a specific centrality range, this method becomes expensive, since many events outside the desired class must still be evolved.

In this dissertation, where the focus is on selected centrality ranges, it is advantageous to establish a direct relation between the initial energy profile and the final multiplicity. Such a relation allows one to determine the centrality class of an event without performing the full hydrodynamic and hadronic evolution. This is particularly useful because the initial-state models constitute the least computationally demanding part of the hybrid simulation. The relation between the final charged particle multiplicity $\mathrm{d}N_\mathrm{ch}/\mathrm{d}\eta$ and the initial energy density $\epsilon$ is provided by the estimator equation.

Immediately after the collision, the system is out of equilibrium, which prevents the definition of genuine thermodynamic quantities such as temperature or entropy \cite{pottier2009nonequilibrium}. However, one can define effective quantities based on the local energy density, which is well defined at this stage. A macroscopic description can then be constructed, notably through the use of attractor solutions.

In the original \Kompost~work\cite{Kurkela:2018vqr}, a universal scaling function was identified that interpolates between the kinetic theory regime and hydrodynamic behavior. In this framework, the effective temperature $T_\mathrm{eff}$ is defined in terms of the energy density $\epsilon$ as 
\begin{equation}
    T^4_\mathrm{eff}(\tau) = \frac{30}{\pi^2 \nu_\mathrm{eff}} \epsilon(\tau),
\label{eq:effetive_temperature}
\end{equation}
where the effective number of degrees of freedom is taken as $\nu_\mathrm{eff} = 40$, consistent with lattice QCD results \cite{HotQCD:2014kol}.

Assuming that the final multiplicity is proportional to the total entropy and using the thermodynamic relation $sT = \epsilon + P$, together with the conformal EOS $P = \epsilon/3$, the equilibrium entropy density can be expressed in terms of the effective temperature. Using \cref{eq:effetive_temperature}, the entropy per unit rapidity at the hydrodynamic stage can be written as 
\begin{equation}
    (s\tau)_\mathrm{hydro} = \frac{4}{3} \left(\frac{\pi^2}{30} \nu_\mathrm{eff} \right)^{1/4} (\epsilon \tau)_\mathrm{hydro}^{3/4}.
\label{eq:entropy_out_eq}
\end{equation}

The connection between the pre-equilibrium energy density and its hydrodynamic counterpart is provided by the attractor solution \cite{Giacalone:2019ldn}: 
\begin{equation}
    \frac{\epsilon(\tau) \tau^{4/3}}{\epsilon_\mathrm{hydro} \tau_\mathrm{hydro}^{4/3}} = \mathcal{E}\left(\tilde{\omega} = \frac{T_\mathrm{eff}(\tau) \tau}{4\pi \eta/s} \right),
\end{equation}
where, in the free-streaming or early-time limit \cite{Giacalone:2019ldn}, the attractor behaves as $\mathcal{E}(\tilde{\omega}) = C_\infty^{-1}\tilde{\omega}^{4/9}$. This relation allows the equilibrium entropy $(s\tau)_\mathrm{hydro}$ to be expressed in terms of the out-of-equilibrium quantity $\epsilon(\tau)\tau$. The total entropy can then be related to the final charged particle multiplicity through the experimental ratio $S/N_\mathrm{ch}$ and the Jacobian connecting rapidity $\mathrm{y}$ and pseudorapidity $\eta$. The resulting expression, known as the estimator equation \cite{Giacalone:2019ldn}, is 
\begin{equation}
    \left\langle \frac{dN_\mathrm{ch}}{d\eta} \right\rangle = \frac{4}{3}\, \frac{N_\mathrm{ch}}{S}\, C_\infty^{3/4}\, \left(4\pi \frac{\eta}{s}\right)^{1/3}  \left( \frac{\pi^2}{30} \nu_\mathrm{eff} \right)^{1/3}  \int \mathrm{d}^2 \mathbf{x} \, [\tau \epsilon]_0^{2/3}.
\label{eq:estimator_equation}
\end{equation}

Care must be taken with the notation: on the left-hand side, $\eta$ denotes pseudorapidity, while on the right-hand side, $\eta/s$ represents the specific shear viscosity. In this dissertation, the parameters are fixed to \cite{Hanus:2019fnc} $N_\mathrm{ch}/S = 1/7.5$ and $C_\infty = 0.87$. With these values, the estimator equation can be evaluated directly from the initial energy density profile, allowing one to determine the expected final multiplicity, and thus the centrality class, without performing the full hydrodynamic evolution.

\begin{figure}[h]
    \centering
    \includegraphics[width=0.7\linewidth]{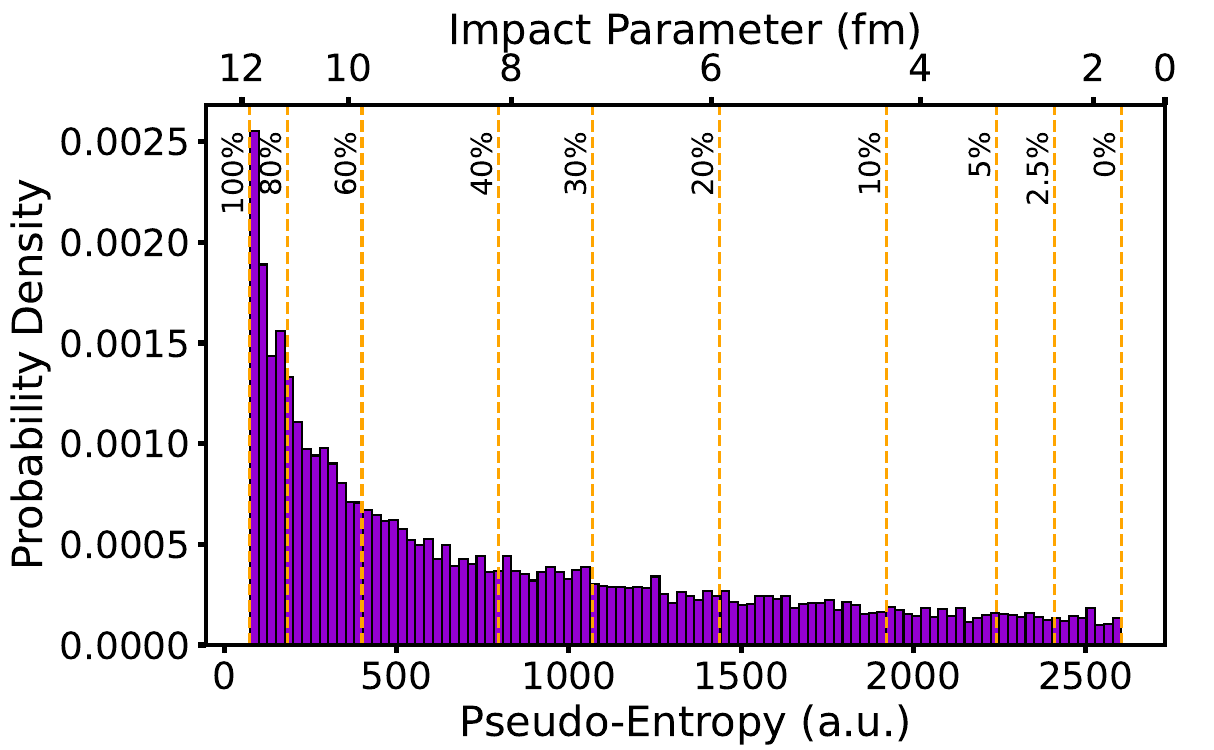}
    \caption{Histogram of the pseudo-entropy, shown in arbitrary units, derived from the energy density profiles produced by an optical Glauber model for $10^4$ Pb-Pb events. The displayed quantity corresponds to the energy density raised to the power $2/3$, integrated over the entire transverse grid without including the normalization $K$-factor. The vertical dotted lines indicate the centrality class boundaries. The upper horizontal axis provides the corresponding mapping to the impact parameter.}
\label{fig:pseudo_entropy}
\end{figure}

\Cref{fig:pseudo_entropy} illustrates a practical method to determine centrality using only the initial energy profile. In this approach, $10^4$ minimum-bias events are generated, in this case for Pb–Pb collisions at $5.02$ TeV. For each event, a quantity referred to as pseudo-entropy, proportional to $\tilde{\epsilon}^{2/3}$, is computed and used to build a distribution. The term pseudo-entropy is used because the actual entropy that gives the multiplicity should have the constants of \cref{eq:estimator_equation} and include the appropriate normalization through the $K$-factor applied to the initial energy density.

By ordering the events according to this pseudo-entropy distribution, centrality classes can be defined. For example, in \cref{fig:pseudo_entropy}, every $10^3$ events corresponds to a centrality interval of $10\%$. Once this classification is established, a newly generated event with the same setup can be assigned a centrality class based solely on its pseudo-entropy value, in a manner analogous to the experimental procedure illustrated in \cref{fig:impact_b}.

However, once \cref{eq:estimator_equation} is available, it becomes possible to determine the final charged particle multiplicity more precisely, provided that all constants and the correct $K$-factor are included. In this sense, the calibration procedure naturally follows from the estimator equation. The calibration involves both the explicit $K$-factor (included in the energy profile $\epsilon$) and an additional normalization factor $C(\eta/s)$, which accounts for small deviations associated with the specific hydrodynamic setup: 
\begin{equation}
    \left\langle \frac{dN_\mathrm{ch}}{d\eta} \right\rangle = C(\eta/s) \frac{4}{3} \frac{N_\mathrm{ch}}{S} C_\infty^{3/4} \left(4\pi \frac{\eta}{s}\right)^{1/3} \left( \frac{\pi^2}{30} \nu_\mathrm{eff} \right)^{1/3} \int d^2 x [\tau \epsilon]_0^{2/3}.
\label{eq:estimator_equation_withC}
\end{equation}
The factor $C(\eta/s)$ is expected to be close to unity, as it represents only a small correction adapting the estimator to the particular hydrodynamic configuration. In contrast, the $K$-factor included in the energy density may assume different values depending on the chosen initial condition model.

The calibration procedure begins by generating a limited sample of events. For the present matching analysis, we produced $10$ events with $b = 0$ fm, and $5$ additional events each with $b = 3$ fm and $b = 6$ fm, initially setting $K = 1$ and $C(\eta/s) = 1$. From the initial energy profiles generated by MC-Glauber, the multiplicity is first estimated using \cref{eq:estimator_equation_withC}. The same $20$ events are then evolved through the full setup described in \cref{sec:Model_s}, and the final multiplicity is extracted from the resulting particle lists.

\begin{figure}[h]
    \centering
    \includegraphics[width=0.49\linewidth]{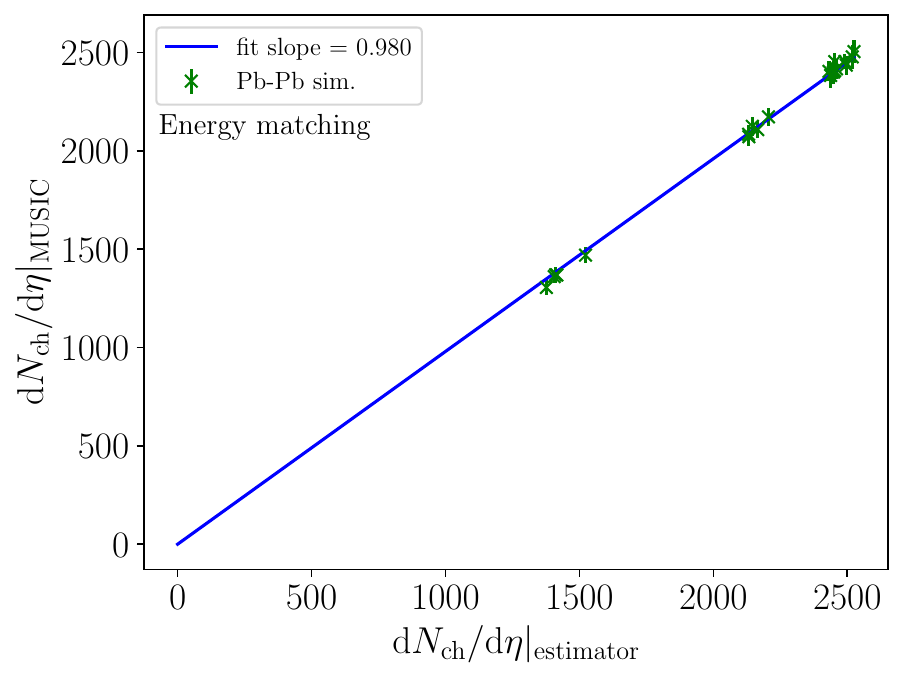}
    \includegraphics[width=0.49\linewidth]{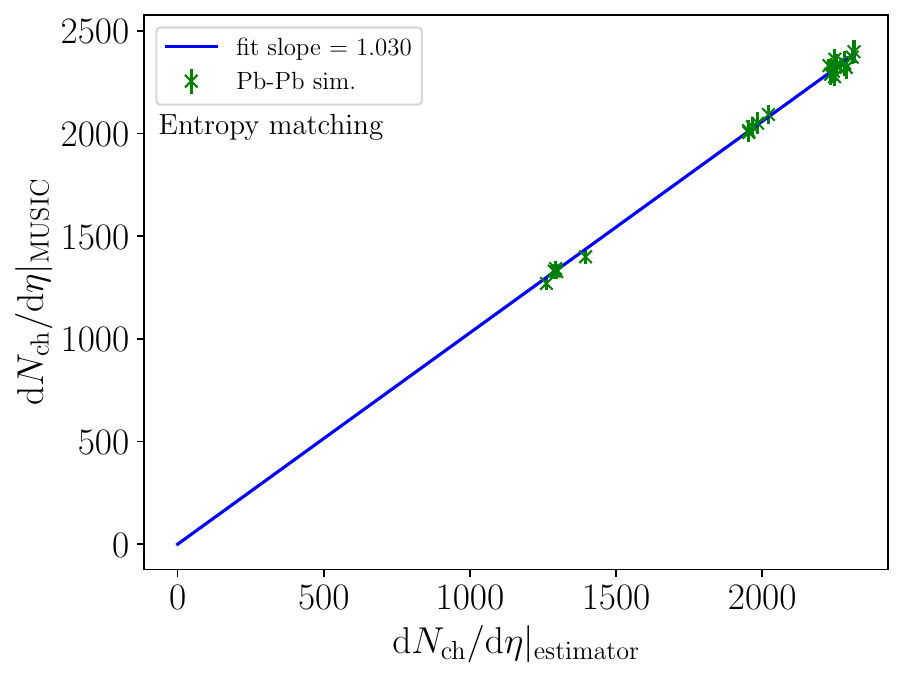}
    \caption{Multiplicities for $20$ simulated events with $b=0$, $3$ and $6$ fm. In the horizontal axis, the multiplicity is calculated by the estimator equation \cref{eq:estimator_equation_withC}, while on the vertical axis, it is obtained from the final particle list after the full evolution. The blue line represents the best linear fit to the green points; its slope defines $C(\eta/s)$. The left panel corresponds to the energy matching, yielding $C(\eta/s) = 0.98$, while the right panel corresponds to entropy matching $C(\eta/s) = 1.03$.}
\label{fig:pre_calibration}
\end{figure}

Both multiplicities are plotted and fitted with a linear function, as shown in \cref{fig:pre_calibration}. The slope of this fit defines the value of $C(\eta/s)$ for the corresponding model. It is important to note that energy matching and entropy matching involve slightly different evolutions and therefore lead to different values of $C(\eta/s)$.

The next step, after including the value of $C(\eta/s)$, is to generate events, vary the $K$-factor, and compare the multiplicity obtained from \cref{eq:estimator_equation_withC} with the corresponding experimental data. For the analysis presented in this chapter, we use Pb–Pb data at $5.02$ TeV from the ALICE collaboration \cite{ALICE:2015juo}. At each iteration of the comparison, the deviation from the experimental values is evaluated. Based on this deviation, the code determines whether the $K$-factor should be increased or decreased in the subsequent iteration. This procedure is repeated until the best agreement with the data is achieved.

\begin{figure}[h!]
    \centering
    \includegraphics[width=0.49\linewidth]{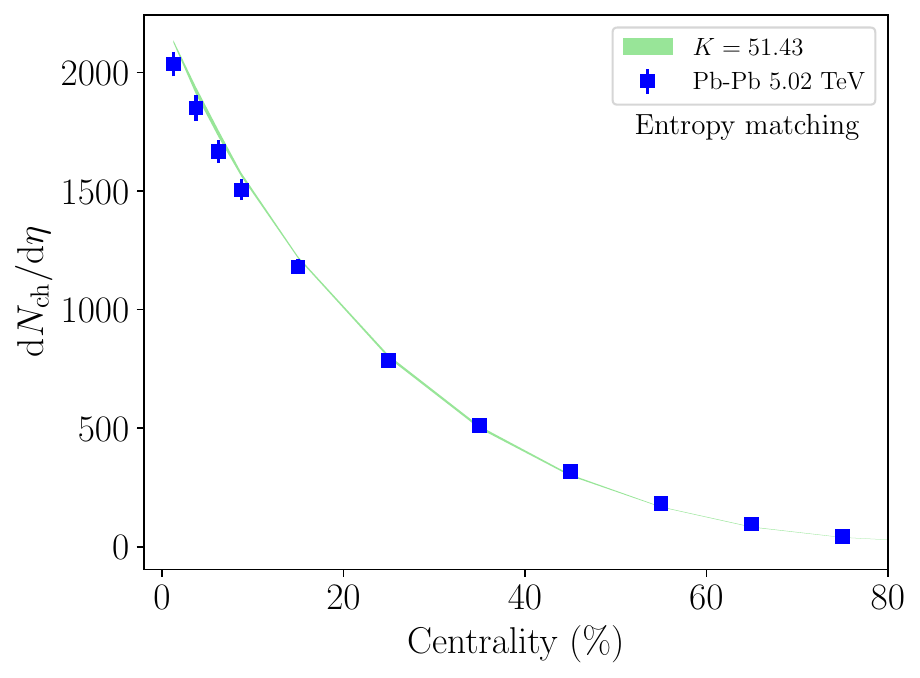}
    \includegraphics[width=0.49\linewidth]{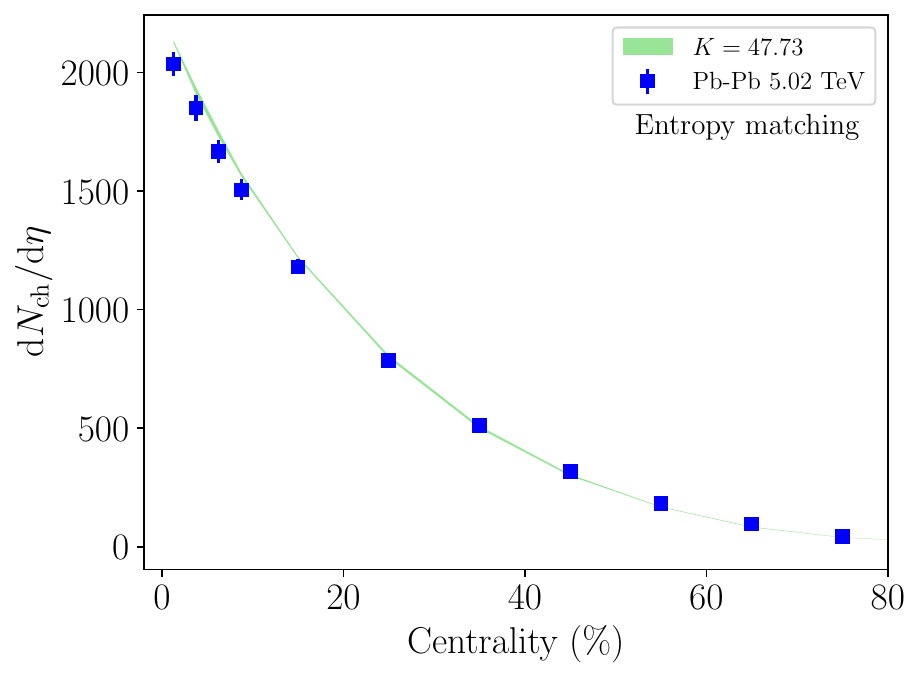}
    \caption{Multiplicity as a function of centrality. The blue points correspond to experimental data for Pb-Pb collisions at $\sqrt{s_\mathrm{NN}}=5.02$ TeV from Ref. \cite{ALICE:2015juo}. The green line represents the prediction from the estimator equation using the calibrated values of $C(\eta/s)$ and $K$. The left panel shows results for energy matching, while the right panel corresponds to entropy matching.}
\label{fig:calibrated_data}
\end{figure}

The resulting optimal normalization is $K \simeq 51.43$ for energy matching and $K \simeq 47.73$ for entropy matching. The reason why entropy matching requires a smaller normalization will be clearer later in this chapter. \Cref{fig:calibrated_data} demonstrates that the calibration procedure is successful, as the simulated multiplicities reproduce the experimental measurements within the considered centrality ranges.

Another validation step consists of generating a small sample of events using the calibrated values of $C(\eta/s)$ and the $K$-factor, computing the multiplicity via the estimator equation, and then evolving the same events through the full hydrodynamic setup to extract the final multiplicity. Repeating the procedure shown in \cref{fig:pre_calibration}, the slope of the resulting fit should be equal to $1$ if the estimator is fully consistent with the complete evolution. For energy matching, the extracted slope is $0.989$, corresponding to an error of approximately $1.1\%$. For entropy matching, the slope is $0.998$, with an error of about $0.2\%$. This confirms that the estimator equation provides an accurate prediction of the final multiplicity, particularly in the entropy matching case, without running the whole evolution.

\begin{figure}[h]
    \centering
    \includegraphics[width=0.85\linewidth]{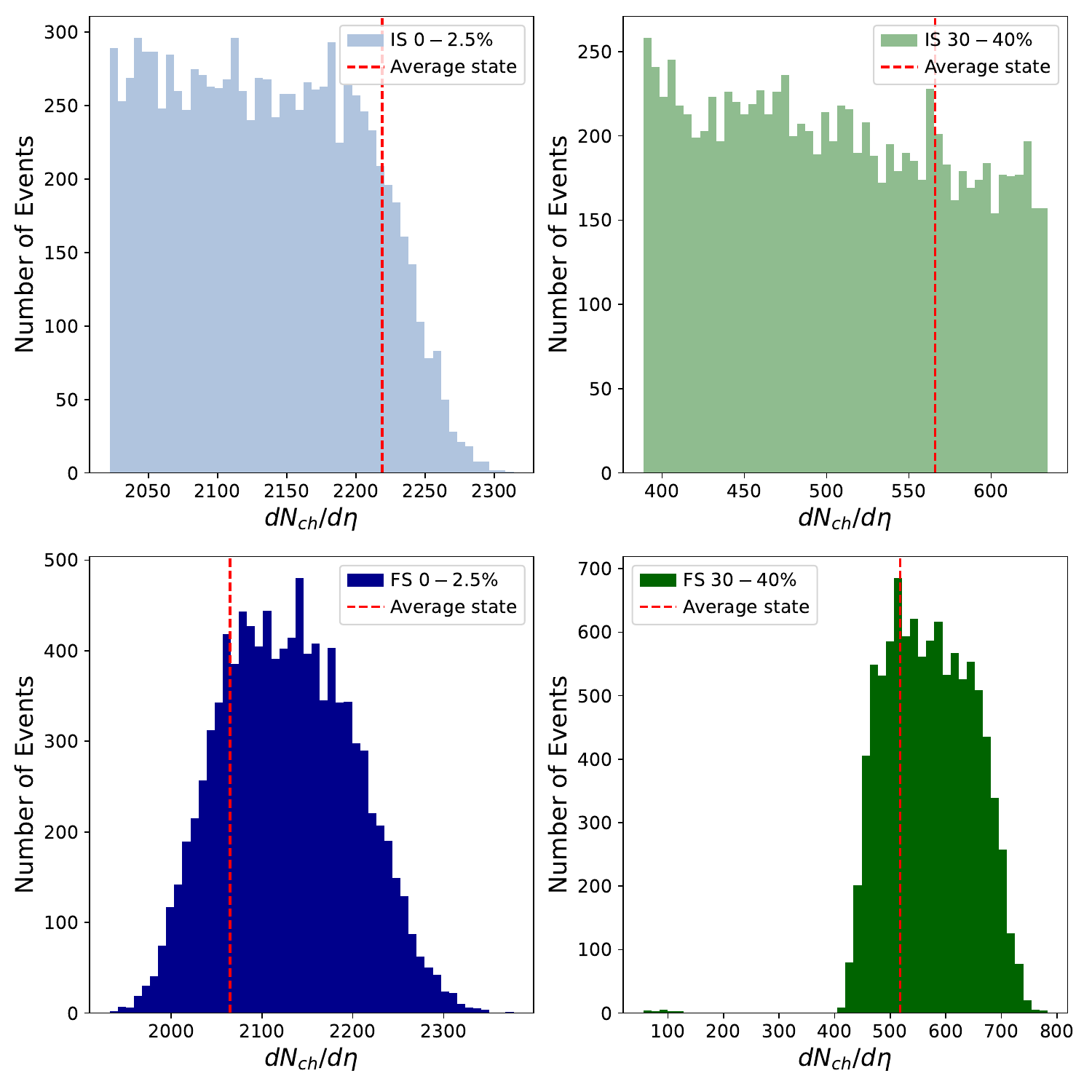}
    \caption{Histograms of the number of particles for $10^4$ events in two centrality classes ($0-2.5\%$ and $30-40\%$), shown at two stages of the evolution: the initial state (IS) and the final state (FS). The vertical red dashed line indicates a smooth average state, obtained by averaging over all initial-state events and subsequently evolving this averaged profile to the final state.}
\label{fig:estimator_cent}
\end{figure}

An additional validation of the estimator \cref{eq:estimator_equation_withC} can be performed after calibration. \Cref{fig:estimator_cent} shows multiplicity histograms for two centrality classes: central $0-2.5\%$ and non-central $30-40\%$ Pb–Pb collisions. The upper panels correspond to initial states (IS). As expected from experimental observations (see \cref{fig:impact_b}), central events exhibit a multiplicity tail, while non-central events show a flatter distribution. The sharp cutoffs at the boundaries of the initial-state histograms are imposed by construction through the estimator equation.

The lower panels display the final states (FS) after the full evolution. The smoother edges indicate a small difference between the estimated initial multiplicity and the final multiplicity. Overall, however, the agreement is very good. In the $30-40\%$ centrality class, there are $18$ events with relatively low final multiplicity ($\sim 100$). These events can be understood by examining the transverse momentum distribution: in such cases, a large amount of initial energy is converted into fewer particles with higher transverse momentum. Nevertheless, this situation is rare, and on average, the estimator formula accurately relates the initial energy density to the final particle multiplicity.

In addition, \cref{fig:estimator_cent} includes the multiplicity obtained from the average state in each centrality class. This profile is constructed by averaging $10^4$ initial-state events from the optical Glauber model \cite{Miller:2007ri}, resulting in a smooth energy density profile that is subsequently evolved. The absence of event-by-event fluctuations in this average configuration explains the smaller multiplicity of the final state compared to the initial state. The role of fluctuations in the dynamical evolution will be explored in more detail in \cref{chap:modesPb} and \cref{chap:3d}.

With this section, we complete an additional component of the framework, namely the normalization of the initial energy profile. After each event is generated, the estimator equation can be applied to determine its corresponding centrality class without requiring the full hydrodynamic evolution. In the following section, we discuss the results from the two different matching procedures. 

\section{Results}
\label{sec:results_s}

The goal of this section is to compare the two matching prescriptions by analyzing selected thermodynamic quantities. Although the matching is performed locally, the observables presented here are integrated over the transverse plane within one unit of rapidity. Only the QGP phase is considered, corresponding to regions with temperatures $T > 155$ MeV.

The results are shown for an average state, defined as a smooth energy profile representing a given centrality class. Since \Kompost~evolves the system from $\tau_0$ to $\tau_\mathrm{hydro}$, each point in the pre-equilibrium phase corresponds to a separate simulation at a specific time. In contrast, \textsc{MUSIC} performs a continuous-time evolution, making the temporal development more straightforward to simulate.

\begin{figure}[h]
    \centering
    \includegraphics[width=0.49\linewidth]{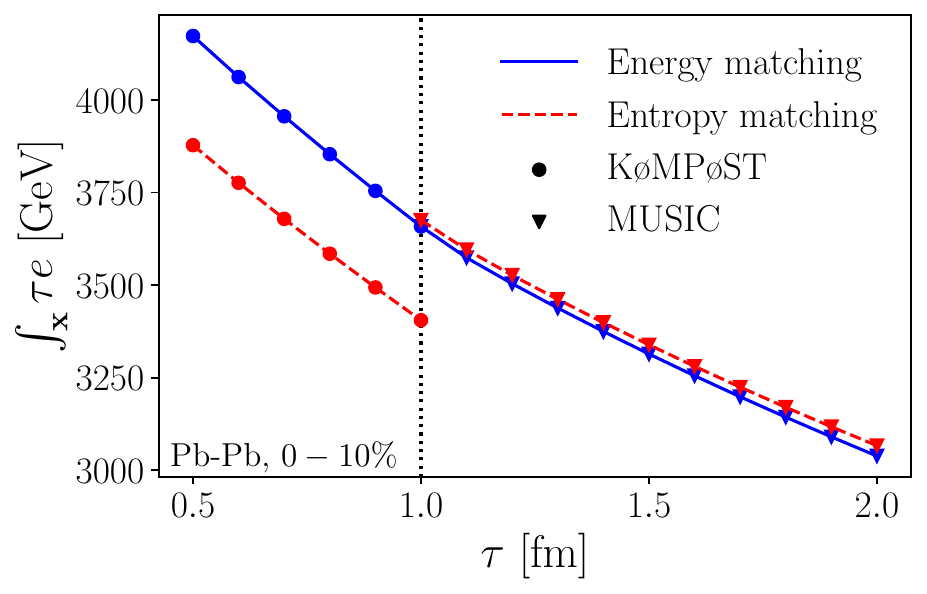}
    \includegraphics[width=0.49\linewidth]{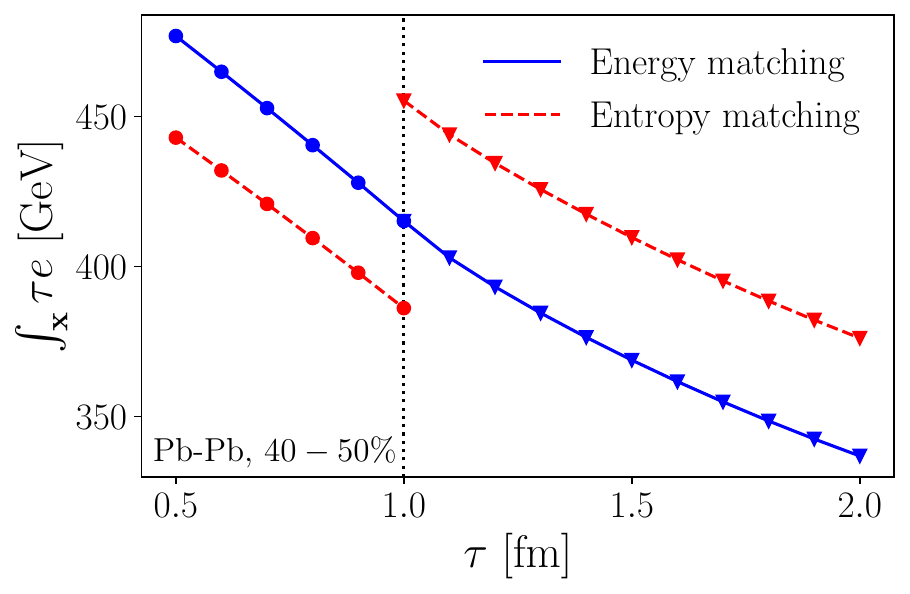}
    \caption{Time dependence of the transverse-plane-integrated energy density for a central average state, $0-10\%$ (left), and a non-central average state, $40-50\%$ (right), in Pb-Pb collisions at $5.02$ TeV. The comparison between energy matching (blue) and entropy matching (red) is shown. The vertical black dotted line indicates the switching time between \Kompost~and \textsc{MUSIC}.}
\label{fig:e_matching}
\end{figure}

\Cref{fig:e_matching} presents the transverse-plane-integrated energy density for two centrality classes ($0-10\%$ and $40-50\%$), in Pb–Pb collisions at $5.02$ TeV, comparing both matching prescriptions. The initial condition is generated at $\tau_0 = 0.2$ fm, and the transition to \textsc{MUSIC} occurs at $\tau_\mathrm{hydro} = 1$ fm, indicated in the plots by a vertical black dotted line. The evolution during the \Kompost~phase is represented by circular markers, while the \textsc{MUSIC} phase is shown using triangular markers. Blue curves correspond to energy matching, and red curves correspond to entropy matching.

As described in \cref{sec:matching}, the blue curve corresponding to energy matching is continuous across $\tau_\mathrm{hydro}$ and decreases monotonically with time, reflecting the expansion and cooling of the system. In contrast, for entropy matching, the energy density exhibits a discontinuity at $\tau_\mathrm{hydro}$, since the matching condition enforces continuity of entropy rather than energy. As a result, energy is not conserved across the transition in this prescription.

Note that, because the $K$-factor differs for each matching prescription, the initial transverse-plane-integrated energy density is also different in the two cases. However, after the matching procedure, both approaches converge to similar final curves, ultimately leading to the same final observables. In \cref{fig:e_matching}, it may appear that only the central events show agreement in the final stage; however, this impression is due to the different vertical scales used in the plots. For both central and non-central events, the difference between the red and blue curves in the late-time regime is approximately $30$ MeV. This small discrepancy indicates that the final observables remain essentially unchanged for the two matching prescriptions, as will be described later.

\begin{figure}[h]
    \centering
    \includegraphics[width=0.49\linewidth]{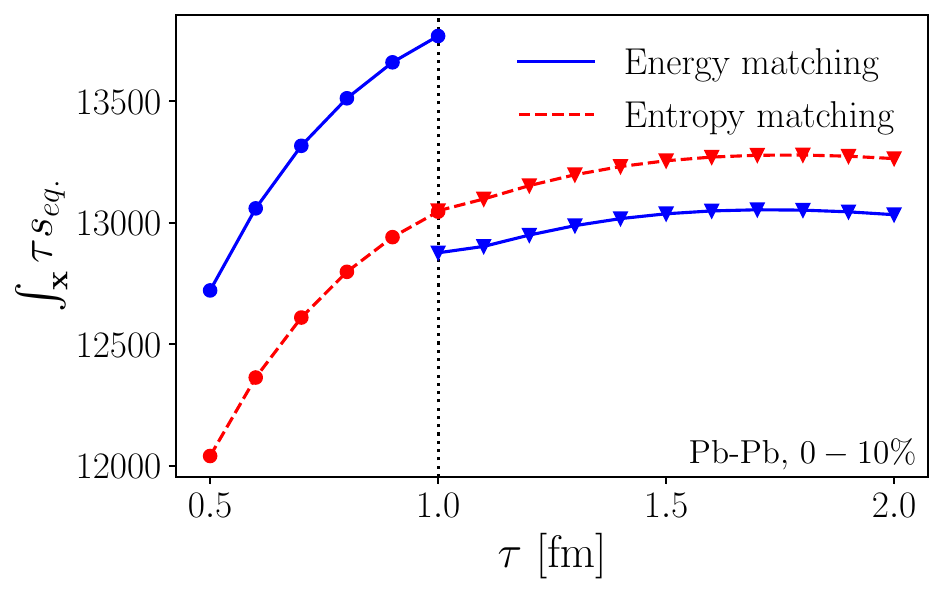}
    \includegraphics[width=0.49\linewidth]{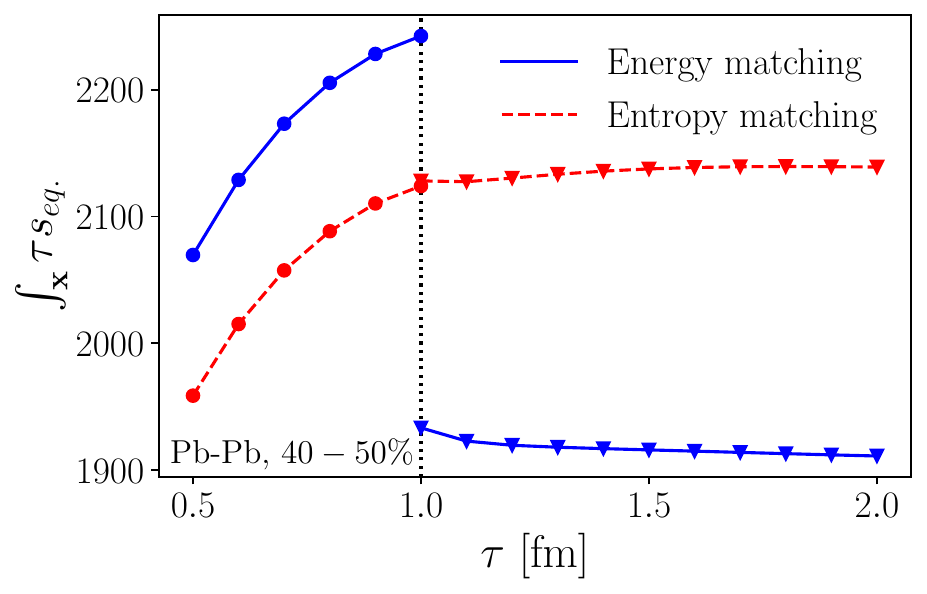}
    \caption{Time evolution of the transverse-plane-integrated equilibrium entropy for a central $0-10\%$ (left) and a non-central $30-40\%$ (right) average state in Pb-Pb collisions at $5.02$ TeV. The comparison between energy matching (blue) and entropy matching (red) is shown. The vertical black dotted line represents the switching time between \Kompost~and MUSIC.}
\label{fig:s_matching}
\end{figure}

\Cref{fig:s_matching} presents the transverse-plane-integrated equilibrium entropy density for central and non-central events. In this case, the equilibrium entropy density is continuous across the switching time for entropy matching, as expected from its definition. In contrast, energy matching introduces a discontinuity in the equilibrium entropy density.

It is important to emphasize that continuity of the equilibrium entropy density does not imply continuity of the total entropy. The reduction of equilibrium entropy observed in the blue curve does not necessarily violate the second law of thermodynamics. The total entropy current $S^\mu$ includes additional contributions from viscous corrections, namely bulk and shear terms, which must be taken into account when assessing entropy production.

\begin{figure}[h]
    \centering
    \includegraphics[width=0.49\linewidth]{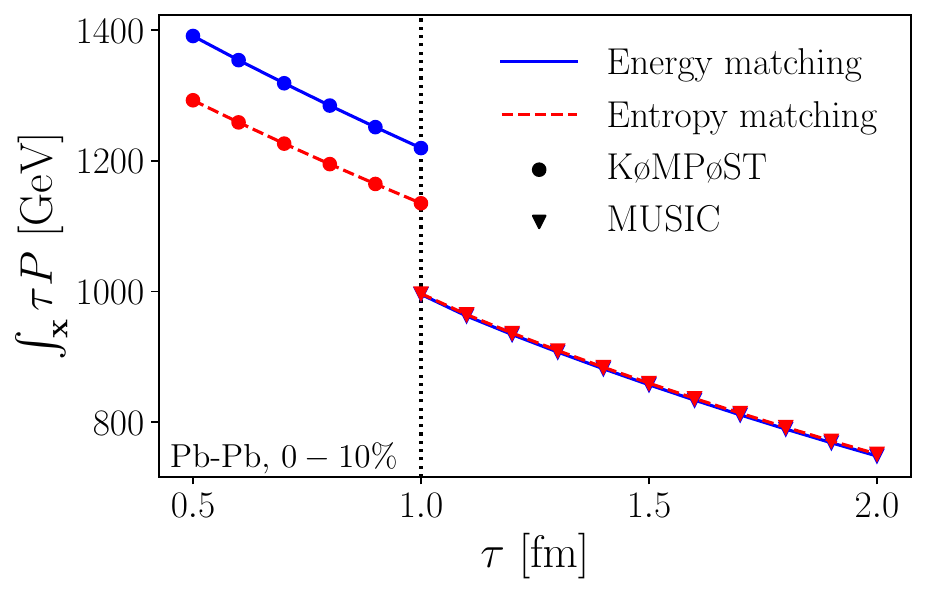}
    \includegraphics[width=0.49\linewidth]{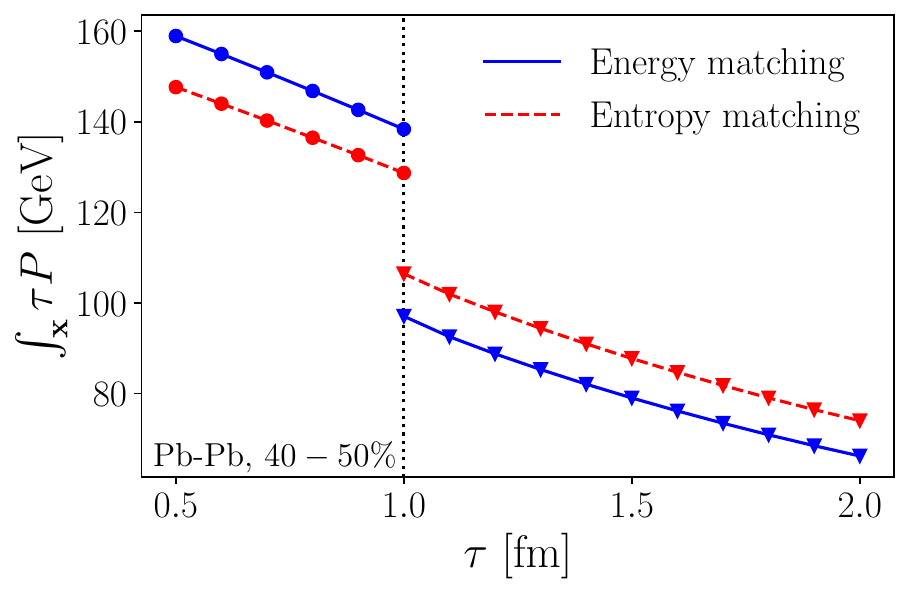}
    \caption{Time dependence of the transverse-plane-integrated thermodynamic pressure for a central $0-10\%$ (left) and a non-central $40-50\%$ (right) average state in Pb-Pb collisions at $5.02$ TeV. The gap between the conformal and nonconformal equations of state is responsible for the discontinuities observed in other thermodynamic variables at the switching time.}
\label{fig:P}
\end{figure}

The difference between the conformal and nonconformal equations of state is illustrated in \cref{fig:P}, which shows the time evolution of the integrated thermodynamic pressure. The gap between the two EOS is smaller in the entropy matching case. This behavior follows directly from \cref{fig:e_matching}, where it was shown that entropy matching leads to a larger energy density at the switching time, resulting in a correspondingly larger thermodynamic pressure.

\begin{figure}[h]
    \centering
    \includegraphics[width=0.49\linewidth]{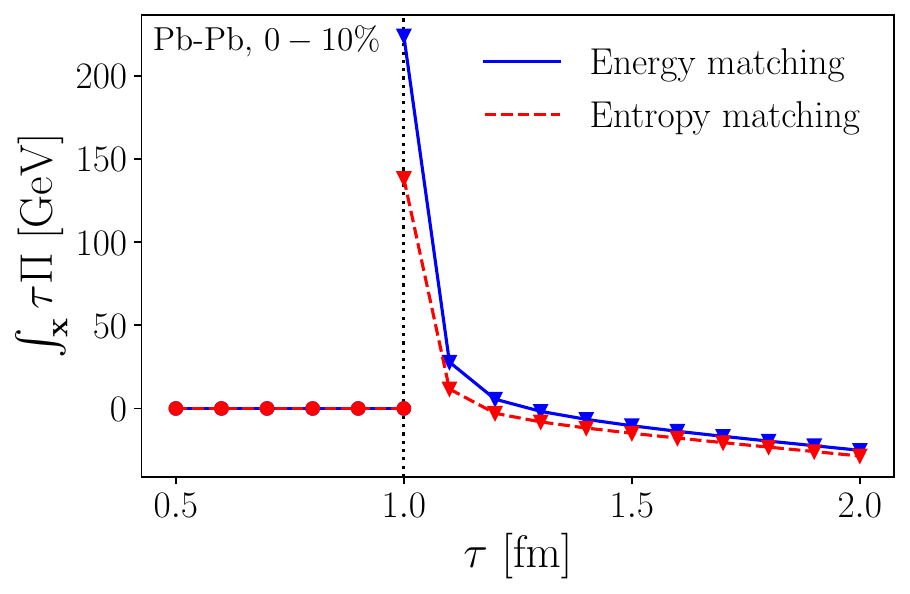}
    \includegraphics[width=0.49\linewidth]{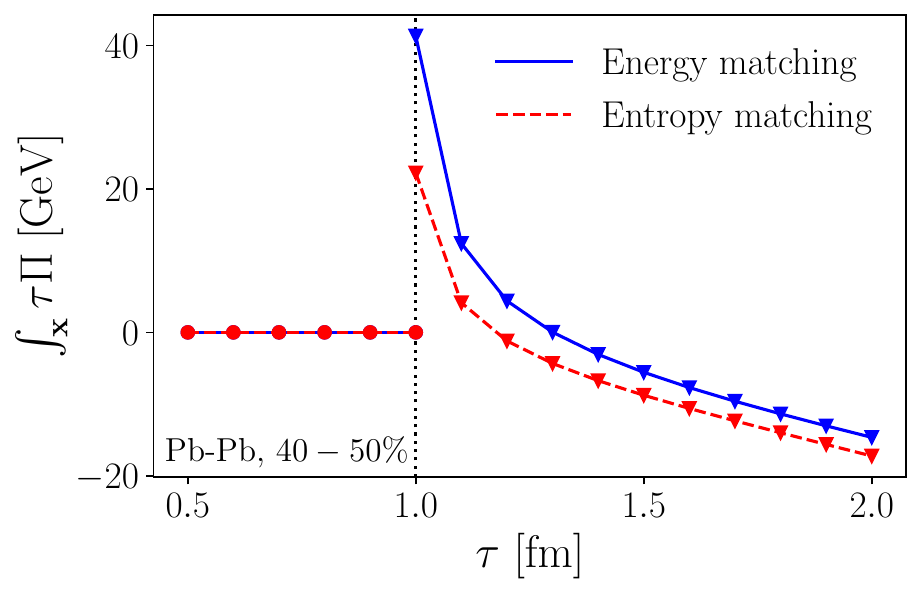}
    \caption{Time evolution of the transverse-plane-integrated bulk pressure for a central $0-10\%$ (left) and a non-central $40-50\%$ (right) average state in Pb-Pb collisions at $5.02$ TeV. Since \Kompost~is based on a conformal description, the bulk pressure in the pre-equilibrium stage is trivial $\Pi_\mathrm{pre} = 0$.}
\label{fig:Pi}
\end{figure}

A larger thermodynamic pressure reduces the initial bulk pressure defined in \cref{eq:bulk_s_matching}. Consequently, the integrated bulk pressure at the switching time is smaller for entropy matching, as shown in \cref{fig:Pi}. This feature is favorable for a hydrodynamic description, which is constructed as a viscous correction on top of ideal hydrodynamics. In other words, the consistency condition $\Pi / P \ll 1$ \cite{landau1987fluid} is better satisfied.

\begin{figure}[h]
    \centering
    \includegraphics[width=0.75\linewidth]{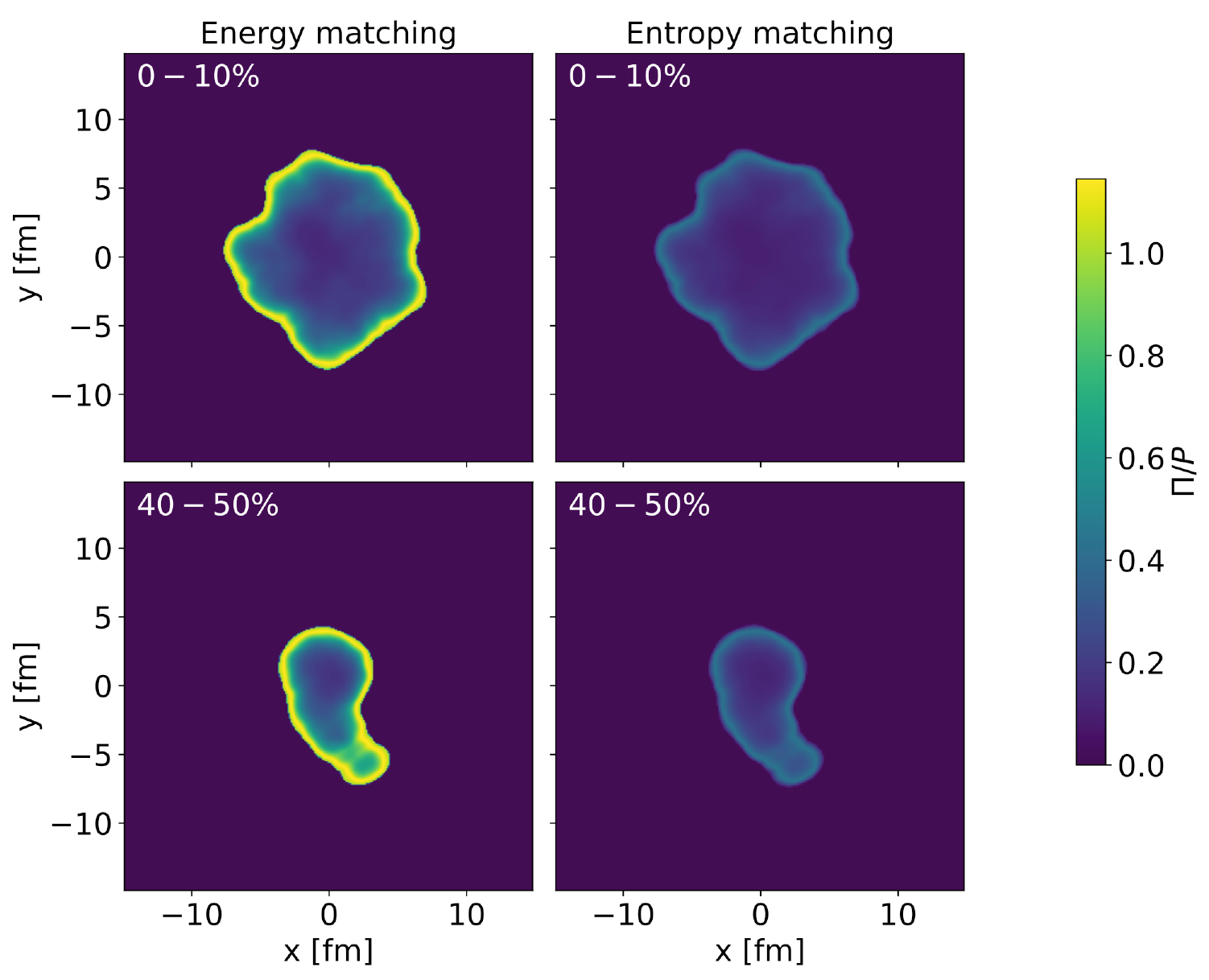}
    \caption{Transverse grid of a random event in the $0-10\%$ (top) and $40-50\%$ (bottom) centrality classes, showing the ratio of bulk pressure to thermodynamic pressure, $\Pi/P$, for energy matching (left) and entropy matching (right).}
\label{fig:bulk_profile}
\end{figure}

For this reason, it is instructive to examine the ratio $\Pi / P$ locally in the transverse plane for a representative event, as shown in \cref{fig:bulk_profile}. In the conventional energy matching, there are regions where $\Pi / P \sim 1$, which challenges the validity of the hydrodynamic approximation. In contrast, entropy matching yields a maximum ratio of approximately $\Pi / P \sim 0.5$, which is more consistent with the assumptions underlying viscous hydrodynamics.

As shown in \cref{fig:Pi}, the bulk pressure at $\tau_\mathrm{hydro}$ is initially large. However, after a few time steps, it rapidly decreases toward zero and even becomes negative. This behavior occurs because the hydrodynamic code identifies such large bulk pressures as nonphysical. Through built-in consistency checks, numerical regulators are applied to suppress this unwanted behavior. Without these regulators, the simulation becomes unstable and fails.

This observation clearly indicates that the transition between the pre-equilibrium stage and the initialization of hydrodynamics remains problematic and requires further theoretical refinement. Although the approach proposed in this work is not a definitive solution, it suggests a possible direction for mitigating these issues.

By construction, the calibration of the $K$-factor enforces a similar final multiplicity for both matching prescriptions. Nevertheless, it is important to examine additional final observables. For this purpose, $5000$ Pb–Pb events are generated for each centrality class considered here and evolved under both matching prescriptions through \textsc{MUSIC}, \textsc{iSS}, and \textsc{SMASH}. The final particle lists are analyzed using \textsc{SPARKX} \cite{Sass:2025opk} within midrapidity, $|\eta| < 0.5$, without any transverse momentum cut.

The event averaged final observables, represented by $\langle ... \rangle$, are presented in \cref{tab:observables}. As expected, the average multiplicity $\langle N_\mathrm{ch} \rangle$ is very similar for both matching prescriptions, with slightly larger deviations in non-central events. This difference arises because the calibration procedure prioritizes central events. As discussed in \cref{subsec:calibration_kfactor}, most of the events used to pre-calibrate the estimator equation are generated with $b = 0$ fm, which naturally favors the description of central collisions.

\begin{table}[h]
\centering
\caption{Final average observables at mid-rapidity for $5000$ Pb-Pb events, comparing the energy-matching and entropy-matching prescriptions.}
\label{tab:observables}
    \begin{tabular}{lcccc}
    \toprule
    & \multicolumn{2}{c}{$0$--$10\%$}  & \multicolumn{2}{c}{$40$--$50\%$} \\
    \cmidrule(lr){2-3} \cmidrule(lr){4-5}
    Observable & Energy & Entropy & Energy & Entropy \\
    \midrule
    $\langle N_{\rm ch} \rangle$ & $1808$ & $1828$ & $259$ & $282$ \\
    $\langle v_1 \rangle$ (\%) & $0.912$ & $0.883$ & $1.476$ & $1.399$ \\
    $\langle v_2 \rangle$ (\%) & $2.652$ & $2.597$ & $7.247$ & $7.309$ \\
    $\langle v_3 \rangle$ (\%) & $1.535$ & $1.530$ & $2.578$ & $2.621$ \\
    $\langle v_4 \rangle$ (\%) & $0.835$ & $0.830$ & $1.563$ & $1.553$ \\
    $\langle v_5 \rangle$ (\%) & $0.506$ & $0.514$ & $1.327$ & $1.274$ \\
    $\langle [p_T] \rangle$ (MeV) & $702.45$ & $697.57$ & $631.21$ & $629.21$ \\
    $\rho([p_T], v_2^2)$ & $0.0911$ & $0.1006$ & $0.1708$ & $0.1773$ \\
\bottomrule
\end{tabular}
\end{table}

The flow coefficients $v_n$, obtained from the expansion in \cref{eq:flow_coefficients}, are calculated using the reaction-plane method, $v_n = \langle e^{in\phi} \rangle$. They are only mildly affected by the different matching prescriptions. The largest deviation is observed for the directed flow $v_1$, where the difference reaches approximately $3\%$.

Although the initial eccentricities weighted by the energy density, $\varepsilon_n$, exhibit a discontinuity at the switching time $\tau_\mathrm{hydro} = 1$ fm in the entropy matching case, this does not significantly alter the linear correlation between $\varepsilon_n$ and $v_n$.

The average transverse momentum $\langle [p_T] \rangle$ would, in principle, be sensitive to differences in the initial bulk pressure. However, because \textsc{MUSIC} employs numerical regulators, the large bulk contributions are suppressed, and only negligible differences are observed in $\langle [p_T] \rangle$. 

The largest deviation, of approximately $10\%$, appears in the correlation between the squared elliptic flow and the transverse momentum,
\begin{equation}
    \rho([p_T], v_2^2) = \frac{\langle [p_T] v_2^2 \rangle - \langle [p_T] \rangle \langle v_2^2 \rangle}{\sqrt{\mathrm{Var}([p_T]) \mathrm{Var}(v_2^2)}}.
\end{equation}
This observable is more sensitive to event-by-event fluctuations and to bulk-pressure effects, which explains the larger difference between the two matching prescriptions.

\section{Summary}
\label{sec:summary_matching}

The main goal of this chapter was to analyze the challenges in connecting different physical models within state-of-the-art heavy-ion simulations and to propose an alternative method for matching the pre-equilibrium stage to hydrodynamics. The central result is the reduction of the artificial bulk pressure $\Pi$ that appears in the conventional energy-matching approach.

This study highlights that, although individual models are well understood within their domains of applicability, difficulties arise at the boundaries where one physical description transitions into another. In particular, the link between a conformal pre-equilibrium and a nonconformal hydrodynamic evolution introduces nontrivial thermodynamic inconsistencies.

The proposed method is based on matching the local equilibrium entropy density rather than the energy density at the switching time. 
This prescription significantly reduces the initial bulk pressure and consequently lowers the ratio $\Pi/P$, which is important for maintaining the validity of viscous hydrodynamics, especially in systems with large gradients.

Although entropy matching does not conserve energy locally at the switching time, this does not introduce a fundamental inconsistency, since heavy-ion simulations already rely on a global normalization of the initial energy density through the $K$-factor. In this sense, the discontinuity in energy can be interpreted as an alternative matching strategy that prioritizes near conservation of equilibrium entropy, which is more directly connected to the final particle multiplicity used for calibration.

The results show that final observables are only minimally affected by the choice of matching prescription once a consistent calibration is performed. While the analysis was carried out using \Kompost~and \textsc{MUSIC}, the entropy-matching idea is general and applicable to any conformal pre-equilibrium model coupled to hydrodynamics.

Beyond the analyses presented in this dissertation, we further employed the entropy-matching framework in Ref.~\cite{Andronic:2025ylc} to investigate transport properties in out-of-equilibrium systems. In that study, we performed a global analysis combining deep inelastic scattering (e+p) with small and large collision systems (p+p, p+Pb, and Pb+Pb) to constrain the shear viscosity during the early-time evolution. Although the physics focus of that work lies beyond the scope of this dissertation, it illustrates the broader applicability of the entropy-matching approach developed here.

In addition to proposing a new matching prescription, this chapter established the calibration framework and the practical use of the estimator equation \ref{eq:estimator_equation_withC}. This provides the methodological foundation for the main focus of this dissertation: the study of initial-state fluctuations and their hydrodynamic evolution, that is explored in the next chapters.

\chapter{Mode-by-mode decomposition in centrality classes}
\label{chap:modesPb}
\DoPToC
\noindent

This chapter is based on the published article \cite{Krupczak:2025gwg}, ``Mode-by-mode evolution of Pb–Pb collisions at 5.02 TeV in a hybrid model'' published in European Physical Journal C (2025), with adaptations and extensions.

This chapter focuses on presenting the theory of mode-by-mode decomposition of initial-state profiles, primarily building on the formalism developed in Ref. \cite{Borghini:2022iym}, which considers the simplest case of a fixed impact parameter. The objective here is to extend this framework to describe mode-by-mode decomposition within centrality classes, analyze the evolution of these modes, and establish their connection to initial- and final-state observables in Pb–Pb collisions \cite{Krupczak:2025gwg}.

To this end, \cref{sec:review_modes} introduces the motivation for using mode decomposition as a tool to study heavy-ion collisions. This section also reviews an alternative approach based on a fixed basis of modes. In \cref{sec:theory_modes}, the theoretical framework for constructing an uncorrelated basis to describe fluctuations is presented. This includes the decomposition of events into modes (\cref{subsec:decomposition}) and the connection between these fluctuations and measurable observables (\cref{subsec:observables}). An essential component of this analysis is the numerical implementation, described in \cref{sec:numerical}, where the simulation chain employed in this work is detailed. The results for the centrality-dependent analysis in Pb–Pb collisions are presented in \cref{sec:results_modes}. Additional technical details of this chapter are provided in \cref{appA}, \ref{appB}, and \ref{appC}.

\section{Motivation for mode-by-mode decomposition}
\label{sec:review_modes}

In the hybrid model for heavy-ion collisions presented in \cref{fig:HIC_numerical}, the largest uncertainties arise from the initial-state and pre-equilibrium phases, collectively referred to as the initial stages. These uncertainties stem from the absence of fully first-principles descriptions in this regime.

These uncertainties are particularly important because the modeling of the initial stage is known to have a direct impact on the final observables of the simulation \cite{Drescher:2006ca,Socolowski:2004hw,Liu:2015nwa}. As transport coefficients are generally extracted under specific assumptions about the initial conditions, uncertainties associated with the initial-state description propagate throughout the hybrid framework \cite{Novak:2013bqa,Bernhard:2019bmu,Nijs:2020ors,JETSCAPE:2020shq} and consequently affect the interpretation of the extracted medium properties.

A central objective in the heavy-ion community is therefore the so-called ``backward engineering'' approach \cite{Schukraft:2012ah}, in which the properties of the initial state are reconstructed from experimentally measured final observables. Achieving this goal requires a precise understanding of the full dynamical evolution, including how small perturbations in the initial state propagate through the system. In this context, the study of fluctuations plays a fundamental role.

\begin{figure}[h]
    \centering
    \includegraphics[width=0.8\linewidth]{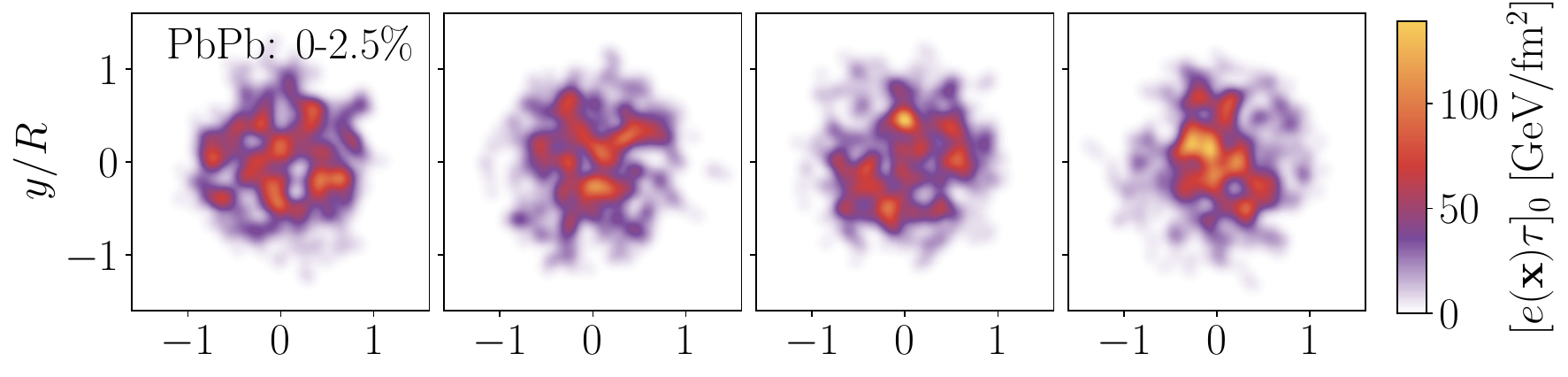}
    \includegraphics[width=0.8\linewidth]{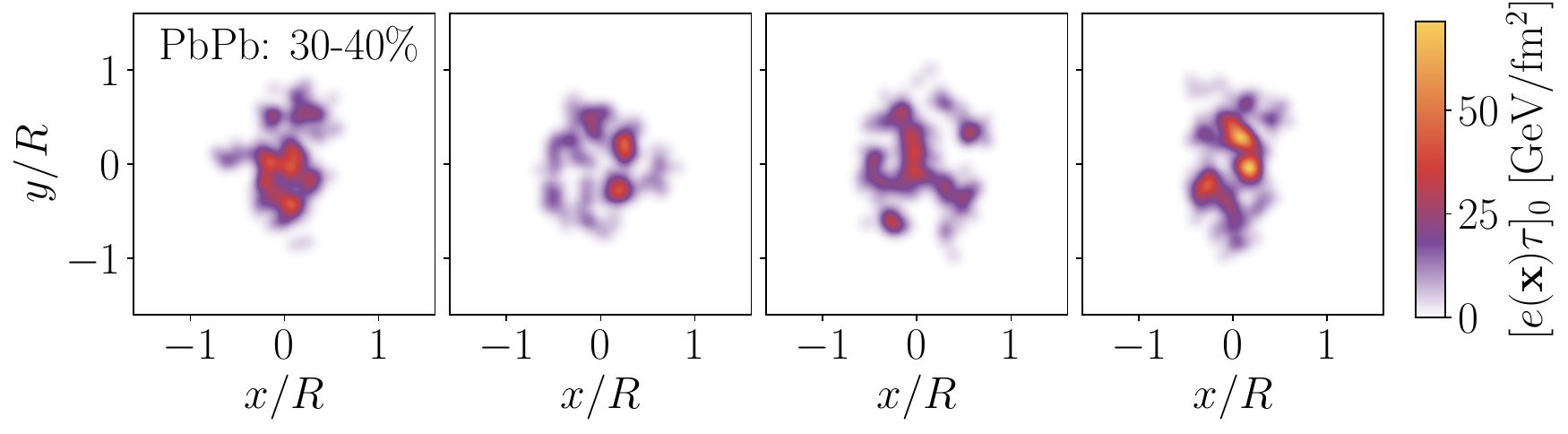}
    \caption{Energy density profiles for Pb-Pb collisions at $5.02$ TeV, shown in the transverse grid for random events in the central class $0-2.5\%$ (top) and the non-central class $30-40\%$ (bottom). The variations in energy deposition between events illustrate event-by-event fluctuations. The transverse coordinates $x$ and $y$ are normalized by the system radius $R$.}
\label{fig:random_profiles}
\end{figure}

Event-by-event fluctuations are illustrated in \cref{fig:random_profiles}, which displays several Pb–Pb events within the same centrality class. Although the overall geometric configuration is approximately fixed by centrality, each collision deposits energy in a unique manner across the transverse plane. These fluctuations originate from variations in nucleon and hotspot positions, as well as from the stochastic nature of the energy deposition process.

The idea of mode decomposition is to describe the initial energy-density profiles in terms of a fluctuation basis, referred to as modes. Instead of evolving each event individually, one evolves only the modes and subsequently reconstructs the final events from combinations of these evolved modes.\footnote{Because each mode contributes only a small fraction of the total energy, the hydrodynamic evolution is carried out for the fluctuation associated with a given mode on top of the background, referred to as the average state.} In principle, this approach offers a computational advantage, since the number of relevant modes can be significantly smaller than the number of event-by-event simulations required for comparable statistics.

To achieve small statistical uncertainties in a conventional event-by-event framework, it is generally necessary to evolve thousands of events. In contrast, within a mode-based approach, one would evolve a reduced set of modes and reconstruct the required statistics from many combinations of these evolved components. As will be shown in this chapter, this objective is not straightforward to realize in practice; nevertheless, it provides a strong motivation for investigating mode decomposition as an alternative framework.

Before introducing the prescription adopted in this dissertation, it is useful to review the previously developed mode-by-mode decomposition based on a fixed basis \cite{Floerchinger:2013vua}. In that approach, the initial enthalpy $w = \epsilon + P$ is decomposed as 
\begin{equation}
    w(\tau_0, r, \phi, \eta) - w_\mathrm{BG}(\tau_0, r) = \int \frac{\mathrm{d}k_\eta}{2\pi} \sum_{m=-\infty}^{\infty} e^{ik_\eta \eta + im\phi} w^{(m)}(\tau_0, r, k_\eta),
\end{equation}
where the initial enthalpy $w(\tau_0, r, \phi, \eta)$, which depends on the radial, azimuthal, and longitudinal coordinates, is separated into a background contribution $w_\mathrm{BG}$ and fluctuations. The background contains only radial dependence and is assumed to be boost-invariant and azimuthally symmetric. It corresponds to a smooth, circular profile obtained by averaging over many events, including variations in the direction of the impact parameter. Because only radial dependence remains, the background evolves effectively in $1+1$D \cite{Floerchinger:2018pje}.

The azimuthal and longitudinal dependencies are encoded in the fluctuation term via a Fourier expansion. Assuming longitudinal boost invariance eliminates the $k_\eta$ dependence. The radial structure of the modes can then be expressed in terms of Bessel functions $J_m$:
\begin{equation}
    w^{(m)}(\tau_0, r) = w_\mathrm{BG}(\tau_0, r)\sum_{l=1}^\infty J_m(k_l^{(m)}, r) \tilde{w}^{(m)}_l,
\end{equation}
where $\tilde{w}_l^{(m)}$ forms an orthonormal basis, with $l$ running up to some maximum value $l_\mathrm{max}$ determined by the desired spatial resolution. In practice, values of $l \sim 20$ already provide a satisfactory average description \cite{Floerchinger:2013vua,Floerchinger:2013rya}.

A key advantage of this approach is that the basis is fixed and the evolution of the background state is relatively simple. However, because the background profile depends only on the radial coordinate, it is necessarily circular. While this description is well suited for central collisions with $b = 0$ fm, it becomes less appropriate for non-central events, where elliptical geometries play an essential role. In these cases, the modes must generate the entire azimuthal asymmetry as corrections to the circular background, causing the average coefficients $\langle \tilde{w}_l^{(m)} \rangle$ to become nonzero in order to account for the missing azimuthal structure \cite{Floerchinger:2013vua}. Within this formulation, observables are linear in the fluctuation modes \cite{Floerchinger:2013rya}. It is also worth noting that resonance decays and hadronic rescatterings after particlization are not included in this framework.

In contrast, our formalism described in the next section does not rely on a fixed basis. Whenever the collision system, energy, or setup changes, the mode basis must be recalculated. However, by fixing the direction of the impact parameter, the average state can already incorporate the elliptical geometry characteristic of non-central collisions. In this case, the modes represent small corrections on top of an anisotropic background, allowing the formalism to be extended beyond linear order, for example, to second order. This feature facilitates a more systematic quantification of the statistical importance of different types of fluctuations.

\section{Theoretical Framework}
\label{sec:theory_modes}

This section is divided into two parts. In \cref{subsec:decomposition}, the basic formalism for decomposing the energy density of a given event into an average background state plus fluctuation modes is presented. Then, in \cref{subsec:observables}, the theoretical framework connecting these fluctuation modes to measurable observables is discussed.

\subsection{Mode-by-mode decomposition}
\label{subsec:decomposition}

Different from previous approaches to mode decomposition, which were based on a radially symmetric average state plus a fixed set of modes, the present formalism allows for an average state that is not necessarily radially symmetric and for a set of modes that depends on the specific system under consideration \cite{Borghini:2022iym, Borghini:2024ekn, Krupczak:2025gwg}. The initial energy density profile is written as
\begin{equation}
    \Phi^{(i)}(\mathbf{x}) = \bar{\Psi}(\mathbf{x}) + \sum_l c_l^{(i)} \Psi_l(\mathbf{x}),
\label{eq:modes_decomposition}
\end{equation}
where $\bar{\Psi}$ represents the average state, $\Psi_l$ are the fluctuation modes, and $c_l^{(i)}$ are the expansion coefficients. The index $l$ labels the mode, and the index $i$ labels the event. In this formulation, each energy profile from \cref{fig:random_profiles} is represented as a column vector $\Phi^{(i)}(\mathbf{x})$, expressed as the sum of a common average state and a linear combination of modes. The event-by-event variation is entirely encoded in the coefficients $c_l^{(i)}$.

Although $\Phi^{(i)}$ may represent an energy density profile as in \cref{fig:random_profiles}, the formalism is more general and can also be applied to other profiles, such as baryon or electric charge densities. The only requirement is that the profile consists of real numbers. In this dissertation, we focus exclusively on the energy density profile, as it is the fundamental quantity governing the evolution in heavy-ion collisions. While the energy density is naturally represented as a two-dimensional grid, here it is treated as a column vector whose dimension is determined by the number of grid points.

The expansion coefficients in \cref{eq:modes_decomposition} are required to satisfy the following relations:
\begin{equation}
\begin{aligned}
    \langle c_l \rangle &= 0, \\
    \langle c_l c_{l'} \rangle &= \delta_{ll'},    
\end{aligned}
\label{eq:properties_cl}
\end{equation}
where $\langle \cdots \rangle$ denotes the average over events. These conditions imply that the set ${c_l^{(i)}}$ is centered, uncorrelated, and has unit variance.

The average state is defined as the ensemble mean over $N_\mathrm{ev}$ events,
\begin{equation}
    \bar{\Psi}(\mathbf{x}) \equiv \frac{1}{N_\mathrm{ev}} \sum_{i=1}^{N_\mathrm{ev}} \Phi^{(i)}(\mathbf{x}).
\label{eq:average_state}
\end{equation}
This average state is expected to be smooth, as fluctuation contributions cancel under averaging. In practice, a reasonably smooth background can often be obtained with $N_\mathrm{ev} \sim 100$ events. However, accurately extracting the fluctuation modes requires a much larger sample. In the present analysis, $N_\mathrm{ev}$ is of order $10^5$, as will be discussed in \cref{sec:results_modes}. No radial symmetry is imposed on $\bar{\Psi}$. To construct it consistently, all events must be aligned with respect to the center of mass and the direction of the impact parameter is fixed in the initial condition model. 

Substituting \cref{eq:modes_decomposition} into \cref{eq:average_state} immediately yields the centering condition for the coefficients:
\begin{equation*}
    \bar{\Psi}(\mathbf{x}) = \langle \Phi^{(i)}(\mathbf{x}) \rangle = \bar{\Psi}(\mathbf{x}) + \langle c_l^{(i)} \rangle \Psi_l(\mathbf{x}) \hspace{0.2cm} \xrightarrow{} \hspace{0.2cm} \langle c_l \rangle = 0~.
\end{equation*}

To determine the modes explicitly, we first assume an orthonormal basis $\tilde{\chi}_l(\mathbf{x})$, with corresponding coefficients $\tilde{c}_l^{(i)}$, which also satisfy $\langle \tilde{c}_l^{(i)} \rangle = 0$. The profile can then be written as
\begin{equation}
    \Phi^{(i)}(\mathbf{x}) = \bar{\Psi}(\mathbf{x}) + \sum_l \tilde{c}_l^{(i)} \tilde{\chi}_l(\mathbf{x}).
\label{eq:modes_decomposition_basis_chi}
\end{equation}
Constructing the ensemble average of the outer product (suppressing spatial indices for simplicity),
\begin{equation}
    \frac{1}{N_\mathrm{ev}} \sum_{i=0}^{N_\mathrm{ev}} \Phi^{(i)}\Phi^{(i) T} = \bar{\Psi}\bar{\Psi}^T + \sum_{ll'} \langle \tilde{c}_l \tilde{c}_{l'} \rangle \tilde{\chi}_l \tilde{\chi}_{l'}^T,
\end{equation}
where we used $\langle \tilde{c}_l \rangle = 0$. The central mathematical object is the correlation matrix
\begin{equation}
    \rho \equiv \frac{1}{N_\mathrm{ev}}\sum_i \Phi^{(i)} \Phi^{(i)T} - \bar{\Psi} \bar{\Psi}^T = \sum_{l, l'} \langle \tilde{c}_l \tilde{c}_{l'} \rangle \tilde{\chi}_l \tilde{\chi}_{l'}^T.
\label{eq:rho}
\end{equation}
The matrix $\rho$ is symmetric, real and represents the event-by-event correlation matrix of fluctuations. Diagonalizing it, $\rho \tilde{\Psi}_l = \lambda_l \tilde{\Psi}_l$, provides an orthonormal set of eigenvectors ${\tilde{\Psi}_l}$ with eigenvalues $\lambda_l$. Using spectral decomposition,
\begin{equation}
    \rho(\mathbf{x}, \mathbf{x'}) = \sum_l \lambda_l \tilde{\Psi}_l(\mathbf{x}) \tilde{\Psi}_l^T(\mathbf{x'}).
\label{eq:spectral_decomposition}
\end{equation}
Comparing \cref{eq:rho} and \cref{eq:spectral_decomposition} gives: $\langle \tilde{c}_l \tilde{c}_{l'} \rangle = \lambda_l \delta_{ll'}$.
Since we impose $\langle c_l c_{l'} \rangle = \delta_{ll'}$, the two sets of coefficients are related by
\begin{equation}
    c_l^{(i)} = \frac{\tilde{c}^{(i)}_l}{\sqrt{\lambda_l}},
\end{equation}
and the corresponding modes are
\begin{equation}
    \Psi_l(\mathbf{x}) = \sqrt{\lambda_l} \tilde{\Psi}_l(\mathbf{x}).
\label{eq:normalization_modes}
\end{equation}

Therefore, each energy density profile is decomposed into an average state plus a linear combination of fluctuation modes ${\Psi_l}$. The modes are obtained from the diagonalization of the correlation matrix $\rho$, which captures the event-by-event autocorrelation of fluctuations after subtraction of the background. The eigenvalues $\lambda_l$ quantify the statistical importance of each mode $\Psi_l$, while the modes themselves form an orthogonal (though not normalized) basis.

It is essential to emphasize that the construction of $\rho$ requires a large number of events in order to accurately capture the full fluctuation structure of the model.

This establishes the basic structure of the mode decomposition of initial energy density profiles. In the next subsection, we describe how this formalism is used to determine the contribution of individual modes to final observables.

\subsection{Mode-by-mode response}
\label{subsec:observables}

This subsection is devoted to describing the theoretical framework for calculating observables from the uncorrelated modes introduced in the previous subsection. The central objective of a mode-decomposition theory is precisely to establish the connection between initial-state fluctuations and observables, which encode the measurable information of a given collision process.

Before proceeding, it is important to clarify the meaning of the term observable in the context of this dissertation. In heavy-ion physics, an observable typically refers to a measurable experimental quantity at facilities such as the LHC or RHIC. In theoretical simulations, an observable usually denotes a final quantity computed from the particle list after the full dynamical evolution of the system. Examples include the charged-particle multiplicity $\mathrm{d}N_\mathrm{ch}/\mathrm{d}\eta$, the transverse momentum $p_T$, and the flow coefficients $v_n$. However, in this dissertation, the term observable is used more broadly. The expression ``initial observables'' refers to quantities computed directly from the initial energy density profile, such as the eccentricities $\varepsilon_n$, which are not experimentally measurable. In contrast, ``final observables'' retain the standard meaning of quantities extracted from the final particle list and comparable to experimental measurements. Thus, whenever the term observable is used, it should not automatically be interpreted as an experimentally measured quantity.

A generic observable $O_\alpha$, evaluated for the decomposition in \cref{eq:modes_decomposition}, can be expanded in a Taylor series in terms of the expansion coefficients:
\begin{equation}
    O_\alpha(\Phi^{(i)}) = \bar{O}_\alpha + \sum_l L_{\alpha, l} c_l^{(i)} + \frac{1}{2} \sum_{l, l'} Q_{\alpha, ll'} c_l^{(i)} c_{l'}^{(i)} + \mathcal{O}(c_l^3),
\label{eq:taylor_expansion_O}
\end{equation}
where the zeroth-order contribution corresponds to the observable evaluated for the average state
\begin{equation}
    \bar{O}_\alpha \equiv O_\alpha(\bar{\Psi}),
\label{eq:Obar}
\end{equation}
the first-order term is defined through the linear response coefficient
\begin{equation}
    L_{\alpha, l} \equiv \frac{\partial O_\alpha}{\partial c_l}\bigg|_{\bar{\Psi}},
\label{eq:linear_response}
\end{equation}
and the second-order term is determined by the quadratic response coefficient
\begin{equation}
    Q_{\alpha, ll'} \equiv \frac{\partial^2 O_\alpha}{\partial c_l \partial c_{l'}}\bigg|_{\bar{\Psi}}.
\label{eq:quadratic_response}
\end{equation}
The expansion is truncated at second order in $c_l$. Given the statistical properties in \cref{eq:properties_cl}, the coefficients $c_l$ are of order unity. As will be demonstrated in \cref{sec:results_modes}, the second-order expansion is sufficient to describe observables within this framework. Importantly, the relevance of each mode is not primarily determined by the magnitude of $c_l$, but by the factor $\sqrt{\lambda_l}$, which encodes the statistical weight of the mode as it was shown in \cref{eq:normalization_modes} with the normalized $\{\tilde{\Psi}_l\}$ basis.\footnote{It has already been mentioned that a large number of events is required to accurately capture fluctuations. One way to verify whether the fluctuations are being resolved with sufficient precision is by examining the eigenvalue spectra. From the modes, we know that degeneracies can occur between modes that differ only by a rotation. However, modes with distinct shapes should have different eigenvalues $\lambda_l$. As the number of events increases, the fake degeneracies are progressively lifted, leading to a clearer separation between the $\lambda_l$.} This feature distinguishes the present formalism from previous fixed-basis approaches, where such a controlled second-order expansion of observables is not possible.

The only assumption underlying the expansion in \cref{eq:taylor_expansion_O} is that the observable $O_\alpha$ is sufficiently smooth to allow the derivatives in \cref{eq:linear_response} and \cref{eq:quadratic_response} to exist. The validity of this approximate linear behavior will be examined in \cref{sec:results_modes}.

Unlike fixed-basis decompositions, where the average state is radially symmetric, the present framework allows the average state to carry the dominant geometric information. Consequently, the zeroth-order term often provides the largest contribution to observables, there are exceptions that will be discussed in the next chapter. The response coefficients in \cref{eq:linear_response} and \cref{eq:quadratic_response} quantify how strongly each fluctuation mode $\Psi_l$ contributes to a given observable $O_\alpha$.

The ensemble average of an observable over many events is given by
\begin{equation}
    \langle O_\alpha \rangle = \frac{1}{N_\mathrm{ev}} \sum_{i=1}^{N_\mathrm{ev}} O_\alpha(\Phi^{(i)}) \simeq \left\langle \bar{O}_\alpha + \sum_l L_{\alpha, l} c_l + \frac{1}{2} \sum_{l, l'} Q_{\alpha, ll'} c_l c_{l'} \right\rangle,
\end{equation}
Using the statistical properties in~\cref{eq:properties_cl}, this simplifies to
\begin{equation}
    \langle O_\alpha \rangle \approx \bar{O}_\alpha + \frac{1}{2} \sum_l Q_{\alpha,ll}.
\end{equation}
Thus, the ensemble average does not depend on the linear response coefficients $L_{\alpha,l}$, but only on the diagonal elements of the quadratic response matrix. In contrast, the covariance between two observables $O_\alpha$ and $O_\beta$ depends only on the linear response coefficients:
\begin{equation}
    \bigg\langle (O_\alpha - \langle O_\alpha\rangle)(O_\beta - \langle O_\beta \rangle) \bigg\rangle \simeq \sum_l L_{\alpha, l} L_{\beta, l}.
\end{equation}
For the special case $\beta = \alpha$, this reduces to the variance,
\begin{equation}
    \mathrm{Var}(O_\alpha) \simeq \bigg\langle (O_\alpha - \langle O_\alpha\rangle)^2 \bigg\rangle \simeq \sum_l L_{\alpha, l}^2.
\label{eq:variance_O}
\end{equation}

With this theoretical foundation established, we now proceed to a more detailed analysis of the implementation of this model.

\section{Numerical setup}
\label{sec:numerical}

This section highlights the specific details of the numerical implementation of the mode-decomposition framework. The original study in \cite{Borghini:2022iym} analyzed the case of fixed impact parameters (for $b = 0$ and $b = 9$ fm). In the present work, a centrality-dependent analysis is performed for Pb–Pb collisions at $\sqrt{s_\mathrm{NN}} = 5.02$ TeV within a boost-invariant scenario.

The energy density profiles $\{\Phi^{(i)}\}$ are generated using the \textsc{MCGlauber} model, described in detail in \cref{subsec:glauber}. The Woods–Saxon parameters used in \cref{eq:woods-saxon} are $R = 6.62$ fm and $a = 0.546$ fm. The geometric collision criterion is based on the inelastic nucleon–nucleon cross section at this energy $\sigma_\mathrm{inel}^\mathrm{NN} = 67.6$ mb. The initial energy density profile is constructed according to \cref{eq:MCGlauber_energy} with $\alpha = 0.2$, multiplied by a global normalization factor $K$.

As discussed in \cref{subsec:calibration_kfactor}, the $K$-factor must be calibrated for the specific model, collision energy, and system under consideration. For this purpose, $20$ initial conditions are generated for impact parameters $b = 0$, $3$, and $6$ fm. Using the estimator equation \ref{eq:estimator_equation}, the corresponding $\mathrm{d}N_\mathrm{ch}/\mathrm{d}\eta$ is computed without prior calibration. The same $20$ events are then evolved through the full hybrid chain.

The pre-equilibrium evolution is performed with \Kompost~from $\tau_0 = 0.2$ fm to $\tau_\mathrm{hydro} = 1.0$ fm. Energy matching is applied at the switching time, and the hydrodynamic evolution is carried out with \textsc{MUSIC}. The shear viscosity is kept constant in both \Kompost~and \textsc{MUSIC}, with $\eta/s = 0.16$. The bulk viscosity in MUSIC is parametrized as in Ref. \cite{Bernhard:2018hnz}:
\begin{equation}
    (\zeta/s)(T) = \frac{0.052}{1 + \left( \frac{T - 0.183}{0.022} \right)^2},
\end{equation}
where the temperature $T$ is expressed in GeV. After hydrodynamics, particlization is performed with \textsc{iSS}, followed by hadronic transport with \textsc{SMASH}, using $25$ oversamplings per hydrodynamic event.

From this full evolution, the final multiplicity $\mathrm{d}N_\mathrm{ch}/\mathrm{d}\eta$ is extracted and compared with the estimator prediction, following the procedure illustrated in \cref{fig:pre_calibration}. For the present setup, the calibration yields $C(\eta/s) = 1.209$ in \cref{eq:estimator_equation_withC}.

Once $C(\eta/s)$ is fixed, new initial profiles are generated with \textsc{MC-Glauber} while varying the $K$-factor until optimal agreement with the ALICE experimental data \cite{ALICE:2015juo} is achieved, without the necessity of a hydrodynamic evolution. With both calibrated constants $C(\eta/s) = 1.209$ and $K = 37.53$, the estimator equation can reliably predict the final multiplicity. Comparing the estimator results with the full simulation shows a discrepancy smaller than $1\%$.

\begin{figure}[h]
    \centering
    \includegraphics[width=0.6\linewidth]{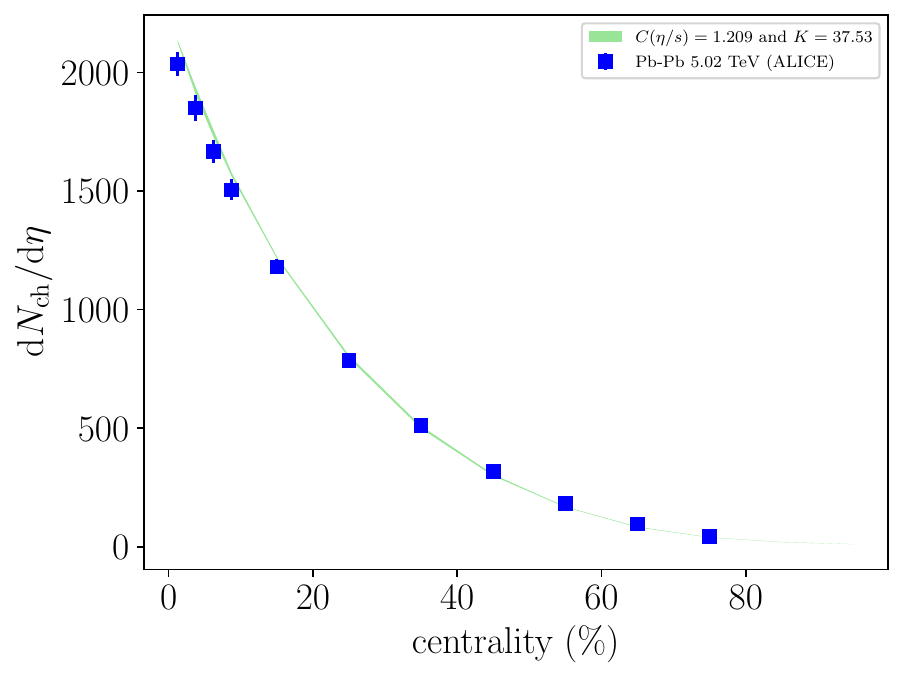}
    \caption{Charged-particle multiplicity as a function of centrality for Pb-Pb collisions at $\sqrt{s_\mathrm{NN}} = 5.02$ TeV. The green curve shows the multiplicity obtained from the calibrated estimator, \cref{eq:estimator_equation_withC}, in good agreement with the experimental data from ALICE \cite{ALICE:2015juo}, shown as blue points with error bars.}
\label{fig:PbPb_calibrated}
\end{figure}

In \cref{fig:PbPb_calibrated}, the calibrated estimator predictions are compared to the experimental data from ALICE. Given the good agreement, the estimator equation can be used to determine final multiplicities without performing the full dynamical evolution for every event.

As explained in the previous theoretical section, the total number of events $N_\mathrm{ev}$ entering \cref{eq:rho} must be large. Since the goal of this work is to perform the analysis for different centrality classes, evolving a large number of events to determine the final multiplicity, following the full procedure used by experimentalists (see \cref{fig:impact_b}), would be computationally expensive. Therefore, the full dynamical evolution is replaced by the estimator equation, as the initial condition models are significantly less computationally expensive than the complete hybrid evolution.

Minimum-bias energy density profiles are generated with the impact parameter fixed along the $x$-direction. The multiplicity is computed using the estimator equation and compared with the experimental centrality limits. If a generated event falls within the desired centrality bin, it is included in the construction of the average state $\bar{\Psi}$ and the correlation matrix $\rho$; otherwise, it is discarded. In this analysis, two centrality classes are considered: $0-2.5\%$ and $30-40\%$, with $N_\mathrm{ev} = 2^{21}$ profiles accumulated for each bin. Due to the computational cost associated with event generation and updates of $\rho$, MPI parallelization is employed to optimize performance.

An important numerical aspect of the implementation is the recentering of the profiles. Since both $\rho(\mathbf{x})$ and $\bar{\Psi}(\mathbf{x})$ are constructed by summing profiles, all profiles must share the same reference center. After generation, each profile is recentered by shifting its center of mass to the center of the grid.

The grid size is another relevant numerical parameter, as the computational cost scales with the number of grid points. In these simulations, a square domain of size $3.2 R_{\mathrm{Pb}} = 21.184$ fm is used, with $N_\mathrm{s} = 192$ grid points per direction, corresponding to a spatial resolution of $0.11$ fm. This resolution is sufficient to capture the fluctuations generated by \textsc{MC-Glauber}, whose characteristic smearing scale is $\sigma = 0.4$ fm. Each initial energy density profile is therefore represented as a column vector $\Phi^{(i)}(\mathbf{x})$ with $N_\mathrm{s}^2 = 192^2$ components. The correlation matrix $\rho$ thus has dimensions $192^2 \times 192^2$. The Eigen library is used to compute the largest $256$ eigenvalues and corresponding eigenvectors.\footnote{See more details in the documentation available at \url{https://libeigen.gitlab.io/}} As will be demonstrated in the next section, this number of modes is sufficient to capture the relevant physical signals considered in this dissertation.

Once $\rho(\mathbf{x})$ has been constructed and diagonalized, yielding the modes $\{\Psi_l(\mathbf{x})\}$ and eigenvalues $\{\lambda_l\}$, an energy–momentum tensor is built for the evolution of both the average state and the individual modes. The construction assumes an ideal, boost-invariant configuration with vanishing longitudinal pressure $P_L = 0$ and transverse pressure $P_T = \epsilon/2$, preserving the traceless property of $T^{\mu\nu}$, as discussed in \cref{sec:hydro}. The energy profiles are interpolated using bilinear interpolation to refine the grid spacing from $0.11$ fm to $0.1$ fm. Additional details of the interpolation procedure are provided in \cref{appA}.

The evolution of the modes is essential for computing final observables and follows the same hybrid structure used in the calibration procedure. The main difference, to be examined in \cref{sec:results_modes}, concerns the treatment of the final particle spectra. One option is to evolve the system through \textsc{SMASH} with a large number of oversamplings. Alternatively, one may extract the full particle spectra directly from a dedicated mode in \textsc{MUSIC}, which includes only decays. The impact of including or neglecting hadronic interactions will be discussed in the next section.

\section{Results and Discussion}
\label{sec:results_modes}

This section presents the results obtained from the numerical implementation of the theoretical framework described in the previous sections. In addition to presenting the results, a more detailed discussion of several relevant aspects is provided.

In \cref{subsec:modes}, we first present the structure of the average state $\bar{\Psi}$ and the fluctuation modes $\{\Psi_l\}$ for the two centrality classes analyzed in this work. This subsection also examines the relative importance of each mode through the analysis of the eigenvalues $\lambda_l$. Furthermore, the statistical properties of the expansion coefficients $c_l$, introduced in \cref{sec:theory_modes}, are explicitly verified. Next, \cref{subsec:ISobservables} discusses the initial observables computed directly from the initial energy density profiles using the Taylor expansion described in \cref{subsec:observables}. Finally, the final observables are analyzed and their connection to the initial observables is established.

It is important to note that these three subsections are closely interconnected. The structure of the modes determines the behavior of the initial observables, which in turn influences the corresponding final observables after the full dynamical evolution.

\subsection{Properties of the decomposed modes}
\label{subsec:modes}

\begin{figure}[h]
    \centering
    \includegraphics[width=0.6\linewidth]{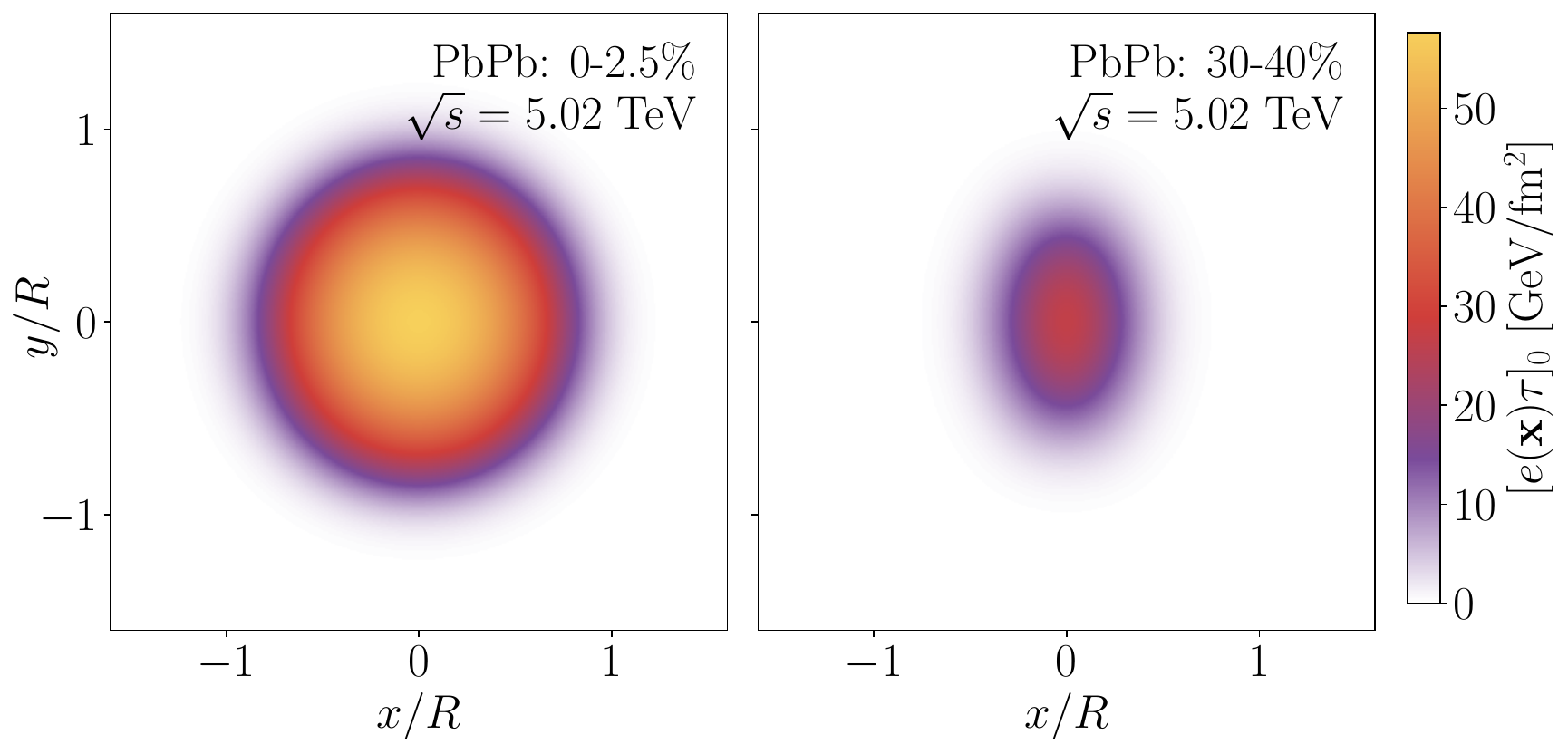}
    \caption{Average state $\bar{\Psi}$ constructed from $2^{21}$ initial profiles in the $0-2.5\%$ (left) and $30-40\%$ (right) centrality classes for Pb-Pb collisions. Both axes are expressed in units of the half-density radius for Pb nucleus $R = 6.62$ fm.}
\label{fig:avg_state}
\end{figure}

Starting this subsection, \cref{fig:avg_state} displays the average state $\bar{\Psi}(\mathbf{x})$, defined in \cref{eq:average_state}, constructed from $N_\mathrm{ev} = 2^{21}$ events in Pb–Pb collisions at $\sqrt{s_\mathrm{NN}} = 5.02$ TeV generated with \textsc{MCGlauber}. The left panel corresponds to the central class $0-2.5\%$, while the right panel shows the non-central class $30-40\%$.

The most evident feature of the average state is its smoothness, reflecting the absence of event-by-event fluctuations. In central collisions, the profile is approximately circular, whereas in non-central collisions it exhibits an elliptical shape, resulting from the fixed direction of the impact parameter. Additionally, the non-central case shows a smaller overall energy density compared to the central case, as expected from the reduced overlap of the colliding nuclei. In summary, the average state represents a smooth, non-fluctuating profile that captures the dominant geometric characteristics of the initial state.

The average state carries the dominant contribution to the total energy, approximately $4014$ GeV for the central class and $700$ GeV for the non-central class. The fluctuation modes provide only a comparatively small contribution to the total energy. The geometric information encoded in $\bar{\Psi}(\mathbf{x})$ is reflected in the initial eccentricities \cite{Ollitrault:1992bk}. For the central case, the elliptic eccentricity decomposed into cosine and sine components yields $\varepsilon_{2,c} = 0.0174$ and $\varepsilon_{2,s} \sim 10^{-6}$, respectively. The dominance of the cosine component indicates a small residual alignment of the impact parameter along the $x$-direction. This behavior becomes more pronounced in the non-central case, where $\varepsilon_{2,c} = 0.3214$ and $\varepsilon_{2,s} \sim 10^{-4}$, reflecting the larger average deformation due to the nonzero impact parameter. In both centralities, the sine component is basically zero, demonstrating the symmetry in the $y$ direction. In contrast, the triangular eccentricity $\varepsilon_3$ is negligible, of order $10^{-6}$ for the $0-2.5\%$ bin and $10^{-4}$ for the $30-40\%$ bin, as expected since triangularity is primarily driven by fluctuations.

\begin{figure}
    \centering
    \includegraphics[width=\linewidth]{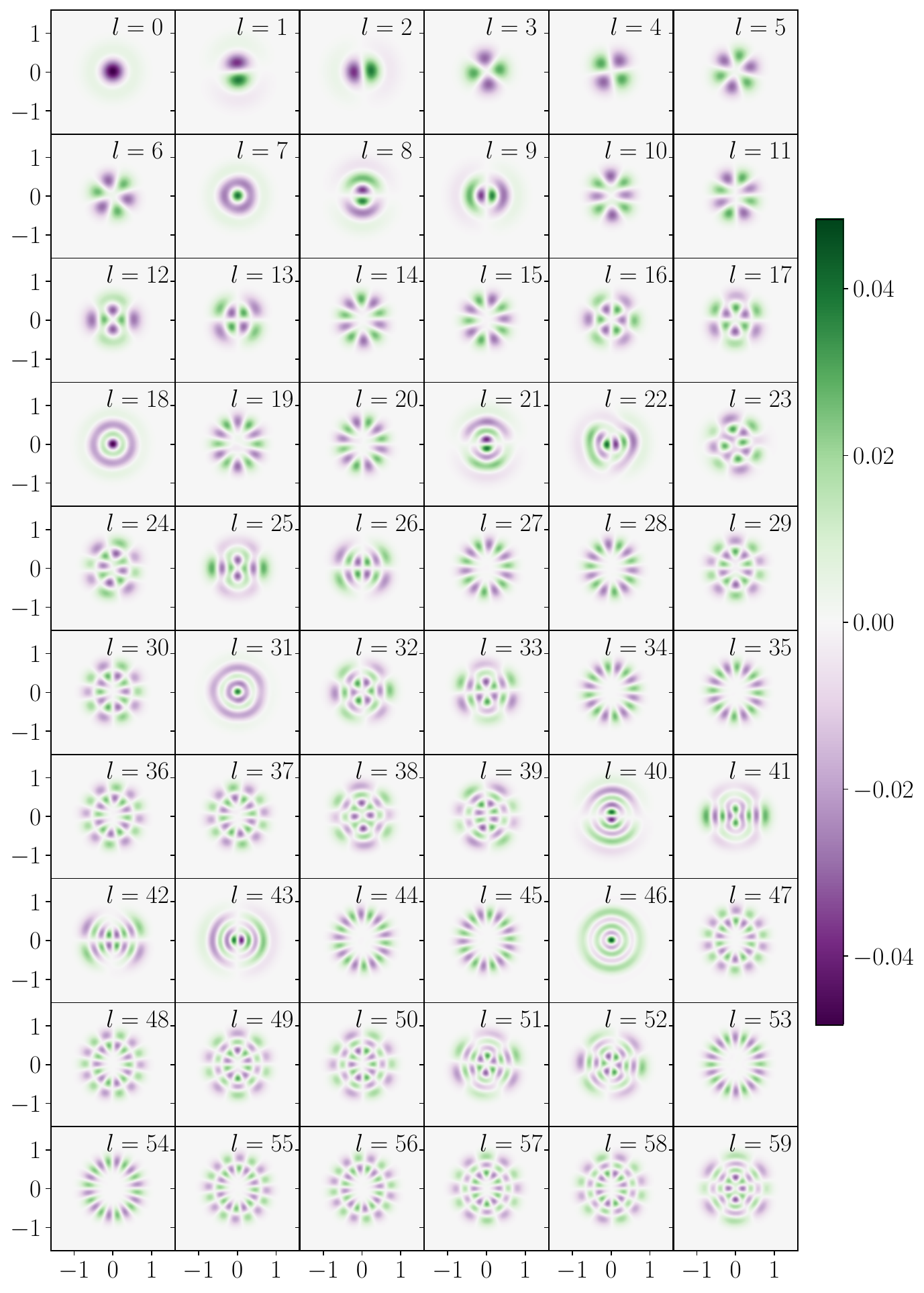}
    \caption{Normalized first $60$ modes in the transverse plane for central events in the $0-2.5\%$ centrality bin. The axes are expressed in units of the half-density radius $R_\mathrm{Pb} = 6.62$ fm, consistent with the average state plot in \cref{fig:avg_state}.}
\label{fig:modes_0-2.5}
\end{figure}

\begin{figure}
    \centering
    \includegraphics[width=\linewidth]{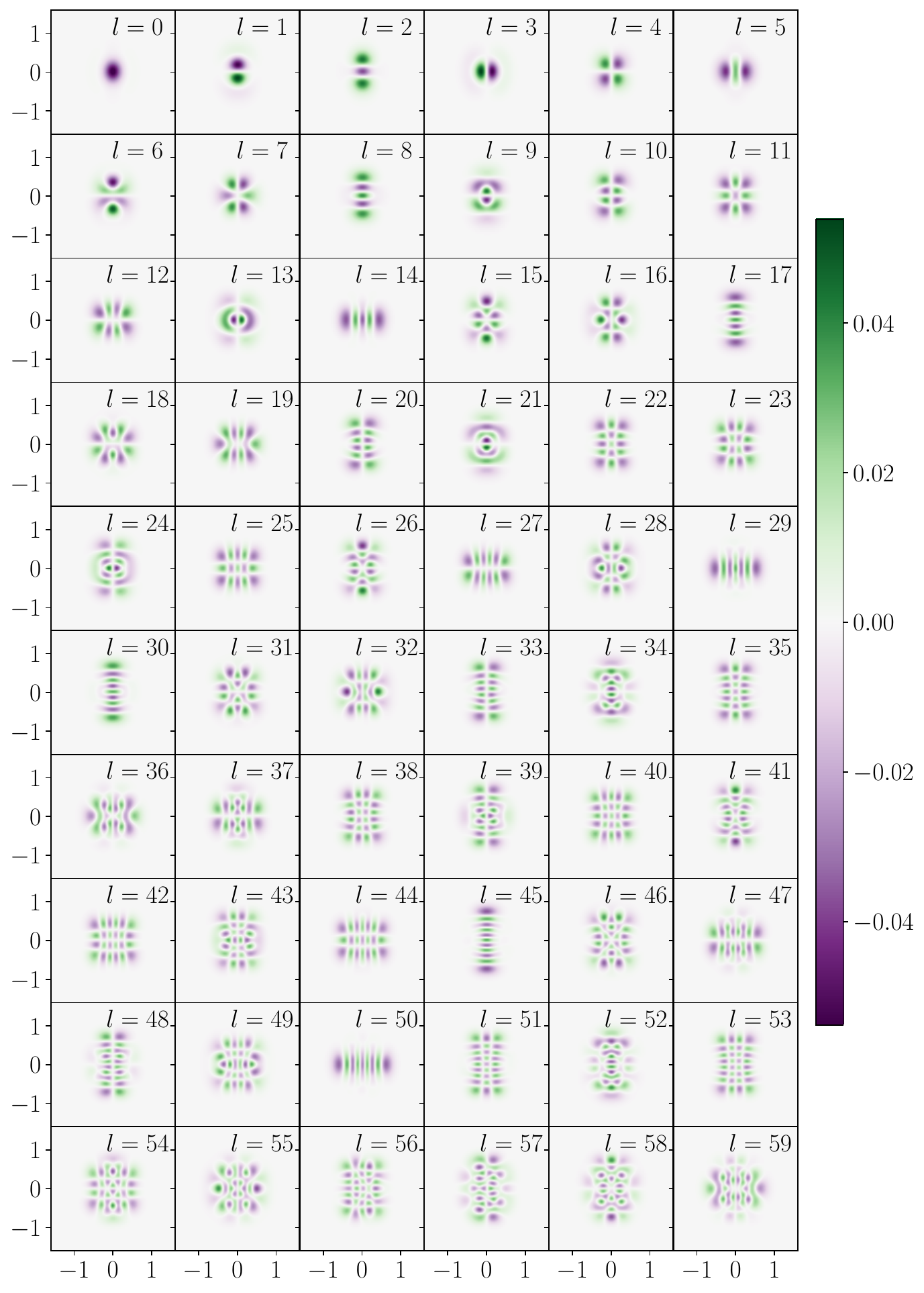}
    \caption{Normalized first $60$ modes in the transverse plane for non-central events in the $30-40\%$ centrality bin. Consistent with \cref{fig:modes_0-2.5}, the axes are expressed in units of the half-density radius $R_\mathrm{Pb} = 6.62$ fm.}
\label{fig:modes_30-40}
\end{figure}

From the ensemble of $N_\mathrm{ev} = 2^{21}$ initial energy density profiles and their correlation with the average state, the matrix $\rho(\mathbf{x})$ defined in \cref{eq:rho} can be constructed. Following the diagonalization procedure described in the theoretical section, the fluctuation modes are obtained. These modes are displayed in \cref{fig:modes_0-2.5} for the $0-2.5\%$ centrality class and in \cref{fig:modes_30-40} for the $30-40\%$ centrality class.

Two important features characterize these figures. First, the modes are ordered according to their statistical importance, starting from index $l = 0$ up to $l = 59$. Only the first $60$ modes are shown here; higher modes ($60 \le l \le 119$) are presented in \cref{appB}. Second, the plots display the normalized eigenmodes $\tilde{\Psi}_l$, which facilitates comparison by keeping all modes on the same scale. The physical modes are given by $\Psi_l = \sqrt{\lambda_l} \tilde{\Psi}_l$; they differ only by a normalization factor and preserve the spatial structure shown in \cref{fig:modes_0-2.5} and \cref{fig:modes_30-40}.

A first qualitative inspection of the central modes shown in \cref{fig:modes_0-2.5} reveals that they exhibit a larger radial symmetry compared to the non-central modes displayed in \cref{fig:modes_30-40}, which show more pronounced elliptical or rectangular structures. This difference reflects the role of the impact parameter in shaping the fluctuation modes.

It is important to note that the modes contain both positive and negative contributions. This is consistent with \cref{eq:modes_decomposition}, where the sign of the expansion coefficients $c_l$ must also be taken into account. Since the average state $\bar{\Psi}$ carries the dominant contribution to the total energy, the addition of a mode with negative regions does not lead to negative total energy density in the combined profile.

The most important mode is $l = 0$, whose shape closely resembles the corresponding average state, reflecting the dominant geometric feature of the event ensemble. In both centrality classes, the eigenvector associated with $l = 0$ appears with an overall negative sign. This sign has no physical meaning; it arises from the arbitrary phase convention in the diagonalization procedure of the Eigen library and does not affect the physical interpretation.

The modes generally follow recognizable symmetry patterns, such as monopole (round), dipole, quadrupole, sextupole, and octupole structures. For instance, in the central case, modes $l = 0$, $7$, $18$, and $31$ display similar radially symmetric shapes and contribute predominantly to the central region of the profile. In contrast, other modes concentrate their contributions toward the periphery, thereby modifying the overall geometry of the system. The detailed shape of each mode is directly connected to the underlying fluctuations in nucleon and hotspot positions, which originate from quantum fluctuations and the intrinsic nuclear structure. This explains why many fluctuation modes exhibit stronger structures in the peripheral region rather than in the center.

In the central case, rotational symmetries are clearly visible. For example, mode $l = 1$ corresponds to a dipole structure that, when rotated by $90^\circ$, results in mode $l = 2$. Similarly, mode $l = 3$ rotated by $45^\circ$ results in mode $l = 4$, and mode $l = 5$ rotated by $30^\circ$ results in mode $l = 6$. These symmetries are not merely visual; they are reflected in the eigenvalues as degenerate pairs. In contrast, such rotational symmetries are less pronounced in the non-central case, and no clear degeneracies are observed in the eigenvalue spectrum. Possible candidates, such as pairs $l = 1$ and $l = 3$, or $l = 22$ and $l = 25$, exhibit different eigenvalues and therefore cannot be considered degenerate states.

Comparing the $0-2.5\%$ and $30-40\%$ centrality classes, the non-central modes retain the general pattern of monopole, dipole, and quadrupole structures but additionally include modes that are absent in the central case, such as $l = 2$, $5$, $6$, $8$, and $14$. These additional structures arise from the larger variation of the impact parameter in non-central collisions, which introduces stronger anisotropies into the system.

The centrality-dependent modes can be compared to the fixed-impact-parameter cases studied in \cite{Borghini:2022iym}. The qualitative structure of the modes in the $0-2.5\%$ centrality class closely resembles the case with $b = 0$ fm, which is expected since most events in this bin correspond to nearly vanishing impact parameter. In contrast, the fixed case $b = 9$ fm shows certain similarities with the $30-40\%$ centrality class, particularly in the appearance of more rectangular structures. However, additional modes such as $l = 2$, $5$, $8$, and others visible in \cref{fig:modes_30-40} arise specifically from the variation of the impact parameter within the centrality bin and are absent in the fixed $b$ configuration.

\begin{figure}[h]
    \centering
    \includegraphics[width=0.6\linewidth]{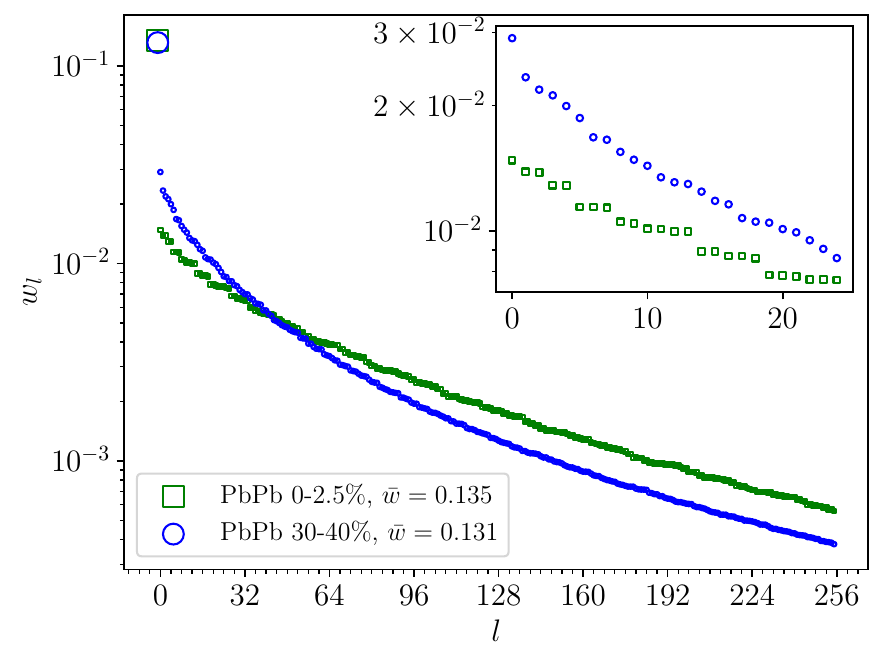}
    \caption{Relative weights defined in \cref{eq:weight_modes} for the first $256$ fluctuation modes in $0-2.5\%$ and $30-40\%$ centrality bins. The largest symbol at $l=-1$ represents $\bar{w}$, defined in \cref{eq:weight_avg}.}
\label{fig:weights_Pb}
\end{figure}

Turning now to a more quantitative description, the magnitude of the modes and their relative importance are determined by
\begin{equation}
    \norm{\Psi_l} = \sqrt{\lambda_l}\,\norm{\tilde{\Psi}_l} = \sqrt{\lambda_l},
\end{equation}
since the eigenvectors $\tilde{\Psi}_l$ are normalized. The eigenvalues $\lambda_l$ therefore quantify the statistical importance of each mode. One may plot $\lambda_l$ as a function of $l$, observing a logarithmic decay from $\lambda_0 \sim 10^6$ to values approaching zero for $l$ of order $10^3$. However, a more informative representation is given by the relative weights \cite{Borghini:2022iym},
\begin{equation}
    w_l \equiv \frac{\norm{\Psi_l}}{\sum_l \norm{\Psi_l} + \norm{\bar{\Psi}}} = \frac{\lambda_l}{\sum_l \lambda_l + \norm{\bar{\Psi}}},
\label{eq:weight_modes}
\end{equation}
where the sum over $l$ includes the first $6900$ modes. This truncation is justified because modes with $l \gtrsim 4000$ contribute negligibly, with eigenvalues of order $\lambda_l \sim 10^{-10}$ or smaller. Similarly, a relative weight can be defined for the average state,
\begin{equation}
    \bar{w} \equiv \frac{\norm{\bar{\Psi}}}{\sum_l \norm{\Psi_l} + \norm{\bar{\Psi}}} = \frac{\norm{\bar{\Psi}}}{\sum_l \lambda_l + \norm{\bar{\Psi}}}.
\label{eq:weight_avg}
\end{equation}
These relative weights are displayed in \cref{fig:weights_Pb} for the two centrality classes analyzed in this work.

The quantity $\bar{w}$ in \cref{fig:weights_Pb} (represented by the larger symbol at $l = -1$) indicates that the average state provides the largest individual contribution, approximately $\sim 13\%$, for both centrality classes. The first fluctuation mode, $l = 0$, contributes about $3\%$ in the non-central case and less than $2\%$ in the central case. This behavior is consistent with the expectation that non-central events contain large fluctuations. In addition to the usual fluctuations arising from uncertainties in nucleon positions, non-central collisions also exhibit geometric fluctuations associated with the variation of the impact parameter.

Comparing the two centrality classes, the $0-2.5\%$ bin displays a flatter decrease of $w_l$ with increasing $l$, indicating a broader distribution of fluctuation strength across modes. In contrast, the non-central case shows a steeper decline, reflecting stronger correlations among the leading modes.

From \cref{fig:weights_Pb}, the degeneracy of certain modes discussed earlier can also be observed. For central collisions, modes $l = 1$ and $l = 2$ share the same eigenvalue, demonstrating degeneracy. However, this is not the case for modes $l = 1$ and $l = 3$ in the non-central bin, which have similar spatial structures but are not degenerate due to the presence of mode $l = 2$ between them in the ordered spectrum. It is worth noting that this lifting of degeneracy appears only in the centrality-based analysis. In the fixed $b = 9$ fm case \cite{Borghini:2022iym}, the corresponding modes remained degenerate.

Before proceeding to the next subsection and discussing the quantitative impact of the modes $\{\Psi_l\}$ on observables, it is important to verify that the statistical properties of the expansion coefficients are indeed satisfied in the numerical implementation.

\begin{figure}[h]
    \centering
    \includegraphics[width=0.49\linewidth]{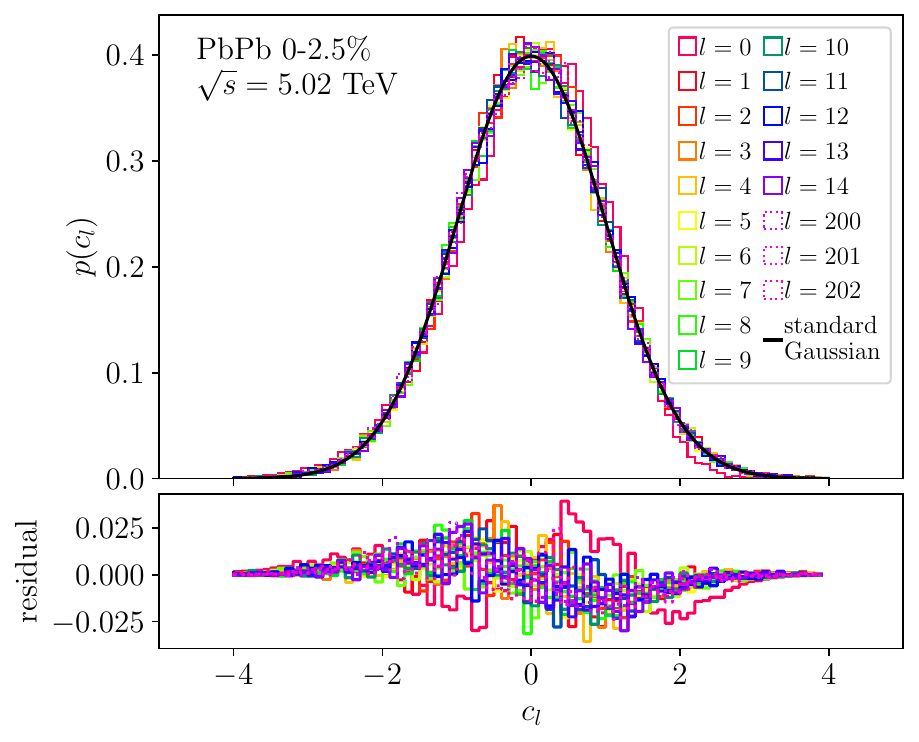}
    \includegraphics[width=0.49\linewidth]{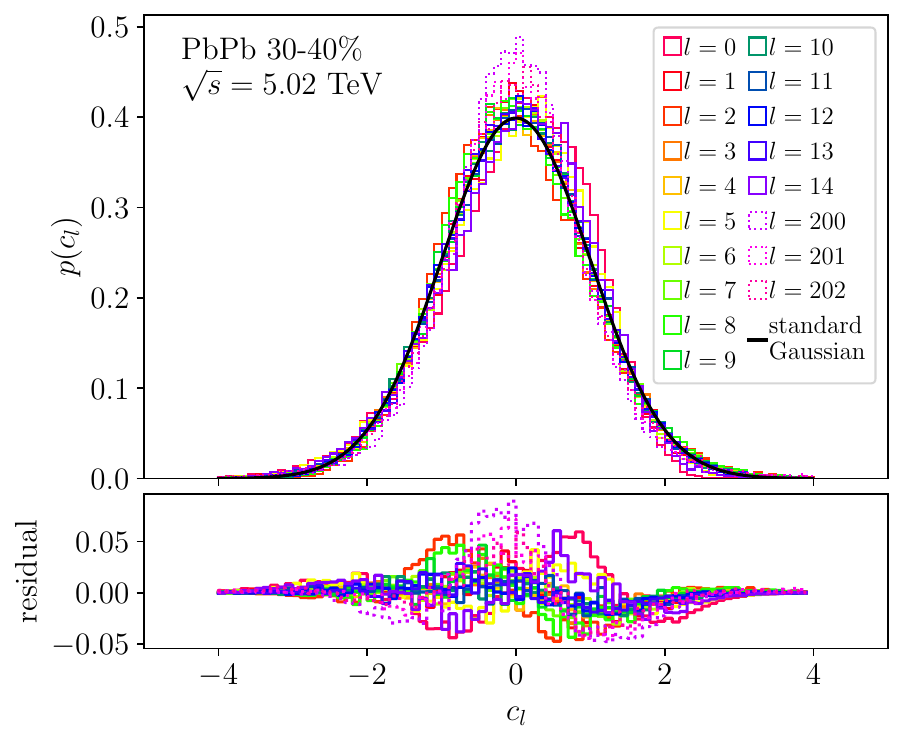}
    \caption{Probability distribution of the expansion coefficients $\{c_l\}$ for the first $15$ modes and for three higher modes. The distributions are compared to a standard Gaussian with unit variance. The residual panel indicates good agreement with the Gaussian shape. The coefficients $\{c_l\}$ were obtained from a decomposition of $2^{15}$ initial profiles for $0-2.5\%$ (on left) and $30-40\%$ (on right).}
\label{fig:coefficients_cl}
\end{figure}

\Cref{fig:coefficients_cl} shows the relative frequency distribution of the expansion coefficients for $2^{15}$ initial profiles, obtained using the decomposition in \cref{eq:modes_decomposition}, for each centrality bin. The distribution of the coefficients $\{c_l\}$, independently of the index $l$, is in good agreement with a standard Gaussian distribution. The maximum residual deviation is approximately $0.025$ for the central bin and $0.05$ for the non-central bin. Mode $l = 0$ exhibits a noticeable skewness in both centrality bins, with the peak of the distribution shifted toward positive values of $c_l$ and an overall negative skewness. This behavior can be understood by comparing \cref{fig:modes_0-2.5} and \cref{fig:modes_30-40}, where mode $l = 0$ is seen to contribute negatively to the energy density. Considering an event constructed as $\Phi = \bar{\Psi} + c_0 \Psi_0$, and recalling that the total energy must remain positive, the expansion coefficient $c_0$ is constrained from taking large positive values. In contrast, negative values of $c_0$ are allowed to enhance the total energy and, consequently, the final multiplicity within the given centrality class. This asymmetry leads to the observed negative skewness in the distribution.

Additional statistical properties, including the first four moments (mean, variance, skewness, and kurtosis), are analyzed in \cref{appC}. Higher modes such as $l = 200$, $201$, and $202$ display narrower distributions, characterized by sharper peaks around $c_l = 0$. These results confirm that the expansion coefficients remain of order unity, validating the use of the Taylor expansion of observables in \cref{eq:taylor_expansion_O}.

\begin{figure}[h]
    \centering
    \includegraphics[width=0.49\linewidth]{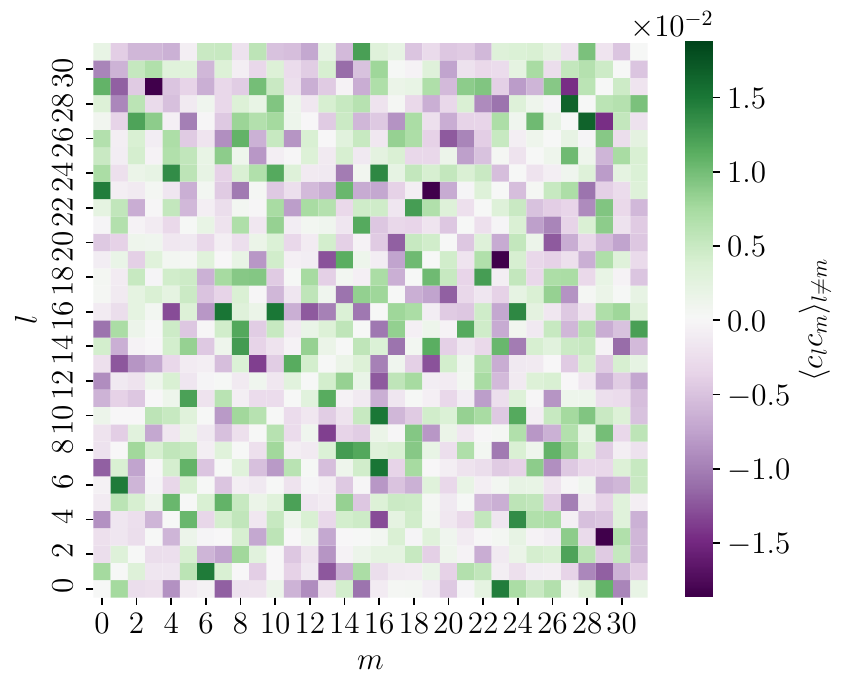}
    \includegraphics[width=0.49\linewidth]{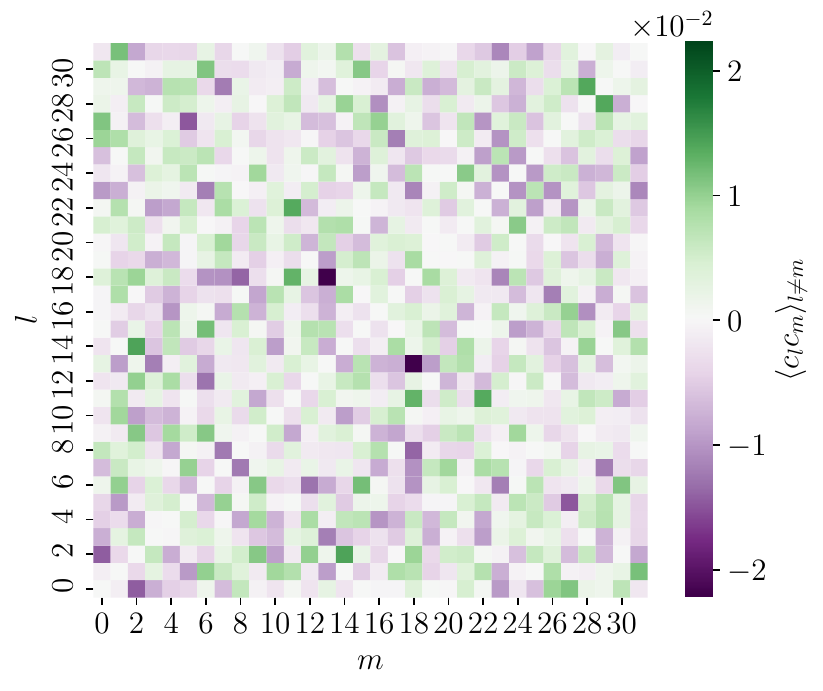}
    \caption{Cross-correlation matrix of the expansion coefficients $\langle c_l c_m \rangle$ for different modes in the $0-2.5\%$ on left and $30-40\%$ on right. The diagonal elements are set to zero in the visualization, since by construction they satisfy $\langle c_l^2 \rangle = 1$.}
\label{fig:correlation_clcm}
\end{figure}

From \cref{fig:coefficients_cl}, it is also evident that $\langle c_l \rangle = 0$, a result further confirmed in \cref{appC}, where the explicit averages are shown to be consistent with zero and the variances satisfy $\langle c_l^2 \rangle = 1$. The uncorrelated nature of the coefficients is demonstrated in \cref{fig:correlation_clcm}, where $\langle c_l c_m \rangle$ for $l \neq m$ is shown to be consistent with zero, with deviations of at most order $10^{-2}$.

This concludes the present section, where the average state and fluctuation modes were introduced and analyzed in light of their theoretical properties. In the following subsection, observables will be computed for each mode in order to quantify their individual contributions.

\subsection{Initial-state observables}
\label{subsec:ISobservables}

From the previous section, it was shown that the expansion coefficients $\{c_l\}$ are sufficiently small to justify the Taylor expansion of observables in \cref{eq:taylor_expansion_O}. The remaining requirement for computing the response coefficients is that the observable be smooth enough to allow the derivatives defining $L_{\alpha,l}$ and $Q_{\alpha,ll'}$ in \cref{eq:linear_response} and \cref{eq:quadratic_response} to exist.

\begin{figure}[h]
    \centering
    \includegraphics[width=0.49\linewidth]{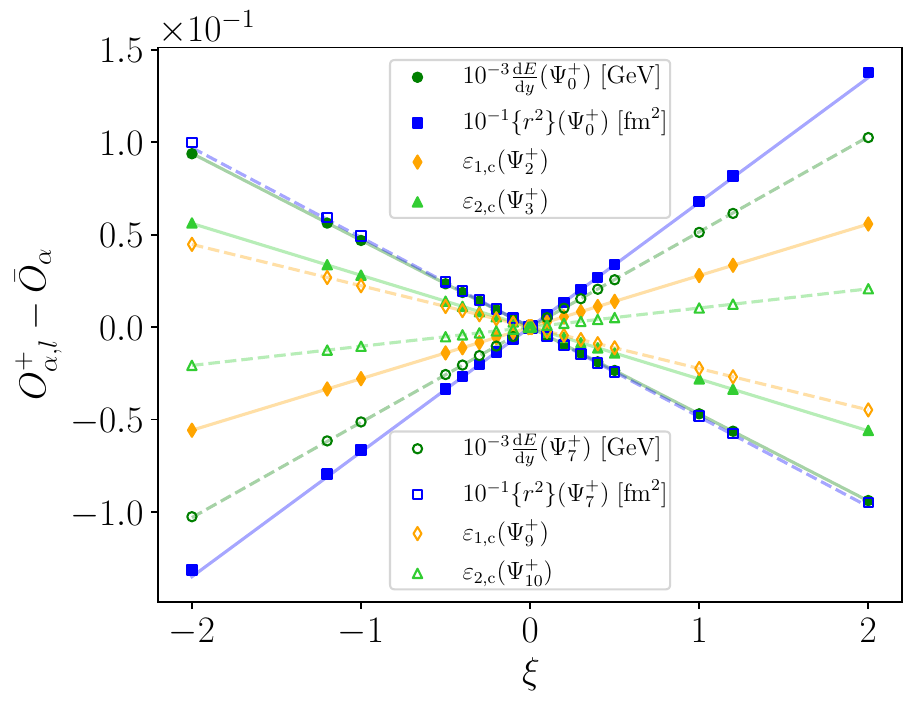}
    \includegraphics[width=0.49\linewidth]{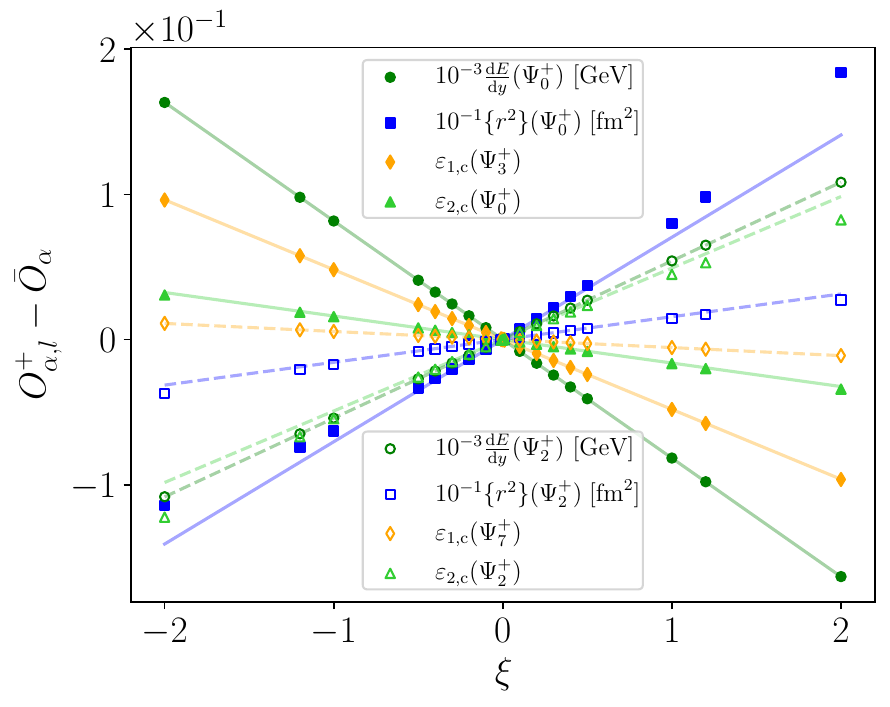}
    \caption{Linearity check of selected initial observables for central (left) and non-central (right) modes as a function of the parameter $\xi$. The filled symbols correspond to the leading modes contributing to each observable, while the open symbols represent the subleading modes. The lines guide the eyes to show the deviations of a linear behavior.} 
\label{fig:linearity}
\end{figure}

To evaluate these derivatives numerically, two auxiliary states are introduced:
\begin{equation}
    \Psi^+_l \equiv \bar{\Psi} + \xi \Psi_l 
    \quad\text{and}\quad
    \Psi^-_l \equiv \bar{\Psi} - \xi \Psi_l,
\label{eq:psi+-}
\end{equation}
which correspond to adding or subtracting a single mode $\Psi_l$ from the average state $\bar{\Psi}$. The parameter $\xi$ is a small constant that controls the perturbation strength and will be discussed below. For these states, the corresponding observables are defined as
\begin{equation}
    O^\pm_{\alpha, l} \equiv O_\alpha(\Psi^\pm_l) = O_\alpha(\bar{\Psi} \pm \xi \Psi_l).
\end{equation}
Using this notation, the linear response coefficient can be written as
\begin{equation}
    L_{\alpha, l} \equiv \frac{\partial O_\alpha}{\partial c_l}\bigg|_{\bar{\Psi}} = \frac{O^+_{\alpha, l} - O^-_{\alpha, l}}{2\xi},
\label{eq:linear_response_open}
\end{equation}
and the quadratic response coefficient as
\begin{equation}
    Q_{\alpha, ll} \equiv \frac{\partial^2 O_\alpha}{\partial c_l^2}\bigg|_{\bar{\Psi}} = \frac{O^+_{\alpha, l} + O^-_{\alpha, l} -2\bar{O}_\alpha}{\xi^2}.
\label{eq:quadratic_response_open}
\end{equation}
From these expressions, it is clear that $\xi$ must be chosen carefully. It cannot be too large, since the derivative approximation requires that the observable behave linearly in $\xi$ between $O_{\alpha,l}^+$ and $O_{\alpha,l}^-$. On the other hand, $\xi$ cannot be too small, because the difference in the numerator must remain numerically resolvable and not dominated by statistical noise. In this work, $\xi$ was tested in the range $[0.001, 0.5]$, as shown in \cref{fig:linearity}. Values $\xi > 0.5$ generally lead to visible deviations from linear behavior for some observables. The choice $\xi = 0.5$ is adopted throughout this chapter; the motivation for this choice, rather than a smaller value, will become clearer in the following subsection.

We now discuss the specific observables considered in this analysis, several of which were already presented in \cref{fig:linearity}. In general, $O_\alpha$ may represent any sufficiently smooth observable. A simple initial observable is the total energy per unit space-time rapidity
\begin{equation}
    \frac{\mathrm{d}E}{\mathrm{d}\eta_s} \equiv \tau_0 \int e(r, \theta) \, r \, \mathrm{d}r \, \mathrm{d}\theta,
\end{equation}
which represents the total energy in the transverse plane for the boost-invariant case. This observable is strictly linear under the addition of fluctuations and therefore does not contain quadratic contributions. Another initial observable is the energy-weighted average squared radius
\begin{equation}
    \{r^2\} \equiv \frac{\int r^2 \, e(r, \theta) \, r \, \mathrm{d}r \, \mathrm{d}\theta}{\int e(r, \theta) \, r \, \mathrm{d}r \, \mathrm{d}\theta},
\end{equation}
which encodes geometric information about the spatial distribution of energy. Because of the ratio structure, quadratic contributions may arise from the denominator.

The most widely used initial observables are the eccentricities,
\begin{equation}
\begin{aligned}
    \varepsilon_1 e^{i \Phi_1} &\equiv -\frac{\int r^3 \, e^{i\theta} \, e(r, \theta) \, r \, \mathrm{d}r \, \mathrm{d}\theta}{\int r^3 \, e(r, \theta) \, r \, \mathrm{d}r \, \mathrm{d}\theta} \hspace{0.6cm}\text{for}\,\, n = 1, \\
    \varepsilon_n e^{i n \Phi_n} &\equiv - \frac{\int r^n \, e^{in\theta} \, e(r, \theta) \, r \, \mathrm{d}r \, \mathrm{d}\theta}{\int r^n \, e(r, \theta) \, r \, \mathrm{d}r \, \mathrm{d}\theta} \hspace{0.6cm} \text{for}\,\, n \geq 2.
\end{aligned}
\label{eq:eccentricities}
\end{equation}
The modulus $|\varepsilon_n|$ is not differentiable with respect to small perturbations and therefore does not satisfy the linearity condition required for the Taylor expansion. In particular, a linearity test, as in \cref{fig:linearity}, would yield a characteristic ``V'' shape for $|\varepsilon_n|$, indicating nondifferentiability. Instead, one considers the complex decomposition
\begin{equation}
    \varepsilon_n e^{in\Phi_n} \equiv \varepsilon_{n, c} + i \varepsilon_{n, s},
\end{equation}
where $\varepsilon_{n,c}$ and $\varepsilon_{n,s}$ denote the real (cosine) and imaginary (sine) components, respectively. These components are smooth and linear under the addition of fluctuations, and are therefore suitable for the response analysis.

The linearity of the observables with respect to $\xi$ was tested in \cref{fig:linearity}. It displays the dependence of $O_{\alpha,l}^+ - \bar{O}_\alpha$ on $\xi$ for some modes. The points correspond to numerical evaluations for different values of $\xi$ entering \cref{eq:psi+-}, and the lines indicate ideal linear behavior. For sufficiently small $\xi$, the dependence is linear, confirming the validity of the derivative approximation. For $\xi > 0.5$, deviations from linearity become apparent for certain observables.

With this established, the linear response coefficient $L_{\alpha,l}$ is determined for each observable discussed above and for each fluctuation mode, using \cref{eq:linear_response_open}. In this way, it becomes possible to quantify how strongly each mode contributes to a given observable.

\begin{figure}[h]
    \centering
    \includegraphics[width=0.49\linewidth]{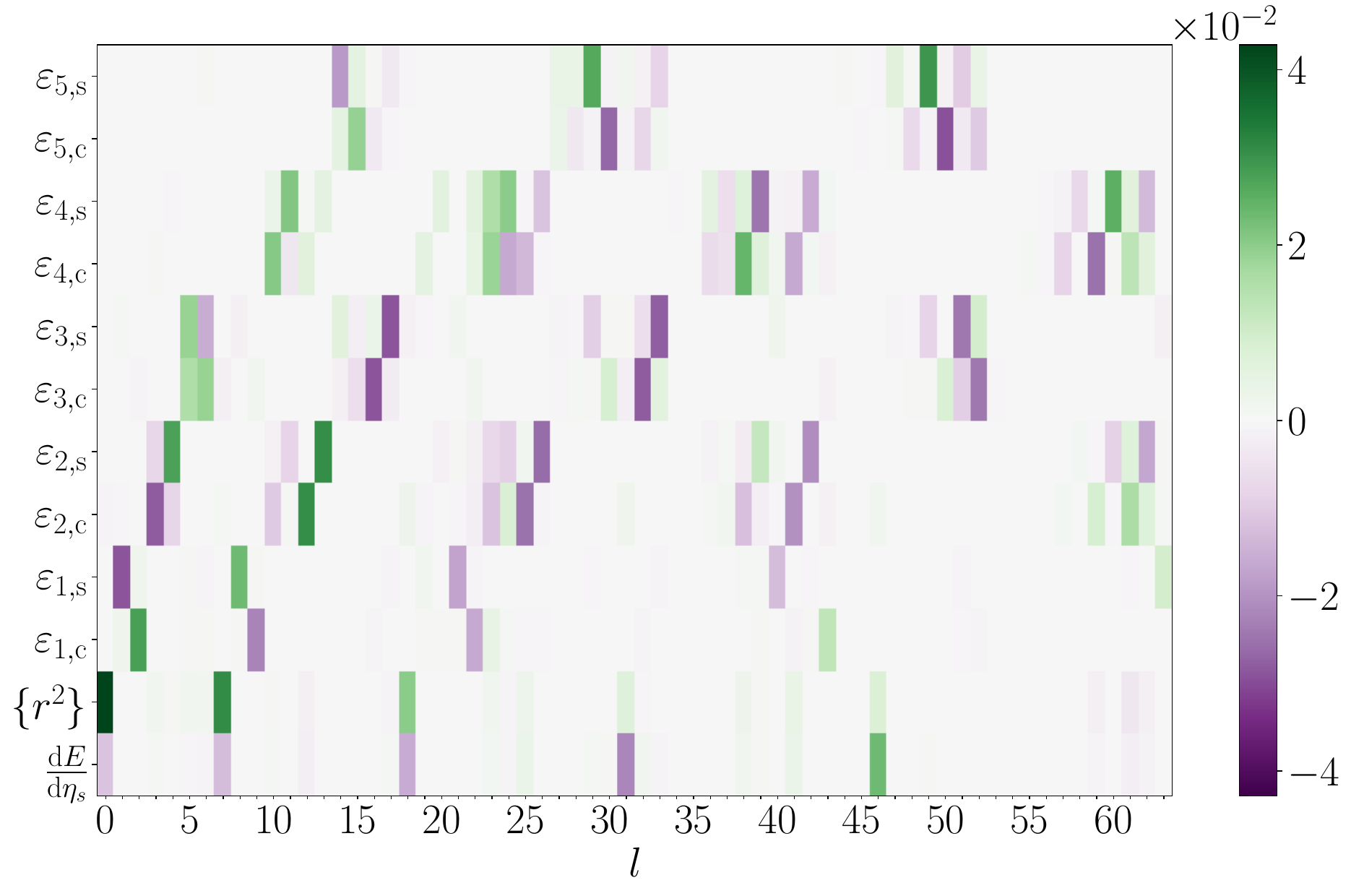}
    \includegraphics[width=0.49\linewidth]{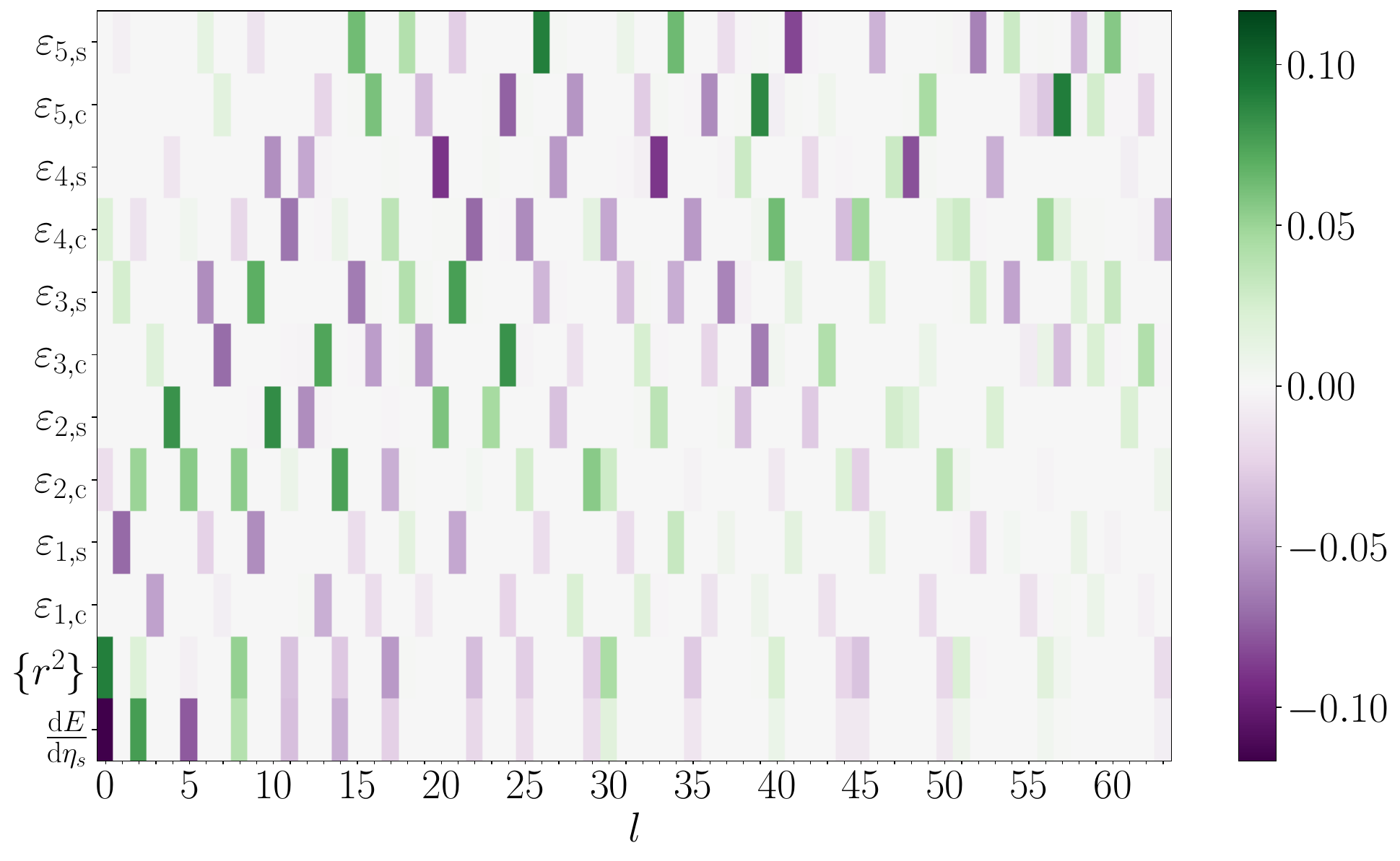}
    \caption{Linear response coefficients $L_{\alpha, l}$ for the initial observables as function of the first $64$ modes, shown for centrality $0-2.5\%$ (left) and $30-40\%$ (right). The coefficients corresponding to $\mathrm{d}E/\mathrm{d}\eta_s$ and $\{r^2\}$ are normalized by their respective average-state values $\bar{O}_\alpha$, ensuring that they are dimensionless.}
    \label{fig:IS_linear_response}
\end{figure}

\Cref{fig:IS_linear_response} displays, on the vertical axis, the initial observables considered in this study as functions of the first $64$ modes. The mode index $l$ follows the same ordering as in the previous figures and corresponds to the ordering of the eigenvalues. The color scale in the plot indicates the magnitude of the contribution of mode $l$ to the observable $O_\alpha$.

As expected from the eigenvalue ordering, the dominant contributions to the observables arise from the lowest-index modes (small $l$), while higher modes contribute progressively less.

Analyzing the left panel of \cref{fig:IS_linear_response}, corresponding to central collisions, the signal appears cleaner and more structured than in the non-central case. In general, each mode contributes predominantly to a single observable or to correlated observables, such as $\varepsilon_n$ and $\varepsilon_{n+2}$ (for $n \ge 2$). This allows one to associate a specific geometric structure of the mode with the corresponding observable.

Using \cref{fig:modes_0-2.5} as a reference for the shapes of the central modes, it is evident that modes such as $l = \{0, 7, 18, 31, 46, \dots \}$, which primarily carry energy, share the same radially symmetric structure as the average state. These radially symmetric modes are precisely those that contribute to the squared radius $\{r^2\}$. The eccentricities appear in pairs corresponding to nearly degenerate states related by rotation, such as $l = 1$ and $l = 2$, or $l = 3$ and $l = 4$. These pairs contribute to the same eccentricity, either through the cosine or sine component. Geometrically, the modes contributing to $\varepsilon_1$, for example $l = \{1, 2, 8, 9, 21, 22, \dots\}$, display a dipole structure. This is expected, since $\varepsilon_1$ characterizes dipolar asymmetry in the energy distribution. Similarly, the elliptic eccentricity $\varepsilon_2$ is associated with quadrupole-shaped modes such as $l = \{3, 4, 12, 13, 25, 26, \dots \}$. The triangular eccentricity $\varepsilon_3$ corresponds to sextupole-shaped modes, followed by octupole structures for $\varepsilon_4$, and so forth.

In this central case, where each mode predominantly contributes to a single observable, the geometric interpretation is particularly transparent. For example, a quadrupole structure naturally leads to elliptic eccentricity. In contrast, for the non-central case (right panel of \cref{fig:IS_linear_response}), the structure is more complex. As noted earlier, non-central modes generally carry a nonzero contribution to $\varepsilon_2$. The right panel shows that individual modes contribute to multiple observables simultaneously. Consequently, the geometric interpretation becomes less straightforward, although the qualitative insights from the central case remain valid.

The modes in \cref{fig:modes_30-40} for the non-central bin exhibit a more intricate structure. This increased complexity arises not only from the intrinsic elliptical geometry of the average state but also from the additional fluctuations associated with the larger variation of the impact parameter. Each mode may now contribute to several observables simultaneously, and the correspondence between geometry and observable is less clean than in the central case. This is consistent with the structure of the average state: in the central bin, it primarily contributes to the total energy, whereas in the non-central bin it contributes to energy, $\varepsilon_2$, and even a small $\varepsilon_4$.

For the non-central modes in \cref{fig:modes_30-40}, those with an elliptical structure (e.g., $l = \{0, 2, 5, 8, \dots\}$) contribute to $\varepsilon_2$ as well as to the total energy, reflecting their similarity to the average state. Dipole modes such as $l = \{1, 3, 6, 9, \dots\}$ still encode the $\varepsilon_1$ contribution, while quadrupole-like modes (e.g., $l = \{4, 10, 12, \dots \}$) contribute more strongly to $\varepsilon_2$. There are also more complex structures, such as mode $l = 11$, which exhibits a rectangular shape and contributes to both energy and $\varepsilon_4$, or mode $l = 28$, which mixes contributions from $\varepsilon_1$, $\varepsilon_3$, and $\varepsilon_5$. These mixed contributions arise from the interplay between the intrinsic elliptical geometry and the event-by-event variation of the impact parameter.

The magnitude of the linear response coefficients is generally smaller in the central case. This reflects the fact that central events exhibit weaker geometric deformation, and therefore fewer corrections to the observables relative to the average state are required. The linear response for higher modes with $l \ge 63$ contributes only marginally to the observables.

\begin{figure}[h]
    \centering
    \includegraphics[width=0.49\linewidth]{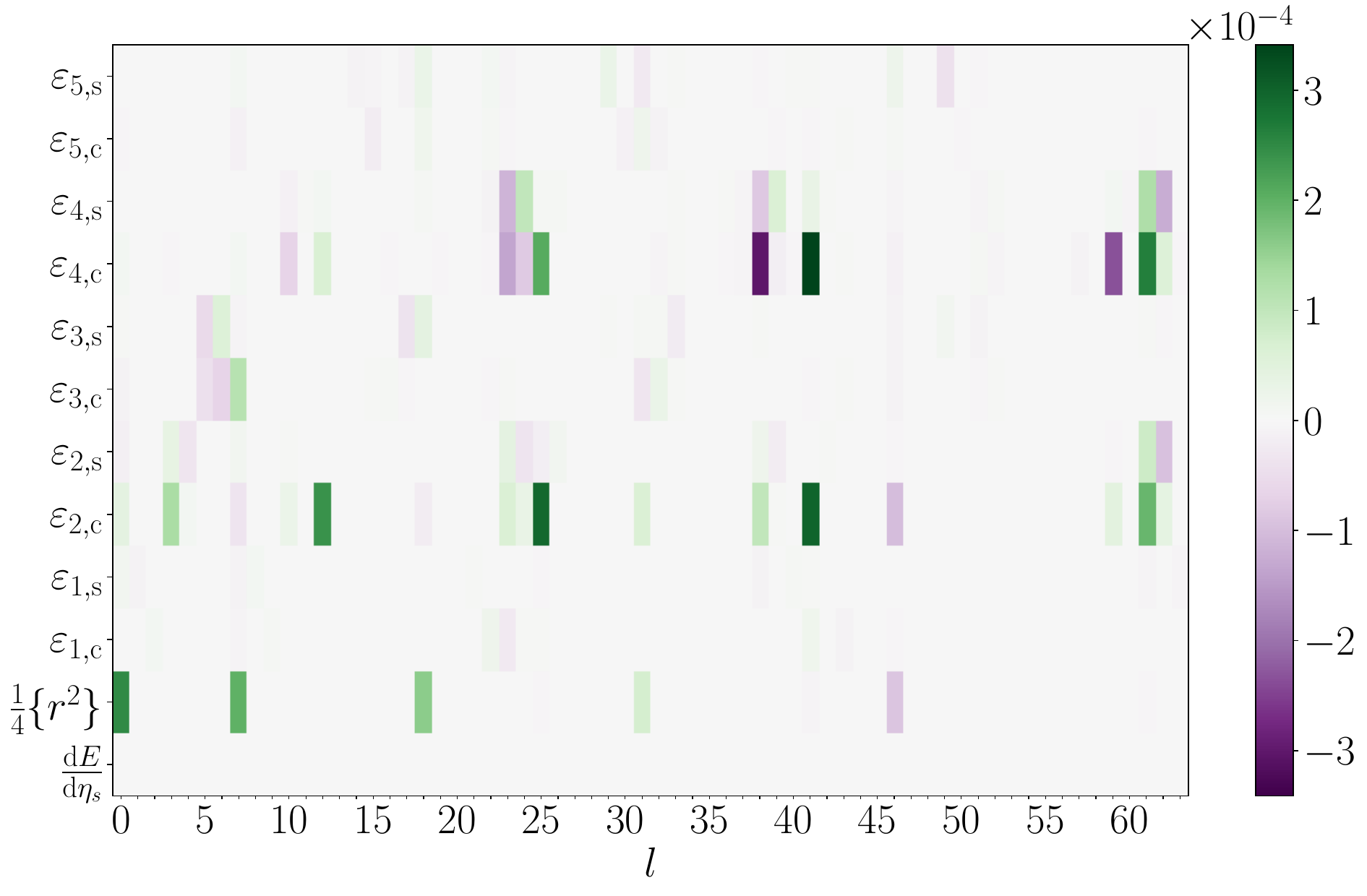}
    \includegraphics[width=0.49\linewidth]{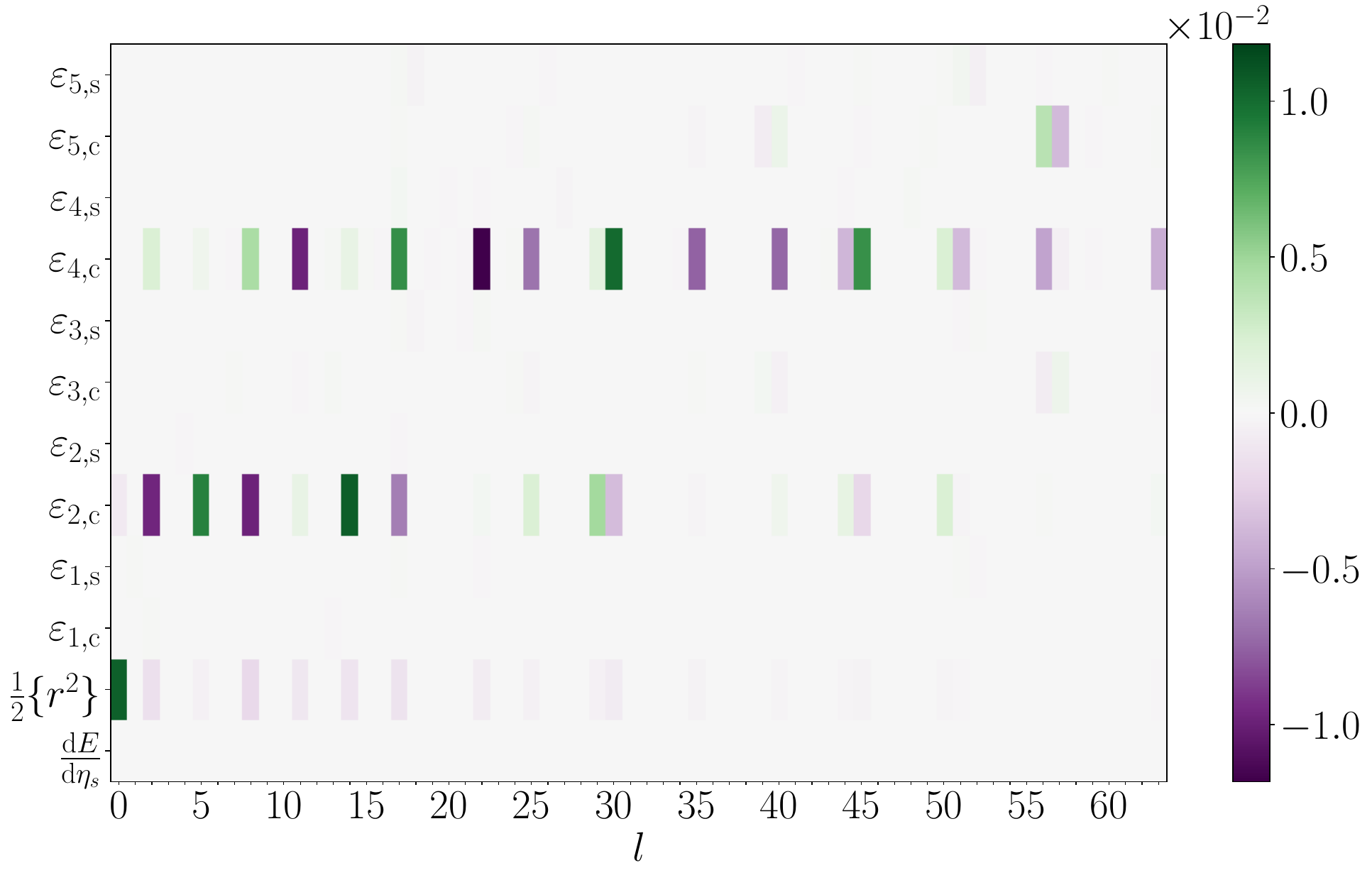}
    \caption{Quadratic response coefficient $Q_{\alpha, ll}$ for the initial observables as function of the first $64$ modes for events in centrality $0-2.5\%$ (left) and $30-40\%$ (right). The coefficients corresponding to $\mathrm{d}E/\mathrm{d}\eta_s$ and $\{r^2\}$ are normalized by their respective average-state values $\bar{O}_\alpha$, ensuring that they are dimensionless.}
\label{fig:IS_quadratic_response}
\end{figure}

Turning to the quadratic response coefficients shown in \cref{fig:IS_quadratic_response} for central (left) and non-central (right) collisions, the expected hierarchy from the Taylor expansion in \cref{eq:taylor_expansion_O} is observed: the second-order contributions are smaller than the linear ones, i.e., $L_{\alpha,l} > Q_{\alpha,ll}$ for both centralities.

As previously discussed, the total energy is strictly linear, and therefore $Q_{\mathrm{d}E/\mathrm{d}\eta_s,l} = 0$. In contrast, the squared radius $\{r^2\}$ exhibits a quadratic contribution in the same modes that appear in the linear response. For the eccentricities, all sine components aligned with the $y$-direction are approximately zero, $\varepsilon_{n,s} \simeq 0$. For the cosine components aligned with the $x$-direction (the direction of the impact parameter), odd harmonics vanish, $\varepsilon_{n,c} \simeq 0$ for odd $n$. The even eccentricities display nonzero quadratic response coefficients, primarily reflecting the underlying elliptical geometry.

In the fixed $b = 0$ case discussed in Ref. \cite{Borghini:2022iym}, no quadratic contributions appear. Here, however, both centrality bins include a distribution of impact parameters, which introduces a quadratic dependence associated with fluctuations in the smaller axis of the ellipse (the $x$-direction in this setup). Since the $30-40\%$ bin corresponds to a more pronounced elliptical geometry, the quadratic response is larger in this case compared to the central bin.

To conclude this section, we introduce a more quantitative measure of the contribution of each mode through the definition of a complex ``eccentricity'' for the individual modes:
\begin{equation}
\begin{aligned}
    \tilde{\varepsilon}_1(\Psi_l) &\equiv -\frac{\int r^3 \, e^{i\theta} \, \Psi_l(r, \theta) \, r \, \mathrm{d}r \, \mathrm{d}\theta}{\int r^3 \, \bar{\Psi}(r, \theta) \, r \, \mathrm{d}r \, \mathrm{d}\theta}  \hspace{0.6cm}\text{for}\,\, n = 1, \\ 
    \tilde{\varepsilon}_n(\Psi_l) &\equiv -\frac{\int r^n \, e^{in\theta} \, \Psi_l(r, \theta) \, r \, \mathrm{d}r \, \mathrm{d}\theta}{\int r^n \, \bar{\Psi}(r, \theta) \, r \, \mathrm{d}r \, \mathrm{d}\theta}  \hspace{0.6cm}\text{for}\,\, n \ge 2.
\end{aligned}
\label{eq:tilde_eccentricities}
\end{equation}
The difference from the standard eccentricity definition lies in the denominator, which uses the average state $\bar{\Psi}$ instead of the mode itself. This choice ensures that the denominator remains nonzero and provides a meaningful normalization. It is also useful to define a ``zeroth eccentricity''
\begin{equation}
    \tilde{\varepsilon}_0(\Psi_l) \equiv \frac{\int \Psi_l(r, \theta) \, r \, \mathrm{d}r \, \mathrm{d}\theta}{\int \bar{\Psi} \, r \, \mathrm{d}r \, \mathrm{d}\theta} = \frac{\frac{\mathrm{d}E}{\mathrm{d}\eta_s}(\Psi_l)}{\frac{\mathrm{d}E}{\mathrm{d}\eta_s}(\bar{\Psi})},
\end{equation}
which represents the relative energy content of mode $l$ per unit space-time rapidity, normalized to the corresponding energy of the average state. This quantity coincides with the linear response coefficient for the total energy,
\begin{equation}
    L_{dE/d\eta_s, l} = \frac{\frac{\mathrm{d}E}{\mathrm{d}\eta_s}(\Psi^+) - \frac{\mathrm{d}E}{\mathrm{d}\eta_s}(\Psi^-)}{2\xi} = \frac{\frac{\mathrm{d}E}{\mathrm{d}\eta_s}(\bar{\Psi}) + \xi \frac{\mathrm{d}E}{\mathrm{d}\eta_s}(\Psi_l) - \frac{\mathrm{d}E}{\mathrm{d}\eta_s}(\bar{\Psi}) + \xi \frac{\mathrm{d}E}{\mathrm{d}\eta_s}(\Psi_l)}{2\xi} = \frac{\mathrm{d}E}{\mathrm{d}\eta_s}(\Psi_l),
\end{equation}
where, in the linear approximation, the contributions from $\bar{\Psi}$ cancel. Since in \cref{fig:IS_linear_response} and \cref{fig:IS_quadratic_response} the response coefficient for the energy is normalized by the average-state contribution, the quantity $\tilde{\varepsilon}_{0,l}$ corresponds directly to the plotted linear response coefficient for the energy. This direct correspondence does not hold for the higher-order eccentricities.

\begin{figure}[h]
    \centering
    \includegraphics[width=\linewidth]{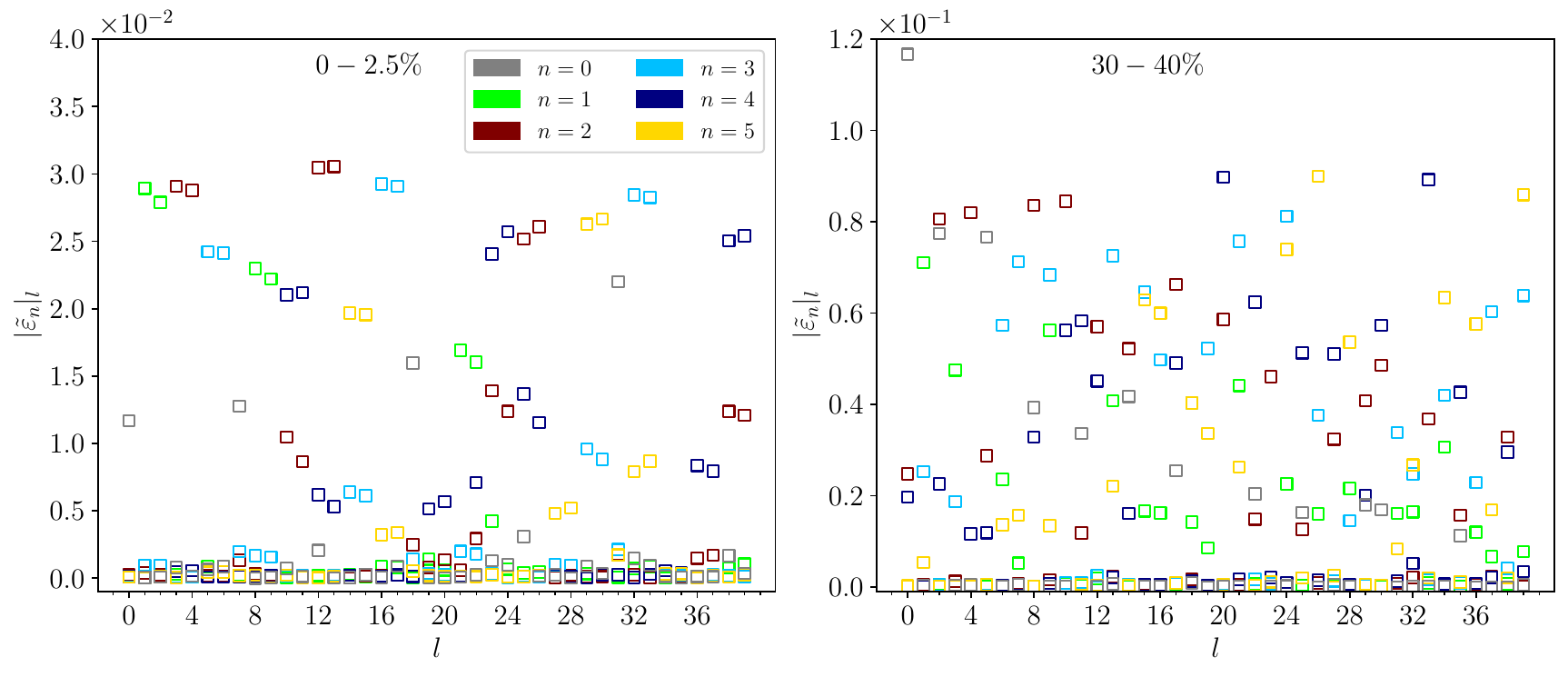}
    \caption{Relative energy given by $|\tilde{\varepsilon}_0|_l$ and the modulus of the spatial anisotropies $|\tilde{\varepsilon}_n|_l$ with $1 \le n \le 5$ that are defined in \cref{eq:tilde_eccentricities} in terms of the modes (here only the first $40$ modes are represented). Central modes $0-2.5\%$ are on the left and non-central $30-40\%$ are on the right side.}
\label{fig:tilde_eccentricities}
\end{figure}

\Cref{fig:tilde_eccentricities} shows $|\tilde{\varepsilon}_n|_l$ for the first $40$ modes. This representation conveys the same qualitative information as the linear-response plots, but in a more geometrically intuitive form. In the central case, nearly degenerate mode pairs are clearly visible, whereas this degeneracy is absent in the non-central bin. The non-central distribution again exhibits a more intricate structure, with stronger mixing between different harmonic contributions.

For the central bin, the relative energy $|\tilde{\varepsilon}_0|_l$ is concentrated in modes $l = 0, 7, 18, 31$, consistent with \cref{fig:IS_linear_response}. A new quantitative insight is that each of these modes contributes approximately $1-2\%$ of the total energy of the average state. In contrast, for the non-central bin, the relative energy is distributed among approximately $12$ of the first $40$ modes. Mode $l = 0$ contributes about $12\%$ of the average-state energy, while modes $l = 2$ and $l = 5$ contribute roughly $8\%$. The remaining relevant modes contribute smaller fractions, typically in the range $1-4\%$. This enhanced contribution is a new feature that arises from the variation of the impact parameter within the centrality bin.

For comparison, in the fixed-impact-parameter analysis of Ref. \cite{Borghini:2022iym}, the largest mode contributions were approximately $1.5\%$ for $b = 0$ fm and $\sim 4\%$ for $b = 9$ fm. The significantly larger contribution of $12\%$ observed here emerges from fluctuations associated with the impact-parameter distribution. Since in \textsc{MCGlauber} the fluctuations from nucleon positions are intertwined with fluctuations in $b$, a simplified toy model based on the Optical Glauber framework is introduced in \cref{subsec:optical} to disentangle these effects and analyze them in more detail.

So far, the fluctuation modes have been determined and linked to the initial observables, establishing a direct connection between the geometry of each mode and its physical interpretation. The next subsection presents the final observables.

\subsection{Final-state observables}
\label{subsec:FSobservables}

In the first subsection, the uncorrelated basis of fluctuation modes was introduced; in the second part, these modes were linked to specific initial observables through the corresponding response coefficients. In the present subsection, the average state $\bar{\Psi}$ and the modes $\{\Psi_l\}$, more precisely the states $\{\Psi_l^+\}$ and $\{\Psi_l^-\}$, are evolved within a hydrodynamic framework and the response coefficients for final observables are computed. This allows us to establish a direct connection between final observables and the fluctuations in the initial state.

The final observables considered here are among the most fundamental quantities in heavy-ion collisions. All are chosen to be linear in the sense required for the Taylor expansion in~\cref{eq:taylor_expansion_O}. The first basic observable is the charged-particle multiplicity per unit pseudorapidity
\begin{equation}
    \frac{\mathrm{d}N_\mathrm{ch}}{\mathrm{d}\eta} = \int \frac{\mathrm{d}N_\mathrm{ch}}{p_T \, \mathrm{d}p_T \, \mathrm{d}\varphi_p \, \mathrm{d}\eta} \, p_T \, \mathrm{d}p_T \, \mathrm{d}\varphi_p,
\end{equation}
where the differential momentum spectrum of charged hadrons, $\mathrm{d}N_\mathrm{ch} / (p_T \, \mathrm{d}p_T \, \mathrm{d}\varphi_p \, \mathrm{d}\eta)$, is evaluated in the region $|\eta| \le 0.5$. Here, $p_T$ denotes the transverse momentum and $\varphi_p$ the azimuthal angle of the transverse momentum. From this distribution, one can also compute the average transverse momentum
\begin{equation}
    [p_T] \equiv \frac{\int p_T \, \frac{\mathrm{d}N_\mathrm{ch}}{p_T \, \mathrm{d}p_T \, \mathrm{d}\varphi_p \, \mathrm{d}\eta} \, p_T \, \mathrm{d}p_T \, \mathrm{d}\varphi_p}{\int \frac{\mathrm{d}N_\mathrm{ch}}{p_T \, \mathrm{d}p_T \, \mathrm{d}\varphi_p \, \mathrm{d}\eta} \, p_T \, \mathrm{d}p_T \, \mathrm{d}\varphi_p}.
\end{equation}
Another important observable is the set of flow coefficients defined through the Fourier expansion introduced in \cref{eq:flow_coefficients}
\begin{equation}
    v_n e^{in\Psi_n} \equiv \frac{\int e^{in\varphi_p} \, \frac{\mathrm{d}N_\mathrm{ch}}{p_T \, \mathrm{d}p_T \, \mathrm{d}\varphi_p \, \mathrm{d}\eta} \, p_T \, \mathrm{d}p_T \, \mathrm{d}\varphi_p}{\int \frac{\mathrm{d}N_\mathrm{ch}}{p_T \, \mathrm{d}p_T \, \mathrm{d}\varphi_p \, \mathrm{d}\eta} \, p_T \, \mathrm{d}p_T \, \mathrm{d}\varphi_p}.
\end{equation}
For the same linearity reasons discussed for the initial eccentricities, the flow coefficients are decomposed into cosine and sine components $v_n e^{in\Psi_n} = v_{n,c} + i v_{n,s}$. This decomposition ensures that the response analysis can be consistently applied to the final-state observables.

It is important to emphasize that the final observables are always computed from the final momentum spectra of particles in each event. The evolution follows the standard hybrid setup described in \cref{sec:numerical}, the main difference in the present work is that the evolution is not performed event-by-event; instead, it is carried out mode-by-mode.

Specifically, an ensemble of $N_\mathrm{ev} = 2^{21}$ events is first generated for a given centrality class. After performing the mode decomposition, only the fluctuation modes are evolved. Since the modes individually carry only a small fraction of the total energy (see $\tilde{\varepsilon}_0$ in \cref{fig:tilde_eccentricities}), they are not evolved independently. Rather, each mode is coupled to the average state, and the evolution is performed for the perturbed configurations $\Psi_l^+$ and $\Psi_l^-$ defined in \cref{eq:psi+-}, for each mode $l$ under consideration.

For the first analysis, the evolution of the modes is performed using \Kompost~and \textsc{MUSIC} until the system reaches the switching temperature of $155$ MeV. At this stage, \textsc{MUSIC} is used in a dedicated mode to compute the full momentum spectra for $320$ hadronic species, allowing only resonance decays and neglecting hadronic rescattering. Although neglecting final-state interactions is not fully realistic compared to detector measurements, the motivation for this choice will be discussed later. It will be shown that including hadronic scatterings does not qualitatively alter the conclusions regarding the role of the modes.

\begin{figure}[h]
    \centering
    \includegraphics[width=0.49\linewidth]{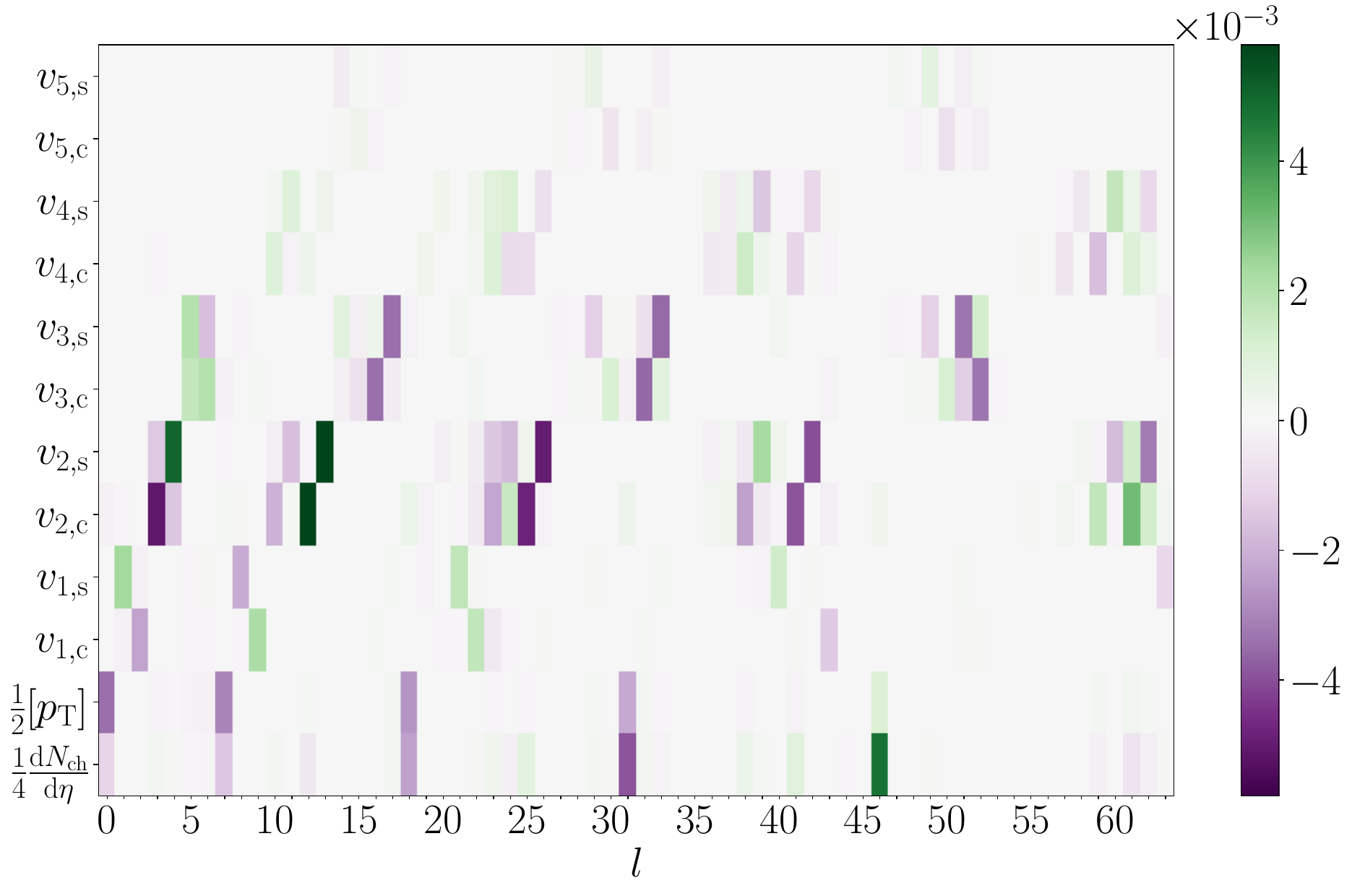}
    \includegraphics[width=0.49\linewidth]{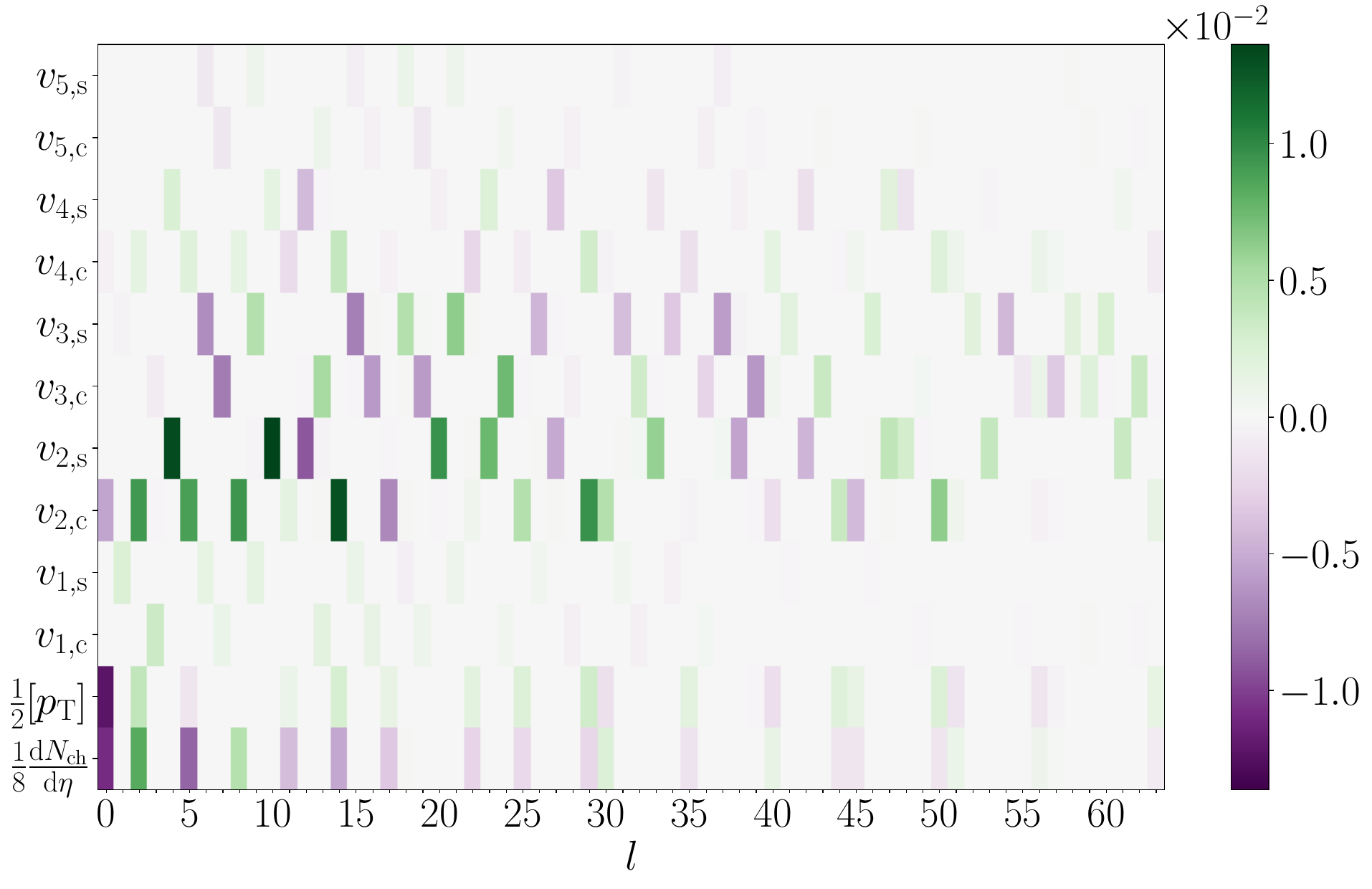}
    \caption{Linear response coefficients $L_{\alpha, l}$ for the final observables as function of the first $64$ modes in centrality $0-2.5\%$ (left) and $30-40\%$ (right). The coefficients corresponding to $\mathrm{d}N_\mathrm{ch}/\mathrm{d}\eta$ and $[p_T]$ are normalized by their respective average-state values $\bar{O}_\alpha$ and an additional constant factor (as indicated in the label), ensuring that they are dimensionless and displayed on a comparable scale in the plot.}
\label{fig:FS_linear_response}
\end{figure}

Using the spectra obtained from the states $\Psi_l^+$ and $\Psi_l^-$, the linear and quadratic response coefficients are computed via \cref{eq:linear_response_open} and \cref{eq:quadratic_response_open}, employing the final observables defined above. The response coefficients associated with the charged multiplicity $\frac{\mathrm{d}N_\mathrm{ch}}{\mathrm{d}\eta}$ and the average transverse momentum $[p_T]$ are normalized by their respective values for the average state. This normalization ensures appropriate scaling in the case of multiplicity and renders the momentum response dimensionless. The linear response coefficients for the final observables are displayed in \cref{fig:FS_linear_response}. As in the case of the initial observables, the non-central case exhibits a more complex structure, with a larger number of modes contributing and stronger mixing between different observables.

The most instructive comparison is between the initial-state signal in \cref{fig:IS_linear_response} and the final-state observables in \cref{fig:FS_linear_response} for the same centrality bin. The overall structure of the signal is preserved: the pattern observed in the final state closely mirrors that of the initial state. This demonstrates that fluctuations encoded in the initial conditions propagate through the full hydrodynamic evolution and leave a measurable imprint on the final observables.

However, the magnitude of the linear response coefficients in the final state is systematically smaller than in the initial state. The suppression arises from viscous effects in the hydrodynamic evolution. Since the evolution includes both shear and bulk viscosity, the response coefficients are effectively damped by approximately one order of magnitude in each centrality bin.

Examining the correspondence in more detail, in the central bin the radially symmetric modes that contribute to the initial energy $\mathrm{d}E/\mathrm{d}\eta_s$ and to the squared radius $\{r^2\}$ (e.g., $l = \{0, 7, 18, 31, \dots\}$) are the same modes that generate contributions to the final multiplicity $\mathrm{d}N_\mathrm{ch}/\mathrm{d}\eta$ and average transverse momentum $[p_T]$. Because multiplicity is directly related to the total energy, their linear response coefficients exhibit the same qualitative pattern. The relationship between $\{r^2\}$ and $[p_T]$, however, is inverse. A mode that increases the squared radius (i.e., spreads energy over a larger area) reduces the pressure gradients, leading to a smaller average transverse momentum. Consequently, the linear response coefficients for $\{r^2\}$ and $[p_T]$ typically exhibit opposite signs. This anticorrelation is also present in the non-central bin, although it may be partially obscured due to the more complex geometric structure of the background state.

The modes responsible for generating initial spatial eccentricities are directly mapped to the corresponding momentum anisotropies. This correspondence is particularly clear for $v_2$ and $v_3$. Higher-order eccentricities such as $\varepsilon_4$ and $\varepsilon_5$ are already small in the initial state and become further suppressed by viscous damping during the hydrodynamic evolution.

The only exception is the directed flow coefficient $v_1$, which exhibits a sign opposite to that of the corresponding initial $\varepsilon_1$. The origin of this inversion is not fully understood within the present framework. Since $v_1$ is more accurately described in fully three-dimensional models, this observable will be mentioned again in the following chapter.

\begin{figure}[h]
    \centering
    \includegraphics[width=0.49\linewidth]{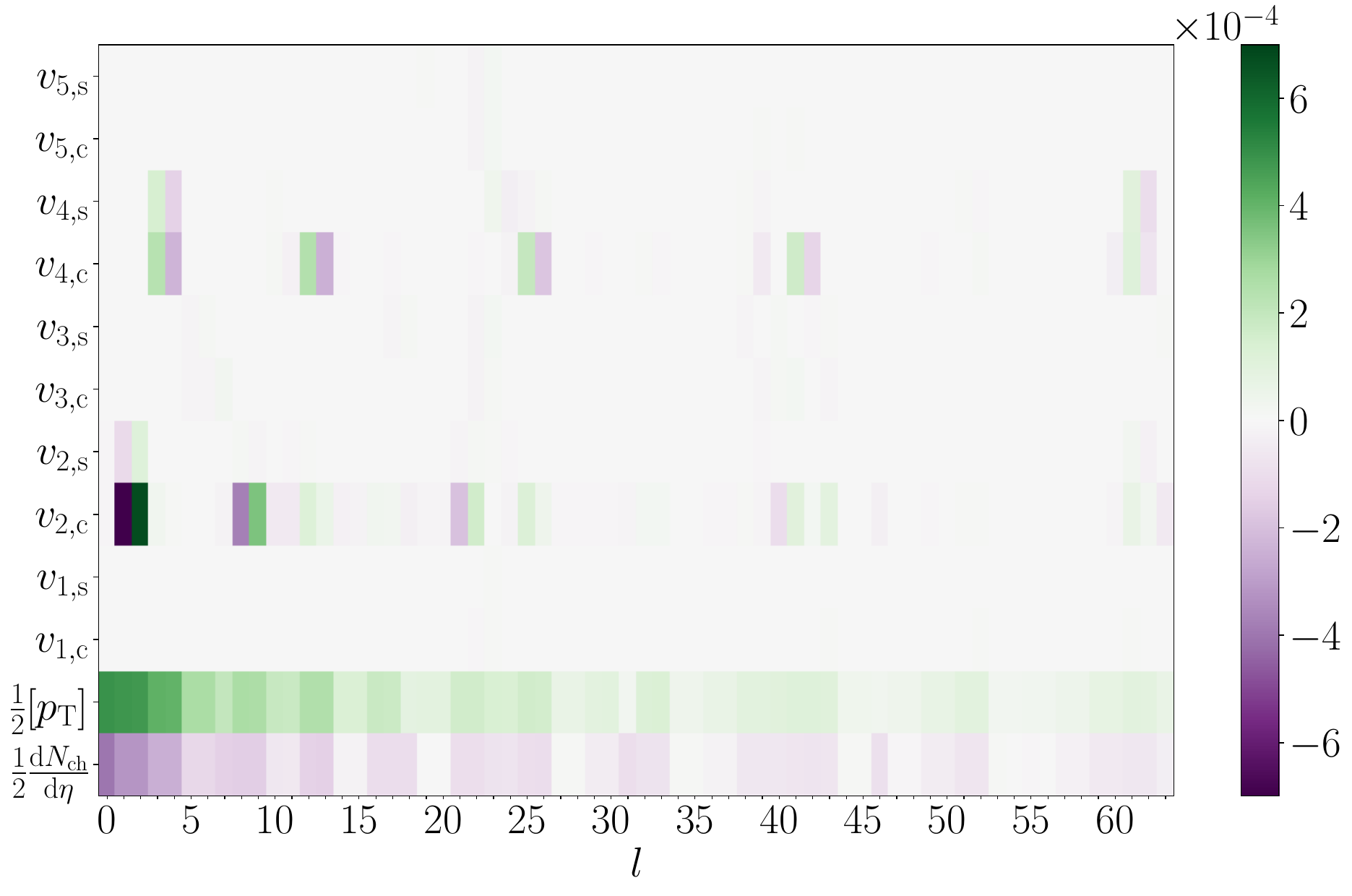}
    \includegraphics[width=0.49\linewidth]{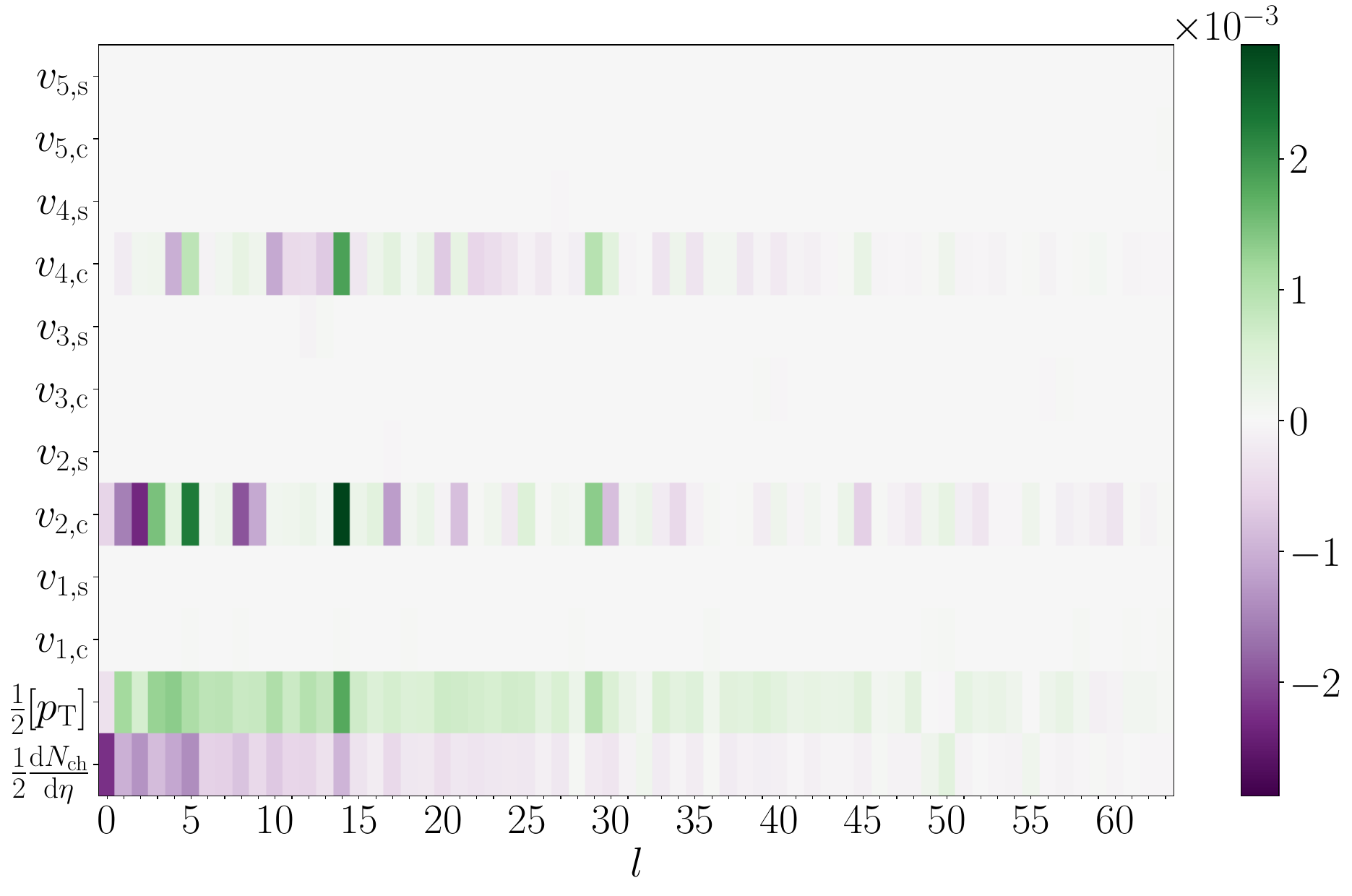}
    \caption{Quadratic response coefficients $Q_{\alpha, ll}$ for the final observables as function of the first $64$ modes in centrality $0-2.5\%$ (left) and $30-40\%$ (right). The coefficients corresponding to $\mathrm{d}N_\mathrm{ch}/\mathrm{d}\eta$ and $[p_T]$ are normalized by their average-state values $\bar{O}_\alpha$ and an additional constant factor (as indicated in the label), ensuring that they are dimensionless and displayed on a comparable scale in the plot.}
\label{fig:FS_quadratic_response}
\end{figure}

\Cref{fig:FS_quadratic_response} presents the quadratic response coefficients $Q_{\alpha,ll}$ for the final observables. For the final state, nearly all modes contribute to multiplicity and transverse momentum at quadratic order. As shown in Ref. \cite{Borghini:2022iym}, fluctuations in multiplicity imply that the quadratic contribution must be negative.

Comparing the quadratic response for final observables with the linear response of the initial state reveals the expected hierarchy $v_{2n,c} \propto \varepsilon_{n,c}^2$, corresponding to a second-order correction to the linear relation $v_n \propto \varepsilon_n$. In other words, dipole modes as $l = 1$ and $l = 2$, which generate $\varepsilon_1$ at linear order, produce a quadratic contribution to $v_2$. Similarly, quadrupole modes as $l = 3$ and $l = 4$, which generate $\varepsilon_2$ at linear order, induce a quadratic contribution to $v_4$. This structure is clearly visible in the $0-2.5\%$ centrality bin.

For the $30-40\%$ bin, the pattern is more intricate. Because most modes already contain an intrinsic elliptical deformation, the quadratic contributions to $v_2$ and $v_4$ arise from a mixture of effects. As in the quadratic response for the initial state, these contributions predominantly affect the cosine components aligned with the impact parameter direction.

The main objective of the mode-by-mode decomposition is to replace the full event-by-event evolution with a computationally more efficient framework. In this approach, only a limited set of modes is evolved, and the final observables are reconstructed from their contributions. Up to this stage, only the hydrodynamic evolution was considered, which is deterministic and therefore does not require large event statistics. However, once hadronic scatterings are included, the situation changes substantially. The evolution becomes probabilistic, and reliable determination of final observables requires sufficient statistical sampling. In practice, although evolving a reduced set of modes is computationally cheaper at the hydrodynamic level, the inclusion of hadronic interactions necessitates a significant increase in statistics.

For the second analysis, after the particlization hypersurface is constructed at $T = 155$ MeV, the code \textsc{iSS} (see \cref{subsec:iss}) samples particles and their momenta. These particles are then propagated through \textsc{SMASH}, where each oversampling in \textsc{iSS} generates an independent \textsc{SMASH} event. Consequently, the total statistical sample is given by the product of the number of hydrodynamic events and the number of oversamplings.\footnote{For minimum-bias simulations, typically $\sim 10^4$ events with $\sim 50$ oversamplings already provide sufficient statistics.} Since the mode-by-mode approach involves fewer initial configurations than a standard event-by-event simulation, a larger number of oversamplings is required to compensate for the reduced number of hydrodynamic evolutions.

\begin{figure}
    \centering
    \includegraphics[width=0.5\linewidth]{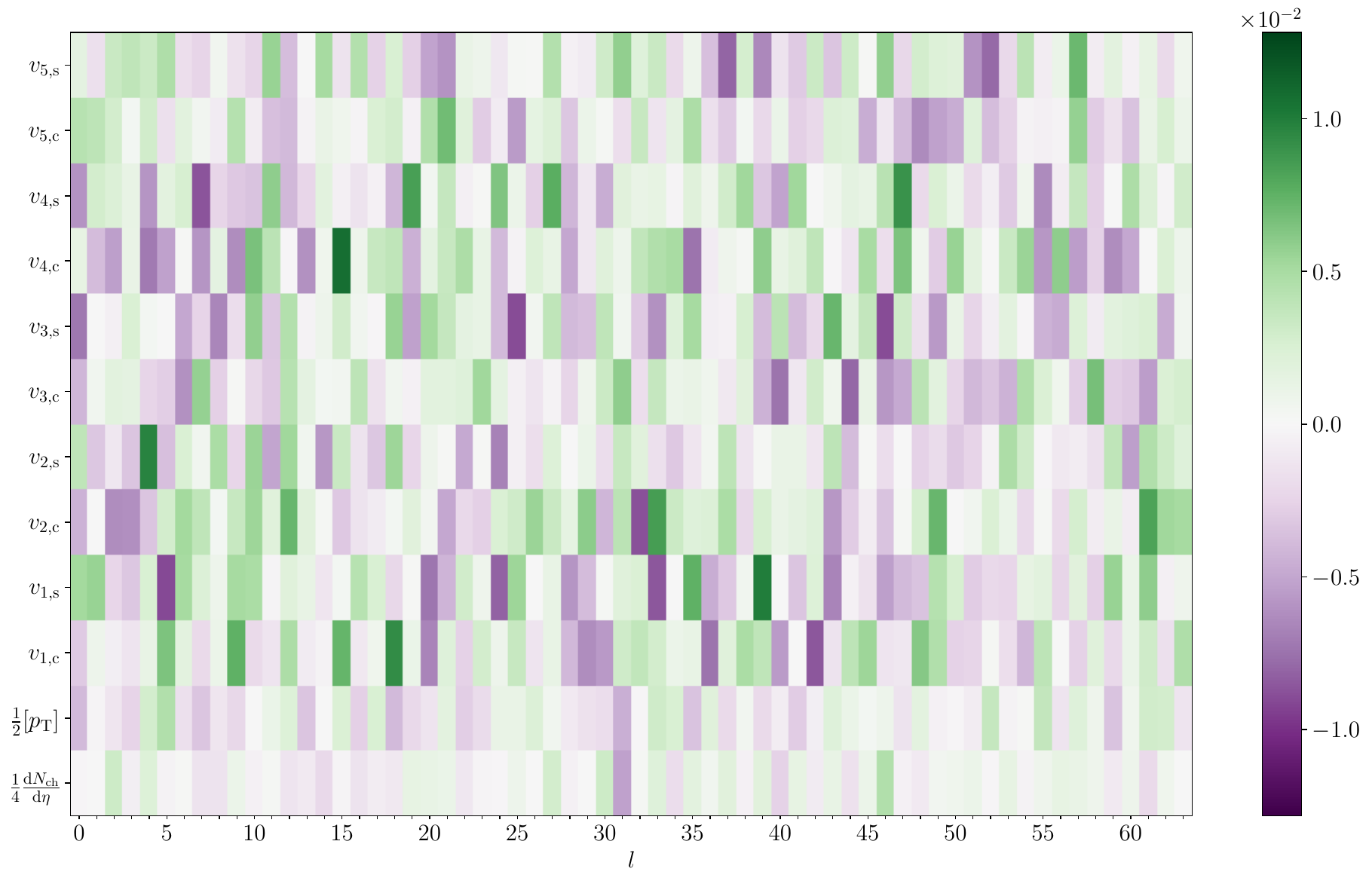}
    \caption{Linear response coefficients for central bin ($0-2.5\%$) for final observables after evolution in SMASH. Although this evolution is more realistic, as it includes hadronic scatterings, the statistical noise exceeds the signal. This is reflected in the corresponding noise plot, where the response structure is no longer clearly distinguishable.}
\label{fig:noise_L}
\end{figure}

An example of insufficient statistics is shown in \cref{fig:noise_L}, where the \textsc{SMASH} evolution, including hadronic scatterings, produces a visibly noisy signal. To control this effect, it is necessary to estimate the number of oversamplings required to reduce the statistical noise below the magnitude of the response coefficient. The statistical noise $\varsigma$ can be approximated as
\begin{equation}
    \varsigma \propto \frac{1}{\sqrt{M}},
\label{eq:noise}
\end{equation}
where $M = N \Sigma$ is the total number of sampled particles, given by the product of the event multiplicity $N$ and the number of oversamplings $\Sigma$.

In the scenario illustrated in \cref{fig:noise_L}, central events produce approximately $N \sim 2000$ particles. For a typical choice of $\Sigma \sim 50$, the total number of sampled particles is $M \sim 10^5$. According to \cref{eq:noise}, this leads to a statistical noise of order $10^{-3}$, which is comparable to the magnitude of the linear response coefficients shown in \cref{fig:FS_linear_response}. This explains the significant instability observed in the signal.

To reduce the statistical noise, the number of oversamplings must be increased. For non-central collisions, where the linear response signal is of order $10^{-3}-10^{-2}$, the statistical noise should be limited to approximately $\varsigma \sim 10^{-4}$. Since non-central events produce about $N \sim 500$ particles per event (see \cref{fig:PbPb_calibrated}), this requirement implies that $\Sigma \sim 10^4$ oversamplings are needed.

\begin{figure}[h]
    \centering
    \includegraphics[width=0.5\linewidth]{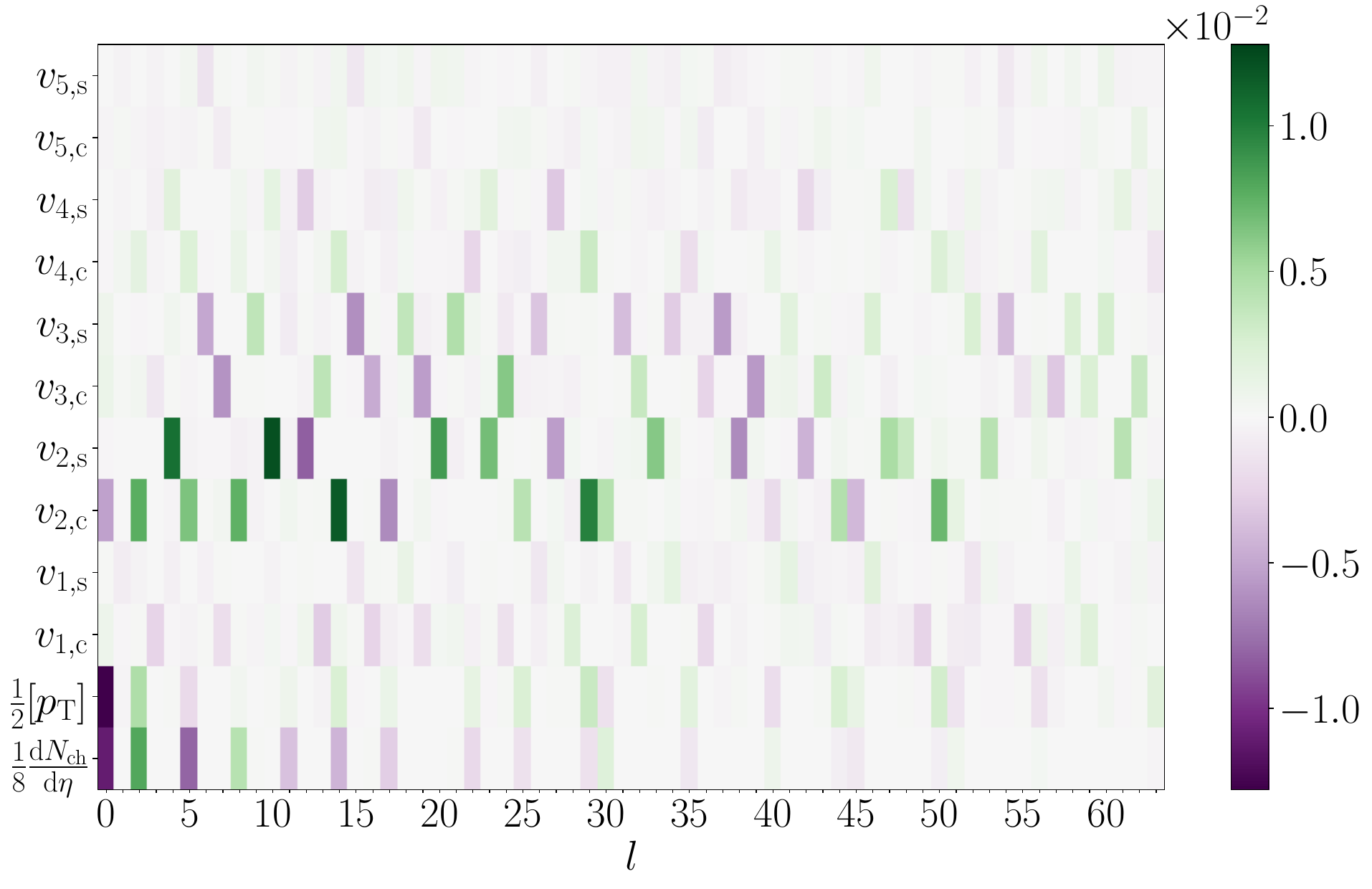}
    \caption{Final linear response coefficients for the first $64$ modes in the non-central bin ($30-40\%$) including hadronic scatterings. The coefficients corresponding to the charged multiplicity and transverse momentum are normalized by their respective average-state values $\bar{O}_\alpha$.}
\label{fig:FS_linear_response_SMASH}
\end{figure}

The results obtained with $20000$ oversamplings are shown in \cref{fig:FS_linear_response_SMASH} for the $30-40\%$ centrality class. In this case, the signal is no longer dominated by statistical noise, at least for the first $64$ modes. Higher modes, whose contributions are smaller and whose quadratic responses correspond to weaker signals, may still be partially affected by residual noise.

A comparison between the results obtained without hadronic scatterings (\cref{fig:FS_linear_response}) and those including \textsc{SMASH} (\cref{fig:FS_linear_response_SMASH}) reveals no significant qualitative differences. The main exception is a modification in the $v_1$ signal, the origin of which is not yet fully understood. A slight quantitative reduction of the response coefficients is also observed. For example,
\begin{equation}
    L_{v_{2,c}, l=5} = 8.9 \times 10^{-3} \hspace{0.2cm} \textrm{(MUSIC)}~, \hspace{0.5cm} L_{v_{2,c}, l=5} = 6.4 \times 10^{-3} \hspace{0.2cm} \textrm{(SMASH)}.
\end{equation}
Despite these modest effects, the large number of oversamplings required to suppress statistical noise results in a computationally expensive approach for \textsc{SMASH}. This justifies why the previously presented final results were obtained using only MUSIC output, including resonance decays but neglecting hadronic scatterings.

Although central collisions produce a larger number of particles, they also exhibit smaller response coefficients. As a result, the oversampling requirements remain comparable to those of non-central collisions. However, the higher particle multiplicity in central events significantly increases the run time of \textsc{SMASH} simulations. For this reason, only non-central results including hadronic scatterings are shown.

Finally, the noise estimate in \cref{eq:noise} applies directly to the response coefficients defined in \cref{eq:linear_response_open}. All calculations were performed using $\xi = 0.5$, corresponding to the largest value consistent with the linearity condition. Reducing $\xi$, for example, to $\xi = 0.1$, would decrease the response coefficients by a factor of five, thereby requiring a proportional increase in oversampling to maintain the same signal-to-noise ratio. Conversely, values $\xi > 0.5$ violate the linearity assumption, which motivates the choice $\xi = 0.5$.

This concludes the block, where the connection between final observables and the initial fluctuation modes has been established. The next subsection introduces a simplified toy model to isolate and analyze the fluctuations arising solely from the variation of the impact parameter within a given centrality bin.

\subsection{Illustrative example: Optical-Glauber model}
\label{subsec:optical}

The purpose of the mode-by-mode decomposition model is to understand the fluctuations inherent in initial-state models. Throughout this chapter, and in comparison with the fixed-impact-parameter analysis of Ref. \cite{Borghini:2022iym}, it has been observed that additional modes, such as $l = 2$, $5$, and $8$ in the non-central case, appear when centrality bins are considered. These additional modes are primarily associated with geometric fluctuations arising from variations in the impact parameter $b$ within a centrality class.

The goal of this subsection is to isolate and better understand the contribution of fluctuations originating solely from variations in $b$. To achieve this, the initial energy density profiles are generated using the Optical Glauber model. As described in \cref{subsec:glauber}, the Optical Glauber model is simpler than the \textsc{MCGlauber} approach. It does not require explicit nucleon positions; instead, it is based on continuous Woods–Saxon density distributions, and the total energy profile is obtained from the superposition of these smooth densities. As a result, fluctuations associated with the random positions of participant nucleons are absent, leaving only the geometric fluctuations induced by the variation of the impact parameter within a given centrality bin.

The numerical setup closely follows that used in the previous analysis: the same grid, collision energy, and system parameters are employed. The key difference is that here, no calibration of the initial energy profile is performed, since no full evolution to final-state observables is considered. This omission does not affect the present analysis, as it would only introduce an overall multiplicative normalization factor in the energy.

Centrality classification is still performed using the estimator equation \cref{eq:estimator_equation}. Although the missing normalization changes the absolute scale of the multiplicity, it does not affect the ordering used to define centrality bins. Therefore, a pseudo-energy is used instead of the fully calibrated energy profile. The final multiplicity (not needed in this subsection) would simply differ by a constant factor.

For each centrality bin, $1000$ events are generated and classified using the estimator equation.\footnote{A conceptually similar procedure is used in parametric models such as $\mathrm{T}_{\mathrm{R}} \mathrm{ENTo}$, where events are ordered by total energy to determine centrality classes, similar to what is shown in \cref{fig:pseudo_entropy}.} The extracted impact-parameter ranges are
\begin{equation}
    b_{0-2.5\%} \le 2.02~\mathrm{fm} \hspace{1cm} 7.25~\mathrm{fm} \le b_{30-40\%} \le 8.35~\mathrm{fm}.
\end{equation}
Although the variation in impact parameter is larger in absolute terms for the central bin (approximately $2$ fm) than for the non-central bin (approximately $1$ fm), the corresponding change in geometry is much more significant in non-central collisions. For large systems such as Pb–Pb, collisions with $b = 0$ fm and $b = 1$ fm both produce nearly circular overlap regions. In contrast, the difference between $b = 7$ fm and $b = 8$ fm leads to a substantial change in the elliptical geometry. This explains why non-central events exhibit stronger geometric fluctuations.

Using the ensemble of $1000$ events per bin, the average state $\bar{\Psi}^{\mathrm{opt.}}$ and fluctuation modes $\{\Psi^{\mathrm{opt.}}_l\}$ are constructed according to \cref{eq:modes_decomposition}, where the superscript “opt.” denotes profiles generated with the Optical Glauber model. The average state $\bar{\Psi}^{\mathrm{opt.}}$ closely resembles the corresponding average states from the \textsc{MCGlauber} model shown in \cref{fig:avg_state}. The main difference is the absence of the calibrated energy scale. Additionally, the Optical Glauber average state can exhibit a slightly larger radius due to the continuous superposition of thickness functions, which differs from the discrete participant-based energy deposition in the \textsc{MCGlauber} approach.

\begin{figure}[h]
    \centering
    \includegraphics[width=0.49\linewidth]{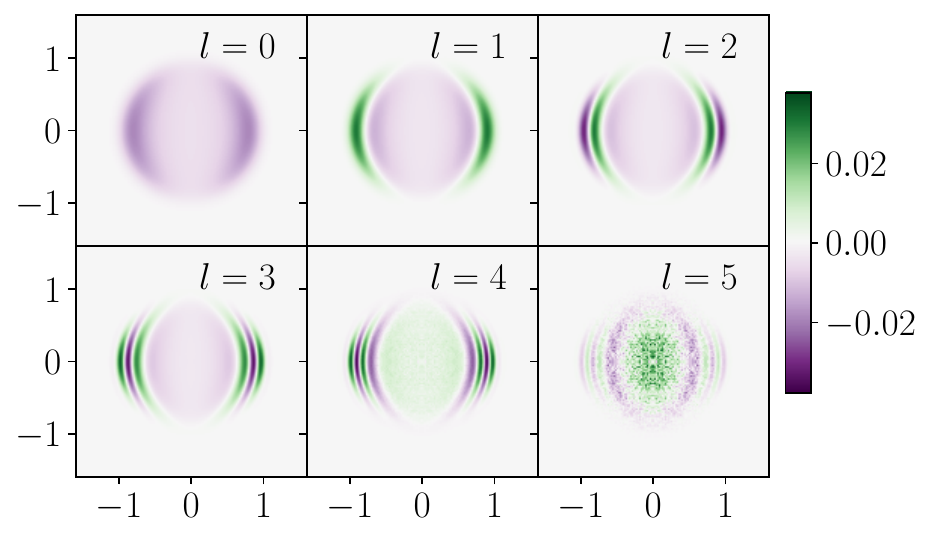}
    \includegraphics[width=0.49\linewidth]{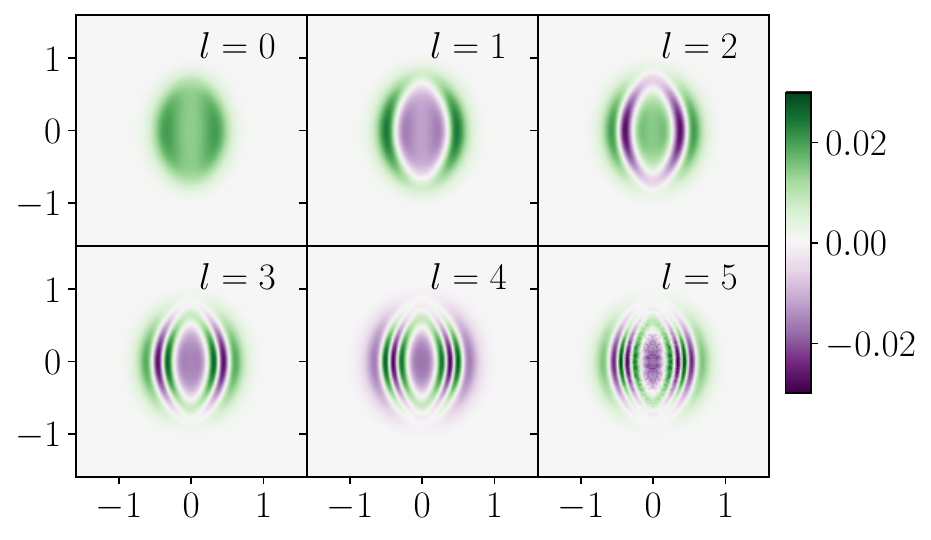}
    \caption{Six most important modes for the Optical Glauber model in Pb-Pb collisions at $\sqrt{s_\mathrm{NN}} = 5.02$ TeV. The left panel corresponds to the central bin ($0-2.5\%$) and the right panel to the non-central case ($30-40\%$).  As in the previous figures of this chapter, the transverse axes are expressed in units of the half-density radius $R_\mathrm{Pb} = 6.62$ fm, and the plots display the normalized modes $\tilde{\Psi}_l$.}
    \label{fig:optical_modes}
\end{figure}

The primary focus of this subsection is the analysis of the fluctuation modes ${\Psi^{\mathrm{opt.}}_l}$, displayed in \cref{fig:optical_modes}. Only the first six modes are shown for each centrality class. From mode $l = 5$ onward, the structures already appear noisy, and the corresponding eigenvalues $\lambda_l$ indicate that for $l \ge 6$, the contribution drops below $10^{-8}$. This demonstrates that geometric fluctuations alone are captured by a very small number of dominant modes.

Due to this rapid decrease in mode importance, the average state carries most of the total energy $\bar{w}_{0-2.5\%} = 0.89$ and $\bar{w}_{30-40\%} = 0.68$. The first fluctuation mode contributes $w_{l=0} \simeq 0.10$ in the central bin and $w_{l=0} \simeq 0.31$ in the non-central bin. All higher modes contribute less than $1\%$ of the total energy.

Analyzing the qualitative structure of the ``geometric'' modes, one observes that they introduce only small corrections to the overall profile shape. Most of the deformation is concentrated along the $x$-direction, which coincides with the direction of the impact parameter $b$. As previously discussed, the overall sign of a mode has no physical meaning, since it arises from the arbitrary phase convention in the diagonalization procedure.

For the non-central case, the leading mode carries a larger relative weight than in the central case, i.e., $w_{l=0}^{30-40\%} > w_{l=0}^{0-2.5\%}$. The remaining modes provide only minor corrections near the system boundary and contribute very little to the total energy.

\begin{figure}[h]
    \centering
    \includegraphics[width=0.37\linewidth]{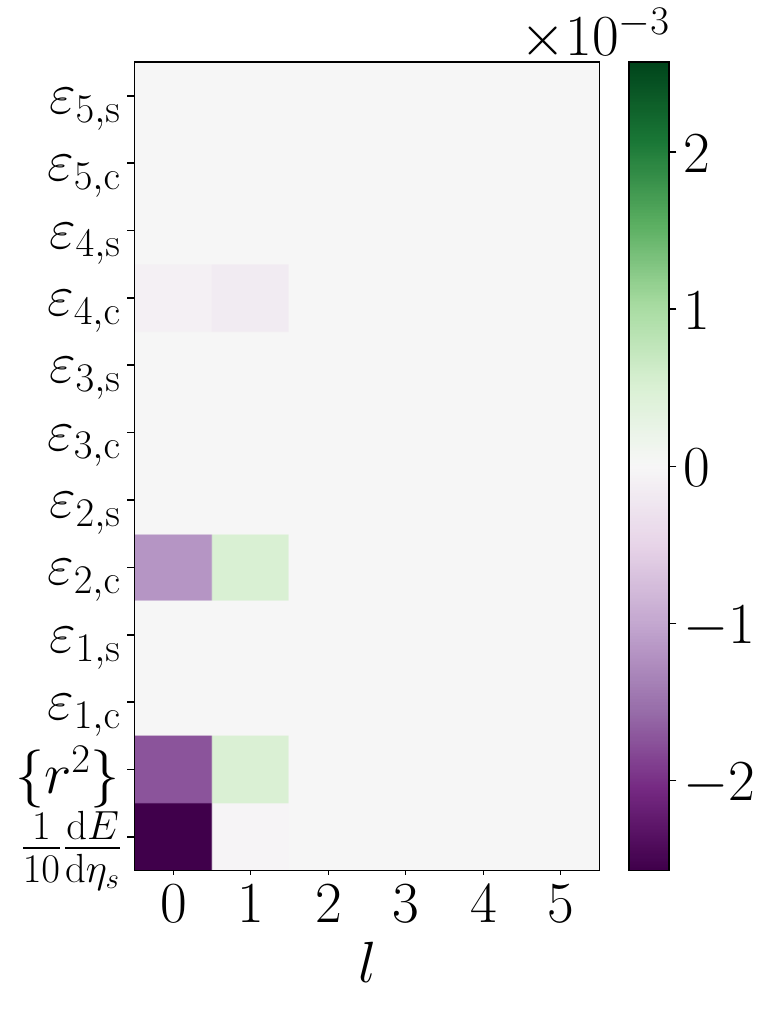}
    \includegraphics[width=0.39\linewidth]{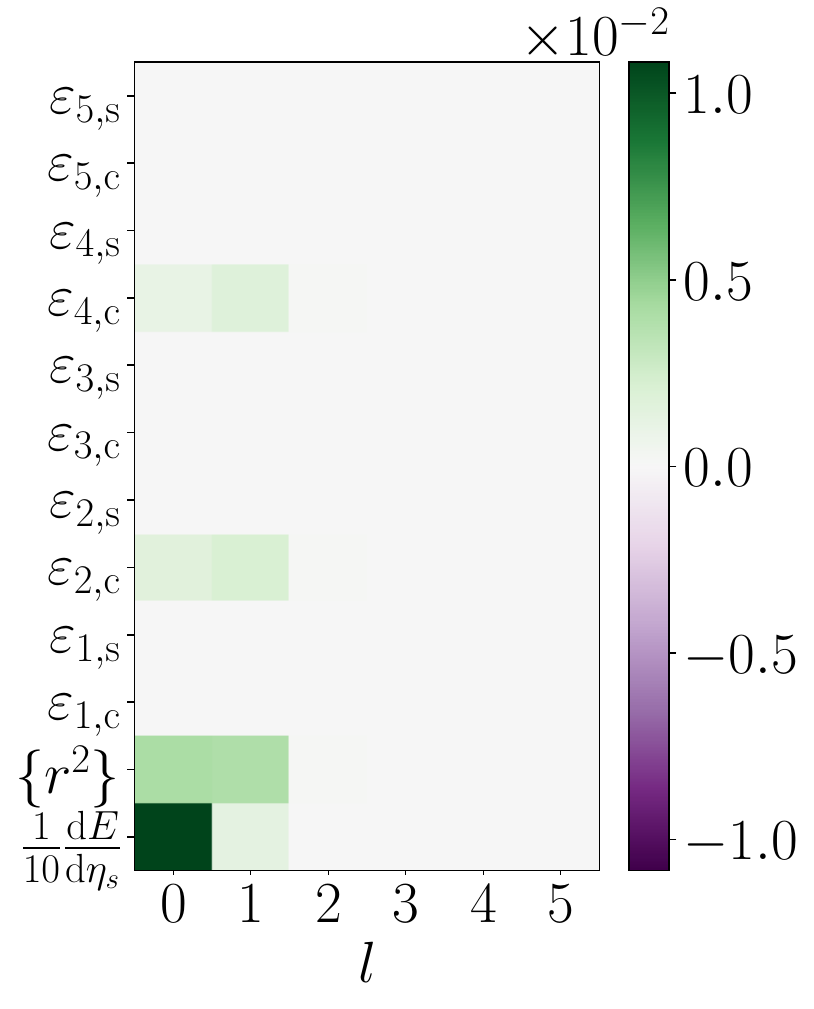}
    \caption{Initial linear response coefficients $L_{\alpha, l}$ for the first $6$ modes in the central bin ($0-2.5\%$, left) and in the non-central bin ($30-40\%$, right).}
    \label{fig:optical_linear_response}
\end{figure}

The corresponding initial linear response coefficients $L_{\alpha,l}$ are shown in \cref{fig:optical_linear_response}. As expected, the dominant contribution arises from mode $l = 0$, which carries the largest fraction of the energy and induces a small but finite correction to the cosine component of the elliptic eccentricity $\varepsilon_{2,c}$. The contributions from the higher modes are negligible in comparison and therefore have no significant impact on the observables. Note that since there is no ``geometric'' fluctuation in the direction $y$, no contributions for the $\varepsilon_{n,s}$ should be found.

Thus, this subsection demonstrates that the ``geometric'' fluctuations arising from the variation of the impact parameter $b$ within a given centrality bin are not the dominant source of event-by-event fluctuations. Although such geometric effects are present and can be systematically isolated, their quantitative impact remains limited compared to other fluctuation sources. The dominant contribution to the observed fluctuations arises from uncertainties in the nucleon positions and from the stochastic nature of the energy deposition process during the collision.

\section{Summary}
\label{sec:summary_PbPb}

The main goal of this chapter was to decompose initial energy density profiles into fluctuation modes and to study their properties in order to improve the description of initial conditions in heavy-ion simulations. The central result is that the fluctuations present in the initial state survive the entire hydrodynamic evolution and leave a measurable imprint on final observables. This finding highlights the importance of accurate initial condition modeling, since the input to the hybrid evolution chain directly influences the final results.

The analysis focused on Pb–Pb collisions in two centrality classes, $0-2.5\%$ and $30-40\%$, allowing a comparison between geometric effects and fluctuations arising from variations in the impact parameter $b$. The first part of the chapter presented the theoretical framework of the mode-by-mode decomposition, centered on two key elements: the average state $\bar{\Psi}$, which captures the dominant geometric features of the collision, and a set of fluctuation modes $\{\Psi_l\}$, which encode the event-by-event deviations from this average state.

The numerical implementation of this framework, along with the associated computational challenges, was described in detail. Particular attention was given to the determination of observables from the isolated modes. Both initial and final observables were analyzed, and the corresponding linear and quadratic response coefficients were computed. The statistical limitations associated with extracting reliable signals from the hadronic stage of the evolution were also discussed.

In the final part of the chapter, a simplified toy model based on the Optical Glauber framework was introduced to isolate the contribution of fluctuations arising solely from variations of the impact parameter within a centrality bin.

The next chapter continues the development of this theoretical framework by extending the analysis to a three-dimensional, non-boost-invariant setup and applying it to intermediate-size systems.

\chapter{Mode-by-mode decomposition in a 3D intermediate system}
\label{chap:3d}
\DoPToC
\noindent

This chapter is based on the submitted article \cite{Krupczak:2026byr}, ``Event-by-event fluctuations of elliptic flow in ultrarelativistic O+O collisions'' available on arXiv and submitted to the European Physical Journal C, with adaptations and extensions.

This chapter extends the work presented in the previous chapter. It continues to build on the mode-decomposition framework developed in \cite{Borghini:2022iym}, previously applied to centrality classes in \cite{Krupczak:2025gwg}. Here, the formalism is extended to a three-dimensional (3D) energy density profile, moving beyond the two-dimensional boost-invariant approximation.

In addition, this chapter focuses on oxygen–oxygen collisions. Oxygen represents an intermediate-sized system that is more strongly influenced by fluctuations, making it an ideal scenario for the application of fluctuation modes.

To accomplish this extension, \cref{sec:motivation_OO_3D} introduces the motivation for a 3D treatment and explains how the mode-decomposition theory is generalized to include longitudinal structure. This section also discusses why oxygen–oxygen collisions are particularly interesting compared to larger systems and provides a brief overview of recent theoretical and experimental studies of light-ion collisions. \Cref{sec:numerical_3D} describes the numerical setup used for the 3D hybrid model implemented in this analysis. The results are presented in \cref{sec:results_3D}. First, the 3D average state is analyzed in \cref{subsec:3D_average_state}, followed by the presentation of the 3D modes in \cref{subsec:3D_modes}. The properties of the modes and expansion coefficients are discussed in \cref{subsec:3D_properties}, including both initial and final response coefficients. Finally, \cref{subsec:modes_application} applies the 3D modes to a study of the longitudinal dependence of the linear response between eccentricities and flow coefficients. In this subsection, the number of modes required to capture the relevant physics is also discussed, illustrating the effectiveness of the mode-decomposition approach. Additional technical details are provided in \cref{appA} and \ref{appB}.

\section{Motivation for a 3D extension to Oxygen-Oxygen collisions}
\label{sec:motivation_OO_3D}

From the previous chapter, the behavior and evolution of fluctuations in the transverse plane (2D case) were analyzed in detail. A natural extension of this study is to investigate mode-by-mode decompositions in a fully three-dimensional (3D) scenario, where the longitudinal dependence is explicitly included.

In addition, smaller systems provide a particularly suitable environment for the modes approach, as they are more strongly driven by fluctuations \cite{Nagle:2018nvi}. In this context, oxygen–oxygen collisions represent an attractive candidate, especially given the recent focus on light-ion collisions in the heavy-ion community and the oxygen run at the LHC in July 2025. This section introduces the relevant background and motivation for extending the mode-decomposition framework to such systems, which will be complemented by the detailed analysis presented in the subsequent sections.

The mode-by-mode formalism described in \cref{sec:theory_modes} is sufficiently general to be applied in a 3D context. The essential modification is that the initial energy density profiles $\Phi^{(i)}(x, y, \eta_s)$ in \cref{eq:modes_decomposition} must now be generated with explicit longitudinal dependence. Consequently, both the average state $\bar{\Psi}(x, y, \eta_s)$ and the fluctuation modes $\Psi_l(x, y, \eta_s)$ acquire full 3D structure.

A key practical consideration is that the inclusion of the longitudinal dimension significantly increases the computational cost. In the 2D case, the average state $\bar{\Psi}$ has dimension $N_s^2$, where $N_s$ denotes the number of grid points in each transverse direction, and the correlation matrix $\rho$ has size $N_s^2 \times N_s^2$. In the 3D case, $\bar{\Psi}$ has dimension $N_s^2 N_L$, where $N_L$ is the number of longitudinal slices, and the matrix $\rho$ becomes $(N_s^2 N_L) \times (N_s^2 N_L)$. This rapid growth demonstrates that the number of longitudinal slices must remain moderate to keep the problem computationally tractable. Aside from this increase in dimensionality and computational complexity, the theoretical framework remains unchanged, and all analyses performed in the 2D case can, in principle, be extended to the 3D scenario. 

The primary motivation for adopting a 3D approach is to access information about fluctuations in the longitudinal direction and to investigate how non-boost-invariant structures propagate through the evolution.

The assumption of boost invariance has long been employed in the heavy-ion community due to the significant simplifications it provides. It was first introduced by Bjorken \cite{Bjorken:1982qr} in 1983. The central idea is that, at sufficiently high collision energies, the strong Lorentz contraction of the incoming nuclei and the extremely short overlap time lead to a system that is approximately invariant under Lorentz boosts along the beam (longitudinal) direction. As a consequence, physical quantities become independent of the space-time rapidity in Milne coordinates (see \cref{eq:Milne_coord}) and depend only on the proper time. This results in an approximately flat distribution of energy density around mid-rapidity. Under this approximation, the longitudinal dynamics effectively decouple, and the evolution is dominated by the transverse degrees of freedom.

This picture provides a good approximation for the mid-rapidity region in high-energy collisions and has been widely and successfully used in hydrodynamic descriptions of the QGP \cite{Moreland:2018gsh, JETSCAPE:2020mzn}. Nevertheless, it is only an approximation, and its range of validity is limited. As the heavy-ion community aims for increasingly precise theoretical predictions and experimental measurements, it becomes necessary to go beyond this simplification and investigate observables within fully 3D hydrodynamic frameworks. This is particularly important for observables that are sensitive to the longitudinal structure or to study longitudinal correlations, which cannot be captured in boost-invariant simulations. Moreover, current experimental programs are extending the exploration of the QCD phase diagram to lower collision energies, where baryon stopping becomes increasingly important and the system exhibits a pronounced dependence on rapidity. In this regime, the boost-invariant approximation is no longer adequate, making a realistic description of the longitudinal dynamics essential.

The full 3D hybrid evolution has recently become an active topic of research in the field. Following the physical stages summarized in \cref{fig:HIC_numerical}, several components of the hybrid framework have been extended to three dimensions. For the initial state, 3D versions of parametric models such as $\mathrm{T}_{\mathrm{R}} \mathrm{ENTo}$ \cite{Soeder:2023vdn} have been developed, while more physically motivated approaches include \textsc{McDipper} \cite{Garcia-Montero:2023gex} and \textsc{IP-Glasma} \cite{Schenke:2016ksl}.

In the pre-equilibrium stage, the standard 2D free-streaming approximation is no longer valid in a 3D framework. A physically consistent 3D extension of \Kompost~has recently been introduced \cite{Du:2025hyk}. Hydrodynamic codes, on the other hand, can be generalized more straightforwardly to 3D, allowing the evolution of non-boost-invariant profiles. After particlization, the dimensionality does not introduce conceptual complications, as the hadronic transport stage naturally operates in full 3D.

These developments demonstrate that a phenomenological 3D description of heavy-ion collisions is now technically feasible and increasingly mature. This progress provides strong motivation for extending the mode-decomposition framework beyond the boost-invariant approximation.

Beyond the extension of the framework to 3D, this chapter also focuses on intermediate-size systems, where fluctuations play a more prominent role. Small systems are typically formed in proton–nucleus collisions, such as p–Pb or p–Au, whereas intermediate systems correspond to collisions between light ions, such as O–O or Ne–Ne. Both types of systems involve a smaller number of participants and therefore generate smaller fireballs, in which fluctuations have a comparatively stronger impact.

As discussed earlier in this dissertation, the initial state constitutes the primary source of uncertainty in the hybrid description. Proton–nucleus collisions are particularly challenging, as they depend sensitively on the internal structure of the proton \cite{Mantysaari:2017cni,Demirci:2021kya}, which remains insufficiently constrained. To reduce this model dependence, and motivated by the recent experimental interest in light-ion collisions \cite{ALICE:2025luc}, this chapter focuses on oxygen–oxygen collisions at $\sqrt{s_\mathrm{NN}} = 5.36$ TeV.\footnote{The center-of-mass energy per nucleon pair in nucleus–nucleus collisions depends on the proton beam energy and on the ratio $Z/A$, where $Z$ is the atomic number and $A$ the mass number. While the positive charge $Z$ enables acceleration in the electromagnetic field of the collider, the mass number $A$ limits the achievable bending in the magnetic field, resulting in a scaling factor $Z/A$ relative to the proton beam energy. For the LHC, the maximum proton–proton energy is $13.6$ TeV, corresponding to $6.8$ TeV per proton beam. For oxygen ($Z=8$, $A=16$), this would allow up to $(8/16)\times6.8 = 3.4$ TeV per nucleon beam, or $\sqrt{s_\mathrm{NN}} = 6.8$ TeV. However, the LHC oxygen run was performed at $\sqrt{s_\mathrm{NN}} = 5.36$ TeV, matching the energy used for Pb–Pb collisions, where $(82/208)\times6.8 = 2.68$ TeV per nucleon beam. This choice also facilitates direct comparisons with p-p and Pb–Pb data collected at similar energies.}

In addition to exhibiting stronger fluctuations, intermediate systems are interesting for several further reasons. Hydrodynamic models aim to describe collective observables in small systems; however, the pronounced jet energy loss observed in Pb–Pb collisions, attributed to the presence of the QGP medium, is not observed in p–Pb collisions \cite{ALICE:2012mj}. This raises the question of the limits to which hydrodynamic descriptions can be extended \cite{Gardim:2022yds}. Intermediate systems, such as O–O and Ne–Ne collisions, provide an opportunity to explore this transition region. Recent preliminary results indicate that hydrodynamics remains applicable in these systems \cite{ALICE:2025luc} and even suggest the presence of moderate energy-loss effects in O–O and Ne–Ne collisions \cite{Modak:2026ymk}.

\begin{figure}[h]
    \centering
    \includegraphics[width=0.4\linewidth]{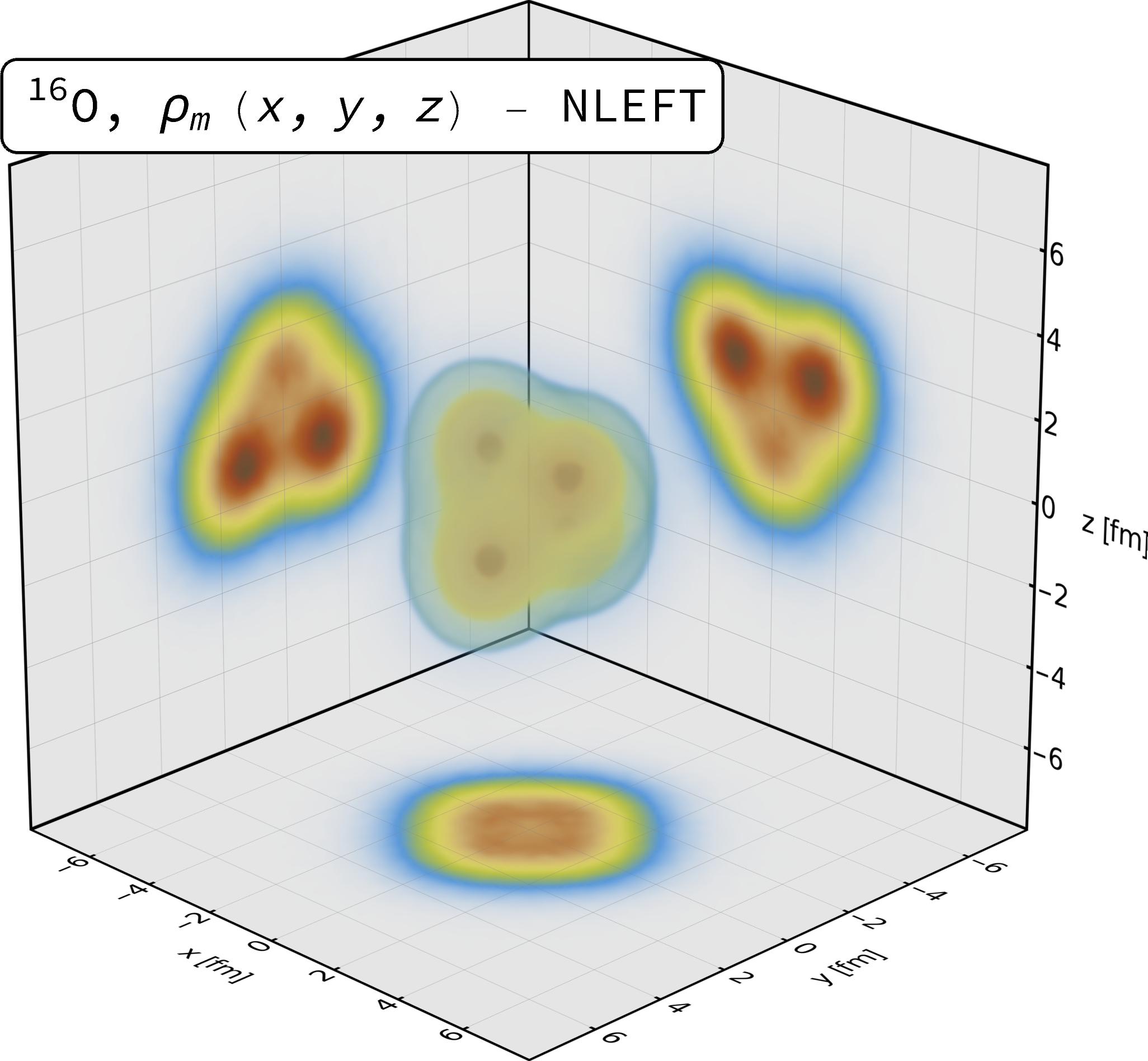}
    \caption{Oxygen $\alpha$-cluster structure determined using the NLEFT method. Image from Ref. \cite{Giacalone:2024luz}.}
    \label{fig:O_NLEFT}
\end{figure}

Another important motivation for studying light-ion collisions is that their nuclear structure can be described with significantly greater precision than that of heavy nuclei. In particular, recent works have determined the nuclear structure of oxygen and neon using \emph{ab initio} calculations that solve the Schrödinger equation for Hamiltonians constructed within effective field theories of low-energy QCD. The Nuclear Lattice Effective Field Theory (NLEFT) framework \cite{Giacalone:2024luz,Blaizot:2025bfu} determines the nucleon positions by solving the full Hamiltonian on a lattice. For oxygen, this approach predicts a tetrahedral configuration supplemented by $\alpha$-cluster structures, as illustrated in \cref{fig:O_NLEFT}. The oxygen nucleus is not perfectly spherical; its intrinsic deformation is now explicitly incorporated into modern modeling efforts \cite{Mehrabpour:2025rzt}, thereby reducing a significant source of initial uncertainty.\footnote{For heavier nuclei such as lead, first-principles calculations are considerably more challenging due to the large number of nucleons ($208$ compared to $16$ for oxygen). Consequently, the precise geometry of the Pb nucleus remains less constrained. Although heavy nuclei are often approximated as spherical, several studies explore the possibility of small deformations to better reproduce specific observables \cite{Carzon:2020xwp,Xu:2025cgx}.}

A second theoretical framework for determining nuclear structure is the Projected Generator Coordinate Method (PGCM) \cite{Frosini:2021fjf}. Unlike NLEFT, PGCM is not based on solving the full nuclear Hamiltonian. Instead, it constructs the nuclear wave function through shape degrees of freedom and nucleon correlations. The central idea is to establish correlations between nucleons in a manner conceptually similar to how correlations between hadrons are studied in the final state of heavy-ion collisions. These correlations can generate deformations in the nucleus and determine its overall geometric configuration. Although the PGCM approach carries larger theoretical uncertainties compared to NLEFT, the qualitative predictions of both methods are in reasonable agreement. In this dissertation, however, only the NLEFT results are used as input for the oxygen nuclear structure.

The same theoretical methods have also been applied to neon nuclei, which exhibit a characteristic ``bowling-pin'' shape. Studying the ratio of observables between O–O and Ne–Ne collisions can help reduce systematic uncertainties \cite{Giacalone:2024luz}, making this combination of systems particularly valuable for investigating QGP properties in intermediate-size systems. The mode-decomposition framework developed here could be applied in an analogous way to Ne–Ne collisions.

After this motivation, the next section describes the numerical setup used in this analysis, followed by the presentation of the results.

\section{3D numerical setup}
\label{sec:numerical_3D}

This section describes the numerical chain used in the analysis presented in this chapter. As mentioned previously, the study focuses on O–O collisions at $\sqrt{s_\mathrm{NN}} = 5.36$ TeV.

The initial energy density profiles are generated on a 3D grid using \textsc{McDipper} \cite{Garcia-Montero:2023gex,Garcia-Montero:2025bpn}, described in \cref{subsec:mcdipper}. In this setup, thickness fluctuations are implemented through the weights $p_i$ in \cref{eq:thickness_fluc}, sampled from a log-normal distribution with width $\sigma = 0.637$ \cite{Garcia-Montero:2025bpn}. Additionally, hotspot fluctuations are incorporated according to \cref{eq:hotspots_fluc}, using $N_q = 3$ constituent quarks per nucleon.

These transverse fluctuations account for uncertainties in the internal nucleon structure, which influence the spatial distribution of the initial energy deposition in the transverse plane, as demonstrated in Ref. \cite{Garcia-Montero:2025bpn}. It is important to emphasize that these subnucleonic fluctuations are implemented only in the transverse plane (2D). The longitudinal dependence of the energy density arises from the unintegrated gluon distribution $\Phi(x,\mathbf{x},\mathbf{q})$, which depends on the momentum fraction $x = \frac{\mathbf{p}}{\sqrt{s_\mathrm{NN}}} e^{\pm \mathrm{y}}$, where $\mathrm{y}$ denotes the momentum rapidity.

\textsc{McDipper} is used in read-in mode, where the nuclear structure of the oxygen nuclei is taken from $14745$ configurations with positive weights provided by NLEFT calculations \cite{Giacalone:2024luz}. Given the relatively small size of the oxygen nucleus (with a radius of $R_\mathrm{O} = 2.608$ fm), the transverse grid is defined with $N_s = 96$ points per direction, while the longitudinal direction is discretized into $N_L = 9$ slices. The transverse box size is set to $x_T = 8.4$ fm, and the space-time rapidity range is $\eta_s \in [-6,6]$. This corresponds to grid spacings of $\Delta x_T = 0.1$ fm and $\Delta \eta_s = 1.5$. Although the longitudinal grid resolution is relatively coarse, the integrals within \textsc{McDipper} are evaluated with higher internal precision, and only the resulting energy density is projected onto the coarser 3D grid used in the mode decomposition.\footnote{In the calibration stage, a finer longitudinal resolution was used, $\Delta \eta_s = 0.2$, followed by interpolation to $\Delta \eta_s = 0.25$. The choice of only $9$ longitudinal slices in the present analysis is dictated by memory constraints associated with constructing the 3D correlation matrix $\rho$, as discussed in the previous section. Tests using sparse matrix representations were performed; however, since the matrix $\rho$ is not sufficiently sparse, this approach proved more computationally expensive than storing the full matrix.}

After generating the initial conditions, the natural next step would be to evolve the system through a pre-equilibrium stage to account for its out-of-equilibrium dynamics. However, at the time this research was conducted, there was no 3D pre-equilibrium model. Now, a 3D version of \Kompost~has become publicly available \cite{Du:2025hyk}, although its applicability to smaller systems, where fluctuations play a dominant role, has not yet been well established because the model relies on a linearized approximation. Consequently, the present analysis was performed without incorporating the full 3D pre-equilibrium evolution.

Thus, the initial energy density produced by \textsc{McDipper} is directly passed to the hydrodynamic code \textsc{MUSIC}. The switching time between the two stages is set to $\tau_0 = 0.6$ fm. At this time, a bilinear–quadratic interpolation is applied to refine the longitudinal grid resolution; further details are provided in \cref{appA}. In this step, only the longitudinal grid is modified, changing from $\Delta \eta_s = 1.5$ to $\Delta \eta_s = 0.25$, corresponding to $N_L = 57$ longitudinal slices. Simultaneously, the box size is increased to accommodate the subsequent hydrodynamic expansion. The transverse box size is increased to $15$ fm, and the longitudinal extent is extended to $14$.

The initial energy–momentum tensor is constructed in a similar way as in \cref{eq:Tmunu_input_kompost}
\begin{equation}
\begin{aligned}
    T^{\mu\nu}(x, y, \eta_s) &= \operatorname{diag}\left(\epsilon(x, y, \eta_s), P_T(x, y, \eta_s), P_T(x, y, \eta_s), \frac{P_L(x, y, \eta_s)}{\tau^2}\right) \\
     &= \operatorname{diag}\left(\epsilon, \frac{\epsilon}{2}, \frac{\epsilon}{2}, 0\right).
\end{aligned}
\end{equation}
This choice corresponds to isotropic pressure in the transverse plane and vanishing longitudinal pressure. The corresponding initialization of the viscous terms is
\begin{equation}
\begin{aligned}
    \pi^{\mu\nu} &= \operatorname{diag}\left(0, \frac{\epsilon}{6}, \frac{\epsilon}{6}, -\frac{\epsilon}{3\tau^2} \right) \\
    \Pi &= \frac{\epsilon}{3} - P(\epsilon),
\end{aligned}
\end{equation}
such that the Landau matching condition is satisfied. \textsc{MUSIC} then evolves the system, but now without assuming boost invariance. The shear viscosity is kept constant at $\eta/s = 0.16$, while the bulk viscosity is taken to be temperature dependent according to Ref. \cite{Denicol:2009am}, as previously used in \cref{chap:entropy_matching} (see \cref{eq:bulk_viscosity_parametrization} with the parametrization).

The particlization hypersurface is defined at $T = 155$ MeV. In the previous chapter, it was discussed that evolving the modes through \textsc{SMASH} is computationally expensive and does not provide qualitatively new information compared to using only resonance decays (see \cref{subsec:FSobservables}). For this chapter, no \textsc{SMASH} evolution is included. However, the special mode in \textsc{MUSIC} that directly produces and decays the particle spectrum is not available in the 3D evolution setup. Therefore, the procedure adopted here is to use \textsc{iSS} to convert the fluid into particles and then allow only decays of these particles, without hadronic rescattering. Since \textsc{iSS} is computationally efficient, a large number of oversamplings is employed: $4 \times 10^5$ in standard cases and up to $8 \times 10^5$ for higher-precision analyses.

As in the previous chapters, the initial condition model requires calibration. The main difference here is that in \textsc{McDipper}, the normalization factor is not a global scaling of the total energy, but rather a normalization of the gluon contribution, as expressed in \cref{eq:energy_mcdipper}. For consistency with the original formulation of \textsc{McDipper}, this factor is denoted by $K_g$. Nevertheless, the calibration procedure follows the same logic as before.

Initially, $40$ events are generated with impact parameters $b = [0, 1, 1.5, 2, 2.5]$ fm. For these events, the estimator equation \cref{eq:estimator_equation_withC} is applied in the central slice $\eta_s = 0$, using preliminary values $C(\eta/s) = 1$ and $K_g = 1$, to estimate the final multiplicity. These same events are then evolved through \textsc{MUSIC} and \textsc{iSS}, and the final multiplicity $\mathrm{d}N_\mathrm{ch}/\mathrm{d}\eta$ is extracted in the pseudorapidity range $[-0.5, 0.5]$. Plots analogous to \cref{fig:pre_calibration} are used to determine the optimal linear parameter $C(\eta/s)$ that best matches the estimator prediction with the fully evolved multiplicity.

With the calibrated value of $C(\eta/s)$, a larger sample of events is generated while varying $K_g$ until the multiplicity obtained from the estimator equation matches the available experimental data \cite{Modak:2026ymk}.\footnote{These recent data are not yet available on HEPData with full error bars.} Only the central slice is used to determine the multiplicity, since it represents the charged-particle yield per unit pseudorapidity. In practice,
\begin{equation}
    \frac{\mathrm{d}N_\mathrm{ch}}{\mathrm{d}\eta} \propto \int \mathrm{d}^2x \left[ \tau_0^{2/3} \int_{-0.5}^{0.5} \epsilon(x, y, \eta_s)^{2/3} \mathrm{d}\eta_s \right].
\end{equation}
Because the region $[-0.5, 0.5]$ in pseudorapidity is approximately boost invariant, the result is insensitive to whether a finer resolution ($\Delta \eta_s = 0.2$) or a coarser resolution ($\Delta \eta_s = 1$) is used. For this reason, during calibration, a longitudinal grid with $N_L = 61$ slices is employed, in contrast to the mode analysis, where $N_L = 9$ slices are used to reduce memory consumption.

\begin{figure}[h]
    \centering
    \includegraphics[width=0.51\linewidth]{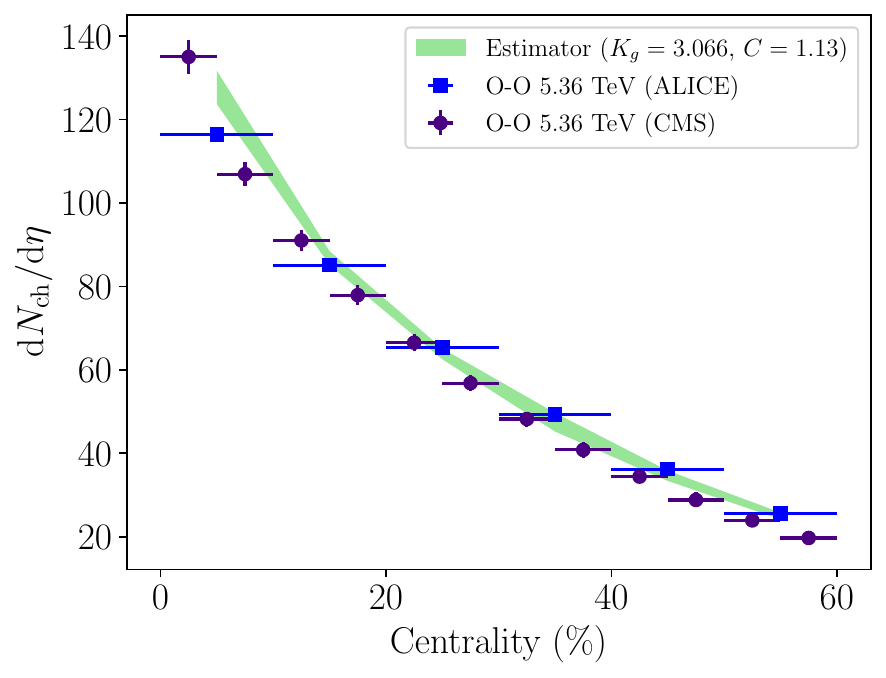}
    \caption{Final charged-particle multiplicity as a function of centrality, obtained from the estimator equation (\cref{eq:estimator_equation_withC} using the calibrated values of $K_g$ and $C(\eta/s)$). This green curve is compared with the experimental data from ALICE \cite{Modak:2026ymk}, shown as blue points. The good agreement supports the validity of the calibration procedure. As additional information, the CMS \cite{CMS:2026sai} data are added in purple, although these data were not used in the calibration procedure.}
\label{fig:calibration_OO}
\end{figure}

The calibration procedure may be iterated to improve precision. The final calibrated parameters are $C(\eta/s) = 1.13$ and $K_g = 3.066$, as shown in \cref{fig:calibration_OO}. It is important to note that the $K_g$-factor in \textsc{McDipper} does not act as a purely global normalization. Since gluons dominate the energy deposition in this regime, it is essential that $K_g$ remains within a physically reasonable range and does not artificially enhance the gluon contribution beyond its expected role.

The calibration procedure adopted in this dissertation aims to reproduce the experimental multiplicities across all centrality classes considered ($0-60\%$). For this reason, the estimator equation provides a more accurate description of the data than the calibration shown in Fig. 1 of Ref. \cite{Modak:2026ymk}, where \textsc{McDipper} was calibrated without including the factor $C(\eta/s)$ and considering only the most central bin. 

In addition, the calibrated green curve in \cref{fig:calibration_OO} also describes the recent CMS data with error bars \cite{CMS:2026sai} in terms of centrality and pseudorapidity (purple points in \cref{fig:calibration_OO}), although these data are not used in the calibration procedure.

With the parameters fixed as described in this section, events can be generated consistently, and the 3D fluctuation modes can be constructed. The next section is dedicated to presenting and discussing the results.

\section{Results and discussions}
\label{sec:results_3D}

This section presents the results and discussion of the fluctuation modes in the 3D framework, focusing on the properties that can be inferred from oxygen–oxygen collisions. First, \cref{subsec:3D_average_state} examines the average state and its properties, which are essential for understanding the subsequent analysis. \Cref{subsec:3D_modes} then introduces the fluctuation modes within a fully 3D description. In \cref{subsec:3D_properties}, the properties of these modes are analyzed following an approach similar to that used in the previous chapter, establishing the connection between initial and final linear response coefficients. Finally, \cref{subsec:modes_application} presents an application of the 3D modes to clarify the connection between $\varepsilon_n$ and $v_n$. This subsection also investigates the limits and applicability of the mode-decomposition framework.

\subsection{3D average state and its properties}
\label{subsec:3D_average_state}

\begin{figure}[h]
    \centering
    \includegraphics[width=\linewidth]{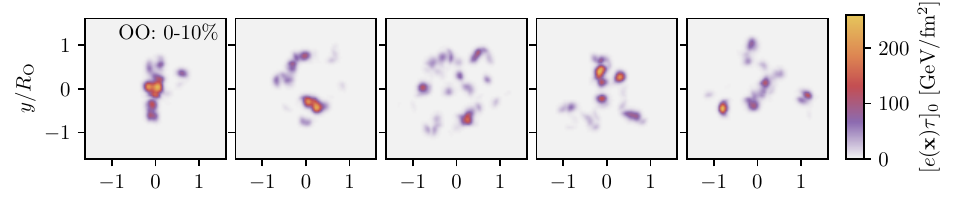}
    \includegraphics[width=\linewidth]{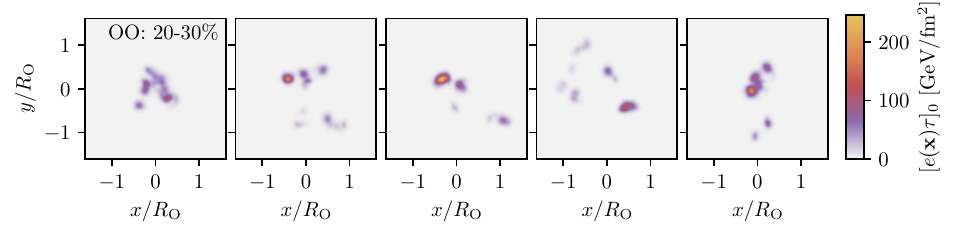}
    \caption{Random events for oxygen-oxygen collisions at $\sqrt{s_\mathrm{NN}} = 5.36$ TeV. The upper panels correspond to central events ($0-10\%$) and the lower panels to non-central ($20-30\%$). The five energy density profiles are shown in the transverse plane, with axes in units of the system radius ($R_\mathrm{O} = 2.608$ fm), evaluated at midrapidity. The similar energy scales and the absence of a pronounced geometric difference between the two centrality classes indicate that oxygen collisions do not exhibit the same geometric centrality dependence observed in large systems.}
\label{fig:random_OO}
\end{figure}

Before analyzing the average state $\bar{\Psi}$ for oxygen–oxygen collisions, it is instructive to develop some intuition about the structure of random energy profiles $\Phi^{(i)}$ in this system. \Cref{fig:random_OO} displays five representative events for the central ($0-10\%$) and non-central ($20-30\%$) bins. This can be compared with the Pb–Pb events shown in \cref{fig:random_profiles}, where centrality clearly modifies both the geometric structure and the overall energy scale of the profiles. In contrast, for oxygen collisions, there is essentially no pronounced geometric difference between the two centrality classes. The overall energy scales are also relatively similar. What effectively determines centrality in such an intermediate system is not a strong geometric shape but rather the total deposited energy or the number of hotspots. In the central case, multiple concentrated energy spots are typically observed, whereas in the non-central case, only one prominent high-density region may appear.

Without information about the total deposited energy, one might inspect the first subplot in the $0-10\%$ panel of \cref{fig:random_OO} and incorrectly assume it corresponds to a non-central collision due to its elliptical appearance. However, this is not the case. This observation provides an initial indication that, for intermediate systems, centrality does not correspond as directly to geometric deformation as it does in large systems such as Pb–Pb collisions.

\begin{figure}[h]
    \centering
    \includegraphics[width=\linewidth]{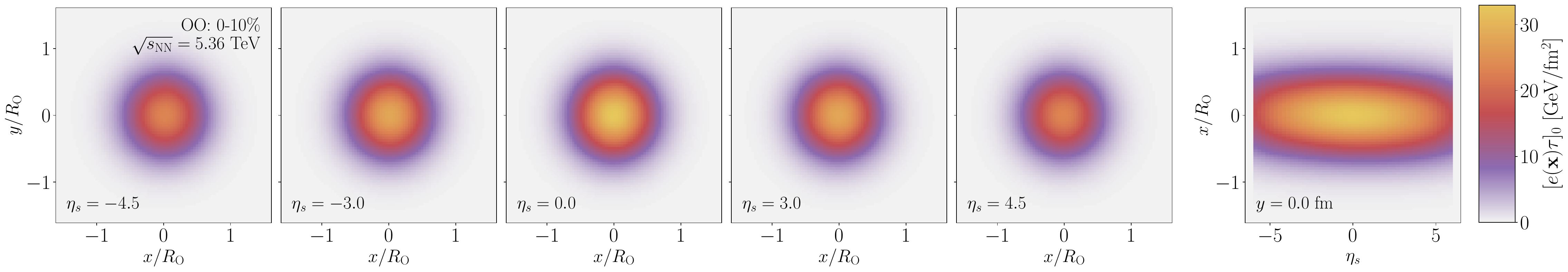}
    \includegraphics[width=\linewidth]{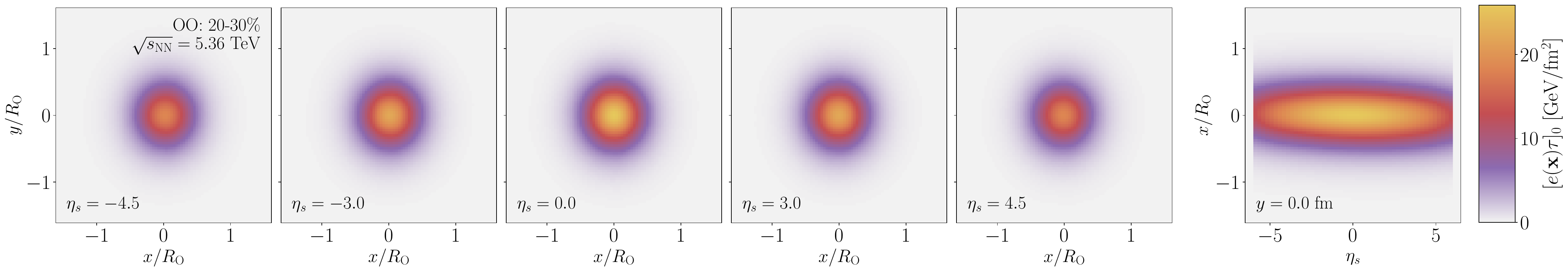}
    \caption{Average state for the $0-10\%$ (top) and $20-30\%$ (bottom) centrality classes. The transverse plane is shown in units of the system radius ($R_\mathrm{O} = 2.608$ fm) for selected slices ($\eta_s = [-4.5, -3, 0, 3, 4.5]$). The last column displays the $x-\eta_s$ plane at $y=0$ fm. No pronounced geometric difference is observed between the two average states, and no significant longitudinal deformation is present.}
\label{fig:avg_state_OO}
\end{figure}

Using $2^{17}$ events in each centrality class, the 3D average state is constructed and displayed in \cref{fig:avg_state_OO} in terms of longitudinal slices. The first point to note is that the average states for both centrality classes exhibit similar energy scales and are more radially symmetric compared to those in large systems. This behavior is consistent with the absence of strong geometric deformation observed in the random events shown in \cref{fig:random_OO}. 

Another important aspect is the energy-density scale. The average state has a smaller energy scale than the individual random events, indicating that the localized high-energy spots visible in single events are effectively ``washed out'' in the averaging procedure. This contrasts with large systems, where the average state typically retains a similar overall energy scale between individual events and the average profile.

Examining the 3D structure, one observes that in the transverse plane the shape remains nearly boost invariant across different longitudinal slices, with the primary variation being a gradual decrease in energy as $|\eta_s|$ increases. In other words, there is no pronounced longitudinal tilt or deformation. Such behavior is expected at high collision energies, where the interaction time is extremely short and does not generate strong longitudinal asymmetries.

\begin{figure}[h]
    \centering
    \includegraphics[width=0.5\linewidth]{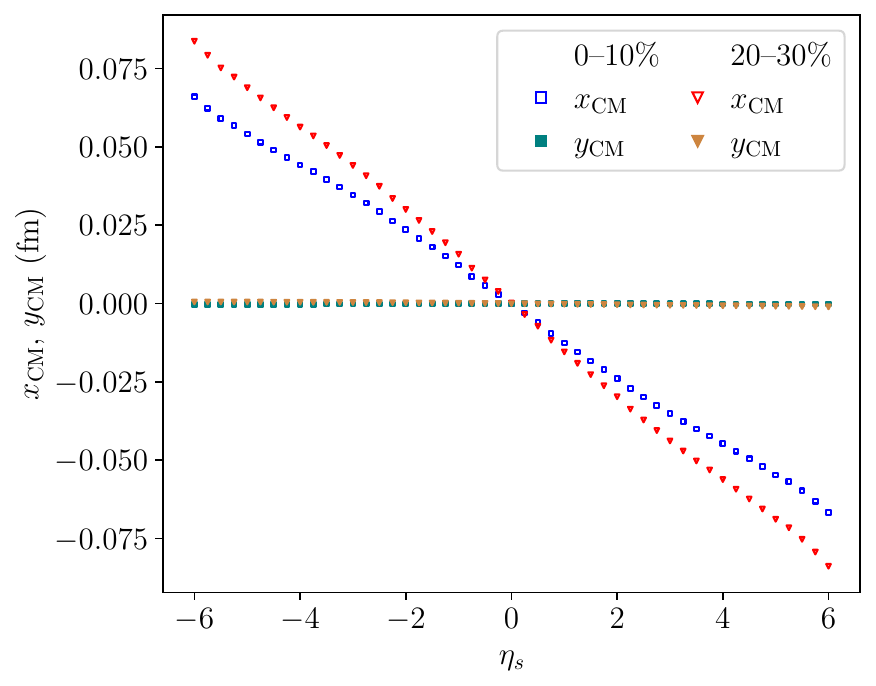}
    \caption{Center-of-mass position (in fm) in the $x$ and $y$ directions as a function of the longitudinal coordinate $\eta_s$ for the average state in the central and non-central bins. The odd dependence observed in $x_\mathrm{CM}$ reflects the small longitudinal tilt of the average state.}
\label{fig:CM}
\end{figure}

Although the longitudinal tilt is not visually prominent, it can be quantified, as shown in \cref{fig:CM}, which displays the center-of-mass position in each slice as a function of $\eta_s$. When constructing the average state, it is imposed that, for each event, the center of mass in the central slice $\eta_s = 0$ is located at the origin of the grid. Consequently, by definition, $x_{\mathrm{CM}}(\eta_s = 0) = y_{\mathrm{CM}}(\eta_s = 0) = 0$ fm. Due to symmetry in the $y$-direction, $y_{\mathrm{CM}}(\eta_s) = 0$ for all slices. However, in the $x$-direction (aligned with the impact parameter), a residual longitudinal structure appears from the asymmetry between target and projectile away from $\eta_s =0$. As a result, the tilt in the average state has a preferred orientation. In this sense, longitudinal deformation can be interpreted as a geometric feature of the collision and should be incorporated into the average state, in analogy with how the dominant elliptical deformation is captured in $\bar{\Psi}$ for large systems. Although the tilt is visible in $x_\mathrm{CM}$, its magnitude is smaller than $1$ fm and therefore not easily seen by eye in the $x-\eta_s$ plane in \cref{fig:avg_state_OO}.

Non-central events are expected to exhibit a larger longitudinal tilt due to the increased number of spectators, which enhances the deformation of the three-dimensional QGP profile. A clearer intuition for this effect can be obtained from lower collision energies (of the order of GeV), where the longitudinal asymmetry is significantly more pronounced \cite{Shen:2020jwv}.

\begin{figure}[h]
    \centering
    \includegraphics[width=0.32\linewidth]{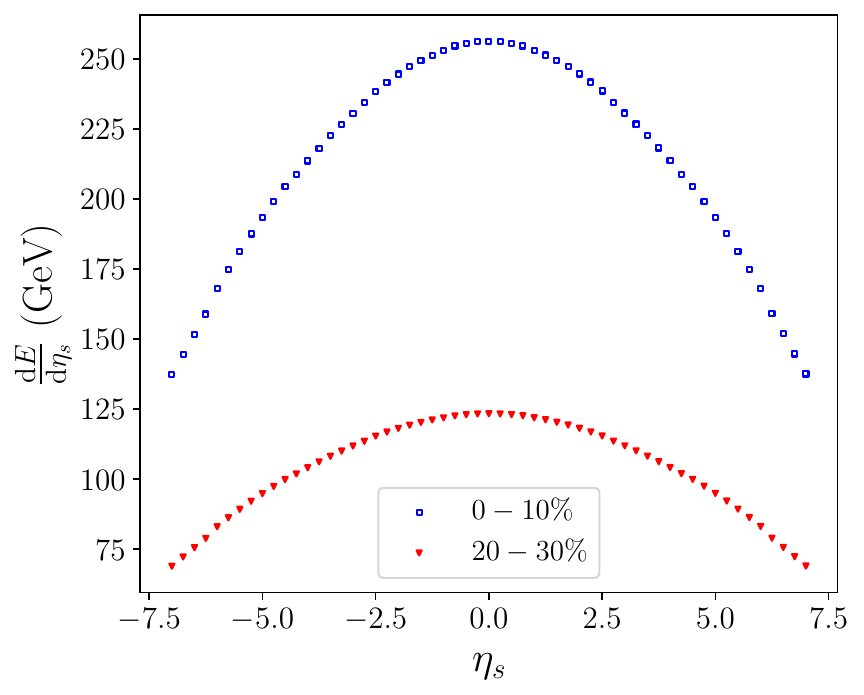}
    \includegraphics[width=0.32\linewidth]{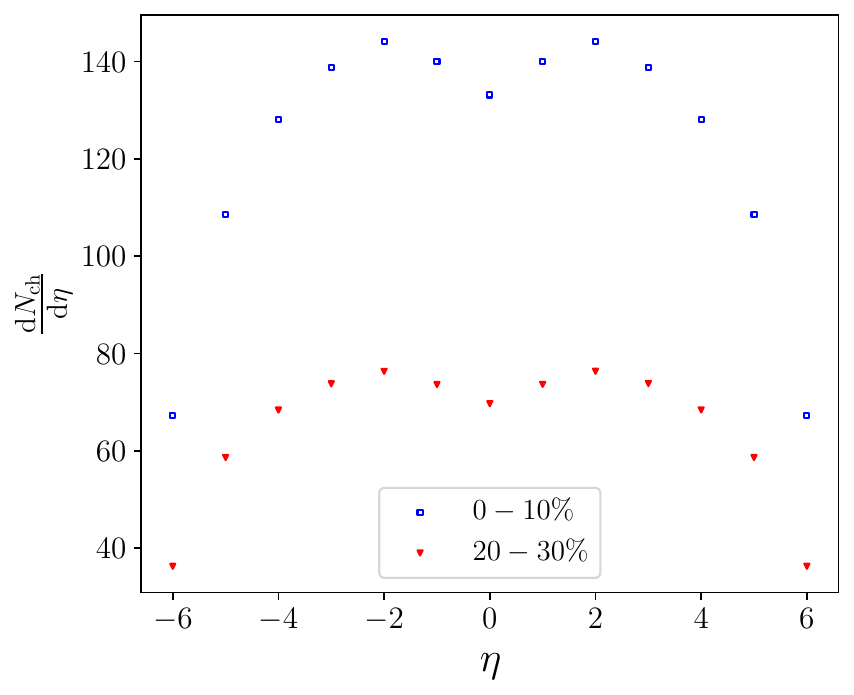}
    \includegraphics[width=0.32\linewidth]{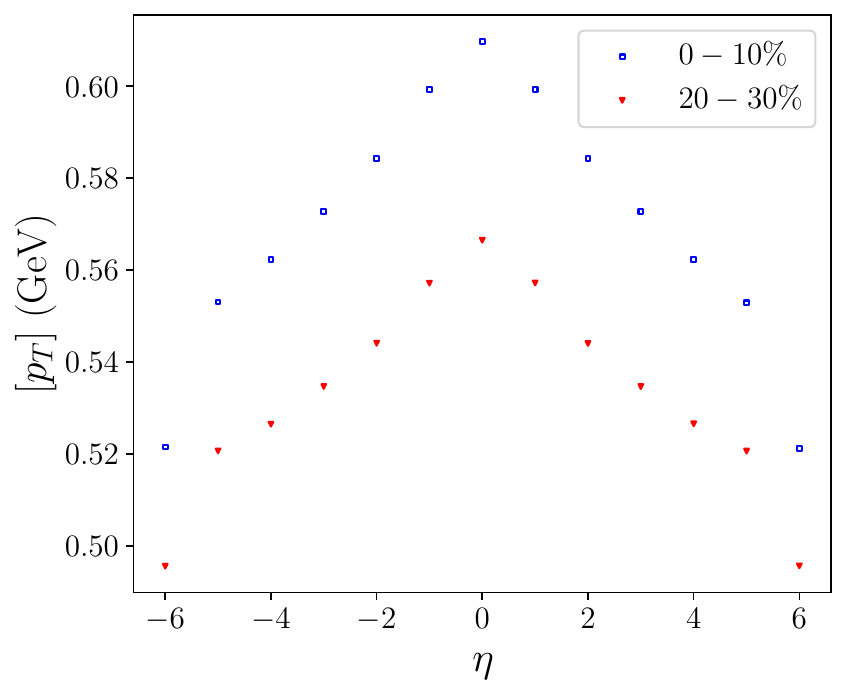}
    \caption{Observables of the average states ($0-10\%$ and $20-30\%$ centrality) as functions of the longitudinal variable: total energy $dE/d\eta_s$ (left), final charged-particle multiplicity $dN_\mathrm{ch}/d\eta$ (middle), and average transverse momentum $[p_T]$ (right).}
\label{fig:observables_psi_OO}
\end{figure}

Regarding the initial and final observables associated with the average state, \cref{fig:observables_psi_OO} displays their longitudinal dependence. The average state $\bar{\Psi}$ carries an energy that, after hydrodynamic evolution, is translated into charged-particle multiplicity. Comparing the longitudinal dependence of energy (in $\eta_s$) and multiplicity (in $\eta$), one observes a difference in shape arising from the Jacobian between space-time rapidity and pseudorapidity, as discussed in \cref{eq:jacobian_yeta}. The multiplicity obtained from the average state is in good agreement with the experimental data used in \cref{fig:calibration_OO} for both centrality classes. Furthermore, this multiplicity allows a comparison between systems: central O–O collisions produce approximately $\sim 120$ charged particles, which corresponds to Pb–Pb collisions in the $60-70\%$ centrality bin, as can be verified from the experimental data shown in \cref{fig:PbPb_calibrated}.

\begin{figure}[h]
    \centering
    \includegraphics[width=0.43\linewidth]{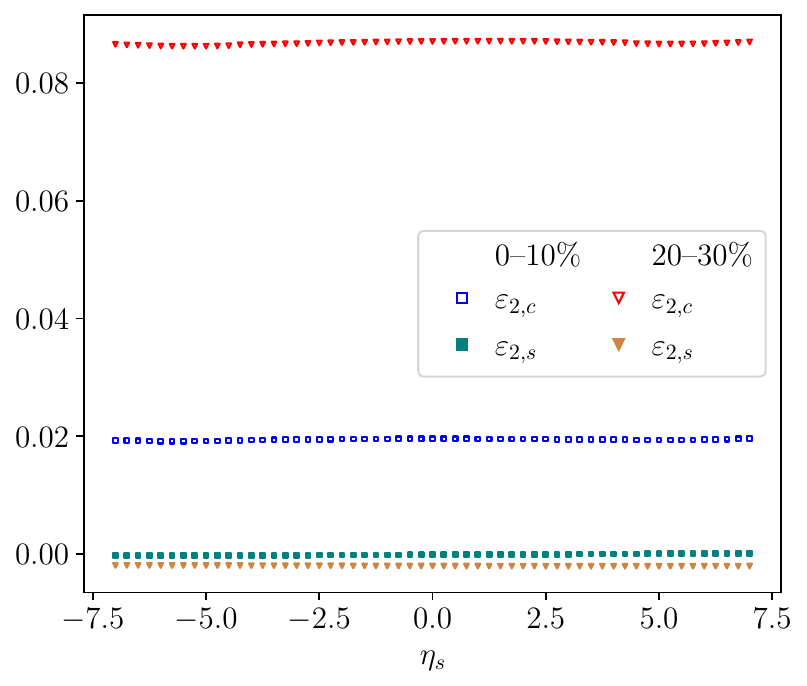}
    \includegraphics[width=0.43\linewidth]{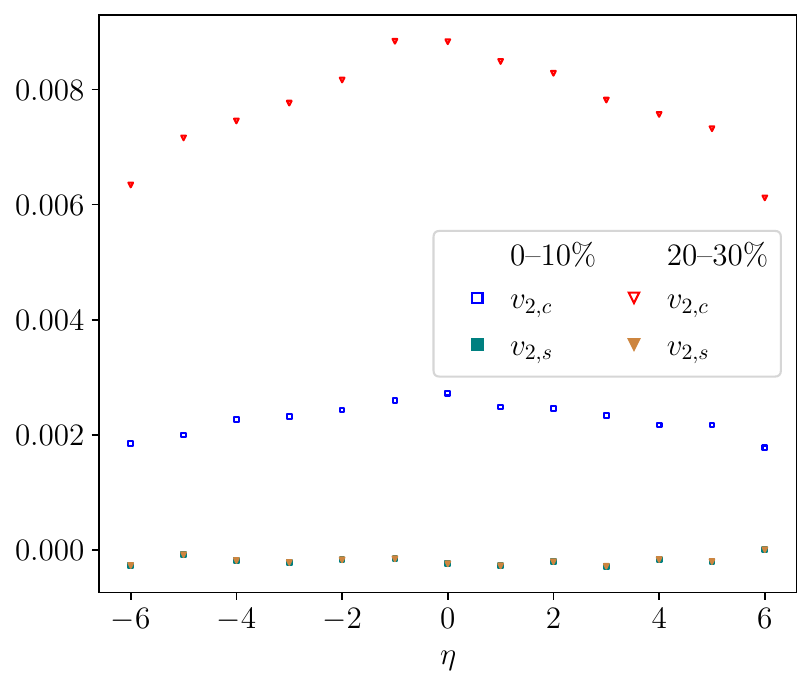}
    \caption{Initial eccentricity $\varepsilon_2$ (left) and final flow coefficient $v_2$ (right) of the average state as functions of the longitudinal variable. Both observables are shown in terms of their cosine and sine components for the two centrality bins.}
\label{fig:e2_psi_OO}
\end{figure}

Since the average state encodes the dominant geometric features of the system, we determine the elliptic eccentricity $\varepsilon_2$ and the corresponding final flow coefficient $v_2$ for the two average states, as shown in \cref{fig:e2_psi_OO}. From the shape of $\bar{\Psi}$ in \cref{fig:avg_state_OO}, no pronounced elliptical deformation is visible by eye; however, a small elliptic signal is present. As expected, this signal is more pronounced in the non-central case.

When calculating the eccentricity in each longitudinal slice, the local center of mass is used as the reference frame \cite{Shen:2020jwv}. This ensures that the coefficient does not incorporate the longitudinal tilt observed in \cref{fig:CM}, resulting in an eccentricity that remains approximately constant as a function of $\eta_s$. After the non–boost-invariant hydrodynamic evolution, this behavior remains largely unchanged.

Although the average state carries a nonzero contribution to $v_2$, this contribution is small compared to the experimental measurements in oxygen–oxygen collisions, where $v_2 \sim 0.06$ \cite{ALICE:2025luc}. This contrasts with large systems, in which the average state typically provides the dominant contribution to $v_2$, while fluctuations provide only subleading corrections. In the present case, as will be demonstrated in the following subsections, the observed flow is not primarily generated by the geometric contribution of the average state, but is instead significantly influenced by fluctuations, which play a dominant role in intermediate systems.

Besides the elliptic eccentricity $\varepsilon_2$, the 3D average state also generates a small but non-vanishing directed eccentricity $\varepsilon_1$. To clarify its structure, it is useful to decompose $\varepsilon_1$ into odd and even components in $\eta_s$:
\begin{equation}
\begin{aligned} 
    \varepsilon_1^{\mathrm{odd}}(\eta_s) &= \frac{\varepsilon_1(\eta_s) - \varepsilon_1(-\eta_s)}{2}, \\
    \varepsilon_1^{\mathrm{even}}(\eta_s) &= \frac{\varepsilon_1(\eta_s) + \varepsilon_1(-\eta_s)}{2}. 
\end{aligned} 
\label{eq:e1_odd_even}
\end{equation}
The odd component, $\varepsilon_1^{\mathrm{odd}}$, captures the dipolar structure generated by the longitudinal tilt of the system and therefore encodes information analogous to the center-of-mass shift shown in \cref{fig:CM}. In contrast, the even component, $\varepsilon_1^{\mathrm{even}}$, is primarily associated with transverse fluctuations and does not correspond to a genuine longitudinal geometric deformation \cite{Gardim:2011qn}.

Within the present framework, the geometric contribution to $\varepsilon_1^{\mathrm{odd}}$ is already contained in the average state $\bar{\Psi}$. As will be discussed in \cref{subsec:3D_properties}, the fluctuation modes encode only the even component, whereas the odd directed eccentricity emerges as a property of the averaged geometry itself.

The magnitude of $\varepsilon_1^{\mathrm{odd}}$ in oxygen–oxygen collisions at this energy remains very small, of the order of $10^{-3}$. Consequently, after hydrodynamic evolution, the directed flow $v_1$ generated from the average state becomes essentially negligible. A stronger connection between $\varepsilon_1$ and $v_1$ would be studied at lower collision energies, where longitudinal asymmetries and the fireball tilt are more pronounced. This is particularly interesting because, although the geometric interpretation of $\varepsilon_n$ linked to $v_n$ for $n \geq 2$ is well established in large systems, the role of $\varepsilon_1$ and $v_1$ remains more subtle, especially in fully 3D scenarios \cite{Luzum:2010fb,Bozek:2010bi}.

Having characterized the properties of the average state, we now proceed to the analysis of the fluctuation modes themselves.

\subsection{3D modes in oxygen-oxygen collisions}
\label{subsec:3D_modes}

\begin{figure}
    \centering
    \includegraphics[width=\linewidth]{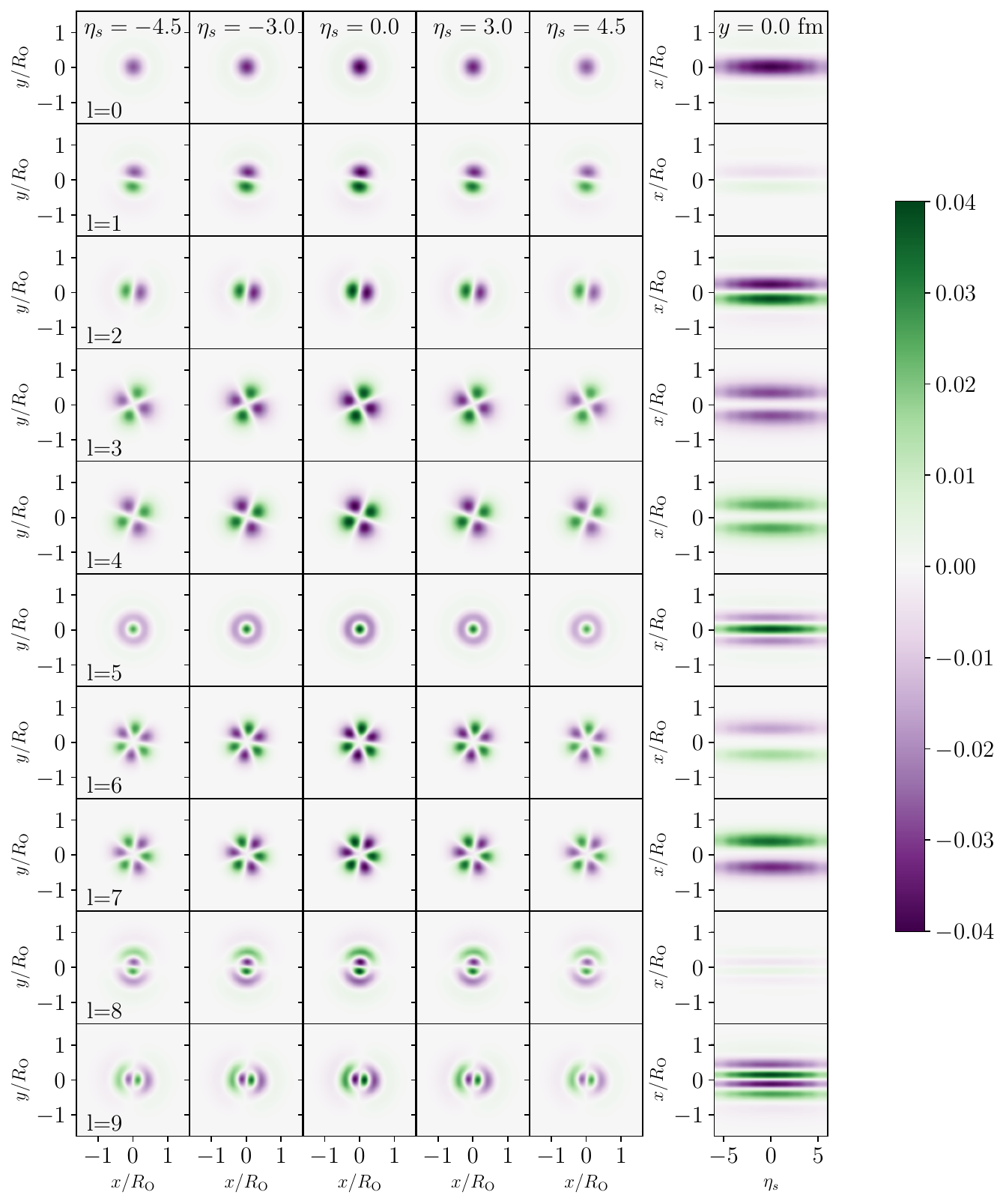}
    \caption{First $10$ normalized fluctuation modes $\tilde{\Psi}_l$ for the central bin ($0-10\%$) shown in the transverse plane in units of the system radius ($R_O = 2.608$ fm) for $\eta_s=\pm 4.5$, $\pm3$ and $0$. The last column represents the $x-\eta_s$ plane for $y=0$ fm.}
\label{fig:modes_3D_0-10}
\end{figure}

\begin{figure}
    \centering
    \includegraphics[width=\linewidth]{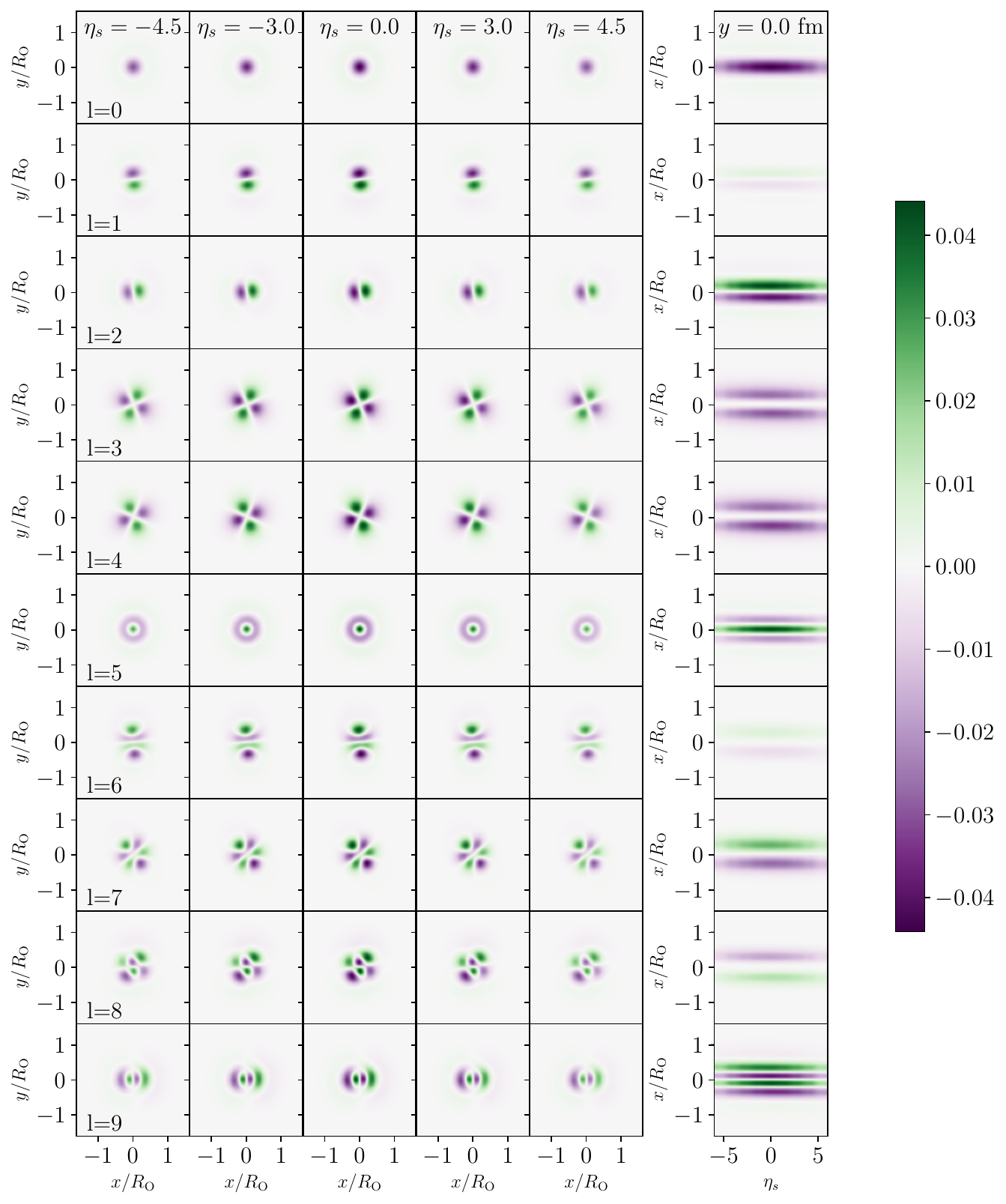}
    \caption{First $10$ normalized fluctuation modes $\tilde{\Psi}_l$ for the non-central bin ($20-30\%$) shown in the transverse plane in units of the system radius ($R_O = 2.608$ fm) for $\eta_s=\pm 4.5$, $\pm3$ and $0$. The last column represents the $x-\eta_s$ plane for $y=0$ fm.}
\label{fig:modes_3D_20-30}
\end{figure}

After discussing the average state and its properties, this section presents the fluctuation modes within a fully 3D framework. \Cref{fig:modes_3D_0-10} shows the first $10$ modes for the $0-10\%$ centrality class, while \cref{fig:modes_3D_20-30} displays the first $10$ modes for the $20-30\%$ centrality class. The modes are visualized in the transverse plane for selected longitudinal slices, and the last column illustrates the longitudinal dependence at $y = 0$ fm.

Note that the 3D modes exhibit transverse structures similar to the 2D modes discussed in the previous chapter. They follow recognizable symmetry patterns: monopole (round), dipole, quadrupole, and higher multipole shapes, with the corresponding rotations of $90^\circ$, $45^\circ$, and other characteristic angles depending on the symmetry class. As will be shown in the next subsection, the observables associated with each mode reflect these characteristic geometric symmetries. In the non-central oxygen case, the deformation remains modest and is not as strongly elliptical as in non-central Pb–Pb collisions.

The longitudinal structure reveals that the fluctuation modes largely replicate their transverse pattern across different $\eta_s$ slices, with a reduced amplitude away from midrapidity. That behavior is expected, given the implementation of fluctuations in \textsc{McDipper}. The rotational symmetries, for example, in $l=1$ and $l=2$ for both centrality cases, are also clearly visible in the $x-\eta_s$ plane.

\begin{figure}[h]
    \centering
    \includegraphics[width=0.95\linewidth]{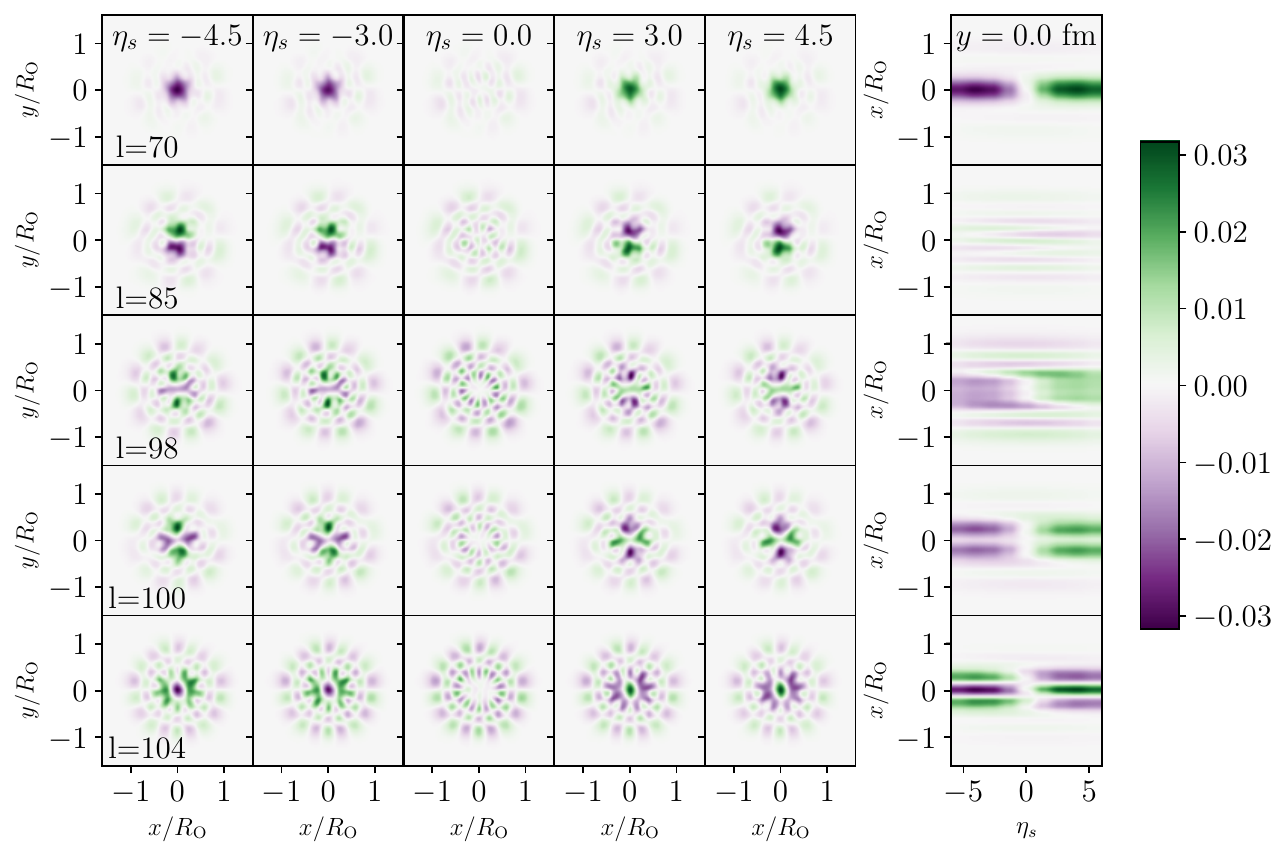}
    \caption{First $5$ normalized fluctuation modes with an odd structure for the central bin ($0-10\%$) shown in the transverse plane in units of the system radius ($R_O = 2.608$ fm) for $\eta_s=\pm 4.5$, $\pm3$ and $0$. The last column represents the $x-\eta_s$ plane for $y=0$ fm. The index $l$ in each line represents which modes are displayed.}
\label{fig:high_modes_0-10}
\end{figure}

A distinctive feature of the 3D decomposition is the appearance of modes with an odd structure in $\eta_s$. \Cref{fig:high_modes_0-10} displays the first five such modes for central collisions ($0-10\%$). These modes are characterized by a change of sign at $\eta_s=0$, which is clearly visible in the $x-\eta_s$ projection. Some of these modes preserve an even transverse background while adding an odd longitudinal component. Although they correspond to relatively high indices $l= \{70, 85, 98, 100, 104\}$, their transverse geometry still resembles the familiar round, dipole, or quadrupole patterns.

\begin{figure}[h]
    \centering
    \includegraphics[width=0.95\linewidth]{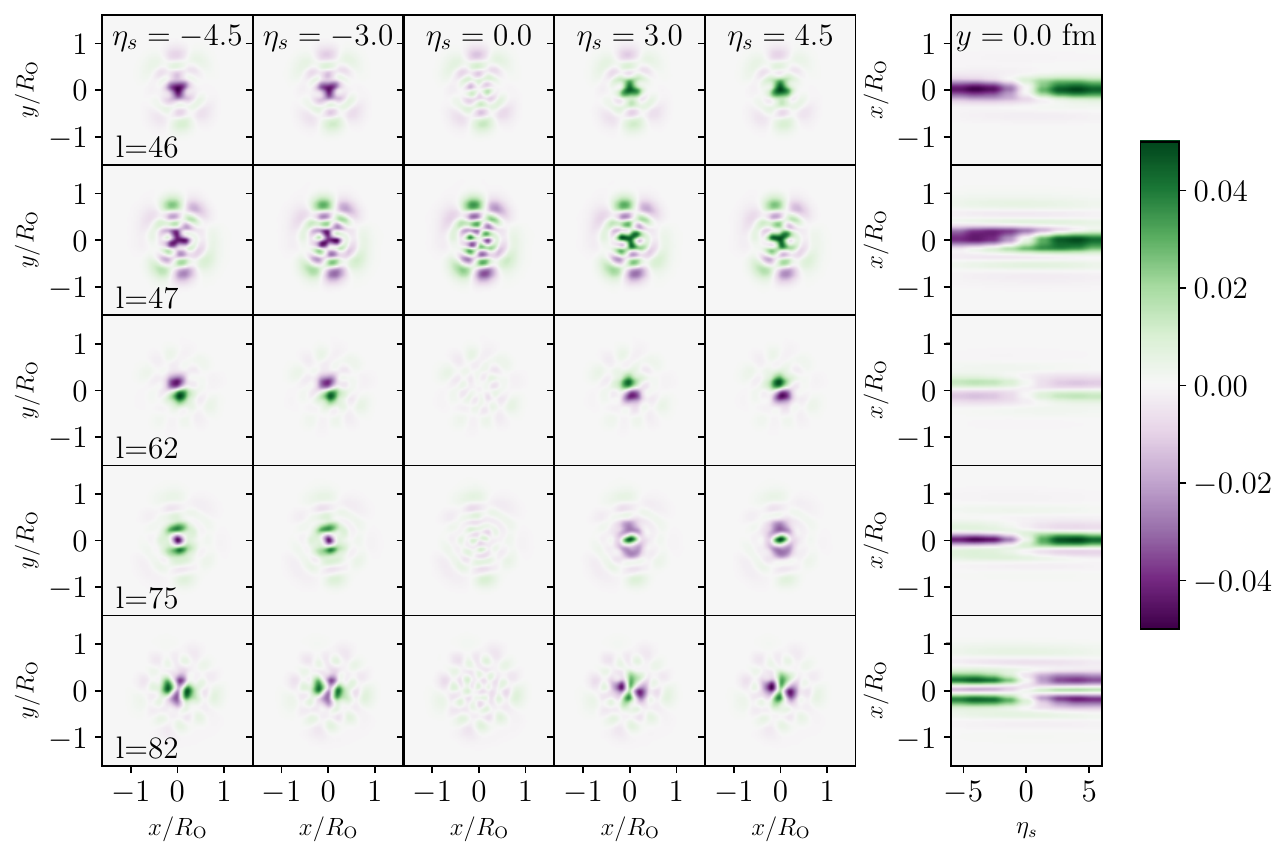}
    \caption{First $5$ normalized fluctuation modes with an odd structure for the non-central bin ($20-30\%$) shown in the transverse plane in units of the system radius ($R_O = 2.608$ fm) for $\eta_s=\pm 4.5$, $\pm3$ and $0$. The last column represents the $x-\eta_s$ plane for $y=0$ fm. The index $l$ in each line represents which modes are displayed.}
\label{fig:high_modes_20-30}
\end{figure}

In the $20-30\%$ centrality bin, shown in \cref{fig:high_modes_20-30}, odd modes appear at lower indices $l= \{46, 47, 62, 75, 82\}$ compared to the central case. This indicates that longitudinally asymmetric structures become relevant earlier in the eigenvalue spectrum for non-central events. As will be discussed in \cref{subsec:modes_application}, these odd modes play a key role in describing longitudinal decorrelations. In this sense, non-central events more easily capture such longitudinal effects.

For completeness and comparison with the 2D case, \cref{appB} presents the transverse-plane structure of the first $60$ modes at $\eta_s=0$. The following subsection builds upon this discussion by analyzing the quantitative properties of these 3D modes and examining their response coefficients.

\subsection{Properties of the 3D modes}
\label{subsec:3D_properties}

This subsection presents and discusses several properties of the fluctuation modes introduced in the previous subsection. As mentioned earlier, the average state in oxygen–oxygen collisions does not capture the majority of the information contained in event-by-event configurations. This already suggests that fluctuations play a comparatively more significant role in intermediate systems.

\begin{figure}[h]
    \centering
    \includegraphics[width=0.6\linewidth]{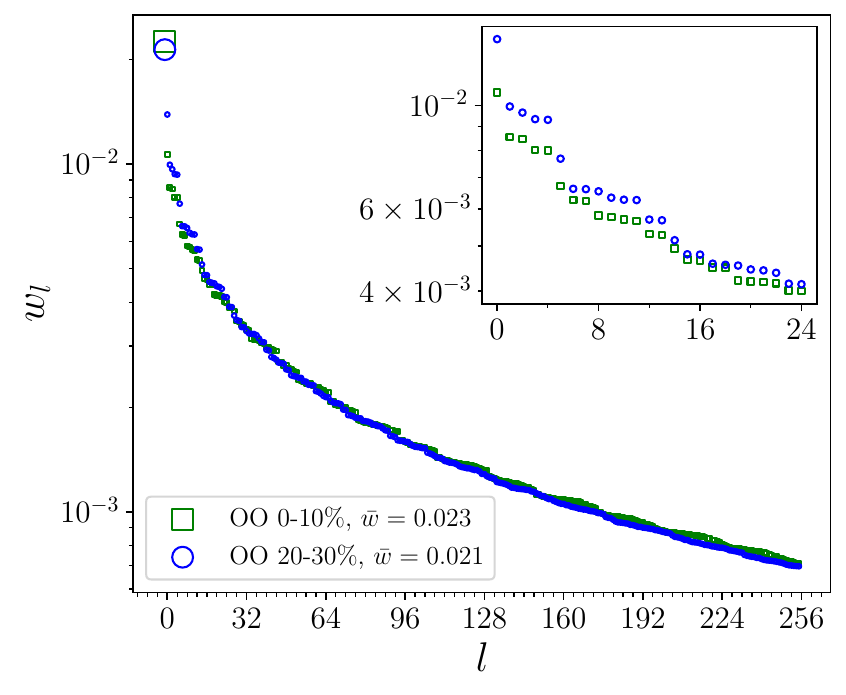}
    \caption{Weights for the first $256$ modes for both centrality bins ($0-10\%$ in green and $20-30\%$ in blue). The weight of the average state is shown by the larger symbol at $l=-1$. The curves exhibit a comparable scale in both centrality classes.}
\label{fig:weights_OO}
\end{figure}

To quantify the relative importance of the fluctuations, the weights defined in \cref{eq:weight_modes} and \cref{eq:weight_avg} are employed. These quantities are displayed in \cref{fig:weights_OO}. Comparing the two centrality classes, one observes that the corresponding curves are nearly identical, indicating that fluctuations contribute with similar relative importance in both centralities. This observation is consistent with the discussion in the first subsection of the results, where it was noted that centrality in intermediate systems does not correspond primarily to geometric deformation, as in large systems. Instead, it is more closely related to variations in the number and distribution of localized energy hot spots.

The weight of the average state ($\bar{w}$) shown in \cref{fig:weights_OO} as $l=-1$, is of order $2\%$, which is significantly smaller than in the Pb–Pb case, where it was approximately $13\%$. Furthermore, the first fluctuation modes have weights of about $\sim 1\%$, resulting in only a small gap between the contribution of the average state and the leading modes. This contrasts sharply with larger systems. From \cref{fig:weights_Pb}, one observes that in Pb–Pb collisions the weight $w_l$ of the first modes differs from that of the average state by nearly an order of magnitude \cite{Krupczak:2025gwg} ($\sim 10\%$) and here this difference is of order $\sim 1\%$. These observations demonstrate that O–O collisions are genuinely fluctuation-dominated systems, rather than merely scaled-down versions of large collision systems.

\begin{figure}[h]
    \centering
    \includegraphics[width=0.49\linewidth]{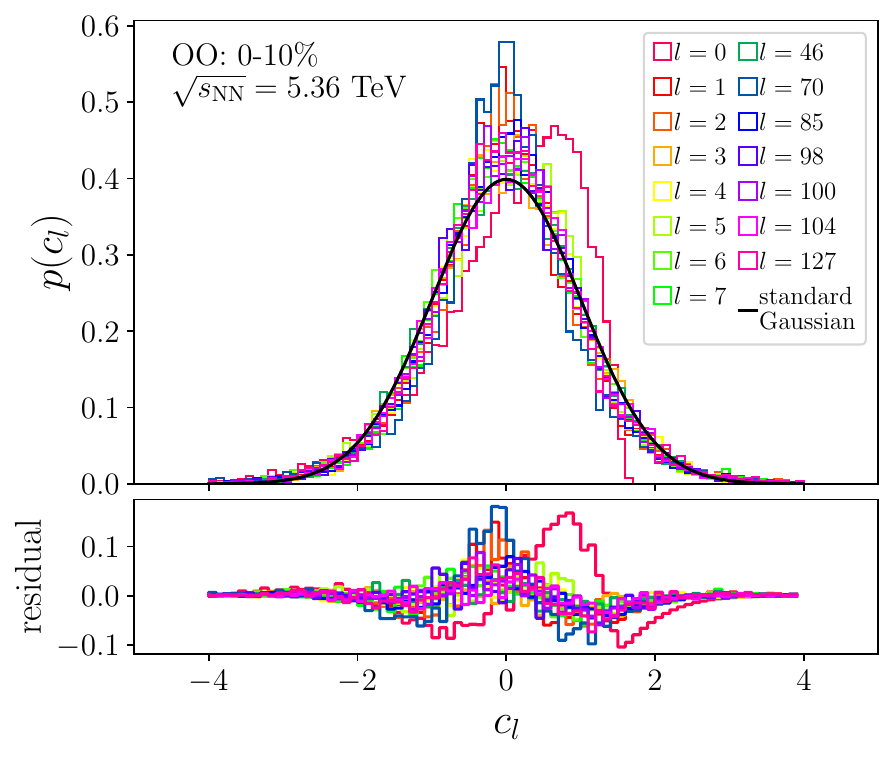}
    \includegraphics[width=0.49\linewidth]{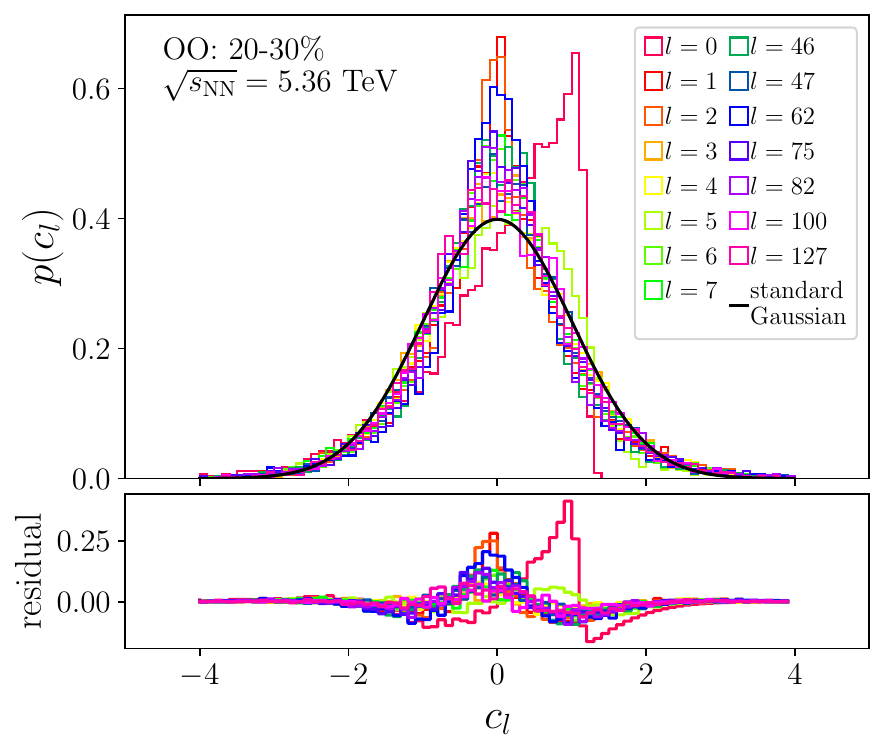}
    \caption{Distribution of the expansion coefficients $c_l$ for $8192$ random events in the $0-10\%$ (left) and $20-30\%$ (right) centrality classes. A standard Gaussian distribution is shown for comparison; all modes except $l=0$ can be approximated by a Gaussian.}
\label{fig:cl_OO}
\end{figure}

Another important property of the modes, already discussed for the 2D Pb–Pb case, concerns the expansion coefficients $c_l$, which are expected to follow an approximately Gaussian distribution in the event-by-event decomposition. \Cref{fig:cl_OO} shows the distributions of $c_l$ for selected modes, including the most significant modes and representative higher even and odd modes. The most noticeable non-Gaussian distribution corresponds to the first mode $l = 0$, which is skewed toward positive values.

From \cref{fig:modes_3D_0-10} and \cref{fig:modes_3D_20-30}, one observes that mode $l = 0$ has a round shape and an overall negative contribution to the energy density. Given that this mode contributes at a level comparable to the average state (see \cref{fig:weights_OO}), and that the total energy must remain positive, the coefficient $c_{l=0}$ cannot take arbitrarily large positive values. Indeed, its distribution is effectively bounded, with $c_{l=0} < 2$. This behavior is observed in both centrality classes.

The remaining leading modes exhibit distributions that agree well with a Gaussian profile, with residual deviations not exceeding $0.25$. Higher-index modes show less Gaussian behavior and are more sharply peaked around $c_l = 0$, similar to the behavior observed in the Pb–Pb case. Overall, the Gaussian approximation remains a good working assumption for the expansion coefficients, and this will be examined further in the following subsection.

\begin{figure}[h]
    \centering
    \includegraphics[width=0.49\linewidth]{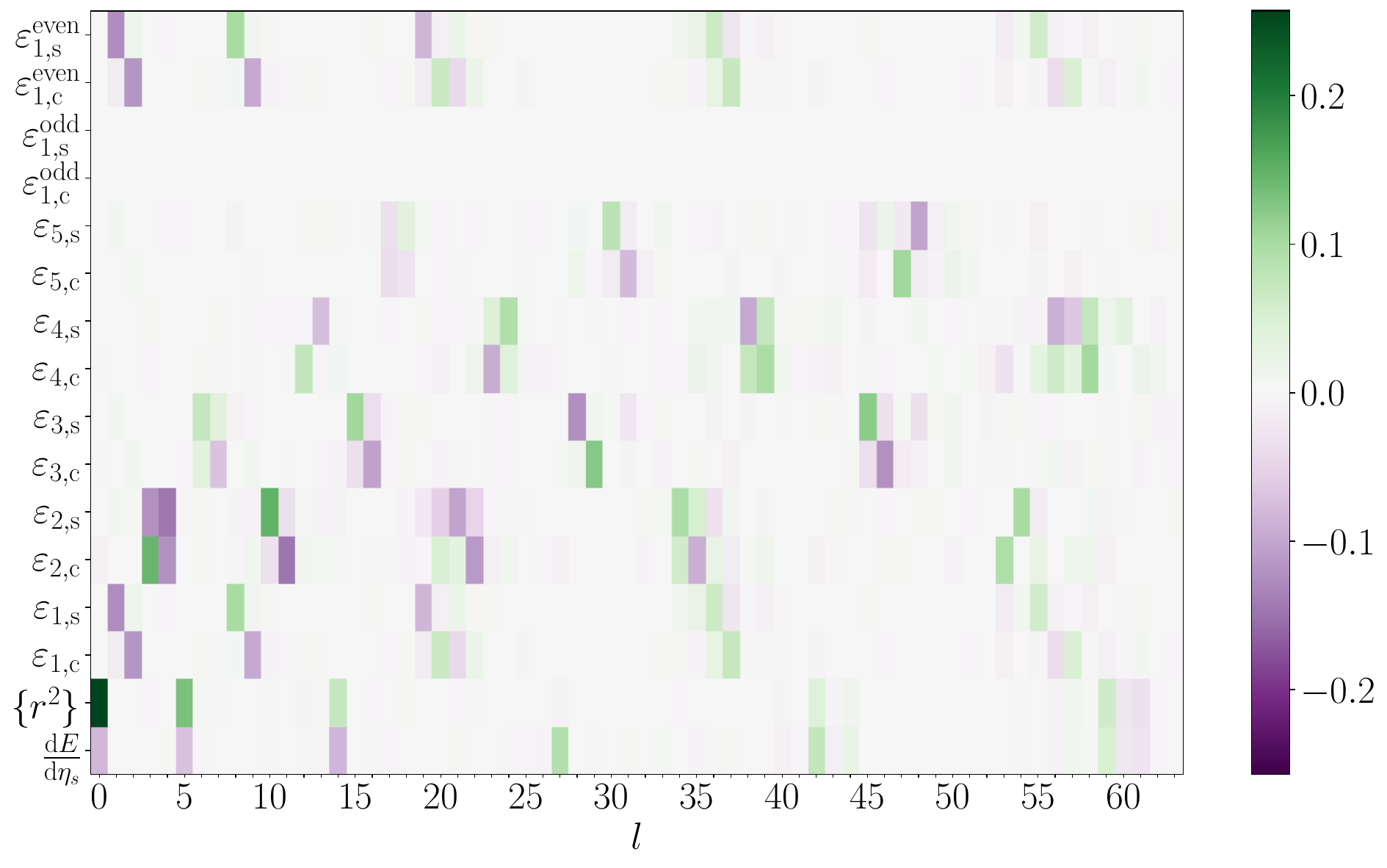}
    \includegraphics[width=0.49\linewidth]{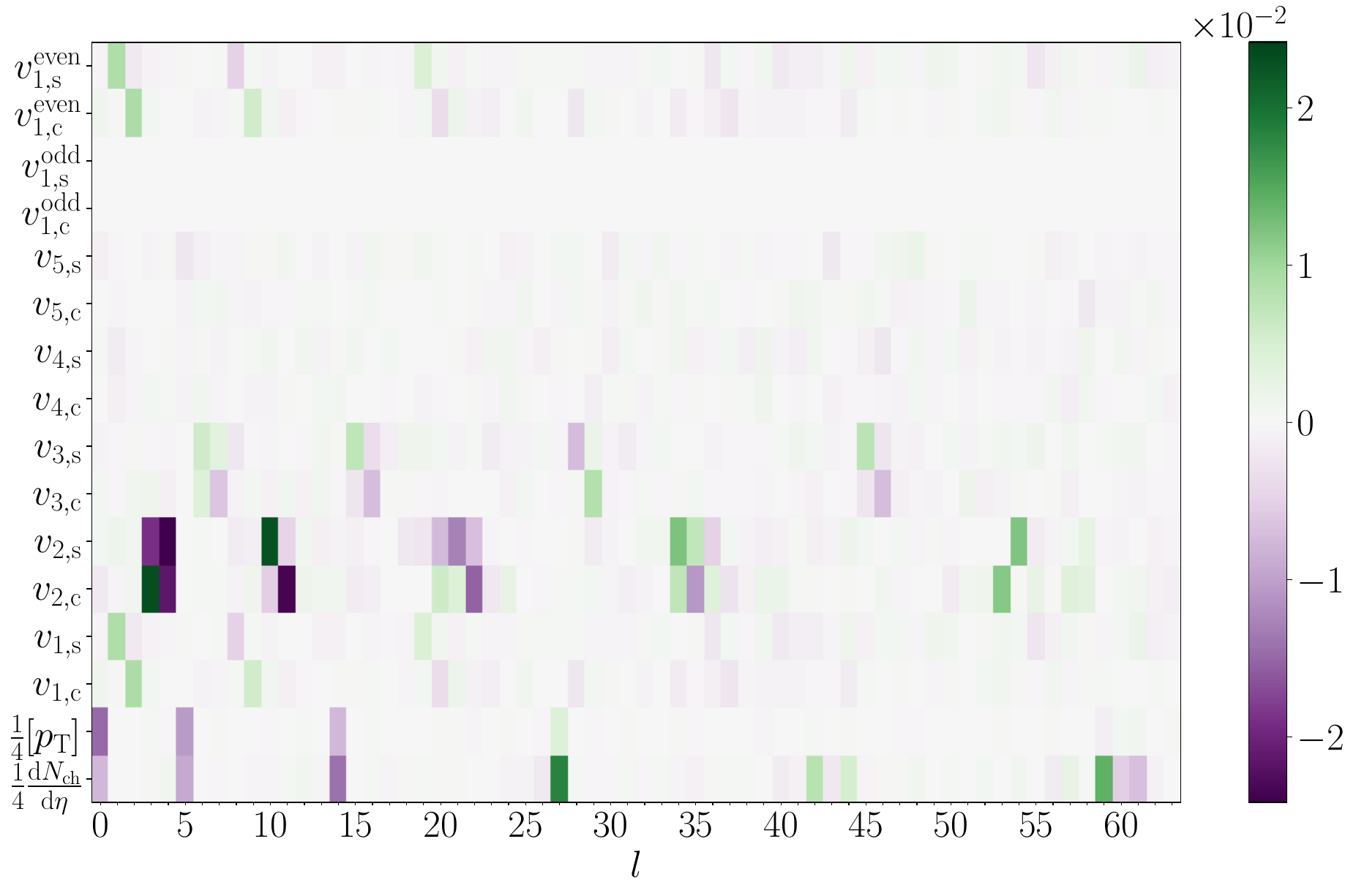}
    \caption{Linear response coefficient $L_{\alpha, l}$ at midrapidity for the first $64$ modes in the $0-10\%$ centrality class, shown for initial observables (left) and final observables (right). $\{r^2\}$, $\mathrm{d}E/\mathrm{d}\eta_s$, $\mathrm{d}N_\mathrm{ch}/\mathrm{d}\eta$, and $[p_T]$ are normalized by the average-state contribution for scale purposes and to render the quantities dimensionless.}
\label{fig:linear_coeff_0-10_OO}
\end{figure}

It is possible to determine the linear $L_{\alpha,l}$ and quadratic response coefficients $Q_{\alpha,ll}$ (see \cref{eq:linear_response} and \cref{eq:quadratic_response}). \Cref{fig:linear_coeff_0-10_OO} shows the initial (left) and final (right) linear response coefficients for the $0-10\%$ centrality bin and considering $\eta_s=0$. With the exception of $v_4$ and $v_5$, whose magnitudes are too small to be resolved clearly, the structure of the initial fluctuations is preserved through the hydrodynamic evolution and manifests in the final observables. Comparing these coefficients with the mode shapes shown in \cref{fig:modes_3D_0-10} reveals the same geometric correspondences discussed in the previous chapter.

In the 3D framework, additional quantities appear in \cref{fig:linear_coeff_0-10_OO}, namely the odd and even components of $\varepsilon_1$ and $v_1$, following the definition in \cref{eq:e1_odd_even}. As discussed in the subsection on the average state, the odd component $\varepsilon_1^{\mathrm{odd}}$ encodes the longitudinal geometric tilt. Since this tilt is fully contained in the average state, the fluctuation modes themselves do not carry this contribution. Consequently, both $\varepsilon_1^{\mathrm{odd}}$ and $v_1^{\mathrm{odd}}$ vanish for all modes. The even component, in contrast, captures transverse dipolar fluctuations. It is present in the fluctuation modes. The same transversal dipole symmetry that appears in $\varepsilon_1$ is therefore reproduced in $\varepsilon_1^{\mathrm{even}}$, confirming that both quantities reflect the same underlying transverse information.

The sign inversion observed between the linear response coefficients of the initial $\varepsilon_1$ components and the final $v_1$ presented in both O-O and Pb-Pb is still an open question. This indicates that directed flow remains a subtle observable whose interpretation is not yet fully understood \cite{Luzum:2010fb,Bozek:2010bi}.

\begin{figure}[h]
    \centering
    \includegraphics[width=0.49\linewidth]{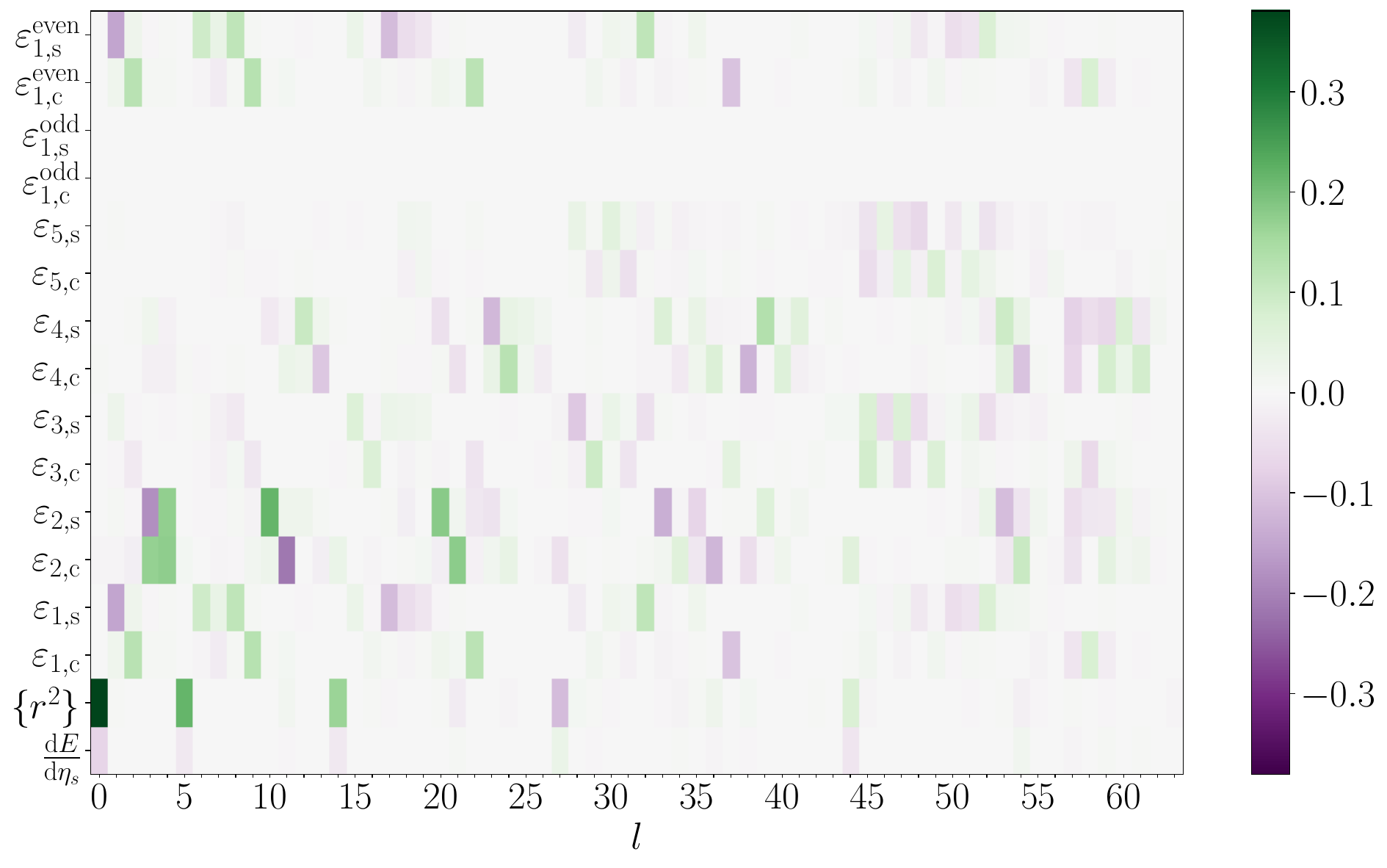}
    \includegraphics[width=0.49\linewidth]{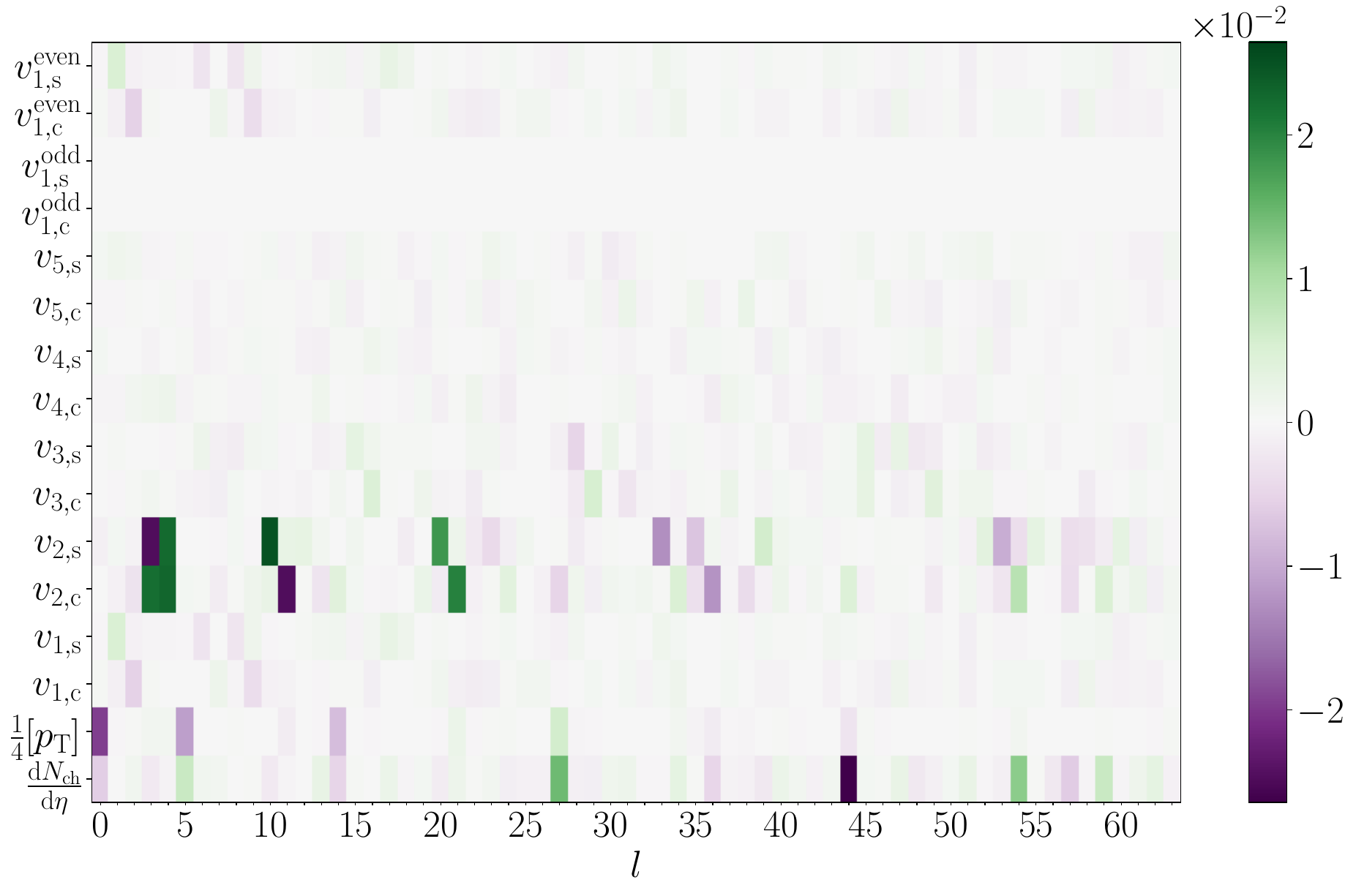}
    \caption{Linear response coefficients $L_{\alpha, l}$ at midrapidity for the first $64$ modes in $20-30\%$ centrality, shown for initial observables (left) and final observables (right). $\{r^2\}$, $\mathrm{d}E/\mathrm{d}\eta_s$, $\mathrm{d}N_\mathrm{ch}/\mathrm{d}\eta$, and $[p_T]$ are normalized by the average-state contribution for scale purposes and to render the quantities dimensionless.}
\label{fig:linear_coeff_20-30_OO}
\end{figure}

For completeness, the linear response coefficients for the non-central $20-30\%$ bin are shown in \cref{fig:linear_coeff_20-30_OO}. In oxygen collisions, even in non-central bins, the response pattern remains relatively clean, reflecting the absence of a strongly elliptical background geometry, unlike the non-central Pb–Pb case. Higher-order eccentricities are more difficult to resolve in the final flow coefficients due to numerical limitations in the particlization stage. As in the central case, the odd components of $\varepsilon_1$ vanish for the modes, while the even components are generated by the same dipole modes that contribute to $\varepsilon_1$.

Another noteworthy difference compared to large systems is that, in oxygen collisions, the magnitudes of the linear response coefficients are of similar order of magnitude for both centrality classes. This further supports the conclusion that centrality in intermediate systems does not correspond to a strong geometric hierarchy, as observed in large systems, but is instead dominated by fluctuations.

It is worth noting that the hadronic decays in this chapter are performed using a large number of oversamplings in \textsc{iSS}. As a result, the final observables retain a residual statistical uncertainty associated with the sampling procedure. This source of uncertainty is absent in the calculations presented in the previous chapter, where \textsc{MUSIC} evaluates the whole particle spectrum directly, without sampling particles. The effect is particularly visible for small signals, such as $v_4$ and $v_5$.

The quadratic response coefficients are not shown here due to limitations in this numerical precision during the particlization stage. Since the following subsection focuses exclusively on linear behavior, the quadratic properties are not further analyzed in this chapter. Nevertheless, their qualitative behavior can be inferred from the Pb–Pb case, where quadratic contributions appear, for example, in $p_T$ and in $v_{n,c}$ for even $n$.

It is also important to note that all observables discussed so far are evaluated at midrapidity. However, the same quantities can be computed at different longitudinal slices. Because the observables depend on the shape of the mode, their qualitative behavior remains consistent across $\eta_s$.

A clearer visualization of how much each mode contributes to the eccentricities is obtained using the definition in \cref{eq:tilde_eccentricities}. In this formulation, the standard eccentricity definition is modified by normalizing with the average state rather than the individual mode contribution, thereby avoiding potential divisions by zero. Although $\tilde{\varepsilon}_n$ is not identical to the physical eccentricity, it is directly proportional to it and therefore provides a meaningful measure of relative importance.

\begin{figure}[h]
    \centering
    \includegraphics[width=0.49\linewidth]{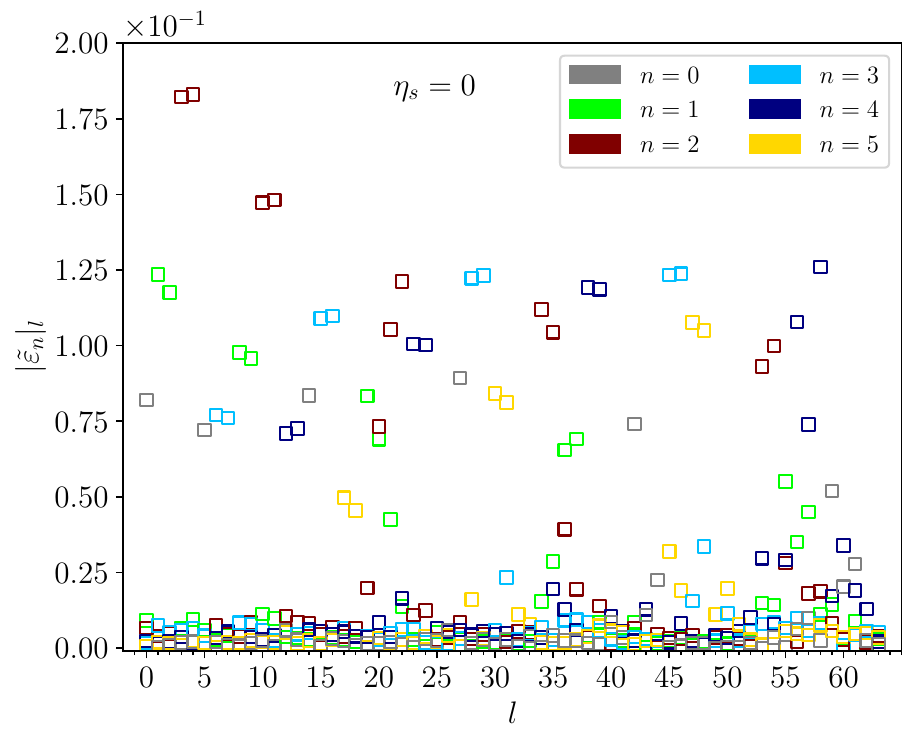}
    \includegraphics[width=0.49\linewidth]{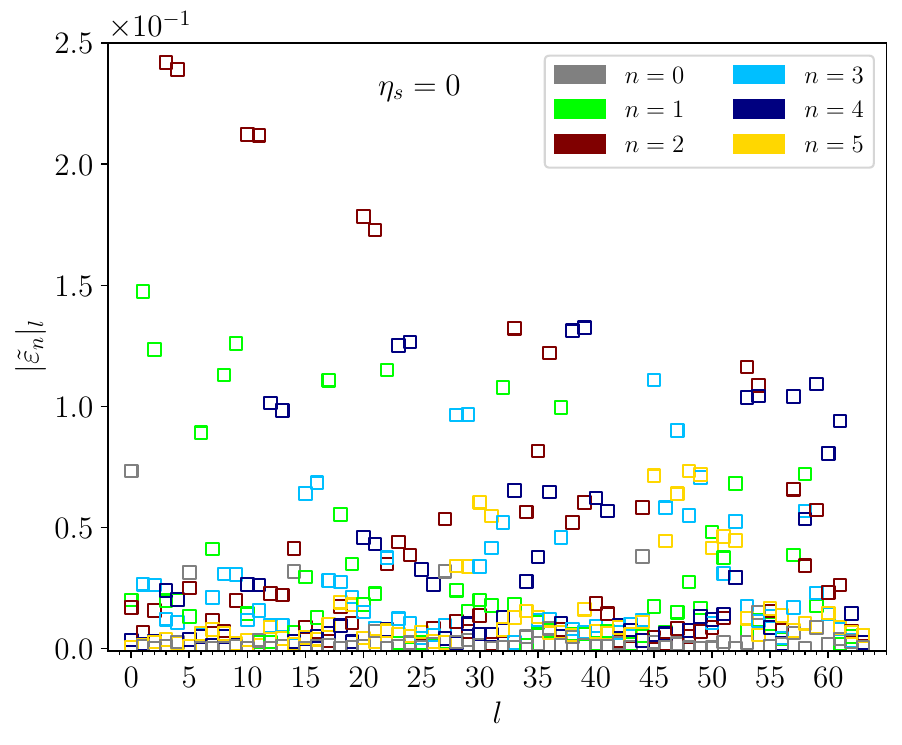}
    \caption{Modulus of the ``eccentricity'' $|\tilde{\varepsilon}_n|_l$ defined in \cref{eq:tilde_eccentricities} for the first $64$ modes in $0-10\%$ (left) and $20-30\%$ (right) centrality classes at $\eta_s=0$.}
\label{fig:tilde_eccentricities_OO}
\end{figure}

\Cref{fig:tilde_eccentricities_OO} shows $\tilde{\varepsilon}_n$ for the first $64$ modes in each centrality class. From this representation, the dominant modes contributing to $\varepsilon_2$ and $\varepsilon_3$ can be readily identified. These modes will be used in the following subsection.

Comparing with the Pb–Pb results in \cref{fig:tilde_eccentricities}, one observes a significant difference in magnitude. In Pb–Pb collisions, the maximum contribution to $\tilde{\varepsilon}_2$ was of order $0.03$ for central and $0.12$ for non-central collisions. In contrast, for O–O collisions, the corresponding values are substantially larger, approximately $0.2$ and $0.25$, respectively. This confirms the picture suggested by the analysis of the average state: the geometric contribution is relatively suppressed, and fluctuations dominate the generation of eccentricities in intermediate systems.

The next subsection focuses on the study of $v_2$ and $v_3$ using a reduced set of dominant modes, in order to test the limits of the mode-by-mode framework and to further elucidate the role of fluctuations in oxygen–oxygen collisions.

\subsection{Application of modes in a 3D scenario}
\label{subsec:modes_application}

Up to this point, the results indicate that O-O collisions are not simply small-scale Pb-Pb collisions, but rather that they behave differently. For larger systems, the geometry dominates; on the other hand, for intermediate/small systems, fluctuations dominate. With this information, one can already understand the centrality dependence of $v_2$ \cite{ATLAS:2025nnt}: for Pb-Pb, this quantity should start small and, with increasing $b$, more elliptic geometric information is translated into $v_2$; for O-O, there is no such strong effect and fluctuations dominate at a similar level across centrality, keeping the dependence weaker. At the same multiplicity, $v_2^\mathrm{PbPb} > v_2^\mathrm{OO}$ because geometrical effects are more pronounced in Pb-Pb collisions. However, for $b=0$ fm, $v_2^\mathrm{PbPb} < v_2^\mathrm{OO}$ because fluctuations play a more dominant role in oxygen collisions \cite{ATLAS:2025nnt}.

With the mode-by-mode decomposition, we can study the properties of these fluctuations. The question now arises of how many modes are necessary for a reasonable description and how well the mode-by-mode description compares with event-by-event approach. In this subsection, we aim to answer these questions for oxygen collisions.

For each centrality bin, $5000$ random events are generated in \textsc{McDipper} and evolved within the hydrodynamical setup using non-boost-invariant \textsc{MUSIC} and \textsc{iSS}. These data will be referred to as ``event-by-event data'' in the plots and serve as a reference for the modes approach.

For the modes, the observables are considered only in linear order, from \cref{eq:taylor_expansion_O}:
\begin{equation}
    O_\alpha \simeq \bar{O}_\alpha + \sum_l^{l_\mathrm{max}} c_l L_{\alpha, l},
\label{eq:obs_linear_modes}
\end{equation}
where the sum is now limited to a selected group of modes, and the expansion coefficients $c_l$ are assumed to follow a Gaussian distribution (to a good approximation, according to \cref{fig:cl_OO}).

One can then sample many $c_l$ coefficients from a Gaussian distribution and calculate the observable based on the average state and the linear response coefficients of the modes considered in \cref{eq:obs_linear_modes}. This method of reconstructing artificial events from the modes will be denoted as ``linear model'' in the plots. Note that, in this approach, only the modes are evolved, unlike in the event-by-event approach. Since generating $c_l$ values is computationally inexpensive, this approach always produces $50000$ artificial events.

Another way to use the modes to determine observables is to obtain the joint probability distribution under the Gaussian approximation \cite{Borghini:2022iym}:
\begin{equation}
    p(\{O_{\alpha_k}\}) = \frac{\exp\left[ -\frac{1}{2} (O_{\alpha_i} - \bar{O}_{\alpha_i}) (\Sigma^{-1})_{ij} (O_{\alpha_j} - \bar{O}_{\alpha_j}) \right]}{(2\pi)^2 \sqrt{\det \Sigma}},
\end{equation}
where the probability is calculated in terms of the cosine and sine components, $p(\varepsilon_{n, c}, \varepsilon_{n, s}, v_{n, c}, v_{n, s})$. The covariance matrix is then determined from the linear response coefficients of the modes (see \cref{eq:variance_O})
\begin{equation}
    (\Sigma)_{ij} = \sum_l^{l_\mathrm{max}} L_{\alpha_i, l} L_{\alpha_j, l}.
\end{equation}
This method will be referred to as ``analytical G.f.a.'' (for Gaussian fluctuation approximation) in the following plots. To obtain the marginal probabilities, one integrates over the other variable:
\begin{equation}
\begin{aligned}
    p(\varepsilon_n) &= \int p(\varepsilon_n, v_n) \mathrm{d}v_n \\
    p(v_n) &= \int p(\varepsilon_n, v_n) \mathrm{d}\varepsilon_n.
\end{aligned}
\end{equation}

\begin{figure}[!t]
    \centering
    \includegraphics[width=0.49\linewidth]{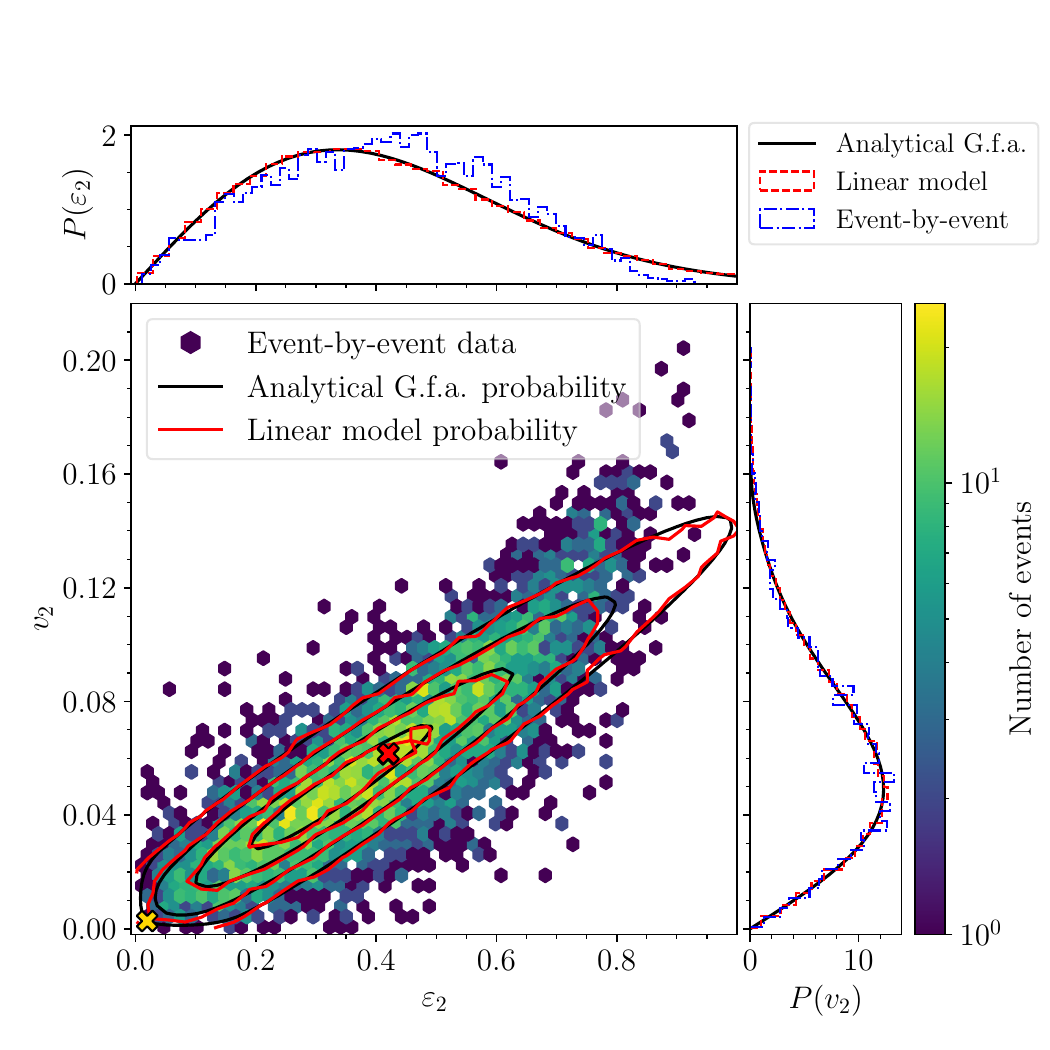}
    \includegraphics[width=0.49\linewidth]{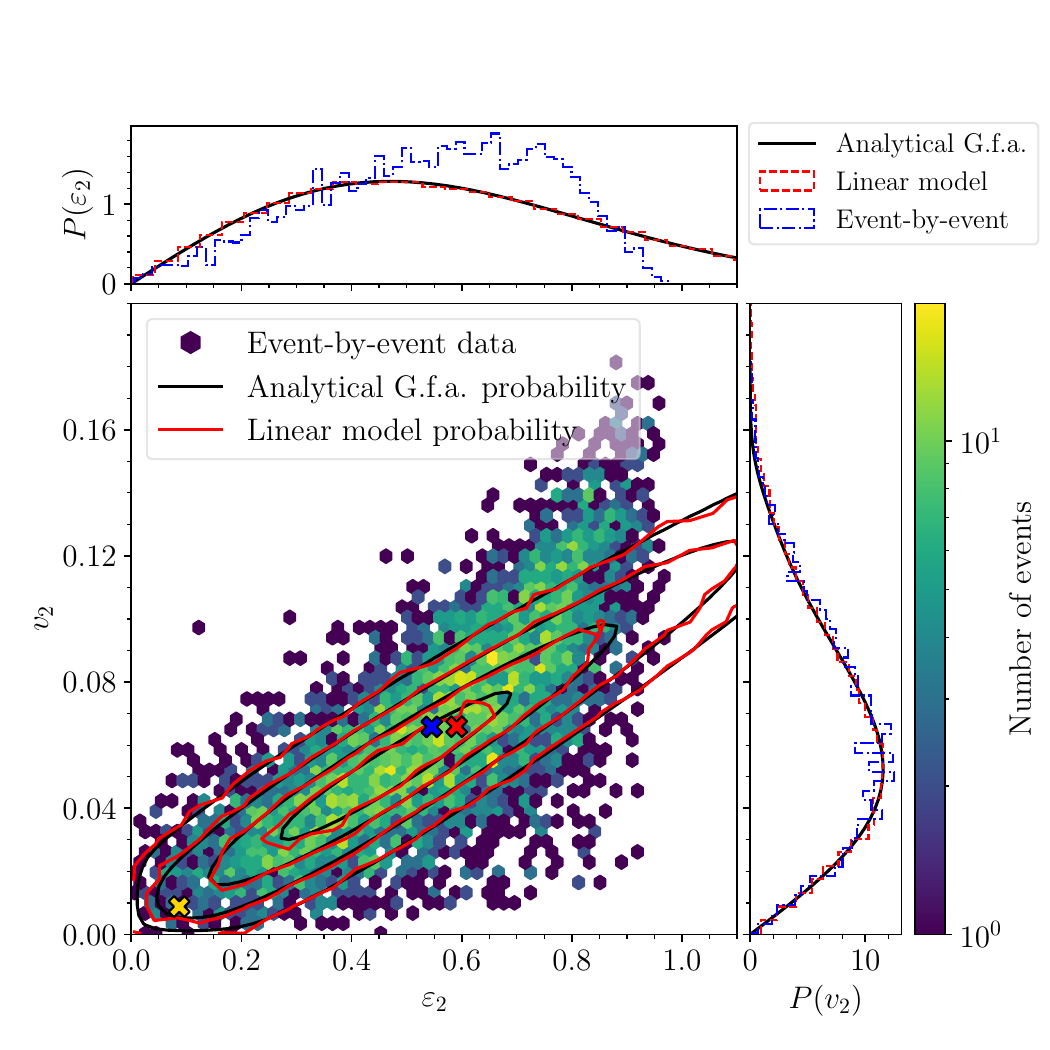}
    \includegraphics[width=0.49\linewidth]{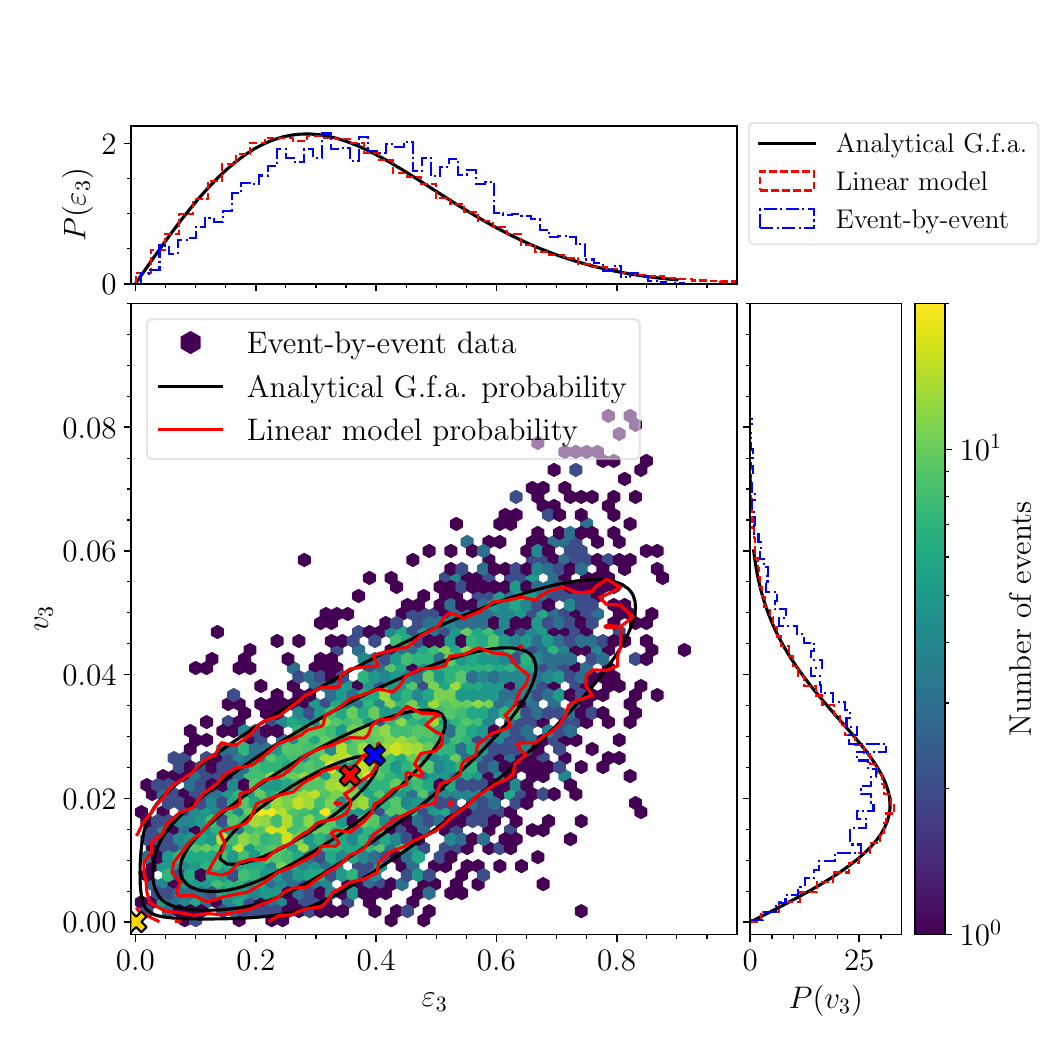}
    \includegraphics[width=0.49\linewidth]{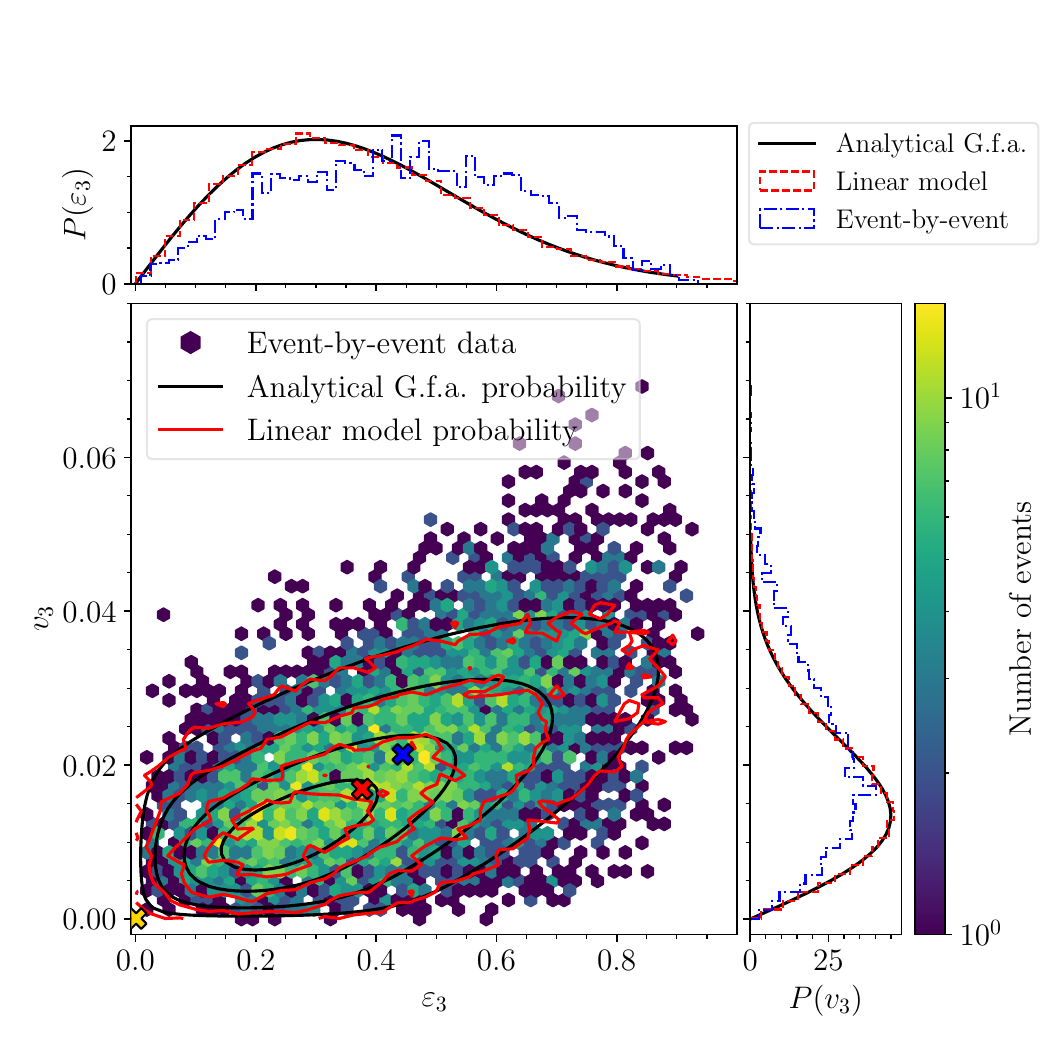}
    \caption{Relation between $v_2$ and $\varepsilon_2$ (top) and $v_3$ and $\varepsilon_3$ (bottom) for the central bin $0-10\%$ (left) and the non-central bin $20-30\%$ (right). The hexagon points show results from the $5000$ event-by-event data. The black curves represent the joint probability obtained from the analytical G.f.a. method described in the text. The red curves represent the results from $50000$ artificial events generated by sampling the expansion coefficients $c_l$ for the linear model. Both curves are shown at probability levels of $20\%$, $50\%$, $80\%$ and $95\%$. The top and right subpanels of each plot show the corresponding marginal probability distribution of each observable. The red cross represents the average over the $50000$ artificial events, the blue cross represents the average over the $5000$ event-by-event, and the yellow cross represents the value obtained from the average state $\bar{\Psi}$, which is much smaller than the event-by-event or modes result.}
\label{fig:prob_vn_en}
\end{figure}

Considering the first $128$ modes, these three approaches are compared in \cref{fig:prob_vn_en}. The top panels show the distributions of $\varepsilon_2$ and $v_2$, while the bottom panels present those of $\varepsilon_3$ and $v_3$. The centrality classes $0-10\%$ and $20-30\%$ are shown in the left and right columns, respectively. The event-by-event data are displayed as hexagonal density maps, while the marginal probability distributions of $\varepsilon_n$ and $v_n$ are shown in the upper and right subpanels, respectively. Overall, good agreement is observed between the mode-by-mode and event-by-event results.

The sampling linear model and analytical G.f.a curves in \cref{fig:prob_vn_en} represent contours containing $20\%$, $50\%$, $80\%$, and $95\%$ of the joint probability distribution. They are expected to be in good agreement because they consider the same modes and rely on the same linear and Gaussian assumptions. However, due to these same assumptions, they fail to describe the pairs $(\varepsilon_n, v_n)$ at large values. This limitation can be seen more clearly in the tails of the marginal probability distributions.

The shape of the event-by-event data naturally suggests the presence of nonlinear corrections beyond the linear response. In particular, the characteristic ``banana'' shape is well described by the inclusion of a cubic term \cite{Noronha-Hostler:2015dbi},
\begin{equation}
    v_2 \simeq \kappa_{2,2} \varepsilon_2 + \kappa_{2,222}\varepsilon_2^3.
\label{eq:cubic_v2}
\end{equation}
The cubic contribution becomes increasingly important for events with large eccentricities, where deviations from a purely linear response are most pronounced. This behavior differs from the results reported in Ref. \cite{Sievert:2019zjr}, which employed the $\mathrm{T}_{\mathrm{R}} \mathrm{ENTo}$ model together with a three-parameter Fermi (3pF) Woods-Saxon distribution for the oxygen nucleus. Since that study used different initial conditions and hydrodynamic frameworks, a direct comparison is not straightforward. Moreover, the analysis in Ref. \cite{Sievert:2019zjr} is primarily restricted to the region of small $\varepsilon_2$, where the cubic contribution is expected to be less significant.

A similar analysis of the correlation between $\varepsilon_3$ and $v_3$ also reveals a characteristic ``banana'' shape, suggesting an analogous cubic correction,
\begin{equation}
    v_3 \simeq \kappa_{3,3} \varepsilon_3 + \kappa_{3,333} \varepsilon_3^3.
\label{eq:cubic_v3}
\end{equation}
While previous studies generally found the linear response to provide an adequate description of triangular flow, the present results indicate that a cubic correction may also become relevant for $v_3$ in intermediate-size systems.

For non-central bins, the agreement between the $128$ modes and the event-by-event basis becomes slightly worse compared to central bins, although the overall behavior is still well described. One interesting feature across the different centrality classes is that the scale of the plots remains fairly similar, which is consistent with the previous discussion that centrality is not characterized by fundamentally different properties, but rather by a different total amount of energy deposition.

Figure \cref{fig:prob_vn_en} also shows the average pair $(\langle \varepsilon_n\rangle, \langle v_n\rangle)$ for the event-by-event data (blue cross) and for the artificial events generated with the linear model (red cross), which are always close to each other. However, when these pairs are compared to the pair $(\varepsilon_n, v_n)$ calculated from the average state $\bar{\Psi}$ (yellow cross), one sees that the latter remains very far away, almost at zero. In fact, these observables are mainly generated by fluctuations. Since the yellow cross lies very close to zero and is therefore difficult to visualize in the plot, \cref{tab:en_vn} presents the corresponding numerical values. In nearly all cases, the average state contributes only a very small fraction of the observable, representing a markedly different scenario from large systems, where the average state already provides a reasonable description \cite{Borghini:2022iym} of $v_2$, for example.

\begin{table}[h]
\centering
\begin{tabular}{c c c ccc}
    \toprule
        \multirow{2}{*}{centrality} & \multirow{2}{*}{$\eta_s$ or $\eta$} & \multirow{2}{*}{observable} & \multirow{2}{*}{$O_\alpha(\bar{\Psi})$} & \multicolumn{2}{c}{sample mean} \\
			& & & & linear model\ & event-by-event \\
    \toprule
        \multirow{4}{*}{$0-10\%$} & \multirow{4}{*}{$0$} 
        & $\varepsilon_2$ & 0.0196 & 0.4208 & 0.4199 \\
        &  & $ v_2$         & 0.0026 & 0.0616 & 0.0614 \\
        &  & $\varepsilon_3$ & 0.0012 & 0.3574 & 0.3976 \\
        &  & $v_3$ & 10$^{-5}$ & 0.0238 & 0.0270 \\
    \hline
        \multirow{4}{*}{$0-10\%$} & \multirow{4}{*}{$4$} 
        & $\varepsilon_2$ & 0.0194 & 0.4151 & 0.4181 \\
        &  & $v_2$         & 0.0022 & 0.0512 & 0.0526 \\
        &  & $\varepsilon_3$ & 0.0015 & 0.3510 & 0.3966 \\
        &  & $v_3$ & 10$^{-5}$ & 0.0195 & 0.0219 \\
    \hline
        \multirow{4}{*}{$20-30\%$} & \multirow{4}{*}{$0$} 
        & $\varepsilon_2$ & 0.0871 & 0.5889 & 0.5456 \\
        &  & $v_2$         & 0.0088 & 0.0657 & 0.0658 \\
        &  & $\varepsilon_3$ & 0.0010 & 0.3758 & 0.4443 \\
        &  & $v_3$ & 10$^{-5}$ & 0.0169 & 0.0214 \\
    \hline
        \multirow{4}{*}{$20-30\%$} & \multirow{4}{*}{$4$} 
        & $\varepsilon_2$ & 0.0869 & 0.5822 & 0.5430 \\
        &  & $v_2$         & 0.0076 & 0.0551 & 0.0559 \\
        &  & $\varepsilon_3$ & 0.0021 & 0.3732 & 0.4433 \\
        &  & $v_3$ & 10$^{-5}$ & 0.0146 & 0.0175 \\
    \bottomrule
\end{tabular}
\caption{$\varepsilon_n$ and $v_n$ calculated from the average state $\Bar{\Psi}$, from the average over $50000$ artificial events generated by sampling the expansion coefficients $c_l$ in the linear model, and from the average over $5000$ event-by-event simulations. Results are obtained using the first $128$ modes. Two centrality classes and two slices in the longitudinal direction are shown.}
\label{tab:en_vn}
\end{table}

\Cref{tab:en_vn} also presents the observables for different slices in the longitudinal direction. Although the observables vary slightly, they remain within the same order of magnitude. Similar plots to \cref{fig:prob_vn_en} could also be produced for different longitudinal slices using $128$ modes, again showing good agreement.

As an application, and to test the mode description against event-by-event simulations, the linear coefficients $\kappa_{2,2}$ and $\kappa_{3,3}$ (see \cref{eq:cubic_v2,eq:cubic_v3}) are computed in each slice in order to extract their longitudinal dependence. 

For the $5000$ event-by-event data set, a cubic fit is performed on the data, according to \cref{eq:cubic_v2,eq:cubic_v3}. However, in the present analysis we focus only on the linear coefficients $\kappa_{2,2}(\eta, \eta_s)$ and $\kappa_{3,3}(\eta_s, \eta)$. From these data, it is also straightforward to estimate the statistical uncertainty using a Jackknife method. 

For the linear model, again $50000$ artificial events are generated based on the linear response coefficients of the first $128$ modes. A linear fit is then performed to extract $\kappa_{2,2}(\eta_s, \eta)$ and $\kappa_{3,3}(\eta_s, \eta)$. Although the data favor a cubic behavior, the linear fit is used here to maintain consistency with this approach, which is based on linear response and is not expected to capture cubic contributions. 

For the analytical method, the linear coefficient is obtained by integrating the joint probability:
\begin{equation}
\begin{aligned}
    \kappa_{2,2}(\eta_s, \eta) &= \frac{\langle v_2(\eta_s, \eta) \rangle}{\langle \varepsilon_2(\eta_s, \eta) \rangle} = \frac{\int p(\varepsilon_2, v_2)\, \mathrm{d}\varepsilon_2}{\int p(\varepsilon_2, v_2)\, \mathrm{d}v_2}, \\
    \kappa_{3,3}(\eta_s, \eta) &= \frac{\langle v_3(\eta_s, \eta) \rangle}{\langle \varepsilon_3(\eta_s, \eta) \rangle} = \frac{\int p(\varepsilon_3, v_3)\, \mathrm{d}\varepsilon_3}{\int p(\varepsilon_3, v_3)\, \mathrm{d}v_3}.
\end{aligned}
\end{equation}
Here again, only the linear approximation is assumed, and the first $128$ modes are used. Note that for the mode-based methods, estimating an uncertainty is more challenging due to the systematic uncertainties associated with the underlying assumptions.

\begin{figure}[h]
    \centering
    \includegraphics[width=0.49\linewidth]{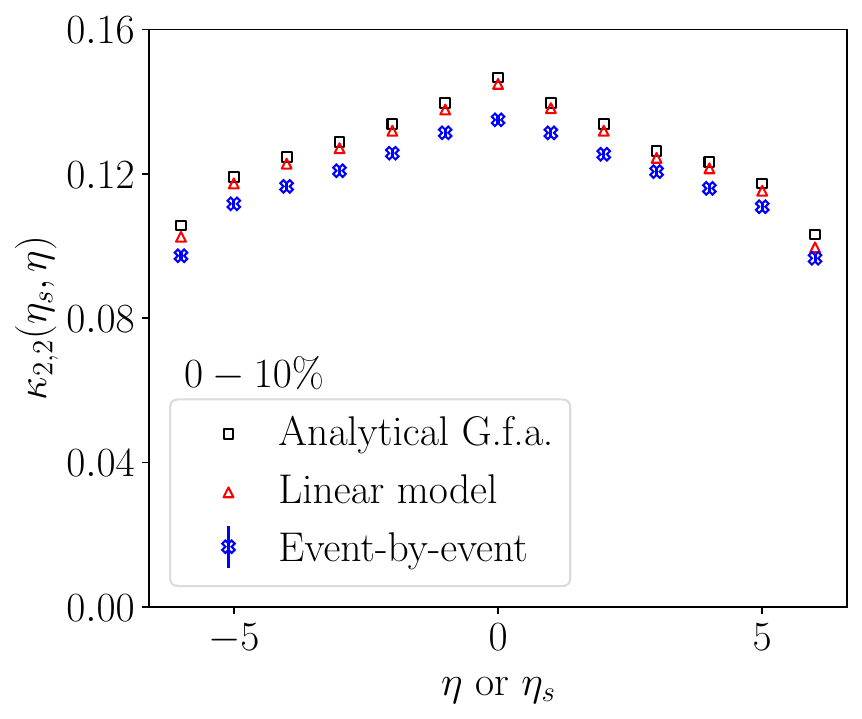}
    \includegraphics[width=0.49\linewidth]{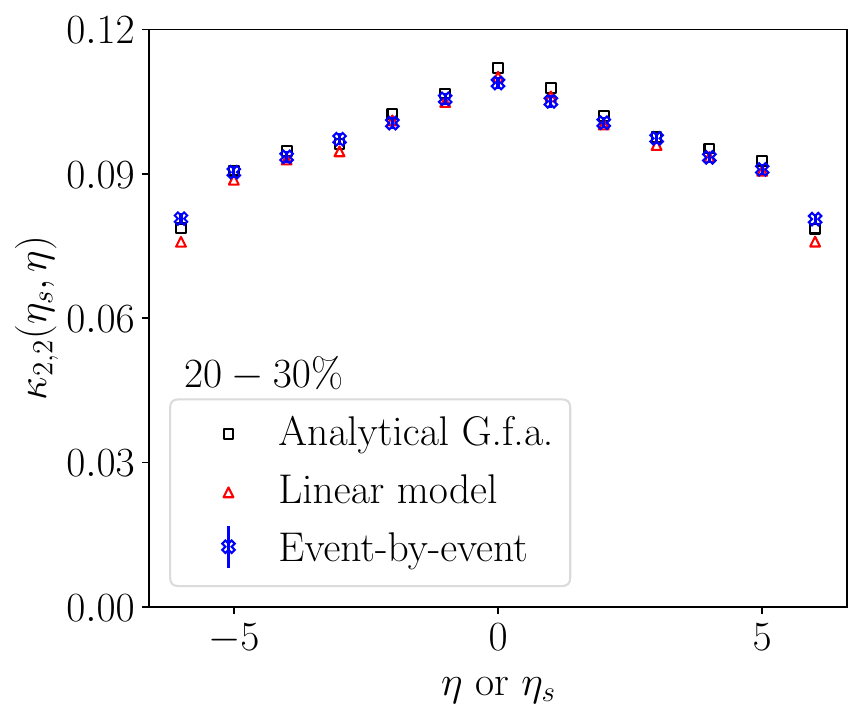}
    \includegraphics[width=0.49\linewidth]{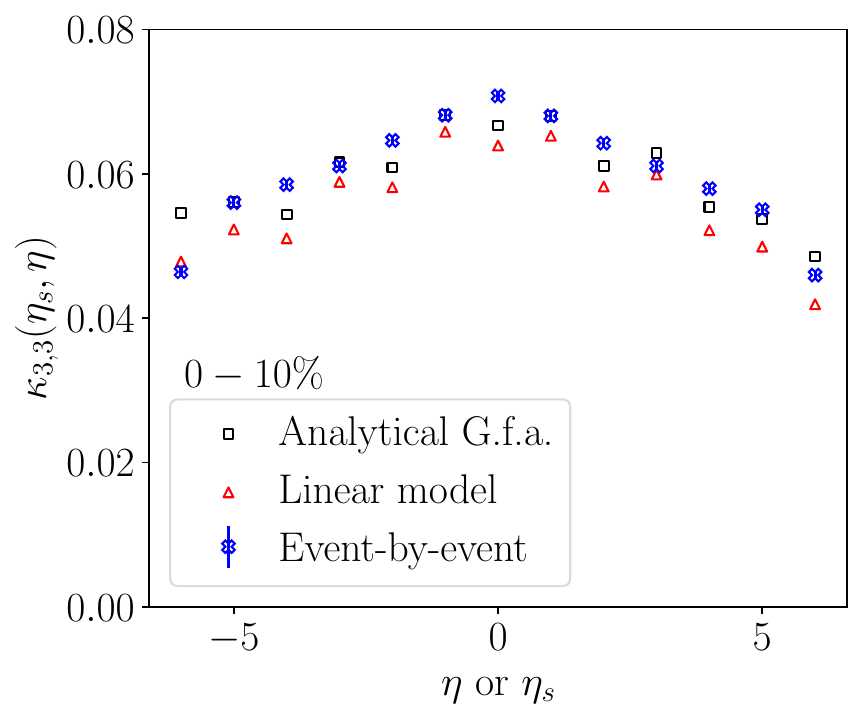}
    \includegraphics[width=0.49\linewidth]{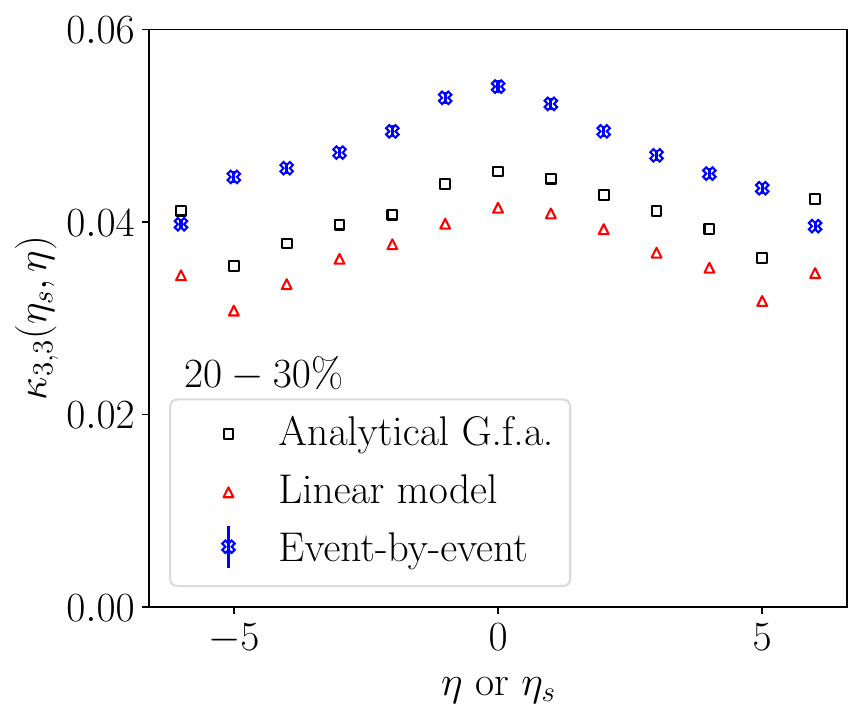}
    \caption{Linear coefficient $\kappa_{2,2}(\eta_s, \eta)$ (top) and $\kappa_{3,3}(\eta_s, \eta)$ (bottom) for central $0-10\%$ (left) and non-central $20-30\%$ (right). The extraction procedures for the three methods are described in more detail in the text.}
\label{fig:kappas}
\end{figure}

The results for the three methods for $\kappa_{2,2}$ and $\kappa_{3,3}$ in both centralities are shown in \cref{fig:kappas}. A clear longitudinal dependence of $\kappa_{n,n}$ is observed. Furthermore, $\kappa_{n,n}$ decreases with decreasing the transversal energy, consistent with previous two-dimensional studies of this coefficient as a function of centrality \cite{Noronha-Hostler:2015dbi}. In this context, higher energy density (or equivalently, more central events or midrapidity regions) corresponds to larger values of $\kappa_{n,n}$.

The ``analytical'' (black points) and ``linear'' (read points) methods are expected to agree with each other, since they rely on the same linear assumptions and use the same set of modes. However, the event-by-event data  (blue points) describe a more realistic scenario, where a cubic fit provides a better representation of the data. In this sense, the black and red points are expected to approach the description given by the blue points. For the central case of $\kappa_{2,2}$, it is straightforward to understand why the coefficients obtained from the linear description are larger than those from the cubic fit: the inclusion of higher-order terms reduces the value of $\kappa_{2,2}$. For the non-central case, the same effect is expected, although the scale suggests that the linear description already provides a reasonable approximation.

For $\kappa_{3,3}$, the assumption in \cref{eq:cubic_v3} is not the only possible choice. The literature also reports contributions involving $\varepsilon_1 \varepsilon_2$ and $\varepsilon_{3,5}$ \cite{Gardim:2014tya}, which could modify the value of $\kappa_{3,3}$ and were not considered in the present analysis.

\begin{figure}[!h]
    \centering
    \includegraphics[width=0.49\linewidth]{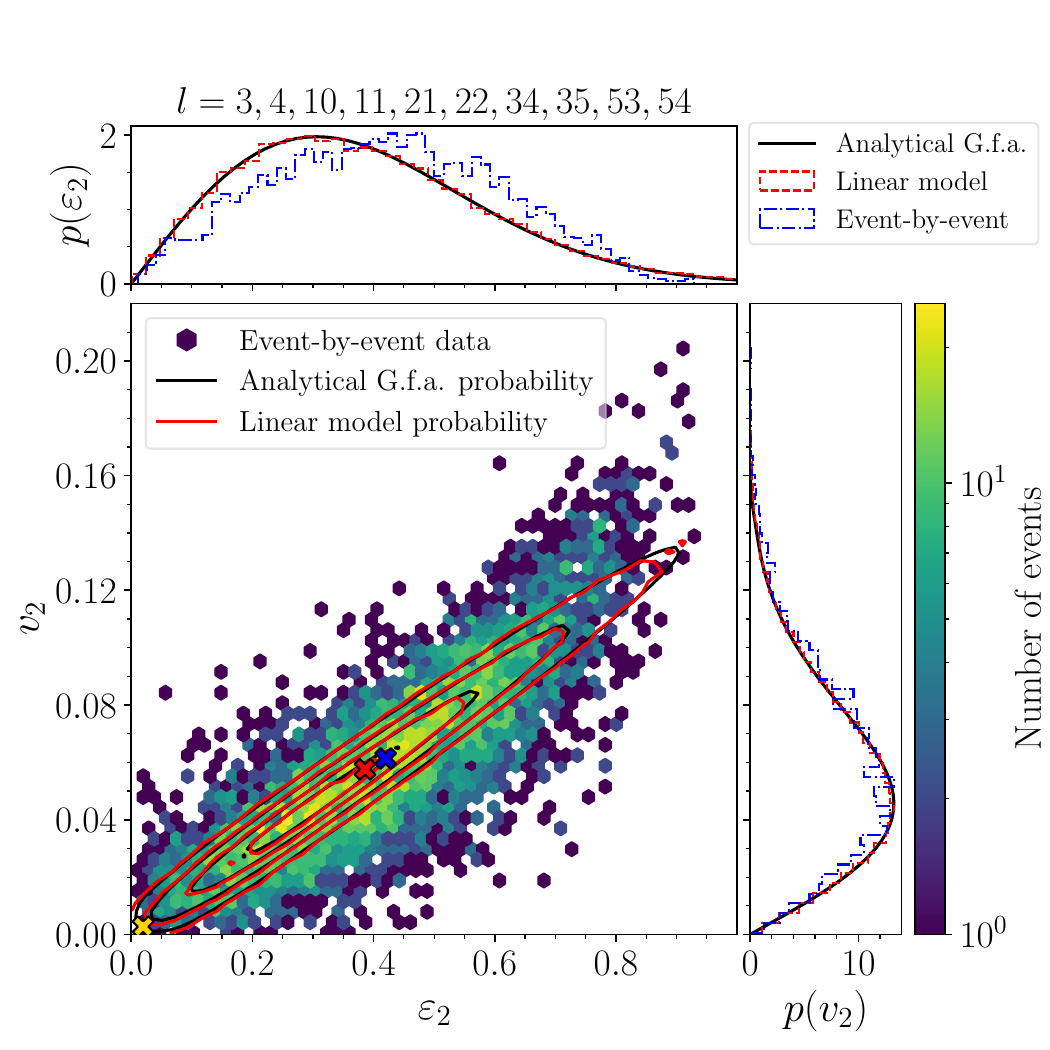}
    \includegraphics[width=0.49\linewidth]{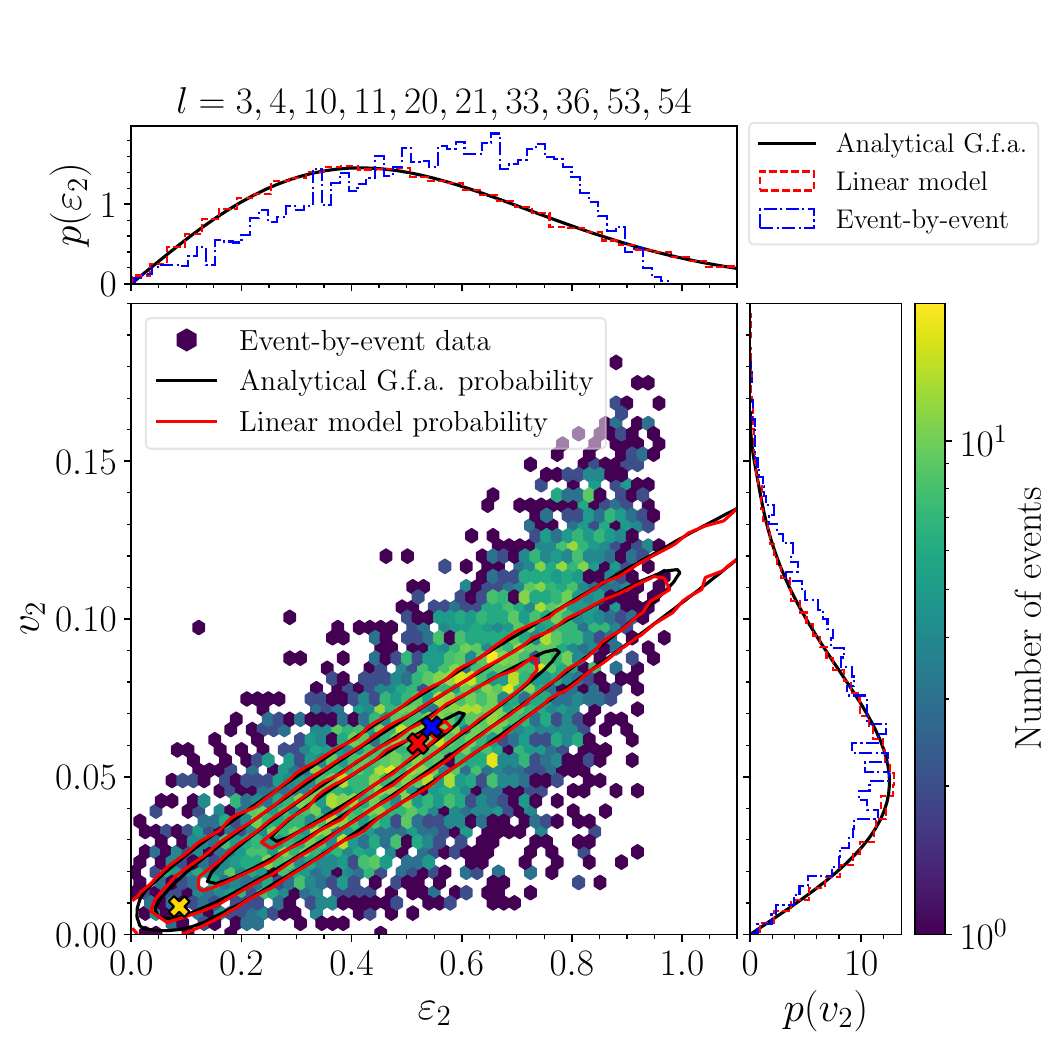}
    \includegraphics[width=0.49\linewidth]{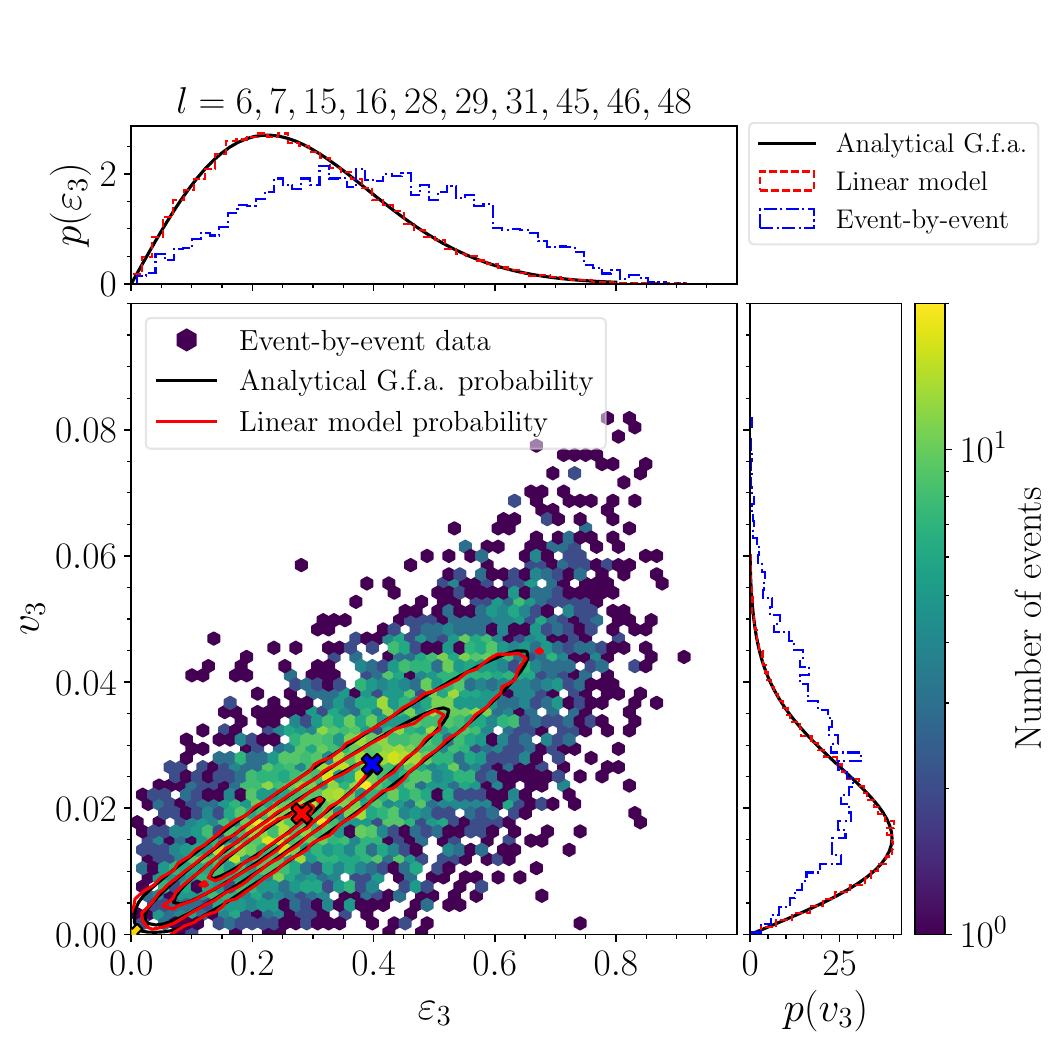}
    \includegraphics[width=0.49\linewidth]{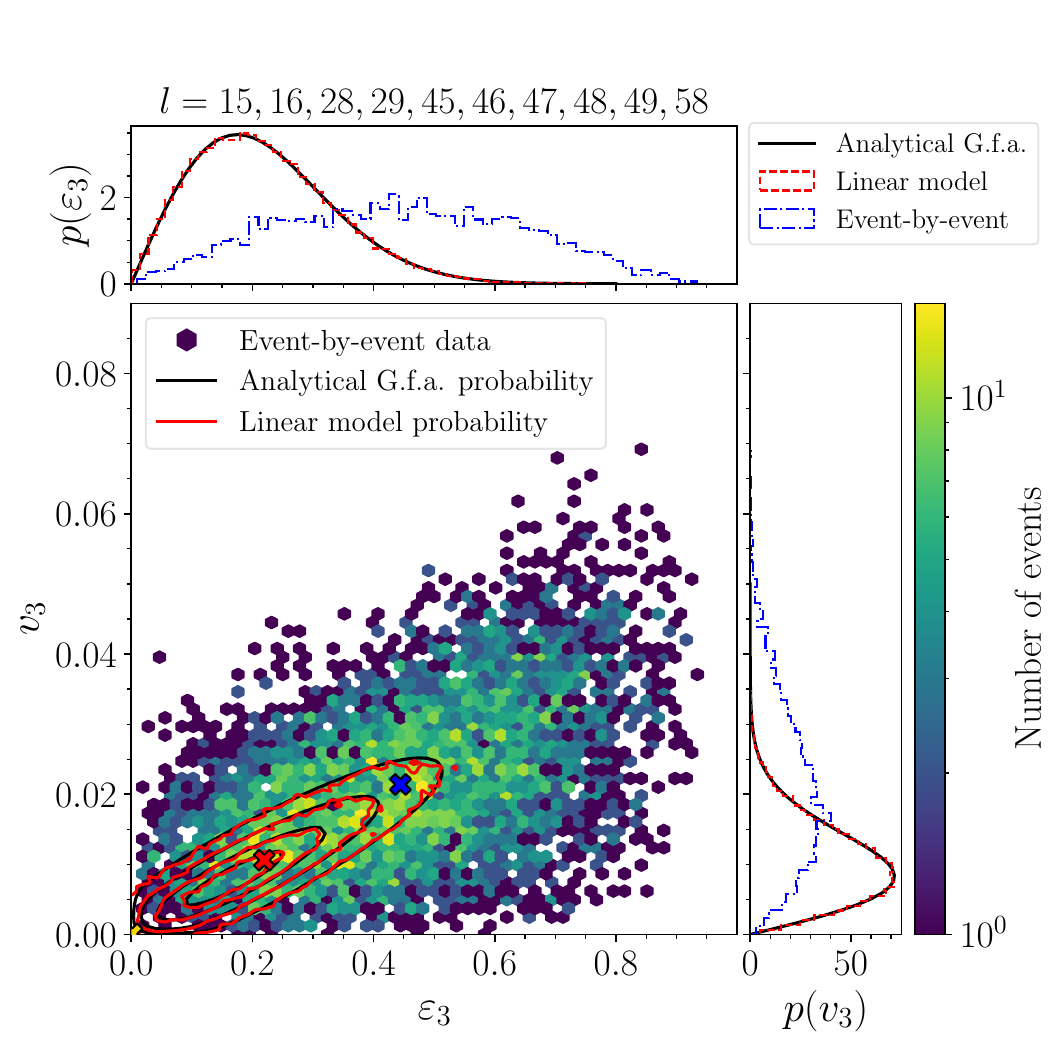}
    \caption{Same as \cref{fig:prob_vn_en}, but the linear model and analytical G.f.a approaches are now performed using only ten selected modes that contribute to the observable, which are indicated in the title of each plot.}
\label{fig:prob_10modes}
\end{figure}

From this mode-by-mode decomposition, a good description is obtained using the first $128$ modes. However, only a subset of these modes contributes significantly to each observable of interest, motivating the possibility of constructing a reduced set of selected modes. Returning to \cref{fig:tilde_eccentricities_OO}, we select, for each centrality class, the ten modes with the largest contributions to $|\tilde{\varepsilon}_2|$ and $|\tilde{\varepsilon}_3|$ among the first 64 modes. Using only these modes, we reconstruct the probability distributions from \cref{fig:prob_vn_en}. The resulting distributions are presented in \cref{fig:prob_10modes}, where each panel is labeled with the corresponding set of ten selected modes. A good agreement is found for $v_2$ while evolving only ten modes, providing a computationally cheaper description. This indicates that, although fluctuations play an important role, only a relatively small number of modes is required to reproduce the overall behavior.

For $v_3$, however, the description becomes worse. The reason is that these ten selected modes do not generate sufficiently large initial $\varepsilon_3$ values (see \cref{fig:tilde_eccentricities_OO}) when compared to the dominant modes contributing to $\varepsilon_2$. Therefore, a larger number of modes is necessary to achieve a better description.

The main difference between the reconstructions using $128$ and $10$ modes lies in how wide the probability levels are. When only $10$ modes are included, the contours describe only the densest region of the distribution, whereas the reconstruction with $128$ modes also captures the less probable events. In other words, increasing the number of modes improves the precision of the reconstruction.

Note that the predicted probability distributions agree reasonably well for low values of $\varepsilon_n$, where the cubic contributions are not significant. To reproduce large values of $v_n$, however, the linear assumption for the modes requires very large values of $\varepsilon_n$, in some cases even exceeding unity (which would be nonphysical for real events). One possible improvement would be to include the quadratic response coefficients to account for these non-linear behaviors. However, this would require reconstructing the model based on $Q_{\alpha, ll}$ while ensuring sufficient numerical precision in their determination.

Up to this point, we have seen how a small set of selected modes can successfully describe several observables, although some require more modes than others to achieve a satisfactory level of precision. However, one observable that requires a significantly larger number of modes is the longitudinal decorrelation, defined as \cite{CMS:2015xmx}
\begin{equation}
    r(\eta_{s1}, \eta_{s2}) = \frac{\langle O_\alpha(-\eta_{s1}) O_\alpha^\ast(\eta_{s2}) \rangle}{\langle O_\alpha(\eta_{s1}) O_\alpha^\ast(\eta_{s2}) \rangle},
\end{equation}
where $O_\alpha$ is a generic observable, and $\eta_{s1}$ and $\eta_{s2}$ represent two different longitudinal slices. In general, the decorrelation is studied for the flow coefficients. Here, we instead consider the elliptic eccentricity $\varepsilon_2$ defined as
\begin{equation}
    r_2(\eta_{s1}, \eta_{s2}) = \frac{\langle \varepsilon_{2,c}(-\eta_{s1}) \varepsilon_{2,c}(\eta_{s2}) + \varepsilon_{2,s}(-\eta_{s1}) \varepsilon_{2,s}(\eta_{s2}) \rangle}{\langle \varepsilon_{2,c}(\eta_{s1}) \varepsilon_{2,c}(\eta_{s2}) + \varepsilon_{2,s}(\eta_{s1}) \varepsilon_{2,s}(\eta_{s2}) \rangle},
\label{eq:decorrelation}
\end{equation}
since at the level of the initial state, there is no statistical noise originating from the final-state evolution.

\begin{figure}
    \centering
    \includegraphics[width=0.49\linewidth]{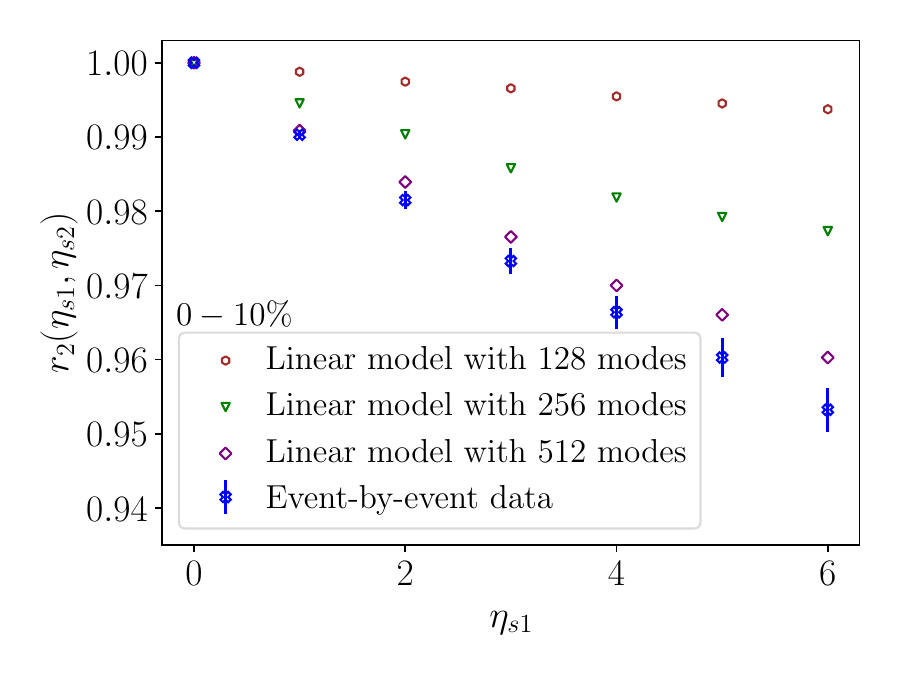}
    \includegraphics[width=0.49\linewidth]{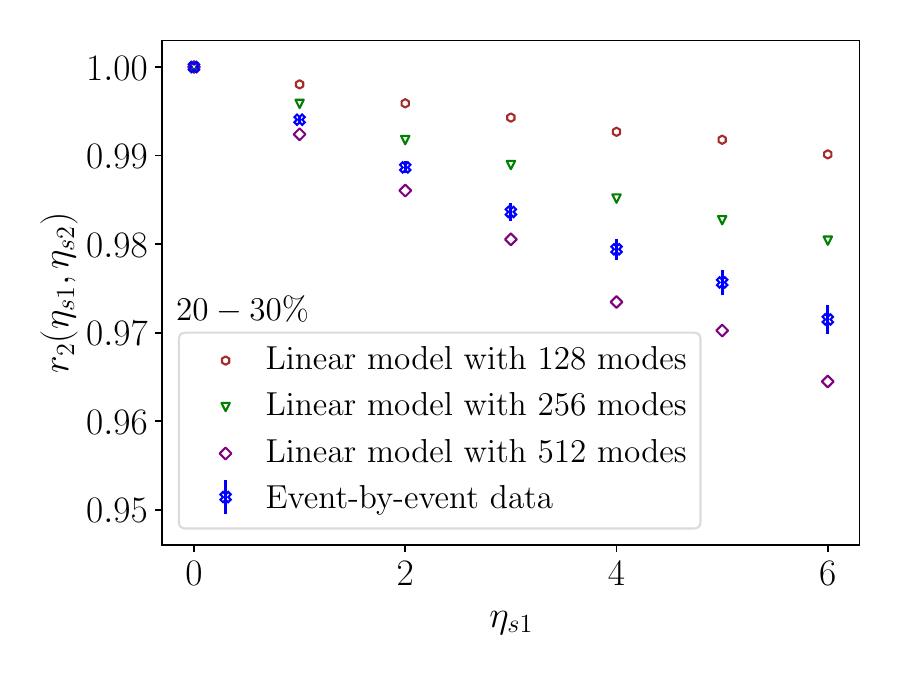}
    \caption{Decorrelation, as defined in \cref{eq:decorrelation}, of the initial eccentricity $\varepsilon_2$ for $\eta_{s2} =5$ in terms of the other slices $\eta_{s1}$. The blue points show the results from $5000$ event-by-event simulations with error bars, while the other colors represent results from artificial events constructed using the linear model with different numbers of modes. The left panel shows results for the $0-10\%$ centrality class, and the right panel shows results for the $20-30\%$ centrality class.}
\label{fig:decorrelation}
\end{figure}

The decorrelation $r_2$ is calculated by fixing the reference slice to $\eta_{s2}=5$ and is shown in~\cref{fig:decorrelation} for the event-by-event simulations, including statistical error bars. For comparison, the same observable is calculated from artificial events reconstructed by sampling the coefficients $c_l$ in the linear model, considering different numbers of modes in the reconstruction. In both centrality classes, the description systematically improves as the number of included modes increases, highlighting the sensitivity of this observable to a broad range of fluctuation modes.

To accurately describe the eccentricity decorrelation $r_2$, a large number of modes, typically of the order of hundreds, is required. The reason is that this observable is particularly sensitive to higher-order modes. Returning to the odd modes shown in \cref{fig:high_modes_0-10,fig:high_modes_20-30}, one can see that they play an important role in this 3D analysis of decorrelations. It is important to emphasize, however, that we are considering collisions at very high energies \cite{Mehrabpour:2025rzt}, where the decorrelation is relatively small and higher-order modes become relevant only when studying high-precision observables.

It should also be noted that the event-by-event decorrelation shown here is based on the \textsc{McDipper} setup. Experimental measurements of flow decorrelation are not yet available at pseudorapidities as large as those considered in this work ($\eta = 6$). Nevertheless, existing measurements already indicate that \textsc{McDipper} does not provide a completely satisfactory description of this observable \cite{Garcia-Montero:2025bpn} around $\eta \simeq 2$. One possible explanation becomes apparent from the present analysis: the odd modes that play a significant role in the decorrelation appear only at relatively high mode numbers. A more realistic description of the longitudinal structure would likely generate these contributions at lower-order modes.

\begin{figure}
    \centering
    \includegraphics[width=0.49\linewidth]{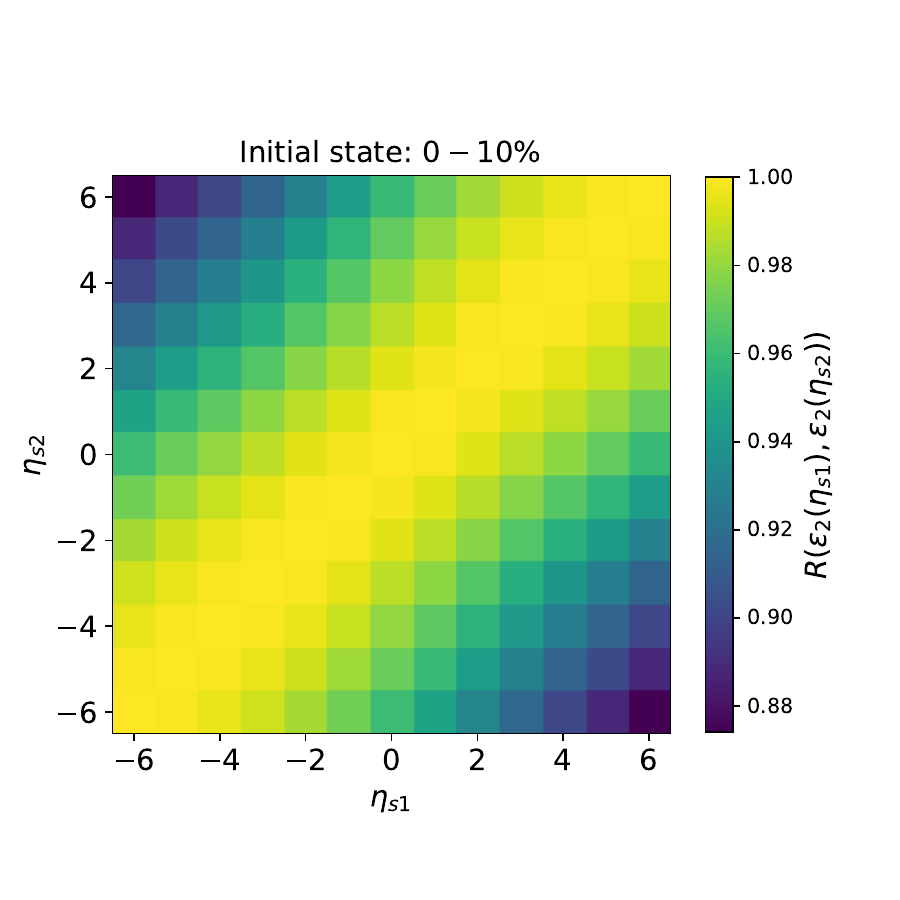}
    \includegraphics[width=0.49\linewidth]{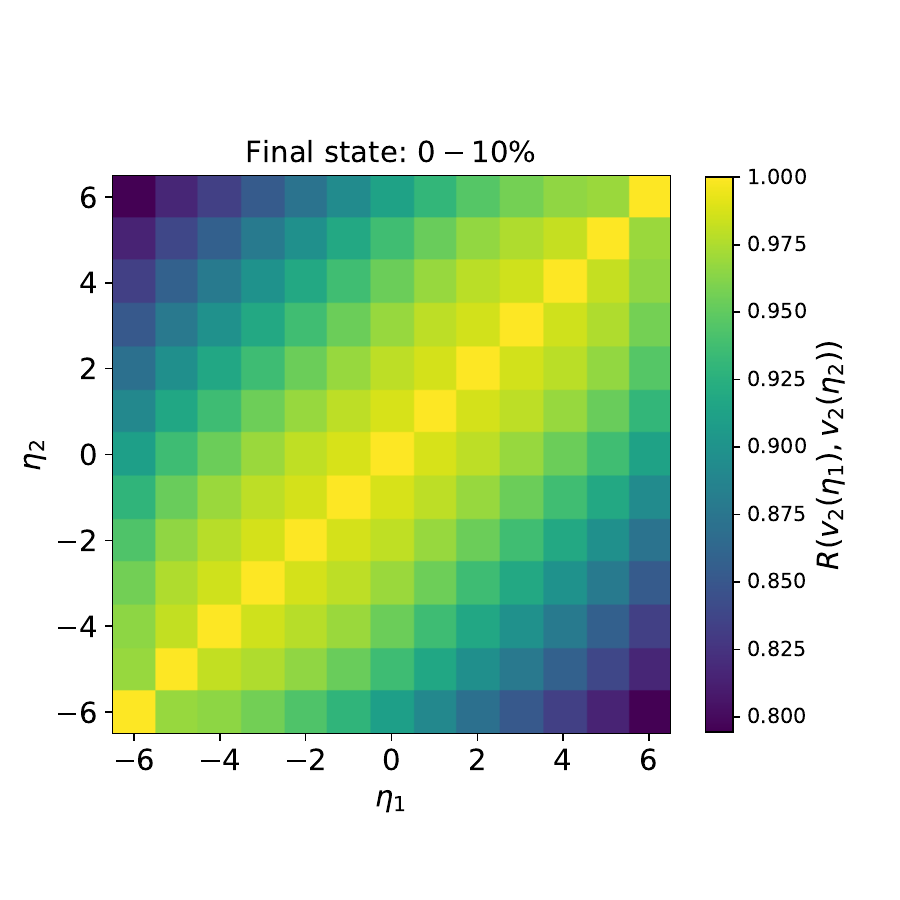}
    \caption{Correlation of observables in different slices along the longitudinal direction. On the left, the correlation between $\varepsilon_2$ in different slices is shown, and on the right, the correlation between $v_2$ in different slices. This is calculated for $5000$ events produced by \textsc{McDipper} in the $0-10\%$ centrality class.}
\label{fig:correlation_R}
\end{figure}

As demonstrated through the average state in \cref{fig:avg_state_OO}, \textsc{McDipper} exhibits only a modest longitudinal dependence. Accurately capturing the initial longitudinal tilt is important for initial-condition models, since it serves as input for the non-boost-invariant hydrodynamic evolution. The correlation between different slices for an observable $O_\alpha$ can be quantified through
\begin{equation}
    R(O_\alpha(\eta_{s1}), O_\alpha(\eta_{s2})) = \frac{\langle O_\alpha(\eta_{s1}) O_\alpha(\eta_{s2}) \rangle - \langle O_\alpha(\eta_{s1}) \rangle \langle O_\alpha(\eta_{s2}) \rangle}{\sigma_{\eta_{s1}} \sigma_{\eta_{s2}}},
\label{eq:correlation}
\end{equation}
where $\sigma_{\eta_{s1}}$ ($\sigma_{\eta_{s2}}$) denotes the standard deviation of $O_\alpha(\eta_{s1})$ ($O_\alpha(\eta_{s2})$). This quantity is shown in \cref{fig:correlation_R} for the $5000$ event-by-event simulated in central events. The left panel displays the initial-state correlations, $R(\varepsilon_2(\eta_{s1}), \varepsilon_2(\eta_{s2}))$, while the right panel shows the corresponding final-state correlations, $R(v_2(\eta_{s1}), v_2(\eta_{s2}))$. The initial-state correlations remain strong in \textsc{McDipper}, and the non-boost-invariant evolution introduces additional decorrelation between slices, although the overall magnitude remains comparable to that of the initial state.

It is worth emphasizing once more that these simulations correspond to oxygen collisions at $\sqrt{s_\mathrm{NN}} = 5.36$ TeV, a very high collision energy at which longitudinal correlations are expected to be stronger than at lower energies of the order of $200$ GeV. Nevertheless, this figure clearly illustrates the importance of the initial-state structure in determining the final-state observables.

\section{Summary}
\label{sec:summary_OO}

The main goal of this chapter was to understand the 3D initial conditions of an intermediate-size system, namely oxygen collisions. The main result obtained was the conclusion that collisions involving intermediate systems do not simply correspond to a scaled-down version of large systems; rather, they produce a system that must be interpreted in a different way. In large systems, the QGP is produced in a fireball whose behavior is largely dominated by geometry. In oxygen collisions, however, the QGP is driven much more by fluctuations than by geometric effects. Our mode decomposition model provided a useful framework to demonstrate and understand how dominant these fluctuations are.

The results presented in this chapter began with the description of the average state $\bar{\Psi}$ in \cref{subsec:3D_average_state}. Already at this stage, it was possible to observe the more spherical shape of these initial conditions. Furthermore, the small tilt present in the longitudinal structure was shown to appear only in the average state, effectively isolating this geometric effect from the fluctuation modes.

The description of the fluctuation modes $\Psi_l$ was presented in \cref{subsec:3D_modes}, while their properties were analyzed in \cref{subsec:3D_properties}. The 3D structure does not exhibit major differences compared to the 2D structure discussed in the previous chapter; the most significant differences appear only at higher mode numbers, where odd longitudinal structures emerge. In terms of their properties, however, \cref{fig:weights_OO} showed that the leading modes in oxygen collisions contribute at a level comparable to that of the average state. This conclusion was further reinforced in \cref{subsec:modes_application}, where the initial and final anisotropies, $\varepsilon_n$ and $v_n$, respectively, were calculated from event-by-event simulations and compared with the corresponding values obtained from the average state. The average-state contribution was found to be too small to reproduce even the qualitative behavior of the event-by-event simulations, demonstrating that fluctuations are important corrections in intermediate-size systems.

This chapter also investigated how far one can go using only a limited set of modes compared to a full event-by-event description. Although fluctuations play a dominant role in oxygen collisions, a carefully selected set of only $10$ modes already provides a reasonable qualitative description of the main observables. Naturally, there are more sensitive observables, such as decorrelations between different longitudinal slices, which are affected by the detailed longitudinal structure of the system. In these cases, a significantly larger number of modes is required.

Overall, this chapter provided a natural comparison with the results obtained for large systems in the previous chapter, a direct application of the theoretical framework developed there, and a demonstration of how it can be used to understand the dynamics of intermediate-size systems.

More generally, these studies of initial-state fluctuations highlight the importance of realistic initial-condition models for achieving an accurate description of final-state observables.

\chapter{General conclusion and outlook}
\label{chap:Conclusion}

\epigraph{``Somewhere, something incredible is waiting to be known'' -- attributed to Carl Sagan}
\noindent

The central objective of this dissertation has been to advance the understanding of the initial stages in simulations of heavy-ion collisions. In the hybrid framework, the initial state and the pre-equilibrium evolution constitute the dominant sources of theoretical uncertainty. Improving their description is therefore essential for strengthening the predictive power of relativistic heavy-ion phenomenology and for extracting reliable properties of the QGP.

The work presented here has addressed this challenge from two complementary directions: first, by discussing the matching between pre-equilibrium dynamics and viscous hydrodynamics; and second, by developing and extending a mode-by-mode framework to systematically analyze fluctuations of the initial energy density in both 2D and fully 3D scenarios. 

In \cref{chap:entropy_matching}, a new prescription for matching the pre-equilibrium stage to viscous hydrodynamics was introduced, based on entropy-density matching. This study went beyond demonstrating the success of kinetic theory in describing the out-of-equilibrium QGP and relativistic dissipative hydrodynamics in modeling the subsequent medium evolution (as reviewed in \cref{chap:entropy_matching}). Instead, it critically examined the limits of these approaches and the inconsistency of their interface. In particular, it highlighted the conceptual tension between conformal pre-equilibrium models and non-conformal hydrodynamic equations of state. The entropy-matching procedure proposed here reduces the artificial bulk pressure generated at the switching time and improves the theoretical consistency of the hybrid framework.

Future work in this direction should focus on developing non-conformal kinetic-theory descriptions, for example by incorporating massive degrees of freedom into the Boltzmann equation or by constructing massive attractor solutions. Such developments would allow for a smoother and more physically consistent transition to hydrodynamics. Additionally, a more systematic study of transport coefficients during the pre-equilibrium phase is needed. Constraints on transport coefficients at high temperatures remain weak, and improved theoretical and phenomenological input in this regime would be valuable.

In \cref{chap:modesPb} and \ref{chap:3d}, the fluctuations of the initial energy density were investigated in detail using a mode-by-mode decomposition. The 2D analysis focused on developing the formalism and characterizing the properties of the modes within centrality bins. A key result was the demonstration that initial fluctuations survive the hydrodynamic evolution and leave measurable imprints on final observables. This finding reinforces the importance of accurately modeling the initial conditions in the hybrid evolution chain.

The 3D extension of the framework, applied to oxygen–oxygen collisions, provided additional insight into intermediate-size systems. In these systems, fluctuations play a dominant role compared to purely geometric effects. The 3D analysis further clarified the longitudinal structure and demonstrated the flexibility of the mode-based approach in non–boost-invariant scenarios.

These studies also revealed limitations of the present framework. For example, the computational cost of simulating the hadronic stage for a reduced set of modes remains significant due to statistical requirements. Moreover, certain sensitive observables require an increasing number of modes for accurate reconstruction. These limitations are not shortcomings but rather indicate directions for future methodological improvements.

Looking forward, several promising research avenues emerge. The study of initial fluctuations should be extended to different initial condition models, collision systems, and beam energies. Deformed nuclei such as Ru or Zr offer natural test cases for studying geometric effects within the mode framework. Lower collision energies would provide an opportunity to investigate stronger longitudinal tilt and direct-flow phenomena in a fully 3D setting. Asymmetric collision systems could further illuminate the role of participant–spectator dynamics in three-dimensional evolution.

Beyond energy density profiles, an extension of the mode decomposition to conserved-charge distributions would be highly interesting, although this requires a more detailed treatment of dissipative hydrodynamics with charge diffusion. A deeper integration of nuclear-structure information into initial condition models, motivated here by the example of oxygen and potentially neon, could also reduce systematic uncertainties in simulations.

Overall, this dissertation has contributed to a more systematic understanding of the initial stages of heavy-ion collisions and provided new tools to analyze their fluctuations. While many open questions remain, the methods and results presented here offer a framework that can be extended to a broad range of problems in relativistic heavy-ion physics.

\appendix

\chapter{Bilinear, trilinear, and bilinear-quadratic interpolation}
\label{appA}
\DoPToC
\noindent

This appendix explains the numerical interpolation procedures used in \cref{chap:modesPb} and \cref{chap:3d}. As discussed in the main text, the mode-by-mode decomposition depends on the number of points in the spatial grid. However, after the decomposition, the average state and the fluctuation modes are converted into energy–momentum tensors and used as input for the pre-equilibrium and hydrodynamic evolution. Since hydrodynamic evolution involves spatial gradients of dynamical quantities, a finer spatial resolution improves numerical stability and smoothness. For this reason, interpolation is applied: the modes are constructed on a relatively coarse grid and then interpolated to a finer grid for the subsequent evolution \cite{press2007numerical}.

The essential constraint of any interpolation procedure is the conservation of total energy. In discretized form, this condition reads
\begin{equation}
    \frac{dE}{d\eta_s} = \sum_{i, j} e(x_i, y_j) \Delta x \Delta y \simeq \sum_{m, n} e(x_m, y_n) \Delta x' \Delta y'~,
\label{eq:total_energy_interpolation}
\end{equation}
where the grid spacing is changed from $\Delta x$ to $\Delta x'$ (and similarly in the $y$-direction). The interpolation must preserve this integral within numerical accuracy.

\section{Bilinear interpolation (2D)}

\begin{figure}[h]
    \centering
    \includegraphics[width=0.5\linewidth]{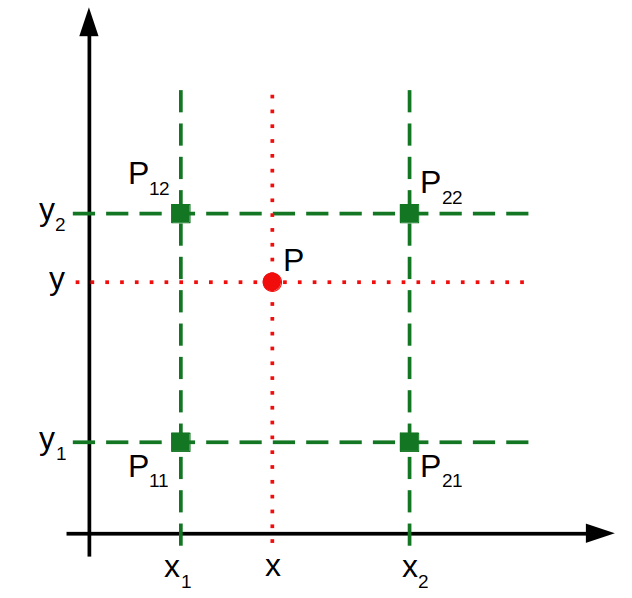}
    \caption{Schematic illustration of bilinear interpolation. The green grid indicates the original coarse grid, where the values at the four neighboring points $P_{11}$, $P_{12}$, $P_{21}$ and $P_{22}$ are known. These values are combined using the interpolation formula described in the text to determine the value at the intermediate point $P$ on the refined grid.}
    \label{fig:bilinear}
\end{figure}

In the 2D case, bilinear interpolation is employed. The basic geometry is illustrated in \cref{fig:bilinear}. The original (coarse) grid is shown in green, where the energy density is known at the four surrounding points: 
\begin{equation}
    P_{11} = e(x_1,y_1), \quad P_{12} = e(x_1,y_2), \quad P_{21} = e(x_2,y_1), \quad P_{22} = e(x_2,y_2). 
\end{equation}

The objective is to determine the energy density at a new point $P = e(x,y)$ on the refined (red) grid. The point $(x,y)$ lies inside the rectangle defined by $(x_1,y_1)$, $(x_1,y_2)$, $(x_2,y_1)$, and $(x_2,y_2)$. Because the refined grid has a smaller spacing, multiple new points may lie within the same coarse cell. For example, another point $P' = e(x',y')$ with $x_1 < x' < x_2$ and $y_1 < y' < y_2$ can be interpolated using the same four surrounding coarse-grid values. Therefore, the relative distances between the interpolation point and the coarse-grid nodes must be taken into account.

The interpolation in the $x$-direction is first performed linearly between the two neighboring grid points:
\begin{equation}
\begin{aligned}
    e(x, y_1) &\simeq \frac{x_2 - x}{x_2 - x_1} e(x_1, y_1) + \frac{x - x_1}{x_2 - x_1} e(x_2, y_1) \\
    e(x, y_2) &\simeq \frac{x_2 - x}{x_2 - x_1} e(x_1, y_2) + \frac{x - x_1}{x_2 - x_1} e(x_2, y_2).
\end{aligned}
\end{equation}
These intermediate results are then used to interpolate in the $y$-direction:
\begin{equation}
    e(x, y) \simeq \frac{y_2 - y}{y_2 - y_1} e(x, y_1) + \frac{y - y_1}{y_2 - y_1} e(x, y_2).
\end{equation}
Combining both steps yields the standard bilinear interpolation formula, and using the simplified notation of points $P$:
\begin{equation}
\begin{aligned}
    P \simeq \frac{1}{(x_2 - x_1)(y_2 - y_1)}\bigg[&(x_2 - x)(y_2 - y) P_{11} + (x - x_1)(y_2 - y)P_{21} + \\
    &(x_2 - x)(y - y_1)P_{12} + (x - x_1)(y - y_1)P_{22} \bigg],
\end{aligned}
\end{equation}
where this expression applies generically to any quantity defined at the grid points, such as energy density, entropy density, or enthalpy. This interpolation is used in \cref{chap:modesPb}.

\section{Trilinear interpolation (3D)}

For a 3D grid, trilinear interpolation is employed. The procedure follows the same logic as in the bilinear case, but now involves eight neighboring grid points: 
\begin{equation}
\begin{aligned}
    P_{111} &= e(x_1,y_1,\eta_1), \quad P_{112} = e(x_1,y_1,\eta_2), \quad  P_{121} = e(x_1,y_2,\eta_1), \\
    P_{122} &= e(x_1,y_2,\eta_2), \quad P_{211} = e(x_2,y_1,\eta_1), \quad P_{212} = e(x_2,y_1,\eta_2), \\
    P_{221} &= e(x_2,y_2,\eta_1), \quad P_{222} = e(x_2,y_2,\eta_2). 
\end{aligned} 
\end{equation}
Here, the third coordinate $\eta$ may represent the space-time rapidity $\eta_s$ or any other longitudinal coordinate.

First, bilinear interpolation is applied separately in the transverse plane at fixed $\eta_1$ and $\eta_2$:
\begin{equation}
\begin{aligned}
    e(x, y, \eta_1) \simeq \frac{1}{(x_2 - x_1)(y_2 - y_1)}\bigg[&(x_2 - x)(y_2 - y) P_{111} + (x - x_1)(y_2 - y)P_{211} + \\
    &(x_2 - x)(y - y_1)P_{121} + (x - x_1)(y - y_1)P_{221} \bigg] \\
    e(x, y, \eta_2) \simeq \frac{1}{(x_2 - x_1)(y_2 - y_1)}\bigg[&(x_2 - x)(y_2 - y) P_{112} + (x - x_1)(y_2 - y)P_{212} + \\
    &(x_2 - x)(y - y_1)P_{122} + (x - x_1)(y - y_1)P_{222} \bigg]. 
\end{aligned}
\end{equation}
A linear interpolation is then performed between the two planes:
\begin{equation}
    P \simeq \frac{\eta_2 - \eta}{\eta_2 - \eta_1}e(x, y, \eta_1) + \frac{\eta - \eta_1}{\eta_2 - \eta_1} e(x, y, \eta_2)~,
\end{equation}
which leads to the full trilinear interpolation formula:
\begin{equation}
\begin{aligned}
    P \simeq \frac{1}{(x_2 - x_1)(y_2 - y_1)(\eta_2 - \eta_1)} \bigg[&(x_2 - x)(y_2 - y)(\eta_2 - \eta) P_{111} + (x - x_1)(y_2 - y)(\eta_2 - \eta)P_{211} + \\
    &(x_2 - x)(y - y_1)(\eta_2 - \eta)P_{121} + (x - x_1)(y - y_1)(\eta_2 - \eta)P_{221} + \\
    &(x_2 - x)(y_2 - y)(\eta - \eta_1) P_{112} + (x - x_1)(y_2 - y)(\eta - \eta_1)P_{212} + \\
    &(x_2 - x)(y - y_1)(\eta - \eta_1)P_{122} + (x - x_1)(y - y_1)(\eta - \eta_1)P_{222}\bigg].
\end{aligned}
\end{equation}

\section{Bilinear–quadratic interpolation (3D)}

The bilinear and trilinear interpolations are linear in each coordinate and therefore approximate the function using first-order terms. A more accurate representation would include quadratic terms, but this increases the computational cost.

In \cref{chap:modesPb}, the grid refinement from $\Delta x = 0.11$ to $\Delta x' = 0.1$ was sufficiently small that bilinear interpolation was adequate. However, in \cref{chap:3d}, the longitudinal refinement was substantial ($\Delta\eta = 1.5$ to $\Delta\eta' = 0.25$), and a quadratic interpolation was applied in the longitudinal direction only.

For this purpose, the energy values at three neighboring slices are used: 
\begin{equation}
    e(x,y,\eta_0)=f(\eta_0), \quad e(x,y,\eta_1)=f(\eta_1), \quad e(x,y,\eta_2)=f(\eta_2),
\end{equation}
with $\eta_0<\eta_1<\eta_2$. The interpolated value is obtained using Lagrange quadratic interpolation:
\begin{equation}
\begin{aligned} 
    P(x, y, \eta) \simeq \frac{(\eta - \eta_1)(\eta - \eta_2)}{(\eta_0 - \eta_1)(\eta_0 - \eta_2)} f(\eta_0) + \frac{(\eta - \eta_0)(\eta - \eta_2)}{(\eta_1 - \eta_0)(\eta_1 - \eta_2)} f(\eta_1) + \frac{(\eta - \eta_0)(\eta - \eta_1)}{(\eta_2 - \eta_0)(\eta_2 - \eta_1)} f(\eta_2),
\end{aligned}
\end{equation}
with $\eta_0 \le \eta \le \eta_2$ and where signs are important. At the boundaries, where three slices are not available, linear interpolation is used.

\chapter{Higher-order modes}
\DoPToC
\label{appB}

\section{Higher modes in Pb-Pb at $\sqrt{s_\mathrm{NN}} = 5.02$ TeV}

\begin{figure}
    \centering
    \includegraphics[width=\linewidth]{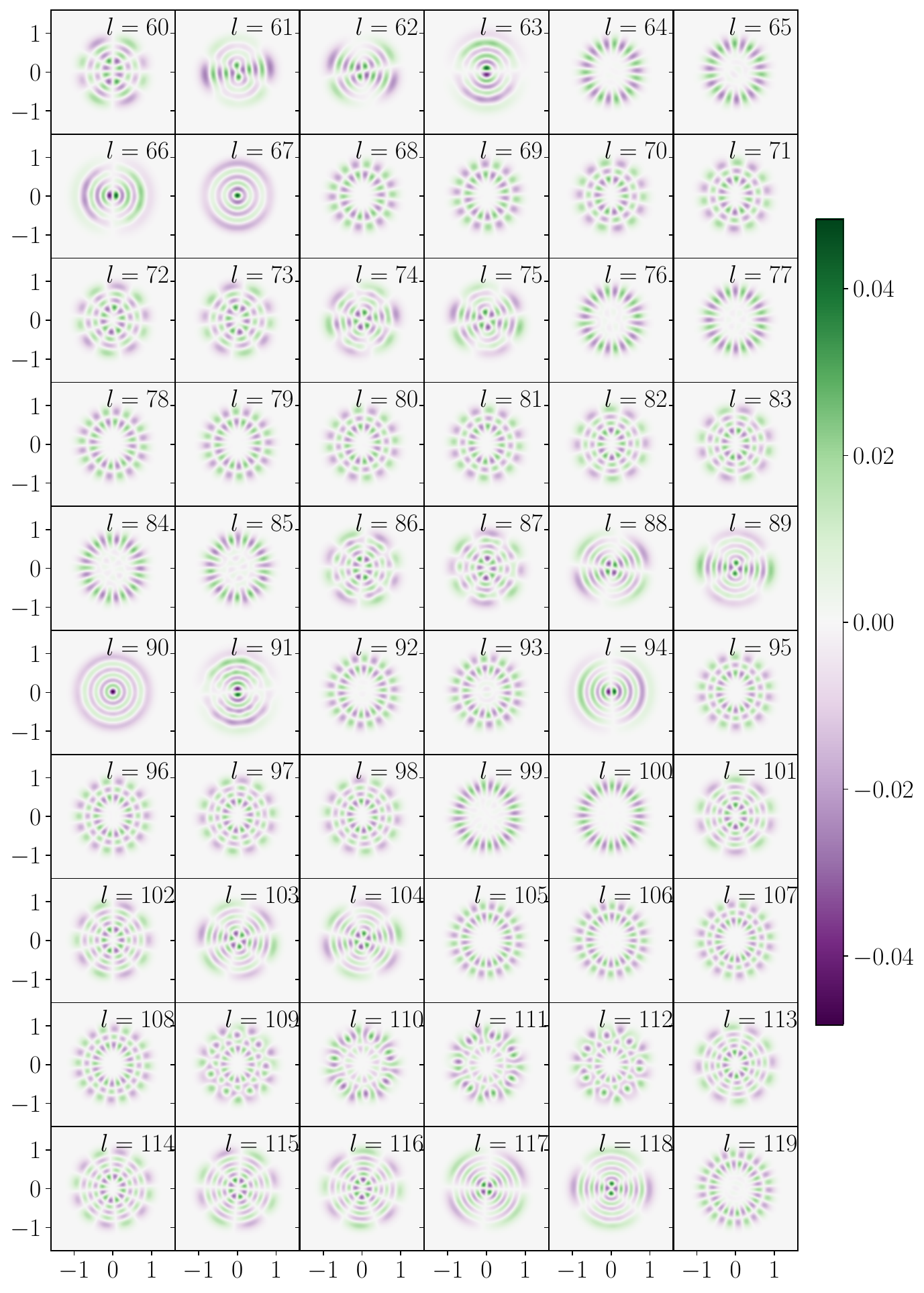}
    \caption{Transverse-plane structure of higher modes for Pb-Pb collisions in the central bin ($0-2.5\%$). The axes are expressed in units of the half-density radius. This figure provides complementary material to \cref{fig:modes_0-2.5}. Further details are discussed in the text.}
\label{fig:extra_modes_0-2.5}
\end{figure}

\begin{figure}
    \centering
    \includegraphics[width=\linewidth]{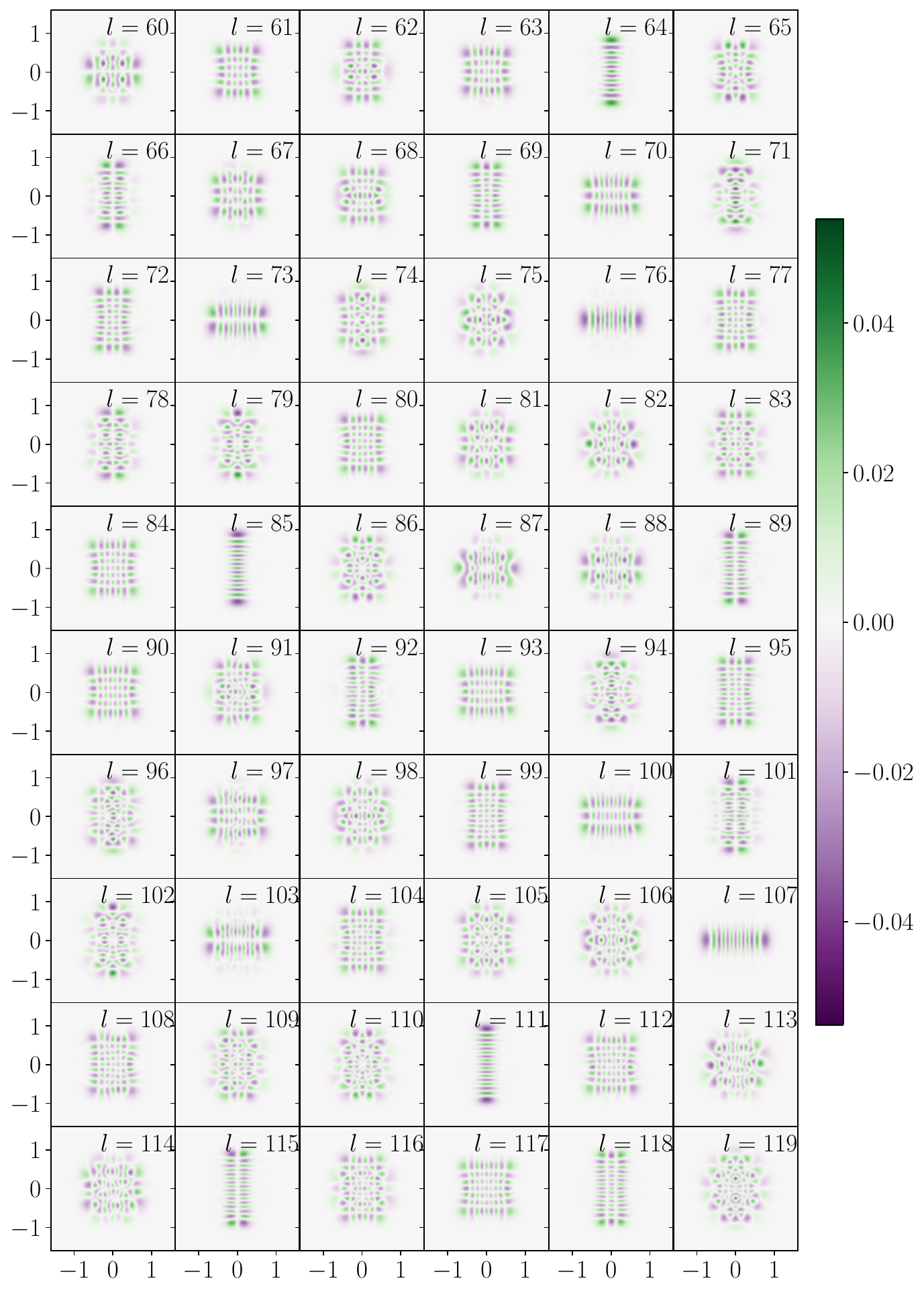}
    \caption{Transverse-plane structure of the higher modes for Pb-Pb collisions in the non-central bin ($30-40\%$). The axes are expressed in units of the half-density radius. This figure provides complementary information to \cref{fig:modes_30-40}.}
\label{fig:extra_modes_30-40}
\end{figure}

This section presents, in \cref{fig:extra_modes_0-2.5} and \ref{fig:extra_modes_30-40}, the higher-index modes $60 \le l \le 119$ as complementary material to \cref{fig:modes_0-2.5} and \ref{fig:modes_30-40}. Although these modes have smaller eigenvalues and therefore reduced physical relevance in the discussion of \cref{chap:modesPb}, they preserve the same symmetry patterns and structural features described in the main text.

\section{Additional modes in O-O at midrapidity}

\begin{figure}
    \centering
    \includegraphics[width=\linewidth]{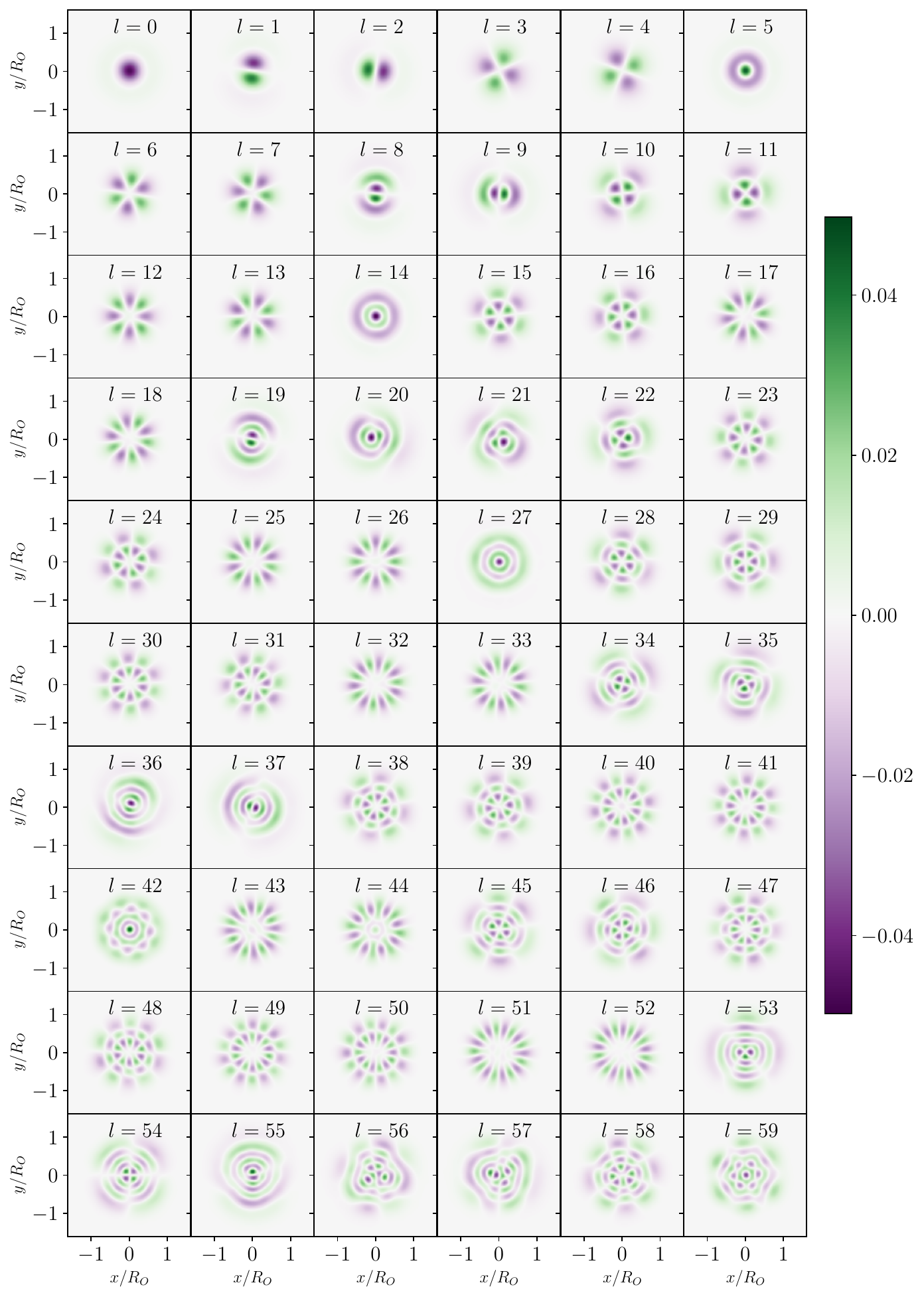}
    \caption{The first $60$ modes for O-O collisions in the $0-10\%$ centrality class. The 3D modes are displayed in the transverse plane at $\eta_s=0$, with the axes expressed in units of the system radius.}
\label{fig:app_OO_modes_0-10}
\end{figure}

\begin{figure}
    \centering
    \includegraphics[width=\linewidth]{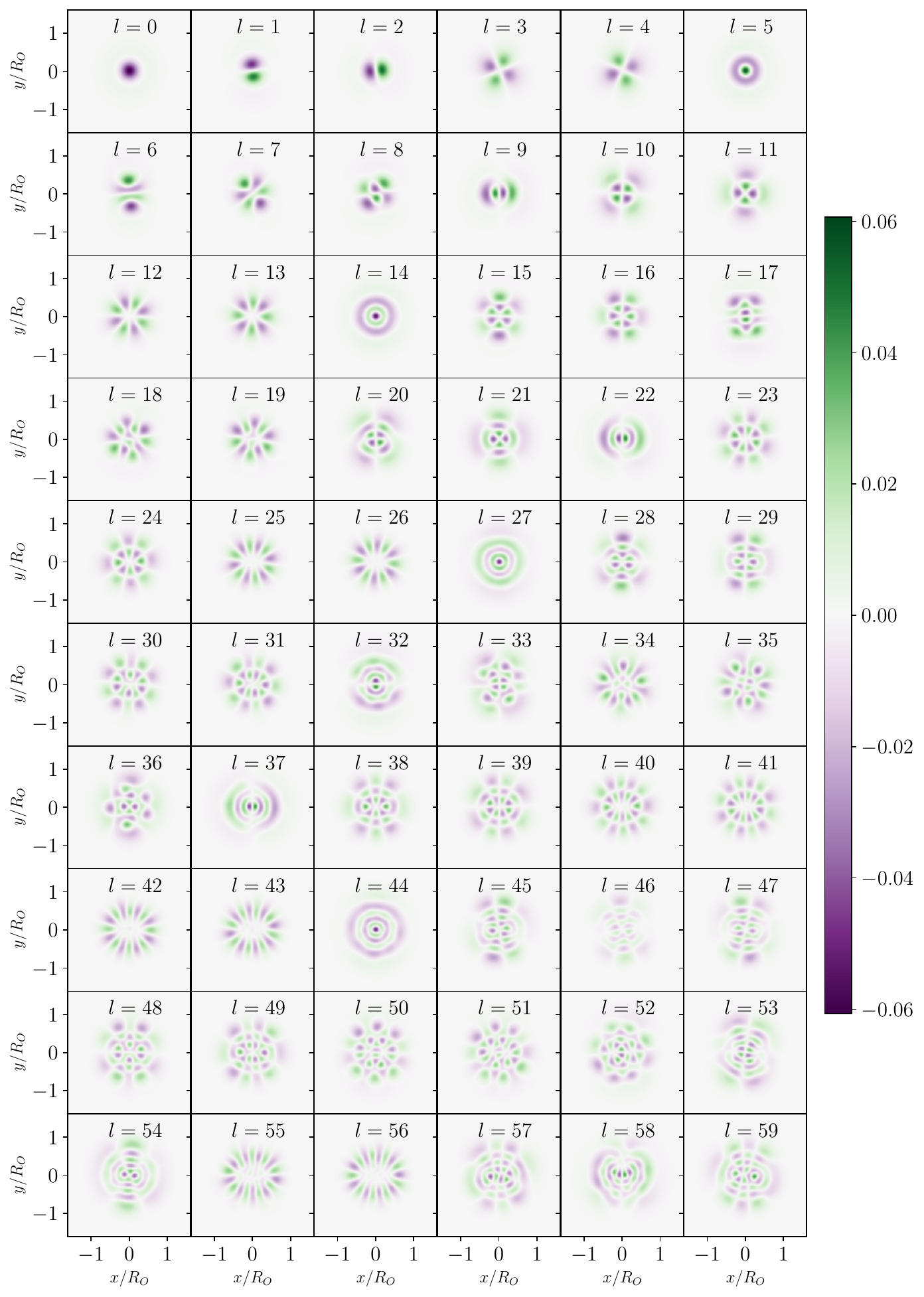}
    \caption{The first $60$ modes for O-O collisions in the $20-30\%$ centrality class. The 3D modes are displayed in the transverse plane at $\eta_s=0$, with the axes expressed in units of the system radius.}
\label{fig:app_OO_modes_20-30}
\end{figure}

This section provides complementary material to \cref{subsec:3D_modes}. The fluctuation modes for oxygen-oxygen collisions at $\sqrt{s_\mathrm{NN}} = 5.36$ TeV, with indices $0 \le l \le 59$ are shown at $\eta_s=0$ in \cref{fig:app_OO_modes_0-10} and \ref{fig:app_OO_modes_20-30} for comparison with the Pb-Pb case. 

The transverse-plane structure at midrapidity reproduces the same geometric patterns discussed in the main text. The longitudinal structure remains consistent with that shown in \cref{fig:modes_3D_0-10} and \ref{fig:modes_3D_20-30}, except for the higher odd modes previously highlighted in \cref{fig:high_modes_0-10} and \ref{fig:high_modes_20-30}.

\chapter{Statistical properties of the expansion coefficients}
\label{appC}

This appendix presents additional quantitative properties of the expansion coefficients $\{c_l\}$. In \cref{fig:coefficients_cl}, the probability distribution of these coefficients was shown for $2^{15}$ events and found to be approximately Gaussian with unit variance. Here, we characterize this distribution more quantitatively using its statistical moments \cite{Lista:2023ith}.

The moments of a distribution include: the mean $\mu$, the variance $\sigma^2$, the skewness $\gamma_1$, which measures asymmetry, and the excess kurtosis $\gamma_2$, which characterizes the shape of the tails and the sharpness of the peak. These quantities are defined as \cite{Lista:2023ith}:
\begin{equation}
\begin{aligned}
    \mu &= \frac{1}{N_\mathrm{ev}} \sum_{i=1}^{N_\mathrm{ev}} c_l^{(i)} = \langle c_l \rangle \\
    \sigma^2 &= \frac{1}{N_\mathrm{ev}} \sum_{i=1}^{N_\mathrm{ev}} (c_l^{(i)} - \langle c_l \rangle)^2 \\
    \gamma_1 &= \frac{1}{N_\mathrm{ev}} \sum_{i=1}^{N_\mathrm{ev}} \left(\frac{c_l^{(i)} - \langle c_l \rangle}{\sigma}\right)^3 \\
    \gamma_2 &= \frac{1}{N_\mathrm{ev}} \sum_{i=0}^{N_\mathrm{ev}} \left ( \frac{c_l^{(i)} - \langle c_l \rangle}{\sigma} \right)^4 - 3.
\label{eq:moments_all}
\end{aligned}
\end{equation}

\begin{figure}[!t]
    \centering
    \includegraphics[width=0.95\linewidth]{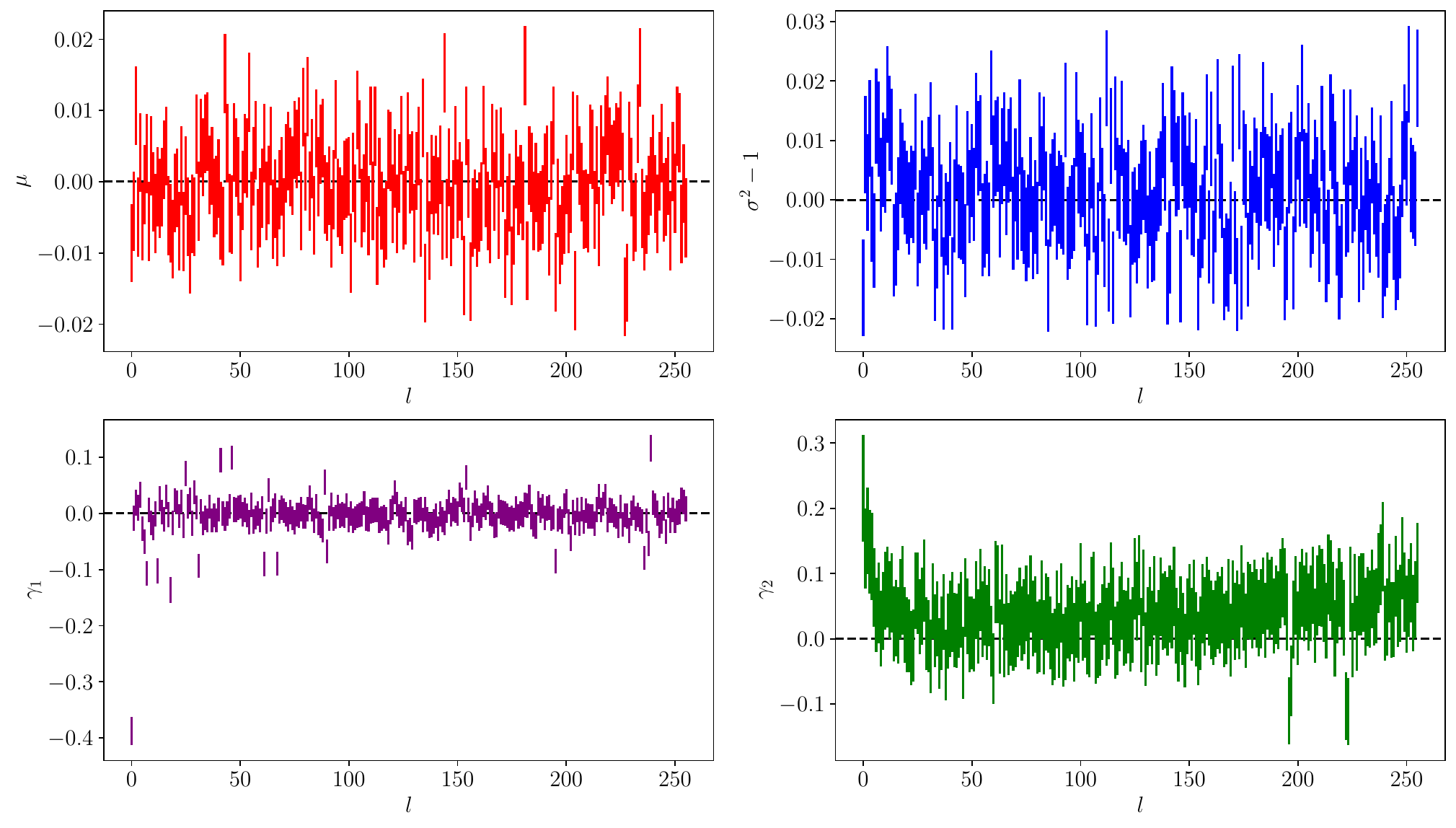}
    \caption{Average $\mu$, variance $\sigma^2$, skewness $\gamma_1$ and kurtosis $\gamma_2$ for the probability distributions of $\{c_l\}$ in the centrality bin $0-2.5\%$.}
    \label{fig:moments_0-2.5}
\end{figure}

\begin{figure}[!t]
    \centering
    \includegraphics[width=0.95\linewidth]{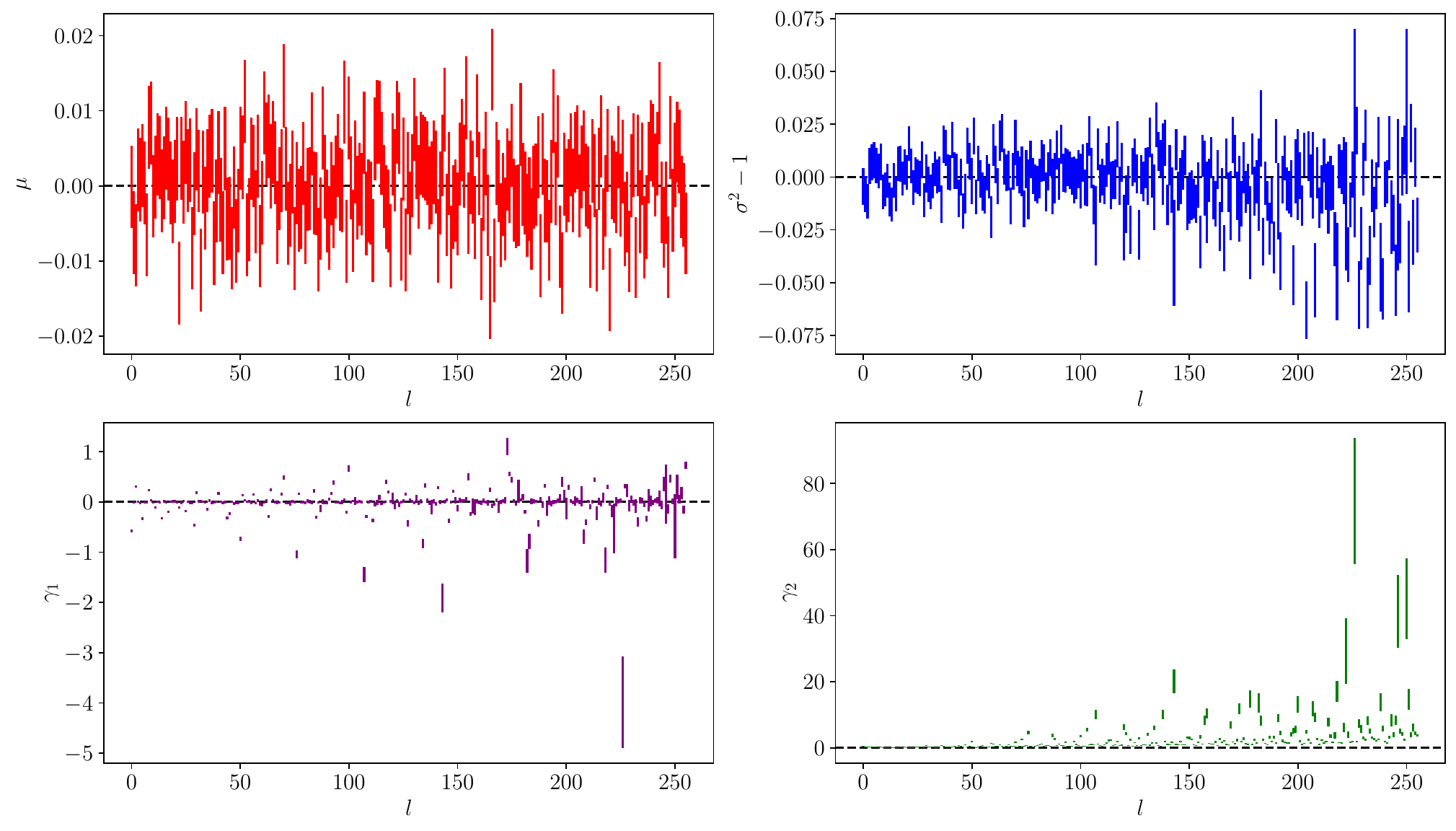}
    \caption{Same as~\cref{fig:moments_0-2.5}, but for the centrality bin $30-40\%$.}
    \label{fig:moments_30-40}
\end{figure}

From the theoretical construction of the mode decomposition, the expansion coefficients satisfy $\langle c_l \rangle = 0$ and $\langle c_l c_{l'} \rangle = \delta_{ll'}$. This behavior is confirmed in \cref{fig:moments_0-2.5,fig:moments_30-40}, where one observes $\mu \simeq 0$ and $\sigma^2 = \langle c_l^2 \rangle \simeq 1$ for all modes.

For a perfectly Gaussian distribution, both the skewness and the excess kurtosis should be approximately zero. In practice, small deviations are observed, consistent with the discussion in \cref{subsec:modes}. In particular, mode $l = 0$ exhibits a noticeable skewness, with a longer tail toward negative values. This asymmetry arises from its negative contribution to the energy density, which constrains the allowed range of the corresponding expansion coefficient.

In the non-central case, higher modes display larger values of kurtosis. This behavior can be associated with the breaking of rotational symmetry in non-central collisions. When comparing with the fixed-impact-parameter study in Ref. \cite{Borghini:2022iym}, a similar trend is observed. For $b = 0$ fm (or in central bins), where the geometry is approximately circular, the distributions of the expansion coefficients for higher modes remain close to Gaussian. In contrast, for $b = 9$ fm (or non-central bins), the loss of circular symmetry modifies the peak and tail structure of the distributions, as also illustrated in \cref{fig:coefficients_cl}.

Overall, these results confirm that the Gaussian approximation provides a good description of the expansion coefficients, with deviations that can be understood in terms of geometric and symmetry considerations.

\backmatter

\chapter{Acknowledgments}

\epigraph{``We are never ready, we only need to be courageous'' -- unknown author}
\noindent

The past three years of my PhD have brought me countless lessons, challenges, joyful moments, and difficult ones. No one does science alone, and I am deeply grateful to all those whose paths have crossed mine throughout this journey.

I would first like to express my sincere gratitude to my PhD supervisor, Professor Nicolas Borghini, for giving me the opportunity to pursue my studies in Bielefeld. I have had an excellent time during my PhD, and I know that much of it is a result of his guidance and support. He was always available to answer my questions and consistently valued my opinions, even when they led to disagreements and discussions between us. Those exchanges helped me grow as a researcher.

I would also like to thank Professor Sören Schlichting, who may not fully realize the impact he has on my life and career. Many opportunities became available to me because of his consideration with me, and for that I am truly grateful.

My special thanks go to Hendrik Roch, affectionately known as my ``numerical advisor''. I learned a lot from you. Thank you for the countless online meetings, repository reviews, and for your patience in helping me navigate the computational side of my research.

I am equally grateful to Professor Cesar Augusto Dartora, my undergraduate research supervisor, who first introduced me to QCD, a field in which I found both intellectual fascination and personal fulfillment. I also thank Professor Tiago José Nunes da Silva, my master's supervisor, who always encouraged me to continue pursuing my goals and consistently believed in my potential, even after I left Brazil.

I would like to thank all the professors and researchers with whom I had the opportunity to discuss science at conferences and meetings, and who took an interest in my work. I am especially grateful to the Brazilian researchers who have always welcomed me with encouragement and kind words. Likewise, I thank the friends I met at scientific events, particularly my fellow Brazilians, whose presence always brought me a sense of warmth and familiarity, no matter where we were.

I am grateful to Irene and Susi for all their administrative support throughout these years. My thanks also go to all members of the Department of Physics at Bielefeld University for creating such a friendly environment.

To my mother and sister: Queria agradecer à minha mãe Luciane e à minha irmã Carla pela resiliência de vida e incondicional apoio. Obrigada por encararem esse longo processo e essa distância comigo, apoiando sempre minhas escolhas por mais dolorosas que possam ser.

Being far from my family, I found in Bielefeld a second support network that became essential to my well-being. Thank you, Nen, for always offering the best life advices and for every hug when I most needed one. Thank you, Utkarsh, for being much more than a friend in German classes and for always keep cheering for me, often even more than I did myself. Thank you, Eduardo and Renata (my clone), for always welcoming me during holidays and making me feel at home. Thank you to Nancy, Henrique, and so many other Brazilians whose friendship, lunches and conversations in Portuguese made life in Bielefeld brighter and easier.

I cannot forget my other friends who, despite the distance, were always present. Whether through long phone calls or voice messages, they helped me through many difficult moments. In particular, I would like to thank my dear friends Rafael, Leonardo, Isa, Yohan, and Thiago for their unwavering friendship throughout these years.

Finally, I would like to thank Germany. It has become much more than a country to me; it has become an important part of my life. Here, I learned how to build a life and meaningful relationships in other languages, I reached from 0 to B1 level in German, I discovered new places, experienced cultural differences, and grew in ways I never expected. This journey was far more than a PhD. It was a life challenge, and I am grateful that I had the courage to accept it. As this chapter comes to an end, I look forward with hope and gratitude, wishing that the next chapters be as meaningful and beautiful as those I have lived so far.

\chapter{Declaration of authorship}
I hereby declare that I have written this present dissertation independently and have only used the materials cited in the text. The artificial intelligence tool provided by Bielefeld University (BIKI) was used only for English grammar and language corrections. No AI tools were used for the scientific content or analysis in this dissertation. 

This dissertation is based primarily on Ref. \cite{Borghini:2024kll, Krupczak:2025gwg} (already published) and \cite{Krupczak:2026byr} (pre-print). For the studies reported in these publications, I was primarily responsible for the simulations, data analysis, and results.

Regarding the computational framework used in this dissertation:
\begin{itemize}
    \item The evolution codes were performed using publicly available codes, which are properly cited throughout the dissertation.
    \item The entropy-matching code used in this work was developed by me and is publicly available on GitHub, as referenced along the dissertation.
    \item For the mode-decomposition framework, I used a pre-version from my collaborators and I developed the centrality classification module, efficiency improvements and implemented the complete 3D extension used in this work. The code is currently not publicly available.
\end{itemize}

\vspace{2cm}

\noindent
\placedate{Bielefeld, October 2, 2026}{place, date}\hfill\signature{Renata Krupczak}

\newpage
~
\vspace*{\fill}
\vspace{0.8\textheight}
\begin{flushleft}
Gedruckt auf alterungsbeständigem Papier ISO 9706
\end{flushleft}


\bibliographystyle{ieeetr}
\bibliography{literature}


\end{document}